In memory of Professor V.N. Rodionov

# MECHANICAL MODEL
# OF
# PERSONAL INCOME DISTRIBUTION

Ivan O. Kitov

Abstract

A microeconomic model is developed, which accurately predicts the shape of personal income distribution (PID) in the United States and the evolution of the shape over time. The underlying concept is borrowed from geo-mechanics and thus can be considered as mechanics of income distribution. The model allows the resolution of empirical and definitional problems associated with personal income measurements. It also serves as a firm fundament for definitions of income inequality as secondary derivatives from personal income distribution.

It is found that in relative terms the PID in the US has not been changing since 1947. Effectively, the Gini coefficient has been almost constant during the last 60 years, as reported by the Census Bureau.

# Content

## §1.1. Introduction

Income distribution is a fundamental process in all economic systems. Conventional economic theories provide a variety of views on the mechanism driving the division of gross domestic product among economic agents. Income distribution at personal level did not deserve the highest attention of the mainstream economists who are focused on households. We do not share this approach and consider *personal* income as a natural and indivisible level for theoretical consideration. Total income of families and households corresponds to a higher level of aggregation and the dynamics of their evolution is prone to all disturbances associated with fluctuations in their composition and average size over time. Therefore, we introduce and elaborate a concept describing the distribution of personal income and its evolution. Because of data availability, quality and time coverage an unavoidable choice for our study is the United States.

Redline of our investigation follows up the answer to the key question: Whether the configuration of personal incomes in the US is the result of distribution of a random part of nominal GDP growing at a rate prone to stochastic external (in economics - exogenous) shocks or there exists a deterministic and fixed hierarchy of personal incomes, which evolution defines the rate of GDP growth? If the distribution is a stochastic process together with the part of GDP related to personal incomes, i.e. with gross personal income (GPI), one should develop a statistical approach. If the distribution is fixed and defines the overall growth of economy one would be able to formulate a deterministic (e.g. mechanical) model. In this Chapter, we are trying to prove that the second answer is valid and the evolution of each and every personal income is predictable, potentially as accurate as in classical mechanics.

We do not feel that economics as a science is currently able to provide adequate concepts and methods to analyze personal incomes in quantitative terms. So, we adapt an interdisciplinary approach, which has already shown its fruitfulness in many scientific and technological areas. This success is achieved not only due to the coincidence of formal description of various physical, chemical, biological, and sociological processes, but also expresses the existence of very deep common roots in the nature. For example, the power law distribution of sizes is observed in economics (Pareto distribution), in frequencies of words in longer texts, in seismology (Guttenberg-Richter recurrence curve), geomechanics (fractured particle sizes), and many other areas. Recent studies associate the power law distribution with a realization of some stochastic processes known as "self-organized criticality" (SOC).

Economics and its numerous applications in real life demand huge amount of numerical data in order to estimate current state of a given economy and future development. Such data have been continuously gathered from the very beginning of capitalism as an economic system, but the 20th century and especially its second part is characterized by a dramatic increase in the number of economic observations and measurements. The resulting data set has become an object of a thorough study not only for professional economists but also for specialists in many other disciplines. There are many examples of successful application of mathematical and physical methods from many adjacent disciplines for understanding economic phenomena and processes.

Personal income distribution (PID) represents one of high-quality sets of quantitative data with a history of more than sixty years of continuous measurement with increasing accuracy. Irrelevant to the nature of these data, even the simplest scatter plot reveals some specific features, which are often observed in physics: growth and fall is well approximated by exponential and power law functions. Some of these functions are the solutions of ordinary differential equation, and thus one can presume that the processes behind the data can be also described by such equations. This makes it very attractive to apply standard methods of analysis and to model the evolution of personal incomes according to 'first principles' adopted in the natural sciences.

Among numerous possibilities, we selected the geomechanical model of a solid with inhomogeneous inclusions proposed and developed by V.N. Rodionov and co-authors (1982) as an analogue of an economy expressed as a set of personal incomes. The economy plays the role of a solid body and personal incomes correspond to inelastic stresses on the inclusions. We expected that some of the already available equations and solutions for a solid would provide an adequate description of incomes, and some of the equations would need modification. The intuition behind such an assumption was based not only on our professional experience in both disciplines but also on a formal equivalence of the PID in the United States and the Guttenberg-Richter recurrence curve.

The original geomechanical model describes the distribution of stresses in solid by separating them into elastic and inelastic components. Inelastic stresses are concentrated only on inhomogeneous inclusions and play an important role in the processes of deformation and fracturing. In the model, the size distribution of inclusions, *d(l)*, is chosen to retain constant the total volume for any size *l*: $d(l) \sim l^{-3}$. In other words, the number of inclusions of a given size *l* (per unit volume) decreases inversely proportional to the size cubed. This is a power law or scale free size distribution. The lower limit of *l* is likely constrained by the characteristic length of atom and the largest size should be substantially smaller then the size of the solid.

The growth rate of inelastic stresses is proportional to the rate of elastic deformation. Inelastic stresses are irreversible and dissipate over time. This is a fundamental property of real

solids – no stress or deformation can be retained forever and even such hard rock as basalt undergoes plastic deformation and dissipation of stored energy. The defining property of the geomechanical model consists in the assumption that the rate of dissipation of inelastic stresses is inversely proportional to the size of inclusion, i.e. the larger is the inclusion the longer time is needed to dissipate the same level of inelastic stress. (When applied to economics this rule says that larger incomes more resistant to decline, i.e. they decay at a lower rate than small incomes.) To simplify relevant mathematics, only one deformation process with a constant rate is usually considered in the geomechanical model. The deformation is caused by some external forces, which provide a constant energy supply.

This geomechanical model has been adapted and modified for the purposes of economic modelling. Formally, the size of inclusion is interpreted as the size of some tool or means, which is used to generate or earn income. Such words as "generate", "produce", "earn" and their synonyms are equivalent in the framework of our model and express the assumption that the sum of all personal incomes is equal to GDP. The proposed model is a *microeconomic* model because it addresses the evolution of personal incomes depending on individual properties and conditions. On the other hand, when aggregated over the whole working age population, the model allows a *macroeconomic* level of consideration. Thus the model is a dual one expressing the fact that by definition Gross Personal Income (GPI) is equal to GDP. Here we assume that GPI is equal to Gross Domestic Income and there is no impersonal income, because any income, personal or corporate, ultimately has its personal owner who can use this income for consumption, saving or investment.

In contrast to the geomechanical model, observations of income force the size of earning means to be distributed uniformly from some nonzero minimum to a finite maximum value. Uniform distributions of sizes are not usual in physics. As a rule, larger objects are less frequent. Because the PIDs measured in the US and their aggregates are well predicted with a uniform size distribution of earning tools, we did not thoroughly analyze alternatives. It could be a good exercise for students, however.

In the microeconomic model, deformation caused by external forces is interpreted as the capability of a person to generate income independent of the size of earning means. As an inherent characteristic of a person it could hardly be changed under normal conditions. This property is related only to money earning and does not depend on other personal talents and deficiencies. In a sense, two persons with equal talent in some profession have quite different salaries. Unlike talent, the capability to earn money is a measurable characteristic expressed in monetary units.

The income earned per year or income rate, as an analogue of inelastic stress concentrated on an inclusion, is proportional to the product of the size of earning means and the capability to earn money. These capabilities (or rates of external deformation) are also distributed uniformly among people of working age. The capabilities and sizes of earning means - both are getting larger as real GDP per capita grows. As a result, the evolution of the system of personal incomes is described by equations, which include some features additional to those in the geomechanical model. The microeconomic model has the same functional dependence between defining variables and similar formal solution. So, in mathematical terms, we are ready to start modelling personal incomes.

Before one starts a quantitative modelling, a thorough investigation of data availability and quality should be carried out. No model can be proved valid or invalid when relevant data do not provide an appropriate resolution. The breakthroughs in the natural sciences always happen at the edge of resolution leaving behind firm knowledge. Following this tradition, §1.2 is fully devoted to the assessment of data quality. The distribution of personal incomes is measured by various institutions, both governmental and private. We rely on the data which have been gathered by the US Census Bureau in the March Supplements of the Current Population Surveys since 1947. Other sources cover shorter periods or have gaps in measurements. Moreover, the Census Bureau provides the dependence on age – a feature most important for an evolutionary model. At the same time, there are numerous and severe deficiencies in the CPS data. The most painful and dangerous for the consistency of quantitative modelling is the incompatibility of data after any new revision to the CPS questionnaire: the unit of income measurement has been randomly changing through time. In physics, metrology was introduced several centuries ago and always serves as a backbone of any empirical investigation.

The microeconomic model is formally introduced in §1.3. This is the final result of an extended empirical investigation. To select some initial model from numerous alternatives, to modify it for matching a bulk of observations, and to estimate empirical parameters and coefficients required time and efforts. In its computer version, the main loop of the model programmed in FORTRAN took around 25 lines. A few subprograms allow different levels of aggregation: from individual income to GPI. The programming is a straightforward one and one can repeat it in no time using defining equations and reported parameters. Real GDP per capita is the driving force of the model. Therefore, we do not need to numerically integrate ordinary differential equations, but to use measured GDP.

To begin with we test the predictive power of the model by estimating the overall PIDs in the United States. This is an intermediate level of aggregation which disregards the dependence of individual incomes on age. Together with predicted PIDs, §1.4 introduces initial conditions for actual modelling. Initial values of defining parameters are obtained by standard trial-and-error method. Since the Pareto law is an empirical one and is obtained directly from observations, the microeconomic model covers only the low income zone. This zone includes 90% of working age population, however.

One of basic results of §1.4 consists in finding of a rigid hierarchy of personal incomes. When normalized to gross personal income and total working age population all PID between 1994 and 2001 collapse to one curve. In other words, the normalized PID is an invariant. In classical mechanics, such invariants (in closed systems) as energy and momentum provide fundamental constraints on possible evolution of the systems and also result in strict links between aggregate variables. These links are usually expressed in homonymic equations of classic mechanics. One can refer to the representations given by Euler and Lagrange, for example. If similar invariants would exist in real economy one could derive numerical conclusions, and likely a sound theory.

Understanding and modelling of age-dependent PID does deserve special attention as demonstrated in §1.5. There are really dramatic changes in the shape of PID: from practically exponential fall in the youngest and eldest age groups to a piecewise function in the mid-age groups. All these features are successfully modelled for the period between 1994 and 2002. The success is even enhanced by the fact that the analysis and prediction was based on the same microeconomic model and parameters as obtained for the overall PID in §1.4. It is a formal quantitative validation of the model – it predicts beyond the set of data used for the estimation of empirical parameters and coefficients.

Therefore, the microeconomic model quantitatively describes the evolution (with age and over time) of each and every personal income as a function of the individual capacity to earn money, the size of earning means, and real economic growth. At this stage, the modelling of age-dependent PID was not accompanied by the explicit prediction of the level of income inequality.

In paragraph 1.6, a different set of data is modelled - the dependence of average and median income on work experience. This data set spans a longer period since 1967. Here we first test the consistency of the model at higher incomes described by the Pareto law. The modelling meets significant difficulties related to the changes in the portion of GPI in GDP and income definition in the CPS questionnaire. The revisions to the CPS and population estimates after decennial censuses create artificial steps in the PIDs. Median income may be a more robust variable due to lower sensitivity to higher incomes. Its dynamics is relatively better predicted by the model. Overall, the dependence of mean and median on work experience and its evolution over time validates the model.

Paragraph 1.7 addresses several problems associated with the Pareto distribution. There is no general understanding and formal model of the processes leading to the power law distribution of personal incomes. This is a challenge for the future. However, there are several quantitative features of the Pareto distribution which can be modelled. Of crucial importance is the dependence of the portion of people in the Pareto distribution on work experience. Apparently, the youngest and eldest age group should be characterized by lower portions than intermediate groups. The model is able to accurately predict this dependence and its evolution through time. It is another point in favour of the model.

Numerous quantitative features related to economic inequality are discussed in §§ 1.8 and 1.9. This type of inequality is an apparently inevitable and multi-dimensional phenomenon in any social system. Due to practical and emotional importance for everyone, inequality attracts high attention of economists, politicians, and ordinary people. The former ones are focused at revealing potential quasi-deterministic or statistical links between economic inequality and numerous micro- and macroeconomic variables. There is no clear understanding whether the economic inequality is a positive or negative factor for such fundamental economic parameters as real economic growth, inflation, and unemployment (Galbraight, 1998).

Income inequality is one of quantitative measures of economic inequality. There are many theories of inequality arising from the distribution of income. Neal and Rosen (2000) presented an almost comprehensive overview of state-of-art in this field. In spite of the efforts associated with the development of a consistent model of income distribution there are some problems yet to resolve. Moreover, modern economic theories do not meet some fundamental requirements applied to scientific theories - a concise description of accurately measured variables and prediction of their evolution beyond the period of currently available measurements.

In §1.8, we model the most popular aggregate measure of income inequality - the Gini coefficient, $G$, for PIDs in the United States. This coefficient is characterized by a number of advantages such as relative simplicity, anonymity, scale independence, and population independence. On the other hand, the Gini coefficient belongs to the group of operational measures: its evolution through time is not theoretically linked to macroeconomic variables and the differences in Gini coefficient observed between various countries are not well explained. These caveats make the Gini coefficient more useful in political and social applications but not in economics as a potentially quantitative (hard) science.

As a rule, the Gini coefficient is estimated from household surveys, and inequality is reported at family and household levels. Such an aggregation is affected by social and demographic processes, which may bias true economic mechanisms driving income inequality.

Theoretically, the indivisible level for the study of income inequality is the personal income. In our framework, personal incomes are presumed to be sensitive only to macroeconomic variables.

Paragraph 1.9 describes the data on personal income distribution in various age groups, presents estimates of relevant Gini coefficients, elaborates on empirical PIDs, and compares the evolution of observed Gini coefficients to that predicted by our model. The age-dependent PID in the youngest group is characterized by large differences from the overall PIDs. Obviously, all individuals start with zero income and the initial part of income trajectory in time, as personal income observations show, is close to an exponential growth. In the mid-age groups, PIDs are similar to the overall PID. In the oldest age group, PID is also different and is closer to that in the youngest group. Accordingly, Gini coefficient undergoes a substantial evolution from the youngest to the oldest age groups.

In §1.10, we join the vivid discussion of increasing economic inequality in the United States. Our quantitative assessment of personal income inequality is quite different from that articulated by many economists. In §1.8, we conducted quantitative estimates of Gini coefficient using personal income distributions, which have been reported since 1947 by the US Census Bureau , and found that this coefficient was practically constant over time. Having a constant Gini coefficient since (at least) 1947, one might find it strange that other researches and media thoroughly discuss increasing inequality during the last 25 years. It was difficult to actually understand why those researches do not use the US Census data despite the Census Bureau (2004) explicitly states:

> Because of its detailed questionnaire and its experienced interviewing staff trained to explain concepts and answer questions, the CPS ASEC is the source of timely *official* national estimates of poverty levels and rates and of widely used estimates of household income and individual earnings, as well as the distribution of that income.

Paul Krugman, the 2008 Laureate of the Sveriges Riksbank Prize in Economic Sciences in Memory of Alfred Nobel, explained why he and other researchers are forced to deny the estimates based on Census Bureau data:

> First, because Census data are based on a limited sample, not the whole population, they're unreliable in tracking the income of small groups – and the really rich are a small group, who just happen to bulk large in the economy. Second, the questionnaire is "top-coded": if the individual interviewed has earnings higher than $999,999, those earnings are recorded simply as $999,999. Since a lot of income growth in the last few decades has taken place among people with multimillion-dollar incomes, the Census data miss an important part of the story.

In practical and theoretical terms, both statements (reasons) are wrong. First, in hard sciences, one is not often able to measure true values of desired variables, but usually measures some portions of them. For example, nobody tries to invent a weighting machine in order to measure the Earth's mass. It is enough to measure gravity acceleration in one point since this acceleration is proportional to the total mass. Therefore, if a portion of a whole object is a representative one and is measured consistently over time, one can carry out a reliable quantitative analysis. A randomly changing portion, as sometimes happens to macroeconomic variables after introduction of new definitions, would, obviously, ruin any such quantitative analysis. So, using surveys of small population samples does create a problem with internal precision, but should not necessary disturb results of overall quantitative analysis. This Chapter provides extensive quantitative results which confirm that the Census Bureau has been collecting high-quality data.

Second, the "top-coded" approach does not harm the estimates of income in the "the richest of the rich" group. This effect is known more than hundred years already. Higher incomes are very accurately distributed according to the Pareto law. As a matter of fact, one does not need to measure any personal income in the high-income group. S/he needs to estimate the number of persons with income above some given (high) threshold. Then, one can use simple mathematical equations to obtain accurate population density at any income level and also total income above any threshold.

The logic of the presentation of a new model in a book cardinally differs from that of scientific research itself. When studying some process or phenomenon, one does not possess complete knowledge about defining relationships and parameters. This state of incomplete knowledge gives birth to numerous questions and problems one has to address during the study. When the research is finished, the obtained model should accurately describe corresponding observations and all (or almost all) wrong hypothesis and irrelevant assumptions are eliminated. As a result, the presentation of a finished model usually skips all unnecessary details and is focused on a comprehensive description of relevant relationships and parameters. Following this tradition, in Chapter 1 we first present our model and then demonstrate how accurate it predicts various quantitative properties of personal income distributions in the US as related to some measured macroeconomic and demographic parameters. The model is validated according to standard procedures involving comparison of predicted and observed data.

*Data sources:*

Bureau of Economic Analysis, (2005). Current-Dollar and "Real" Gross Domestic Product, (Seasonally adjusted annual rates), table, last modified 05.25.05, http://bea.gov/bea/dn/gdplev.xls

Bureau of Economic Analysis, (2008). National Economic Accounts, tables, retrieved on March 30, 2008

Census Bureau, (2000). The Changing Shape of the Nation's Income Distribution, retrieved on February 26, 2007 from http://www.census.gov/prod/2000pubs/p60-204.pdf

Census Bureau, (2002), Technical Paper 63RV: Current Population Survey Design and Methodology, issued March 2002", http://www.census.gov/prod/ 2002pubs/tp63rv.pdf

Census Bureau, (2002). Source and Accuracy of the Data for the March 2002 CPS, http://www.bls.census.gov/cps/ads/2002/S&A_02.pdf

Census Bureau, (2004). Detailed Income Tabulations from the CPS. Last revised: August 26 2004; http://www.census.gov/hhes/income/dinctabs.html

Census Bureau, (2004). Methodology: National Intercensal Population Estimates. Last revised: August 20, 2004 at 07:19:10 AM. http://www.census.gov/popest/archives/ methodology/intercensal_nat_meth.html

Census Bureau, (2004). Historical Income Tables - People. (Table) P-9. Age-People (All Races) by Mean Income and Sex: 1967 to 2001. Last revised: Thursday 13-May-2004 11:31:11 EDT.

Census Bureau, (2004). Historical Income Tables - People. (Table) Table P-10. Age--People (Both Sexes Combined--All Races) by Median and Mean Income: 1974 to 2003, Last Revised: Thursday 13-May-2004 11:31:11 EDT.

Census Bureau, (2004). Guidance on Differences in Income and Poverty Estimates from Different Sources, August 19, 2004

Census Bureau, (2005). U.S. Interim Projections by Age, Sex, Race, and Hispanic Origin. Table. Last revised: 8:40 am on May 13$^{th}$.

Census Bureau, (2005). Changes in Methodology for the March Current Population Survey. Last Revised: May 13, 2005.

Census Bureau, (2007). Population Estimates. Retrieved March 14, 2007 from http://www.census.gov/popest

Census Bureau, (2008). Current Population Reports. Consumer Income Reports from 1946-2006 (P60), http://www.census.gov/prod/www/abs/income.html

Internal Revenue Service, (2007). Selected Income and Tax Items from Inflation-Indexed Individual Tax Returns. All Returns: Sources of Income, Adjustments, and Tax Items in Constant 1990 Dollars, http://www.irs.gov/taxstats/indtaxstats

## §1.2. Measurements of personal income

Before starting any quantitative research one needs to evaluate data consistency and quality. This paragraph addresses several questions associated with the quality of income measurements:

What is personal income?
Are there alternative definitions of personal income?
How is personal income measured?
How large is the uncertainty of income measurements?
What improvements are necessary for successful quantitative modelling?

The history of income measurements in the United States started in 1947, when a question about individual income was first included in the first Annual Social and Economic Supplement, ASEC, (former Annual Demographic Survey, ADS, or March Supplement) to the Current Population Survey (http://www.census.gov). Fully electronic tables presenting personal incomes by detailed socioeconomic characteristics have been published by the US Census Bureau since 1994. Scanned versions of hard copies of income tables since 1947 are also available via the Census Bureau.

The electronic tables include information on the number of people with income in relatively narrow intervals of $2500, starting from zero income and current losses, up to the highest income of $100,000. These tables also provide mean incomes in five-year wide age groups, except that for the youngest group, which spans the ages from 15 to 24 years. The detailed income tabulations from the CPS related to personal income are available since 1994 and aggregated data - for a longer period since 1967. Before 1994, data representation has been changing through time with numerous revisions to income definition and methodology of measurements.

The definition of personal income has been actively discussed by the Canberra Group on Household Income Statistics (2001) under the umbrella of the United Nations. As a matter of fact, there are many definitions of personal income and, thus, there are many measures of income. From the point of view of the orthodox economics, the Haig-Simons-Hicks (HSH) concept could provide the theoretical framework for any definition of personal income. The concept defines the income as the maximum amount that can be consumed in a given time period without any change in real wealth. Technically, two most elaborated definitions of personal income in the US are given by the Bureau of Economic Analysis (BEA) and the Census Bureau; both measure only different portions of the HSH amount. Therefore, there is always some room for improvement in the definitions. The Internal Revenue Service's definition is of lower theoretical interest for quantitative modelling because it excludes too many sources of income for unbiased estimates.

The BEA conducts the estimates of personal income from administrative record and defines it in the following terms:

> **Personal income.** Income received by persons from all sources. It includes income received from participation in production as well as from government and business transfer payments. It is the sum of compensation of employees (received), supplements to wages and salaries, proprietors' income with inventory valuation adjustment (IVA) and capital consumption adjustment (CCAdj), rental income of persons with CCAdj, personal income receipts on assets, and personal current transfer receipts, less contributions for government social insurance.

This is the approach which considers personal income as a portion of GDP for macroeconomic purposes. As one can judge, this is a technical definition and relevant income measurements need enormous efforts for data gathering and processing. Private companies or scientific foundations can hardy repeat such measurements. Therefore, alternative measurements of personal income, as defined by the BEA (and Census Bureau), are not feasible.

The Census Bureau uses the microeconomic approach and measures money (personal) incomes in the ASEC Supplement to the Current Personal Survey (CPS), which are aimed at understanding of income distribution. Under this framework, income is split into categories related to the types of transaction disregarding the sources of income. Overall, the Census Bureau defines personal income as

> … as total pre-tax cash income earned by persons, excluding certain lump sum payments and excluding capital gains.

A striking difference for a researcher carrying out standard analysis consists in the total personal income reported by the BEA and CB. In 2001, the former agency reported $8.678 trillion and the latter published the estimate of gross personal (money) income of $6.446 trillion. The difference is $2.232 trillion or 35%. Disregarding any details of the difference in personal income components following from both definitions one should assess the difficulty met by quantitative modelling. The uncertainty of 35% can make any reasonable model an appropriate one, and also allows denying almost any model as statistically unreliable.

In this study, we use only the data on personal income distribution gathered by the Census Bureau. The reasons for this choice follow from the "Guidance on Differences in Income and Poverty Estimates from Different Sources. August 19, 2004":

- The CPS ASEC provides a *consistent historical time series* of many decades in length at the national level...
- ...*detailed* questionnaire and its experienced interviewing staff trained to explain concepts and answer questions...
- The CPS ASEC provides the most timely and *most accurate* cross-section data for the nation on income and poverty.

Hence, the defining factor consists in the availability of the detailed and accurate age-sex-race tabulation in various income bins since 1947. The BEA reports only the aggregate Gross Personal Income value. For a quantitative model, the detailed data are of crucial importance, but one has to bear in mind the possibility of highly biased personal income estimates to be present. Our working hypothesis is that the extra personal income reported by the BEA is distributed over the working age population in a way, which does not disturb the personal income distribution reported by the Census Bureau. Obviously, this is a crude extrapolation, but it does not contradict our empirical findings. We also disregard personal income definitions related to decennial censuses, the American Community Survey, the Survey of Income and Participation Program, and the Small Area Income and Poverty Estimates. They do not provide either the national level of coverage or the continuous time series.

The next question concerns the procedure of income measurements conducted by the Census Bureau. The CPS is basically a labor force survey covering 60,000 households, and the ASEC uses a sample of about 100,000 addresses per year. Currently, the questionnaire contains 50+ sources of income. For each person of 15 years old and over in the sample, the CPS asks questions on the amount of money income received in the preceding calendar year from each of the following sources: earnings, unemployment compensation, workers' compensation, social security, supplemental security income, public assistance, veterans' payments, survivor benefits, disability benefits, pension or retirement income, interest, dividends, rents, royalties, and estates and trusts, educational assistance, alimony, child support, financial assistance from outside of the household, other income. The income statistics is gathered during one month but covers the whole preceding calendar year. This introduces a bias into such demographic characteristic as age.

The CPS does not measure capital from the sale of property, including stocks, bonds, a house, etc.; withdrawals of bank deposits; money borrowed; tax refunds; gifts; and lump-sum inheritances or insurance payments.

A crucial issue for our analysis is the precision of income measurements. (Because the true personal income is not defined one can measure only internal consistency of measurements, i.e. precision, not the distance between measured and true value, i.e. accuracy. However, we do not make any major difference between the terms.). There are two principal sources of errors in income measurements - sampling error and non-sampling error as presented by the US Census Bureau (2002). The first is related to the difference between the CPS sample and working age population as a whole. The CPS sample may differ in many ways from the population controls as obtained from decennial censuses and further corrections between the censuses for total deaths, migration, and people in the armed forces. When the CPS is projected to the total population in the weighting procedure, the biases may still be present and affect income estimates in the unknown way. Statistical accuracy of this type of error can be estimated and included into analysis.

The non-sampling error includes all other potential sources of error including under-coverage, definition difficulties, variations in interpretation of questions, incorrect information due to unwillingness or inability to recall, etc. Due to the intrinsic nature of these type errors their magnitude is unknown. Small population numbers related to specific income-age-sex-race groups deserve closer attention. In relative terms, the sampling and non-sampling errors are both most prominent where the enumerated population is comparable to the measurement accuracy. For example, the number of very young and very old people with very high incomes is always small. As shown in §1.5, there is no people reported in some income bins near $100,000 in the age group between 15 and 24 years. An obvious problem, which hardly can be resolved by any population survey, is associated with the measurement of very low incomes near $1 per year. Also due to definitional problem, many persons with very low incomes are reported as people without income. So, the uncertainty in these sensitive groups is very high and one should not expect any quantitative modelling to fit relevant data. On the contrary, a good quantitative model validated by accurate data can provide an invaluable assistance in the development of a sound definition and related measurement procedure. This is a standard situation in the natural sciences, where theoretical predictions stimulate more accurate measurements, i.e. determine what and how to measure.

The US Census Bureau also constantly improves the precision of the personal income measurements. This process includes extension of the number of interviewed households, covering some specific groups such as Hispanic origin, etc. The last major change in the procedure occurred in 2002, when the total number of the surveyed units became 99,000. Between 1994 and 2001 there were 60,000 units in the basic CPS and additional 21,000 units for the March Survey, i.e. 81,000 in total. As explicitly stated by the US Census Bureau, these changes make it sometimes difficult to *compare* income data sets obtained in different years. Most prominent revisions were as follows:

| Income year | Change |
|---|---|
| 1947 | Data based on 1940 census population controls. |

| Year | |
|------|---|
| 1961 | Implementation of first hot-deck procedure to impute missing income entries (all income data imputed if any missing). Introduction of 1960 census-based sample design. |
| 1965 | Implementation of new procedures to impute missing data only. |
| 1966 | Questionnaire expanded to ask eight income questions. |
| 1967 | Implementation of a new March CPS processing system. |
| 1974 | Implementation of a new March CPS processing system. Questionnaire expanded to ask 11 income questions. |
| 1975 | These estimates were derived using Pareto interpolation and may differ from published data which were derived using linear interpolation. |
| 1976 | First year medians are derived using both Pareto and linear interpolation. Prior to this year all medians were derived using linear interpolation. |
| 1979 | Implementation of 1980 census population controls. Questionnaire expanded to show 27 possible values from 51 possible sources of income. |
| 1985 | Recording of amounts for earnings from longest job increased to $299,999. |
| 1993 | Data collection method changed from paper and pencil to computer-assisted interviewing. Limits either increased or decreased in some categories. |
| 2000 | There are two versions of the 2000 income data available. One version is based on the traditional sample of about 50,000 households and reflects the use of 1990 census population controls. The second version is based on a sample of 78,000 households, reflecting a 28,000 household sample expansion and the use of Census 2000 population controls. |

The income part of the CPS questionnaire has not suffered major revisions since 1980. There are several important items which would be appropriate to include in the next revision, as proposed by the Canberra Group. Among them are interhousehold transfers and some fringe benefits. In addition, some items such as wages, transfer payments, and self-employment require substantial improvement due to clear misreporting.

Summarizing the quantitative properties of personal income measurements in the United States relevant to our modelling we would like to stress that:

- Personal income definition has suffered severe revisions in the past, but does not meet yet the complete set of requirements associated with variables in the natural sciences. The "right" definition of personal income should provide the full coverage of working age population, not only the most obvious groups such as workers and pensioners.

- The right definition of personal income has to provide the gross personal income equal to GDP.

- The uncertainty of personal income measurements is inherently related to two sources: definition and methodology, the latter being easier to resolve.

- The magnitude of the uncertainty varies with age, race, sex, and income level. Due to non-sampling errors it is difficult to estimate the absolute and relative magnitude of the uncertainty.

- Potentially, the upper limit of the uncertainty might be estimated as the difference between various definitions of income. In that case, the uncertainty is unacceptably high in some age groups and income bins.

- Quantitative properties of personal income distribution and their evolution over time might be helpful for the proper definition of personal income.

After the start of this research, the problem of quality and accuracy of income measurements immediately became a fundamental one. Good measurements are always based on a positive feed back from scientific knowledge to technological implementation and vise versa. Metrology, which embraces both theoretical and experimental determination of uncertainty, is always the first step in any scientific research. In other words, any quantitative theory or model demands the units of measurements, system of units and the development of measuring techniques. Economics, if accepting the challenge to join the natural sciences, needs to demonstrate a sufficient level of uncertainty characterizing major data sets. No surprise that, in line many empirical studies in physics, we start every chapter with sometimes extended but very specific review of accuracy related to studied economic variables. Otherwise, our concept cannot be realized in quantitative models.

## §1.3. Microeconomic model of the personal income distribution and evolution

Historically, the model is the result of a long process. Initially, the idea was to represent the distribution and evolution of personal incomes as defined by geo-mechanical model of a solid with non-linear inhomogeneous inclusions. This geo-model is characterized by a specific sizes distribution, which retains constant total volume for all sizes of inclusions. The initial idea had undergone a number of modifications and corrections before was confirmed in its final form. Because of the geo-mechanical legacy we called the invented a new term for economics presented in the framework of (geo-) mechanics "mechonomics". Later on, this term was applied to the research as a whole.

Therefore, the microeconomic model for the evolution of personal incomes is based on a "first principle" approach expressed as an equation balancing money production and money dissipation. The former is formally equivalent to the production of goods and services. The latter is the result of a number of factors acting against individual production efforts and reducing the amount of money earned below its potential level. Similarly, efficiency of a steam engine is not 100% because of heat losses and friction. Such processes are often referred as dissipative processes or dissipation. In that sense, our model is an analogue to numerous physical models including the model of solid with inhomogeneous inclusions. Thus we use the term "dissipation" instead of economic terms like "depreciation" in order to stress that economic systems are similar to physical systems.

The model presents functional dependence for continuous evolution of all personal incomes in an economic system and attributes real economic growth to personal efforts to earn money. Such an approach links the model to the roots of economic activity because the only source of any income or wealth consists in personal efforts. Due to natural interactions and inevitable economic ranking in any society personal incomes are distributed not in a random, but in predefined and fixed way, and actual PIDs demonstrate a simple functional dependence on real economic growth.

Figure 1.3.1 illustrates schematically some principal features of personal income distribution in the USA, which undoubtedly has to be addressed in any model of income distribution. These features were a part of the empirical intuition behind our model. The Figure displays the dependence of mean income (averaged in 10-year intervals) on work experience in the USA in 2002. The empirical dependence is normalized to its peak value and is approximated by two functions: *(1-exp(-$\alpha_0 t$))* ($\alpha_0$=0.085) in the interval of work experience from 0 to 39 years, and by *(1-exp(-$\alpha_0 T_{cr}$))exp(-$\alpha_1 t$)* ($\alpha_1$=0.06) above $T_{cr}$=39 years. In spite of a good description of the observations, this approximation does not present a correct model because it does not distinguish between people with different incomes and does not provide any driving force for the evolution of personal income. This is an example of the approximation often used instead of "first principle" model. It demonstrates, however, potential simplicity of the processes underlying income distribution, which lead to simple functional dependencies.

The microeconomic model presented in this paragraph takes the advantage of recent accurate measurements of personal income distribution in the USA. There was no such data for the period before 1947, and individual incomes were thought to be unpredictable. This unpredictability is valid for an individual, if to trace his/her income evolution in time. But any individual can only follow one of the predefined trajectories predicted by the model. If somebody suddenly jumps to a new and higher income value some other individual with the income value equal to the new income of the first person has to drop to first person's income value in order to retain fixed PID. One can treat this observation as a manifestation of a conservation law for PID – relative number of positions with given income is conserved. This is similar to the level's degeneracy in quantum physics.

Conventional economic models of income distribution lack these natural roots aiming at artificial division of total personal income into employee compensation and corporate profits with further (again artificial) separation into groups attributed to various types of economic activity: consumption, savings, and investment. Relative importance of these parts separated by force varies with time depending on individual decisions to split a personal income into these three portions. This decision is based, however, only on the individual effort to earn income and can be random in the sense of incomplete information and wrong interpretation. That's why one can not construct a precise model for the economic evolution expressed in terms of these artificial parameters and any economic policy based on a full control of these parameters ultimately fails. There is nothing except the personal income production in any developed economy and the production volume expressed in monetary units is predefined. The PID evolution can be exactly predicted to the extent of the accuracy of population counting and GDP measurements as we demonstrate in the following paragraphs.

Principal assumption of the microeconomic model is that every person above fourteen years of age has a capability to work or earn money using some means, which can be a job, bank interest, stocks, interfamily transfers, etc. An almost complete list of the means is available in the US Census Bureau technical documentation (2002) as the sources of income are included in the survey list. Some important sources of income are not included, however, what results in the observed discrepancy between aggregate (gross) personal income, GPI, and GDI.

Here we introduce the model as described by Kitov (2005a). The rate of income, i.e. net income a person earns per unit time, is proportional to her/his capability to earn money, $\sigma$. The person is not isolated from the surrounding world and the work (money) s/he produces dissipates through interaction with the outside world, decreasing resulting income rate. The counteraction of external agents, which might be people or any other externalities, determines the price of goods and services a person creates. The price depends not on some absolute measure of quality of the goods but on the aggregate opinion of the surrounding people on relative merits (expressed in monetary units) of the producers, not goods. For example, the magic of famous brands provides a significant increase in incomes for their owners without proportional superiority in quality because people appreciate the creators not goods. As a whole, an equilibrium system of prices arises from aggregate opinions on relative merits of each and every person not from the physical quantities and qualities of goods and services. Personal incomes are ranked in some fixed hierarchy and, when expressed in monetary units, the hierarchy is transformed in a dynamic system of prices. Since the hierarchy of incomes is fixed, the amounts and qualities of goods can only reorder individuals not change the final aggregate price of everything produced – GDP.

  Analogously to many cases observed in natural sciences, the rate of dissipation is proportional to the attained income (per unit time) level and inversely proportional to the size of the means used to earn the money, $\Lambda$. Bulk heating of a body accompanied by cooling through its surface is an analogue. For a uniform distribution of heating sources, the energy released in a body is proportional to its volume or the cube of characteristic linear size and the energy lost through its surface is proportional to the square of the linear size. In relative terms, the energy balance or the ratio of cooling and heating is inversely proportional to the linear size. As a result, a larger body undergoes a faster heating because loses relatively less energy and also reaches a higher equilibrium temperature. In line with this consideration, one can write an ordinary differential equation for the changing rate of income earned by a person in the following form:

$$dM(t)/dt = \sigma(t) - \alpha M(t)/\Lambda(t) \qquad (1.3.1)$$

where $M(t)$ is the rate of money income denominated in dollars per year [$/y], $t$ is the work experience expressed in years [y], $\sigma(t)$ is the capability to earn money [$/y$^2$]; and $\alpha$ is the dissipation coefficient also expressed in [$/(y$^2$)]. The size of earning means, $\Lambda$, is expressed in [$/y]. General solution of equation (1), if $\sigma(t)$ and $\Lambda(t)$ are considered to be constant (as shown later these two variables evolve very slowly with time), is as follows:

$$M(t) = (\sigma/\alpha)\Lambda(1 - exp(-\alpha t/\Lambda)) \qquad (1.3.2)$$

In quantitative modelling, we integrate (1) numerically in order to include the effects of the changing $\sigma(t)$ and $\Lambda(t)$. Equations (1.3.2) through (1.3.4) are derived and discussed in detail below to demonstrate some principal features of the proposed model. These equations represent some particular solutions of (1.3.1), where measured change in $\sigma(t)$ and $\Lambda(t)$ in all terms of (1.3.1) is neglected.

One can introduce the concept of a modified capability to earn money as a dimensionless variable $\Sigma(t)=\sigma(t)/\alpha$. Absolute value of the modified capability, $\Sigma(t)$, and the size of earning means evolves with time as the square root of real GDP per capita:

$$\Sigma(t) = \Sigma(t_0)sqrt[GDP(t)/GDP(t_0)]$$

and

$$\Lambda(t) = \Lambda(t_0)sqrt[GDP(t)/GDP(t_0)],$$

where $GDP(t_0)$ and $GDP(t)$ are per capita values at the start point of the modelling, $t_0$, and at time $t$, respectively. Then the capacity of a "theoretical" person to earn money, defined as $\Sigma(t)\Lambda(t)$, evolves with time as real GDP per capita. Effectively, equation (1.3.2) states that the evolution in time of a personal income rate depends only on personal capability to earn money, the size of means used to earn money, and economic growth. The latter factor is common for everybody and does not affect relative distribution of personal incomes just the overall level.

The modified capability to earn money, $\Sigma(t)$, and the size of earning means, $\Lambda(t)$, obviously have positive minimum values among all individuals in a given economy, $\Sigma_{min}(t)$ and $\Lambda_{min}(t)$, respectively. One can now introduce relative and dimensionless values of the defining variables in the following way: $S(t)=\Sigma(t)/\Sigma_{min}(t)$ and $L(t)=\Lambda(t)/\Lambda_{min}(t)$.

From a calibration procedure described below, a fundamental assumption is made that possible relative values of $S(t_0)$ and $L(t_0)$ can be represented as a sequence of integer numbers from 2 to 30, i.e. only 29 different integer values of the relative $S(t_0)$ and $L(t_0)$ are available: $S_1=2,..., S_{29}=30$; $L_2=2,..., L_{29}=30$. The largest possible relative value of $S_{max}=S_{29}=30=L_{max}=L_{29}$ is only 15 (=30/2) times larger than the smallest possible value $S=S_1$ and $L=L_1$ (in the model, the minimum values $\Lambda_{min}$ and $\Sigma_{min}$ are chosen to be two times smaller than the smallest observed values of $\Lambda_1$ and $\Sigma_1$). Because the absolute values of variables $\Lambda_i$, $\Sigma_i$, $\Lambda_{min}$, and $\Sigma_{min}$ evolve in time according to the same law, the relative and dimensionless variables $L_i(t)$ and $S_i(t)$, $i=1,...,29$, retain the same discrete distribution. This means that the distribution of relative capability to earn money and the size of earning means is fixed as a whole over calendar years and also over ages.

Figure 1.3.2 depicts several examples of the evolution of personal income. The curves are normalized to the maximum income possible in the model - $SL=30 \times 30=900$, and dissipation factor $\alpha=0.07$. The predicted income for a person using a means of dimensionless size $L=2/30$ and having a dimensionless capability to earn money $S=2/30$ approaches its maximum possible level of $4/30^2$ (relative to the overall maximum possible of $30^2/30^2=1$) just in few years after the start of work. During the rest of her/his life, the person has the same relative personal income, and the absolute level of the income increases proportionally to the growth of per capita GDP.

In the case where $S=15/30$ and $L=15/30$, a longer time is necessary for a person to approach the maximum potential income equal to $15^2/30^2$. This person reaches 95% of the potential income in 10 to 15 years from the start of work. Then, almost no change in relative income is observed as in the case of the lower income person.

It is interesting to compare two cases with the same potential maximum level of income but different $L$ values. These cases are $S=2/30$, $L=30/30$; and $S=30/30$, $L=2/30$. Corresponding curves in Figure 1.3.2 reach the same level in 40 years and approximately 3 years, respectively. Hence, the earning means' size plays a key role in the evolution of theoretical PID. This parameter defines the change of effective dissipation rate in (1.3.2), because $\alpha$ is constant. Thus, effective time constant in (1.3.2) is $\Lambda/\alpha$. The larger is the effective time constant, the longer time is needed to reach the same relative level of income. So, increasing value of $L$ leads to slower relative income growth.

Now, one can carry out appropriate substitutions in (1.3.2) and normalize the equation to the maximum values $S_{max}$ and $L_{max}$. The normalized equation for the rate of income, $M_{ij}(t)$, for a person with capability, $S_i$ and the size of earning means, $L_j$ is as follows:

$$M_{ij}(t)/(S_{max}L_{max}) =$$
$$= (\Sigma_{min} \Lambda_{min})(S_i/S_{max})(L_j/L_{max})(1 - exp(-(\alpha/\Lambda_{min} L_{max})t/(L_j/L_{max}))) \quad (1.3.3)$$

or

$$M'_{ij}(t) = \Sigma_{min}(t)\Lambda_{min}(t)S'_i L'_j \{1 - exp[-(1/\Lambda_{min})(\alpha' t/L'_j)]\} \quad (1.3.3')$$

where $M'_{ij}(t)=M_{ij}(t)/(S_{max}L_{max})$; $S'_i=(S_i/S_{max})$; $L'_j=(L_j/L_{max})$; $\alpha'=\alpha/L_{max}$; $S_{max}=30$, and $L_{max}=30$. Below we omit the prime indices. The term $\Sigma_{min}(t)\Lambda_{min}(t)$ corresponds to the total (cumulative) growth of GDP per capita from the start point of a personal work experience, $t$ $(t_0=0)$, and is different for different start years. This term might be considered as a coefficient defined for every single year of work experience because this is a predefined external variable. Thus, one can always measure personal incomes in units $\Sigma_{min}(t_0)\Lambda_{min}(t_0)$. Then (1.3.3') becomes a dimensionless one and the coefficient changes from 1.0 as GDP per capita evolves relative to start year.

Equation (1.3.3') represents the rate of income for a person with the defining parameters $S_i$ and $L_j$ and work experience $t$ relative to the maximum possible personal income rate, which is obtained by a person with $S_{29}=30/30=1$ and $L_{29}=30/30=1$ at the same work experience $t$. The term $1/\Lambda_{min}$ in the exponential term evolves inversely proportional to the square root of GDP per capita. This is the key term for the evolution of personal incomes, which accounts for the differences between the start years of work experience. Numerical value of $\alpha/\Lambda_{min}$ is obtained by calibration carried out for the start year of corresponding modelling. This calibration assumes that $\Lambda_{min}(t_0)=1$ (and $\Sigma_{min}(t_0)=1$ as well) at the start point of the modelling and only the dimensionless dissipation factor $\alpha$ has to be empirically determined. In this case, absolute value of $\alpha$ depends on the start year of modelling.

This is a good place to speculate on the sufficient size of economy in terms of population, which follows up from the distribution of earning capability and means size. There are 841 (=29x29) different states or income trajectories available for people in a given economy. It is natural to assume that each state should be occupied by at least 841 persons, each of them having an equal chance to join any other state. So, to fill all possible states 841x841=707281 people are needed. This number of people is related to the same birth year. There are about 50 different years of age in a standard working age population in developed countries. Then the number of people in the sufficient size economy is estimated as ~35,000,000. This is a very crude estimate, but it gives a useful threshold discriminating small and large economies. Working age population is such self-consistent economies as Germany, France, the UK, Italy, Spain scatters around this level. The economies are characterized by a practically full set of internally produced goods and services, which provides economic independence and sustainability. Personal income distribution in these countries should be driven by the uniform distribution of capabilities and earning means. In other words, income distribution in these countries should mimic that in the United States. Thorough investigations are needed in this field, although.

The world's biggest economies, the US and Japan, are essentially bigger than the threshold and must be characterized by the best fit between the modelled and observed personal income distribution. This conclusion follows from the number of people in each single year of age group reaching 4,500,000 and 1,700,000, respectively. The overwhelming number makes the distribution of people among 841 available states more uniform or less volatile.

In smaller economies, say less than 1,000,000 to 5,000,000 people, there are not enough people to fill all income states uniformly and higher disparity between theoretically equal states is observed. As a result, personal income distribution in such smaller countries might be biased relative to that predicted by the model. In many cases social policy aimed at a lower income (economic) inequality. However, these countries are on the brink of self-consistency and sustainability. In many cases they depend on larger neighbours or are focused on few products. Vulnerability is a common consequence.

As Figure 1.3.1 indicates, money earning capacity, $S_i L_j$, drops to zero at some critical time, $T_{cr}$, in any personal work experience history and the solution of (1) becomes as follows:

$$M_{ij}(t) =$$
$$M_{ij}(T_{cr}) \exp(-\alpha_1 (t-T_{cr})/ \Lambda_{min} L_j) = \qquad (1.3.4)$$
$$= \{\Sigma_{min}(t) \Lambda_{min}(t) S_i L_j (1-\exp(-\alpha T_{cr}/\Lambda_{min} L_j))\} \exp(-\alpha_1 (t-T_{cr})/ \Lambda_{min} L_j)$$

The first term is equal to the level of income rate attained by the person at time $T_{cr}$, and the second term represents an exponential decay of the income rate for work experience above $T_{cr}$. The latter variable also evolves in time as the square root of GDP per capita:

$$T_{cr}(t) = T_{cr}(t_0) sqrt(GDP(t)/GDP(t_0)) \qquad (1.3.5)$$

The exponent index $\alpha_1$ is different from $\alpha$ and varies with time. It was found that the exponential decrease of income rate above $T_{cr}$ results in the same relative (as reduced to the maximum income for this calendar year) income rate level at the same age. It means that the index can be obtained according to the following relationship:

$$\alpha_1 = -ln(M_r)/(A_r - T_{cr}) \qquad (1.3.6)$$

where $M_r$ is the relative level of income rate at the reference age $A_r$, both are effectively constant. Thus, when current age reaches $A_r$ the maximum possible income rate $M_{ij}$ (for $i=29$ and $j=29$) drops to $M_r$. Income rates for other values of $i$ and $j$ are defined by (1.4.4). For the period between 1994 and 2002, empirical estimates are as follows: $M_r =0.72$ and $A_r=64$ years. In our model, the exponential roll-off observed for the mean income in Figure 1.3.1 corresponds to a zero-value work applied by individuals with work experience beyond $T_{cr}$ to earn money. People do not exercise any effort to produce income starting from some predefined (but changing) point in time, $T_{cr}$, and enjoy exponential decay of their incomes. It is important that this critical work experience is below the age or retirement and was even lower in the past.

A physical analogue of such exponential decay is cooling of a body, for example - the Earth. When all sources of internal heating (gravitational, rotational, and radioactive decay) disappear, the Earth only will be loosing the accumulated internal heat through the surface before reaching an equilibrium temperature with the outer space. This process of cooling is also described by an exponential decay because the heat flux from the Earth is proportional to the difference of the temperatures between the Earth's surface and the outer space.

The probability for a person to get an earning means of relative size $L_j$ is constant over all 29 discrete values of the size. The same is valid for $S_i$, i.e. all people of 15 years of age and above are distributed evenly among the 29 groups for the capability to earn money. Thus, the capacity for a person to earn money is distributed over working age population as the product of independently distributed $S_i$ and $L_j$ - $S_iL_j = \{2\times2/900, 2\times3/900, \ldots, 2\times30/900, 3\times2/900, \ldots, 3\times30/900, \ldots, 30\times30/900\}$. There are only 841 (=29x29) values of the normalized capacity available between 4/900 and 900/900. Some of these cases seem to be degenerate (for example, 2x30=3x20=4x15= …= 30x2), but as discussed above, all of them define different time histories according to (1.3.3'), where $L_j$ is also present in the exponential term. The discrete spectrum of the capabilities and sizes of earning means between 2 and 29 is obtained by a trail-and-error method. The values were varied in a wide range in order to obtain a good agreement between observed and predicted PIDs. Figure 1.3.3 illustrates the final result of the variation – the oscillations in the observed PIDs for 1994, 1997, and 2001 are well represented by the model in the low-income zone. One can test that other discrete spectra associated with different combinations of maximum $S$ and $L$ do not provide the same accuracy of the description of peaks and troughs and actual roll-off of the observed PIDs. Also the discrete and even distribution of the two defining variables is very simple and natural.

In reality, no individual income trajectory is predefined, but the model puts a strong constrain that it can only be chosen from the set of the 841 predefined individual future for each single year of birth.

It is not possible to quantitatively estimate the value of dissipation factor, $\alpha$, using some independent measurements. Instead, a standard calibration procedure is applied. By definition, the maximum relative value of $L_j$ ($L_{29}$) is equal to 1.0 at the start point of the studied period, $t_0$. The value of $\Lambda_{min}(t_0)$ is also assumed to be 1.0. Thus, one can vary $\alpha$ in order to match predicted and observed PIDs, and the best-fit value of $\alpha$ is used for further predictions. Figure 1.3.4 presents some examples of income evolution for various effective values of $\alpha$ in the range from 0.09 to 0.04. This range is approximately the same as obtained in the modelling the PID for the time period from 1960 to 2002 (Kitov, 2005a). Actual initial value of $\alpha$ is found to be 0.086 for $t_0$=1960. The value of $\Lambda_{min}$ changes during this period from 1.0 to 1.49 according to the square root of real GDP per capita growth. The cumulative growth of real GDP per capita from 1960 to 2002 is 2.22 times.

Because the exponential term in (1.3.2) includes the size of earning means growing as the root square of real GDP per capita, longer and longer time is necessary for a person with the maximum relative values $S_{29}$ and $L_{29}$ to reach the maximum income rate. There is a critical level of income rate, however, which separates two income zones with different properties. This level is called the Pareto threshold of income. Figure 1.3.5 illustrates the increase in time necessary to reach the Pareto threshold depending on the decrease in dissipation factor $\alpha$. For $\alpha=0.1$, the time is about 6 years, and for the current value of 0.057 – 10 years. Effective dissipation will decrease in future according to GDP per capita growth. Overall, this process results in fewer and fewer young people to be able to reach the Pareto distribution, i.e. to become rich.

Below the Pareto threshold, in the sub-critical income zone, observed PIDs are accurately predicted by the microeconomic model for the evolution of individual income. One can crudely approximate the PID in this zone by an exponent with a small negative index. Above the Pareto threshold, in the supercritical income zone, the PIDs are governed by a power (equivalent to the Pareto) law. The presence of the high-income zone with the Pareto distribution allows any person reaching the threshold to obtain any income in the distribution, with rapidly decreasing probability, however.

The mechanisms driving the power law distribution and defining the threshold are not well understood not only in economics but also in physics as well for similar transitions. The absence of any explicit description of the driving mechanisms does not prohibit the usage of well established empirical properties of the Pareto income distribution as measured in the USA – constancy of the exponential index through time and the evolution of the threshold in sync with the cumulative value of real GDP per capita (Kitov, 2005a, 2005c). Therefore we include the Pareto distribution with empirically determined parameters in our model for the description of the PID above actual Pareto threshold. The usage of a power law distribution of incomes implies that we do not need to follow each and every individual income above the Pareto threshold as we did in the sub-critical income zone. All we need to know the number of people in the Pareto zone, i.e. the number of people with incomes above the Pareto threshold, as defined by relationships (1.3.3) and (1.3.4).

The initial dimensionless Pareto threshold is found to be $M_P(t_0)=0.43$ and it evolves in time as real GDP per capita:

$$M_P(t) = M_P(t_0)[GDP(t)/GDP(t_0)] \qquad (1.3.6)$$

When a personal income reaches the Pareto threshold, it undergoes a transformation and obtains a new quality to reach any income with a probability described by the power law distribution.

This approach is similar to that extensively used in natural sciences involving self-organized criticality. Due to the exponential (with a small negative index) character of the growth of income rate the number of people with incomes distributed according to the Pareto law is very sensitive to the threshold value, but people with high enough $S_i$ and $L_j$ can eventually reach the threshold and obtain an opportunity to become rich, i.e. to occupy a position at the high-income zone, as shown in Figure 1.3.6. It is illustrative that nobody with both $S$ and $L$ below 20 can reach the threshold. As measured, only 10 per cent of total population is eventually able to reach the threshold, however. This portion exactly defines the dimensionless threshold in the model.

There is another feature of the observed PIDs, which has to be addressed in the model. Actual income distributions span the range from $0 to several hundred million dollars, and the theoretical distribution extends only from $0 to about $100,000. The power law distribution starts from the Pareto threshold somewhere between $40,000 and $60,000. Above the threshold, the theoretical and measured distributions should diverge. What is the exact threshold? Figure 1.3.7 presents the predicted and observed dependence on income of the cumulative (normalized) number of people with incomes below given level in 1999. The curves start at the point (0,0) and practically coincide up to $54K since our model accurately describes the low-income branch of the PID. This value is the determined absolute value of the Pareto threshold for 1999, which corresponds to the dimensionless Pareto threshold value of 0.951 in 1999 and 0.430 at the start point of the modelling in 1960.

Above the Pareto threshold, the predicted distribution drops with an increasing rate to zero at about $100,000. This limit corresponds to the absence of the theoretical capacity to earn money, $S_iL_j$, above 1.0. The dimensionless units can be converted into actual 2000 dollars by multiplying factor of $120,000, i.e. one dimensionless unit costs $120,000. Actual and theoretical absolute income intervals are different above the Pareto threshold but contain the same portion of total population (~10%). Thus, the total amount of money earned by people in the Pareto distribution income zone, i.e. the sum of all personal incomes, differs in the real and theoretical cases.

Here one can introduce a concept distinguishing below-threshold (subcritical) and above-threshold (supercritical) behavior of the income earners. Using analogues from statistical physical, Yakovenko (2003) associates the subcritical interval for personal incomes with the Boltzmann-Gibbs law and the extra income in the Pareto zone with the Bose condensate. In the framework of geomechanics, as adapted in the modelling of personal income distribution (Kitov, 2005a), one can distinguish between two regimes of tectonic energy release (Rodionov et al., 1982) – slow subcritical dissipation on inhomogeneities of various sizes and fast energy release in earthquakes. The latter process is more efficient in terms of tectonic energy dissipation. The frequency distribution of earthquake sizes also obeys the Pareto power law.

If to sum all personal incomes in the Pareto zone, then the net actual income is 1.33 times larger than that would be earned if incomes were distributed according to the theoretical curve, in which every income is proportional to the capacity to earn money. This affectively means that in average every person in the Pareto zone earns 1.33 times more money than prescribed by the model. Figure 1.3.8 illustrates the concept. Two curves in the Figure correspond to the theoretical and observed total income received by people with incomes below a given value, i.e. the sum of all personal incomes from a given value to zero income. The theoretical curve is not corrected for a 33% increase for each personal income above the Pareto threshold.

This multiplication factor is sensitive to the definition of the Pareto threshold. In order to match the theoretical and observed total amount of the money earned in the supercritical zone one has to multiply every theoretical personal income in the zone by a factor of 1.33. This equalizes the theoretical and observed number of people and incomes in both zones: sub- and supercritical. It seems also reasonable to assume that the observed difference in distributions in the zones is reflected by some basic difference in the capability to earn money.

The model is finalized. An individual income grows in time according to relationship (1.3.3') until some critical age $T_{cr}(t)$. Above $T_{cr}$, the income rate is exponentially decreasing according to (1.3.4). When the income is above the Pareto threshold it gains 33% of its theoretical value in order to fit the overall income above the Pareto threshold. Above the Pareto threshold, incomes are distributed according to a power law with an index to be determined empirically. It is obvious that if a personal income has not reached the Pareto threshold before $T_{cr}$, it never reaches the threshold because it starts to decrease exponentially. A personal income above the Pareto threshold at critical work experience $T_{cr}$ starts to decrease and can reach the Pareto threshold at some point. Then it loses its extra 33% value.

All people above 14 years of age are divided into 841 groups according to their capacity to earn money. Any new generation has the same distribution of $L_j$ and $S_i$ as previous ones, but different start values of $\Lambda_{min}$ and $\Sigma_{min}$, which evolve with real GDP per capita. Thus, actual shape of PIDs depends on the single year of age population distribution. The population age structure is an external parameter evolving according to its own rules. The critical work experience, $T_{cr}(t)$ also grows proportionally to the square root of per capita real GDP. Based on independent measurements of population age distribution and GDP one can model the evolution of the PID below and above the Pareto threshold.

Because of the reverted logic of presentation, results are presented later than the model itself. In the following paragraphs various data sets related to personal income distribution are used for calibration of the model, i.e. for obtaining accurate estimates of defining parameters, and for validation as well. An adequate model has to predict the evolution of observed PIDs in the past, i.e. before the start year of modelling, and also in the future.

## §1.4. Modelling the overall personal income distribution in the US between 1994 and 2002

The essence of a quantitative model consists in the description of a set of variables and the prediction of their behavior beyond the period covered by data. The higher is the accuracy of description and prediction, the more reliable is the model. In order to prove the adequacy, our model has to describe several important aspects of income distribution. Among them are: the evolution of PID over calendar time, the dependence of individual income on work experience, the time history of the number of people in various income bins. The data counted in the CPS are aggregated in various ways. This paragraph is devoted to quantitative modelling of the overall, i.e. aggregated over age in given bins, personal income distribution in the US from 1994 to 2002.

We have formally introduced the model for the evolution of personal income distribution depending on economic growth in §1.3. PID is one of the key economic parameters. Despite some principal uncertainties in the data set on personal incomes, it represents the longest and the most detailed and accurate source of information on the distribution of income (individual, family, household) for a quantitative analysis and modelling. (As discussed in §1.2, other source provide estimates, which either cover a shorter period or less resolved, like the BEA gross personal income estimates.) Original income distributions, i.e. the number of people in given income bins, for even years between 1994 and 2002 are displayed in Figure 1.4.1. The width of corresponding bins is fixed to 2500 current dollars, i.e. it is not corrected for inflation. For the sake of clarity, the numbers of people with income inside original $2500-wide income intervals are aggregated into $10,000-wide intervals. The distributions show an increasing number of people in the fixed bins with income above ~$20,000 and a decreasing number below this value. (Notice the lin-log coordinates.) This is an expected result of population growth, real economic growth, and inflation. The first of the three processes potentially leads to an upward displacement with time of the curves as a whole. The displacement is uniform (in relative terms) when the population added every year is distributed over income in the same way as before, i.e. when the PID of the added population mimics the original PID. The US population grows at a rate of approximately 1% per year due to the excess of births over deaths and positive immigration.

The latter two processes result in the change of the shape of the distribution. Inflation (as represented by GDP deflator) in the US between 1994 and 2002 was measured between 1.2% in 1998 and 2.4% in 2001. The effect of inflation consists in higher nominal incomes potentially obtained without real economic growth. Because population was counted in fixed income bins during the entire period, one can expect that some people from lower income bins were eventually lifted into higher income bins. Moreover, an increasing portion of people were moving into the zone above the highest income of the survey - $100,000. These people find themselves outside the detailed counting scheme and one cannot calculate population density, i.e. the number of people in a given bin divided by its width, because the upper end of the income scale is open. With time, this effect became so prominent that it forced the Census Bureau to introduce new income intervals for higher incomes after 2000. Because of much lower population density at the highest incomes, these intervals are $50,000-wide, i.e. twenty times wider than the standard bins. Here we observed another deficiency in the CB's reporting methodology – it does not retain the resolution of PID uniform through time.

With respect to the evolution of PID in the United States, real economic growth leads to an effect similar to that caused by inflation. The increment in the volume of goods and services produced by the US economy results in an increase in the gross personal income, GPI, which according to our definition must be equal to GDP (GDP=GPI). People earn more and drift with time in the direction of higher incomes. In some rare years of economic contraction, personal incomes drop and some people may fall back into lower income bins.

An aggregate effect of these three processes divide the distributions into two zones – lower and above mean income, as seen in Figure 1.4.1. The mean income increases from $23,278 in 1994 to $32,222 in 2002 (in current dollars). So, the turning point between these two zones is somewhere between these values.

The next step of our analysis is to normalize the original PIDs to several aggregate values. A natural normalization is associated with total population. Such representation suppresses the effect of population change and reduces the original PIDs to population density distributions, (PDD). These PDDs can be considered as probability density functions (pdf) because the integral of a PDD over income, or the area below the PDD curve, is equal to unit. As shown later on, these density distributions better characterize the hierarchy of personal incomes in the US. Figure 1.4.2 illustrates the evolution of the PDD during the period between 1994 and 2002.

As often happens in scientific research associated with empirical data, a trail-and-error method enhanced by some new conceptual assumptions provides a fruitful approach. However, the underlying logic of this new approach is opposite to the logic of the representation of obtained results. One cannot know the final result before the completion of the search, but knows exactly the final result when reports. In fact, only the final result really matters. But it became a standard to present all principal stages of corresponding search procedure and most important outcomes.

In order to find an invariant in the US PIDs, we have conducted a series of corrections to the original PIDs and related PDDs. The information on the growth rate of real and nominal GDP and the changes in total population in the US has been used to reduce the distributions to those in 1994. A well-known procedure of such a reduction is the adjustment for inflation. Since the March Supplement of the CPS gives the number of people with incomes in fixed $2500-wide bins one has first to correct the enumerated distributions for the change in dollar value. This correction can be implemented as the contraction of income scale by a factor, which entirely compensates the extension caused by inflation. For example, in order to correct for 10 per cent inflation, one has to compress the income scale by a factor 1.1. Thus, the bin between $50,000 and $52,500 is transformed into the bin between $45,455 and $47,727, with its center shifted from $51,125 to $46,591. Both bins are effectively equivalent in terms that $50,000 income in a given year has a value equivalent to $45,455 in the previous year, when the rate of price inflation is 10% per year. Hereafter we associate measured values of population density with the centers of relevant income bins. For nonlinear functions approximating actual PDDs, this is not a quantitatively accurate procedure. Estimated population densities should be associated with the incomes, which provide the exact values of the approximated PDDs. This procedure will be described in §1.8, where the Gini coefficient is modelled. However, for the purpose of illustration and comparison the difference between the centers and exact incomes is negligible.

The original PID for 1994 is the starting and reference distribution. The distributions for the following years are corrected for inflation, as represented by GDP deflator. Obviously, the corrected curves should reveal the change over time in real PIDs. Figure 1.4.3 displays some results of the correction for inflation between 1994 and 2002 as applied to the population density distributions. These PDDs demonstrate an increase in the portion of people with higher real incomes. This observation is consistent with the positive growth in real GDP during the studied period. Due to weak real GDP growth in 2001, the curve for 2000 is very close to that for 2002. One can observe the contraction of income scale resulted from the corrections described above - the centers of the bins corrected for inflation are drifting to the center of coordinates with time.

The same correction procedure has been applied to the changes associated with nominal and real economic growth, both total and per capita. When there is no total (working age) population change and inflation, the correction for real economic growth could potentially reveal the changes in the distribution of gross personal income in the same (in number but not in individual representation) population. Figures 1.4.4 and 1.4.5 compare the effects of the correction for the growth in nominal GDP on the original personal income distribution, and the correction for the growth in nominal GDP per capita on corresponding population density distribution. Both corrections address the question of the redistribution of the increasing income generated by inflation and real economic growth over the growing population. Is the increment in nominal and real volume of money distributed evenly (in relative terms) among income groups or some selected groups benefit from the redistribution? This is a political question as well, but first it must be answered using quantitative estimates provided by the Census Bureau.

The curves representing PIDs in Figure 1.4.4 are practically parallel. This indicates that the increase in working age population forces only an upward movement of the 1994 curve. Because these curves are parallel, the relative increase in the number of people in every income bin is the same, and the distribution of the income increment among the newcomers is exactly the same as among experienced people. In other words, the PID in the US is characterized by the existence of a hierarchy, and this hierarchy is rigid over time and generations. This is a fundamental conclusion. It is better illustrated in Figure 1.4.5, where the population density curves for the studied years practically coincide. It means that equivalent portions of total (working age) population always receive equal portions of gross personal income. This fixed income distribution implies the constancy of income inequality between 1994 and 2002. In §1.8, we will extend this conclusion to the period between 1947 and 2006.

It is important that the hierarchy is observed in both branches of actual (and modelled) PIDs: sub- and supercritical one, as characterized by quasi-exponential and power law distributions. Figure 1.4.6 depicts all PDDs between 1994 and 2002 (corrected for the growth in nominal GDP per capita) and introduces two nonlinear trend lines. This graph also illustrates the presence of a rigid hierarchy.

Using the set of defining equations developed in §1.3, we start the modelling of the overall PIDs with some simple examples. Our initial model is constrained to reproduce all aspects of actual observations at any level of detalization, but only aggregated values are important at this stage. The model is characterized by a number of external and internal parameters. The external parameters include the growth rate of GDP: both real and nominal, total and per capita, and the distribution of population over (single year of) age. The internal defining parameters of the model are the initial critical work experience, $T_{cr}(t_0)$, and the initial dissipation factor, α.34The34former parameter can34be34estimated using some independent34 observations and the latter one - only by calibration, i.e. by trial-and-error.

Various estimates of the growth rate of GDP between 1950 and 2002 are presented in Table 1.4.1. Total increase of the nominal and real GDP during this period is 35.7 and 5.7, respectively. Relevant GDP per capita, corrected for the difference between total population and that above 15 years of age, changed by a factor of 17.5 and 2.5, respectively. The latter value indicates that the real GDP per head for working age population changed only by a factor of 2.5 during the fifty two years after 1950. It is 2.28 times less than the total change in the real GDP. Actual economic development is not as fast as seems from some economic news.

Thus, the observed growth of real GDP is half due to the population growth or extensive growth. The proposed model includes the distribution of working age population over age as a key parameter defining fine and aggregated characteristics of personal income distribution. Figure 1.4.7 depicts the single-year-of-age population estimates for some selected calendar years as obtained from the US Census Bureau (2004b). Total population in the United States grew eventually from 152,200,000 in 1950 to 286,200,000 in 2001. Total population of 16 years of age and over grew during the same period from 111,300,000 to 225,670,000.

As found above, the increase in total population results only in a parallel shift of the PIDs for the years between 1994 and 2002. What important is the change in the portion of population with given age through time. Figure 1.4.8 displays the same population distributions as in Figure 1.4.7, but normalized to the largest population for all ages. The age of the peak number evolves over time and was 45 years in 2001. The population peaked at this age gives the largest share of the total population with almost the highest attained average income. In other words, the current age distribution in the US corresponds to a very effective case for income earning – a larger portion of working age population receives almost the maximum possible income, as defined by our model. Effectively, people between 45 and 55 years of age are closing the critical age when the increase in mean income turns to the exponential fall (see Figure 1.3.1).

When the age peak surpasses the (predefined by age structure and the initial level of GDP per capita, as shown in Chapter 2) critical working experience, $T_{cr}(t)$, which is also growing with time, these favourable conditions for income earning will start to deteriorate, but not severely. There will be only a 10% drop in the share of population with the peak income in the next 15 years. Five years after 2002, however, are marvelous for an increase in gross personal income extra to that associated with the age distributions in the past. Figures 1.4.7 and 1.4.8 demonstrate the importance of the changes in the age distribution for our model. As a thought experiment one could imagine that all population has the same year of birth and track the evolution of mean income over time. In the beginning, when all people are 16 years of age and start working career, the mean income is zero. At $T_{cr}$, it finds its peak, and then exponentially decays with time. The age distribution smashes this pure evolutionary pattern by mixing a large number of similar curves with various time shifts.

It is worth noting in this paragraph that there are visible variations in the population counts within 10-year age windows, which are used by the US Census Bureau for averaging personal income readings. This can cause substantial variations in the average income estimates, especially in the youngest age group, where the personal income increases exponentially with age. Fortunately, the estimates of single year of age population are available from 1900, with a varying accuracy. We estimate the accuracy of the population counts for single years of age as 5% to 7%. In wider age intervals, this accuracy is higher and may reach 1% to 2%.

In the model, the population of each single year of age for each calendar year (in the studied period) is divided by the number of different income states. Resulting values represent the portions of total population with the same history of income evolution, because only 841 different combinations of the product of capability to earn money, S, and means to earn money, L, are available. One should bear in mind that combinations ($S=2$, $L=30$) and ($S=30$, $L=2$) are quite different due to the fact that the means size L defines the time constant of dissipation, $\alpha/L$. People in a given group have the same income equal to the product of the current level of GDP per capita (relative to the start year) and time dependent functions in (1.3.3') and (1.3.4). For example, the age group of 50+ year-olds has work experience of ($t-t_0=$) 35+ years. Current value of L is the size of real earning means in the group relative to the initial value. Actually, there are as many age groups as listed in the tables published by the US Census Bureau. Because of the uncertainty associated with the enumeration in elderly group the model includes only ages from 16 to 75 years. Admittedly, the number of elderly is not large and is prone to strong fluctuations due to the changes in the population controls. This limitation can potentially affect the modelling of the overall personal income distribution: some people of working age and their incomes counted by the Census Bureau are not included in the model.

The internal parameters $T_{cr}(t)$ and $\alpha(t)$ depend on time. To estimate their values by trial-and-error method, a series of calculations for various time intervals between 1950 and 2002 has been conducted to fit the measured PIDs. For 1950 as a starting point, the best fit estimates are $T_{cr}=23.5$ years, $\alpha = 0.097$, respectively. According to (1.3.5), these values can be reduced to any other years using the level of GDP per capita (see table 1.4.1).

Having the externally measured age distribution and GDP per capita as well as the above of the internal model parameters, one can predict the overall PID for any year. For the purpose of this paragraph, we chose the period between 1994 and 2002. Technically, the model is very simple and consists of three main steps. First, we calculate personal incomes for all 841 income states (trajectories) and for all ages under modelling, as defined by (1.3.3') and (1.3.4). As a result, the number of different incomes in the model for a given year reaches 841 times the age range (chiefly, 60 years between 16 and 75). Second, each income estimated in step one is multiplied by the number of people with this income in actual economy, i.e. the number of people of relevant age divided by 841. Third, we aggregate over income or age intervals, as appropriate. In the case of the overall PID, we aggregate over all ages in 0.0025 model units bins. Since the model is dimensionless, we introduce an empirically determined calibration factor, which transforms between current and dimensionless spaces. In other words, total personal income in reality, i.e. the integral of PID over income, must be equal to total personal income in the dimensionless space multiplied by calibration factor.

Figure 1.4.9 presents predicted overall PIDs in current dollars for some selected years between 1960 and 2002. (Notice the log-log coordinates.) The evolution of the predicted PIDs reproduces some principal features of the observed distributions: the level of population density decreases at lower incomes and increases at higher income. The point where these two processes meet is somewhere between $20K and $30K. The curves are obtained from relevant dimensionless curves scaled with a factor of $70,000 in 1990. This means that one unit of the dimensionless scale costs $70,000 (current) for the year of 1990. The scaling factor is proportional to nominal GDP per capita. In 2002, the scaling factor was about $105,000 and the predicted personal income distribution in current (2002) dollars extends to $103,000. This value is the largest income which would have been predicted by the model if we do not recall that actual distribution at higher incomes is governed by the Pareto law. The purpose of this paragraph is to predict PID between zero income and the Pareto threshold. As mentioned in §1.3, the model defined by (1.3.3) and (1.3.4) describes the evolution of individual incomes for 90% of working age population. PID beyond the Pareto threshold is described and modelled in §1.7.

Figures 1.4.10 through 1.4.12 compare the observed and predicted distributions for 1994, 1998, and 2001. The observed distributions, aggregated in $10K intervals, are interpolated by splines in order to present continuous distributions. In reality, fine structure of the observed PIDs may differ from these smooth lines. But we consider this way of representation as an adequate one because the uncertainty in actual PIDs is large enough and the smoothed curve is a good visual characteristic of the PID suppressing measurement errors and related fluctuations. So to say, we expect that the shape of real PID would be very close to that of predicted one if the former is measured precisely. The predicted distributions are obtained from a model with the following parameters: start37year $t_0$=1960,37 $T_{cr}$(1960)=26.5 years, $α$=0.087. These calculations were carried out in dimensionless units and current dollars. Therefore, the initial value of the Pareto threshold of 0.43 evolves in time as nominal GDP per capita. The dimensionless width of counting bins is 0.001 and effective current dollar bins expand over time. The conversion factor between dimensionless units of the model and current dollars in 1960 is $10,500. This means that the entire 1960 theoretical PID can be placed between $0 to $10,500; the whole 1980 PID spans from $0 to $40,000 because nominal GDP per capita grew by a factor of 3.8 from 1960 to 1980. There are no people in several highest dimensionless income groups, however, because nobody approached the level 1.0 (dimensionless units) during the period of income growth, i.e. before $T_{cr}$.

The PIDs for 1994 are presented in Figure 1.4.10. The predicted distribution coincides with the observed one in the range between $5K and $35K. The latter value is very close to the Pareto threshold for this year situated somewhere between $35K and $45K. Beginning from the Pareto threshold, the observed and predicted distributions diverge because of different character of decay, as described in §1.3. The observed distribution decays with income according to a power law. The predicted distribution decays slowly just above the threshold, but then the rate of the decay grows very fast and it intercepts the observed distribution near $60K.

Figures 1.4.11 and 1.4.12 illustrate the evolution of the distributions and corresponding Pareto threshold. With time, the threshold moves towards higher incomes. It was around $45K in 1998, and near $52K in 2001. Because the Pareto threshold quickly moves to $100,000, as a result of the observed intensive nominal economic growth in the US, one can expect that in the near future the overall PIDs measured by the CB will not contain the Pareto portion of the distribution in the range from $0 to $100,000. Therefore, the US Census Bureau will not be able to correctly describe personal income distribution. Even now, the distribution contains only a narrow range where the power law rules. As a response, the Pareto zone has been covered with $50,000-wide bins since 2000.

The above analysis of the overall PID demonstrates the existence of some fixed hierarchical structure in the personal income distribution in the United States. The PIDs normalized to the total population above 15 years of age and corrected for nominal GDP per capita effectively coincide for the years between 1994 and 2002.

The structure of the measured personal income distribution can be simulated by using the microeconomic model with some simple assumptions related to the distribution of capabilities to earn money and sizes of earning means. In the lower income zone, the observed and predicted distributions coincide up to the income level interpreted as the Pareto distribution threshold or the minimum possible income in the Pareto distribution. The Pareto part of the actual PIDs is considered to be a result of some processes associated with self-organized criticality and does not need any additional modelling except the prediction of the portion of the total population in the Pareto zone. This portion is of about 10 per cent and its distribution over work experience is also exactly predicted by the microeconomic model, as described §1.7.

The evolution of the overall PID is also well predicted depending on nominal GDP growth from 1994 to 2002. This includes the prediction of the subcritical zone width and relevant change of the PID slope with time. One can easily predict the future PIDs as a function of GDP growth and population changes.

On the other hand, the observed accurate prediction of the US PIDs for years between 1994 and 2002 demonstrates validity of the microeconomic model and general concept inherently related to the personal income as the only source of the economic growth.

### §1.5. Modelling the age-dependent personal income distribution in the US between 1994 and 2002

In §1.4, we revealed the importance of several key parameters defining the model and estimated their empirical values. These parameters provide a deterministic description of the overall PID and its evolution between 1994 and 2002. At this high level of aggregation, the model provides accurate predictions for incomes below the Pareto threshold. At a lower level of aggregation, as represented by age-dependent PIDs, one observes some new effects representing a big challenge to the model. Therefore, in order to answer this challenge, the age-dependent personal income distribution in the United States was modelled using the same (microeconomic) model and the parameters obtained in the overall PID modelling. This is an independent validation of the model: it was not tuned to describe new PID features when the overall PIDs were fitted. In any case, a model with empirically determined parameters accurately (and quantitatively) predicting *new* effects beyond its initial scope is of high scientific value.

In §1.3, the concept of two branches of PID in the USA is introduced: a low-income branch and a high-income branch. The former is accurately described by the model and the latter is a standard power law (Pareto) distribution. In natural sciences, the Pareto distribution is a common observation and is thought to be a result of a number of processes called as a whole "self-organized criticality". There is no economic model available to formally express some processes at micro level leading to the Pareto distribution, as we did in §1.4 for the quasi-exponential distribution at lower incomes. However, one can accurately predict the number of people in this distribution using the developed microeconomic model and their distribution over age. In a sense, the number of people and their distribution over age is the only feature needed to model measured Pareto distribution, because other parameters of the distribution follow up from the definition and properties of power law. Therefore, the age dependence of personal income plays the defining role for the empirics of the Pareto law in economics.

The overall PIDs presented in §1.4 include personal incomes of all Americans above 15 years of age as published by the U.S. Census Bureau. By design, since 1994 the Census Bureau has been reporting population counts in relatively narrow income bins of $2500, also in five to ten-year-wide age groups. It is obvious that each and every personal income undergoes important changes with age or, what is equivalent for the fixed age of workforce joining, with work experience. The starting point for all personal incomes is apparently zero at the age of 15 years. Then personal incomes grow at a decreasing rate to some peak values defined by given *S* and *L*. Between 1994 and 2002, the average income measured in the USA usually reached its relative maximum value at some age between 45 and 55 years and then started to decrease exponentially.

The overall PIDs in §1.4 are not able to express all these complex features and processes due to the absence of age resolution. These processes and features, however, are extremely important from personal point of view as a prediction of potential income trajectory. They are also a big challenge to any model for the personal income distribution and its evolution.

The overall PIDs in the USA demonstrate a very stable social structure in respect to the distribution of gross personal income. However, the observed PIDs change very fast from one age group to another. There are income data sets with high age resolution for the years between 1994 and 2002. These sets include information on the number of people in 5-year-wide age intervals starting with 15 years of age. The first bin is 10-year long, however, and spans the age interval between 15 and 24 years. This widening reflects severe problems in income measurements in this age group, where standard definitions of income do not cover all potential income sources as, for example, intra-family money transfer (see discussion in §1.2 for details).

Figures 1.5.1 and 1.5.2 display the PIDs measured in 1998 in various age groups in absolute values and normalized to total population (in given age group) of 15 years of age and over, respectively. In the youngest age group, the distribution is close to an exponential one, as expressed by a straight line in the lin-log coordinates, with strong variations observed at incomes above $50K. As discussed in §1.2, this is the result of the small size of the survey, which covers only approximately ~100,000 households. When the level of PID in this relatively narrow age group drops by three orders of magnitude, such coverage can not provide adequate estimates. Some bins at higher incomes are not populated at all! In the Figures, the absence of population is manifested by gaps in the curves and zero values in corresponding tables published by the CB. When using the CPS data for quantitative modelling, one should bear in mind that the high-income tails of age-dependent PIDs are not reliable (not populated). In natural sciences, there are many cases when size distributions expressed by power law are characterized by high-amplitude fluctuations at the upper size extreme. These fluctuations are usually artificial as well as those observed in Figures 1.5.1 and 1.5.2 in the youngest group. The gaps induced by low resolution are also partly responsible for the underestimation of average income in the youngest age group.

With increasing age or work experience, the age-dependent PIDs obtain a slowly extending quasi-flat part at lower incomes followed by an exponentially decreasing part. The PID for the age group between 70 and 74 years as a whole is characterized by an exponential roll-off similar to that for the youngest group. This roll-off clearly demonstrates that a significant part of elderly population in the USA loses income very fast with time and not many of them can retain the same income as they had before: compare the eldest group with the group between 60 and 64 years of age. Figure 1.5.2 illustrates this process in relative terms. The curve corresponding to the eldest reported age group lies below the curve for the age group between 25 and 29 years for incomes higher than $30K and well above the curve for incomes below $20K. The middle age groups are characterized by almost identical population density distributions. The observed PIDs for the year of 1998 reveal some complex features of the evolution with age. The PIDs for other years of the studied interval are similar and demonstrate the same principal features.

PID in a given age group also evolves over calendar time. The features associated with this evolution are also of principal importance for the model - is it capable to predict well PID changes in all age groups. Figure 1.5.3 displays two PIDs (current dollars) in the youngest age group aggregated in $10K bins for the calendar years of 1994 and 2002. An exponential regression gives a negative index, increasing from -0.125 in 1994 to -0.095 in 2002. The ratio of the indices is 1.32. This value is very close to that observed for the growth in nominal GDP per capita from 1994 to 2002, equal to 1.34 (Bureau of Economic Analysis 2005). Thus, an adjustment for the change in nominal GDP per capita, like that applied to the overall distributions, should completely merge the curves.

From the above Figures, we would like to highlight three important features of the original PIDs' dependence on work experience. First, income distribution in the youngest age group is exponential over the whole reported range. Second, the distributions normalized to population are characterized by the development of a quasi-constant (slightly decreasing) part, which spans the range from zero income to approximately $30K in the age groups above 25 years. Third important phenomenon is the exponential decrease in the distribution for the eldest age group. This distribution is similar to that in the youngest age group. One can assume that in some older age group, say above 75 years of age, the PID is equivalent to that in the youngest group.

The adjustment for nominal GDP per capita effectively reduces the overall population density distributions to one line. It is instructive to apply the same procedure to the PIDs in various age groups. Because the adjustment might produce different outcome depending on age, some principal cases are presented in Figures 1.5.4 through 1.5.6. In the youngest age group, the adjustment results in the same pattern as observed for the overall distribution. The only difference is higher scattering at large incomes related to low reliability of measurements in this age interval. The largest differences between the reduced PDDs are observed in the age groups from 45 to 49 years and from 50 to 54 years (Figures 1.5.5 and 1.5.6, respectively). These ages are near the critical work experience, $T_{cr}$, where the empirical relationship between mean personal income and work experience turns from growth to fall. This critical age also changes with time as the square root of GDP per capita. So, one can expect the largest disturbance in these PDDs as induced by real economic growth and inflation. In 2002, the critical age was somewhere between 50 and 55 years in the USA.

Having studied some features of the age dependent PIDs, we calculated theoretical distributions in the age groups defined by the U.S. Census Bureau. Since the model predicts each and every individual income for people of 15 years of age and over, it is a simple aggregating procedure to estimate the number of people in any given age and/or income interval. As discussed in §1.4, the model takes into account only 841 distinct individual income histories for every single year of age and then weights individual incomes using total number of people of the same age. People of the same age are effectively divided into 841 equal sub-groups, and the size of these sub-groups changes with age and calendar time. For the sake of simplicity we assume that all people of the same age have the same birthday and first working day. This makes the model a discrete one.

Everybody can choose and follow up one of the 841 available trajectories. The person may also leap from one trajectory to another. In the case when all income positions in a given economy are filled, the change actually must be an exchange. In other words, the leap forces the person who occupied before this new position to drop to the trajectory abandoned by the first person: the trajectories can be only swapped but not created over the fixed number. Actually, much longer swap chains are possible. In any case, population density distribution in any given age group is always fixed despite any finite number of exchanges in income trajectories. The ratio of personal incomes of two persons of different age but the same $L$ and $S$ depends on economic growth during their job careers.

As in §1.4, the best fit defining parameters for 1960 are $T_{cr}$=26.5 years and $\alpha$=0.087, respectively. After aggregation of individual incomes in 10- and 5-year bins we obtained PDDs in all predefined age groups. Figures 1.5.7 through 1.5.10 depict the predicted and observed distributions in the age groups from 15 to 24 years, from 25 to 29 years, from 30 to 34 years, and from 60 to 64 years in 1998. The year of 1998 has an advantage of simple conversion factor between dimensionless units and current units (dollars) equal to 100,000. As has been already shown, the other middle age groups have distributions very similar to those for the age groups from 30 to 34 years and from 60 to 64 years.

In the youngest age group, the curves in Figure 1.5.7 diverge almost everywhere. The predicted distribution lies above the actual one. One can explain this observation using the factors discussed above: the low resolution in this group, the undercount at higher incomes and the absence of adequate income sources in the Census Bureau's questionnaire (West & Robinson, 1999). So, this deviation is expectable and should be resolved somehow in further income surveys. We presume that this deviation is induced by some deficiencies in the current design and methodology of the surveys not by the model. However, both distributions are characterized by the same index of exponential decay from the zero income bin to the bin starting at $40K. The observed distribution has already an emergent part characterized by the Pareto distribution (above $48K) and also has a large number of people with very low income. The latter observation is well predicted. Again, the accuracy of enumeration at lower incomes is under doubt due to the limitations in the questionnaire.

In the middle age groups, evolution of the predicted and observed distribution is described with a reasonable accuracy. Here we meet again the low income counting problem. In the eldest age group among all presented here (from 60 to 64 years of age), the theoretical curve fits the measured one with an excellent precision. Apparently, the CPS covers the elderly population with the highest resolution and defines adequate sources of income. This effect will be also discussed in §1.9, as related to age dependent Gini coefficient. In Figure 1.5.10, as with the overall PIDs, one can distinguish a sub-critical zone, a zone where the theoretical distribution is above the observed and a zone of an opposite behavior. In §1.3, this age group was used for calibration of the earning capabilities.

The observed PDDs suffer from lack of resolution in the most important for our model age group – from 15 to 24 years of age, where the dynamics of income evolution is the fastest and the range of income change in the largest. The range of change is a crucial parameter for the reliability of any model; for a zero-wide-range any functional dependence between variables is void. One can find a link between measured variables only when they are changing. The observed PDD in this age group actually consists of ten very different single year PDDs. Figure 1.5.11 displays the predicted evolution of the single year of age PDDs for 1998 starting from 3 years of work experience. This is the decomposition of an aggregated distribution into single year of age distributions. During the first two years of work the predicted incomes are concentrated in the lowermost income bins and are not worth displaying.

The predicted evolution reveals some important PID features. During the first eight years of work, nobody is able to reach the Pareto threshold of $48K. This might be the reason for the U.S. Census Bureau to aggregate all the personal incomes between 15 years and 24 years of age. Otherwise, there is literally nobody filling the higher income bins in the Pareto distribution range.

There are actually quite a few people starting their job career before reaching the age of 15 years. There is no official statistics for these people, however, and their overall impact is negligible in the PID evolution because they affect only the youngest age group with the lowermost incomes. With age, all these differences in the start year of work should disappear.

In addition to a higher resolution, the predicted PID can be extrapolated years back and ahead. Figure 1.5.12 and 1.5.13 display predicted PIDs in 5-year-wide work experience intervals for calendar years 1980 and 2002. One can observe that the PIDs span very different ranges of income despite the same virtual procedure of counting in $2500 wide intervals.

In 1980, the predicted distributions span the income interval from $0 to $35K. The Pareto threshold was at about $20K according to the increase in nominal GDP per capita by a factor of 2.9 from 1980 to 2002. Actual distribution, if measured, would be characterized by power law above $20K. The actual distribution is also extended to very high incomes well above the level of $35K predicted by the model in the sub-critical range.

In 2002, the Pareto threshold was at the level of $58K and the predicted distributions occupy the whole actual income range of the survey – from $0 to $100K. One of the effects of the observed PID stretching with time is that relatively lower numbers of people occur in the predefined income intervals. This effect results in the observed oscillations in the PIDs. The PIDs in 1980 look much smoother than in 2002 due to smoothing effects of the large numbers. Oscillations in the predicted PID are only induced by the discrete distribution of the capacity to earn money, which was revealed by the model's calibration procedure.

The overall personal income distribution in the USA was modelled for the period between 1994 and 2002 and an excellent match between the predicted and measured PIDs has been found. Defining parameters of the model are obtained and are accurate in prediction of the PIDs' evolution in time. The prediction of the evolution was almost solely based on the assumption that critical time, $T_{cr}$, and $\alpha$ evolve as the square root of the per capita real GDP.

This paragraph addresses a more complicated problem of age structure of the observed PIDs. Complexity of the problem is obvious due to the cardinal changes in the PID shape with age. The model, however, meets this new challenge and accurately predicts not only the overall behavior of the age-dependent PIDs, but also some fine details. Moreover, the model unveils some shortcomings of the income survey methodology and design, which lead to degradation of the observations' accuracy with time.

The data on the age-dependent PIDs are obtained from the U.S. Census Bureau web-site. Two important features of the data should be mentioned here. Definition of income adopted by the Bureau in its questionnaire is very limited and does not comply with broader definitions based on consumption or expenditures. This kind definitions better present the production side of the income definition adopted in the model. The main assumption of the model is that personal income is exactly equal to the total price of the produced goods and services. This assumption effectively balances net expenditures and incomes in the society. Difference between the two definitions under discussion is the largest in the lower income zone where pure money earnings sometimes are not the major source of income. Interfamily money transfer also can be of importance in this zone. The difference is clear when we compare the observed and predicted distributions in this zone. The majority is concentrated in the first income bin from $0 to $2500, where also people with total loss are placed. There are some doubts, however, that a person without any income can survive for a long period of time. If to consider the real expenditures of the person to stay alive as his/her income we effectively have our definition of the personal income. At higher incomes, the difference is obviously lower because principal sources of income have here some monetary form. It is worth noting also that the net income in the poorest group is less than that in the middle and higher income groups and plays only a marginal role in the total economic development.

The second feature of the data is related to the coverage of the age and income intervals. With the PIDs stretching with time over wider and wider income range, smaller and smaller number of people is counted in the designed income bins. In the youngest age group, the number of people in the predefined income bins changes by almost four orders of magnitude with income, i.e. if there are 10,000 people in the lowermost income interval, less than ten people are present in the highest measured income bin. Moreover, in some income bins there is nobody counted in. These measurements do not present true personal income distribution, but reveals only increasing problems in the current survey design. There are two ways to resolve the problem: not to publish the data characterized by very high uncertainty or to increase the population coverage (number of households) to make the data more accurate and representative.

The presented model is a good basis for the development of a new methodology for the income distribution measurements and definitions of economic equality/inequality. The model predicts the evolution of the PID with economic growth and reveals important future changes. For example, the overall economic growth results in longer time necessary to reach higher income (in relative terms). This makes young people relatively poorer - a trend which is observed in the USA. Further increase of $T_{cr}$ and decrease of $\alpha$ will accelerate this process in near future. Also, people reach the peak personal income later in age, but before the retirement age: the former approaching the latter with the overall economic growth.

## §1.6. Modelling the average and median income dependence on work experience in the USA from 1967 to 2002

This paragraph analyses and models a different slice of the overall data set available from the same US Census Bureau web-site (2004b) – the dependence of average income on work experience. The average income has been reported since 1967 as one of aggregated measures of income distribution. Due to some major changes in the income survey procedure applied before and after 1967 the accuracy of relevant measurements has likely been progressively improved, but were accompanied by a significant loss in data compatibility.

Unfortunately for the purpose of this paragraph, personal incomes have been averaged only in 10-year intervals since the beginning. (After 1994, 5-year intervals were introduced, which allowed better resolution of the critical work experience.) Corresponding electronic table contains mean personal incomes for male and female separately. Therefore, additional efforts are necessary for obtaining gender independent estimates. It is possible because the number of people with income is also listed for each gender and each age group. For the youngest age group, data are available only from 1974, and for the eldest age group, from 65 to 74, only from 1987. This difference is induced by major changes in the CPS procedures.

The electronic table presents both current dollars and chained 2001 dollars estimates. Figure 1.6.1 illustrates the dependence of average income on work experience as measured in current dollars. Here and below in this paragraph, we use cubic spline to interpolate the dependence between discrete readings, which are associated with the central points of work experience intervals, i.e. 5, 15, …, 55 years. As discussed before, the centers are not exact points related to mean incomes for exponential dependence, but we neglect the difference in illustrations. There are two striking features in the curves, better demonstrated in the *log-lin* coordinates: quasi-exponential growth and roll-off of the mean income with work experience, and the existence of some critical work experience, $T_{cr}$. This critical point divides the curve into two branches: a part increasing as a function like *(1-exp(-αt))*, and an exponentially decreasing branch beyond the critical work experience. Historically, this plot was behind the first intuition that the character of income distribution is equivalent to several well-known processes in geomechanics and physics. The authors have not found any major research addressing this crucial empirical finding neither in quantitative nor in qualitative terms. Our search can not pretend to be a complete one, but the effect is so clear and prominent that must be a part of each and every (micro-) economic theory related to personal income distribution. In our model, the dependence of mean income on age allows to empirically estimate two key parameters: the dissipation factor, α (and α1), and critical work experience, $T_{cr}$. The former defines the rate of growth before $T_{cr}$. The later may be directly measured from the curves in Figure 1.6.1. However, the width of averaging intervals (10 years) does not provide sufficient age resolution and we use the trial-and-error procedure to fit the curves as a whole for the estimation of $T_{cr}$.

Figure 1.6.2 shows the same average income dependence on work experience expressed in chained 2001 dollars. The only visible difference from the curves in Figure 1.6.1 is in the amplitude of the overall increase in level. The growth in the real mean income is much smaller than in the nominal one and differs between age groups: the mean real income in the age group from 25 to 34 years increased by a factor of 1.27 between 1967 and 2001, from 35 to 44 years - by 1.40, from 45 to 54 years - by 1.5, and from 55 to 64 - by 1.59. This is also a big challenge to any theory of income distribution to explain such a divergence in growth rates.

Another important effect to be mentioned and explained consists in a smaller real mean income change observed between 1967 and 1991 compared to that between 1991 and 2001, despite the fact that real GDP between 1967 and 1991 increased by a factor of 2, and between 1991 and 2001 only by 1.4. This discrepancy can be simply explained as associated with the evolution in the portion of people with income during these years as displayed in Figure 1.6.3. There was a strong increase from 0.8 in 1967 to the level around 0.95 in 1980 in all age groups except the youngest one. The latter group has an almost constant participation factor near 0.75. When corrected for the participation factor the mean income distributions look more consistent with the observed monotonic growth in real GDP as displayed in Figure 1.6.4.

There was a dramatic increase in the number of people with income between 1977 and 1979. Surprisingly for conventional economic theories, this increase did not cause a proportional growth in real GDP during these years. So, the same gross personal income was distributed over a larger number of people. The average incomes in all age groups changed proportionally to the growth in real GDP. In other words, the level of total income does not depend on any implicit or explicit mechanisms of income distribution.

To address the effect of population without income, Figure 1.6.4 presents a natural mean income dependence on work experience, estimated as total income in a given age group divided by total population in this age group, while Figures 1.6.1 and 1.6.2, undoubtedly, show biased dependencies. Apparently, the portion of people with income is arbitrarily defined by income survey procedures and fluctuates with time relative to true number according to the changes in the survey questionnaire. We consider as a reliable one the definition of mean income based on the concept of personal income presented in §1.3. In physical terms, our definition is based on the full size of a closed (economic) system, and thus, naturally obeys all conservation laws. As discussed in Introduction, no definition based on a varying portion of a closed system is valid since it is prone to uncontrolled fluctuations. A textbook example here is the gas laws – one would not be able to reveal any link between pressure, temperature and specific volume when measuring them in open atmosphere or using randomly changing gas volume.

Among all age groups, the largest correction for the people without income has to be applied to the youngest one. This group is also the most affected by the procedure of income estimate because a larger part of a young person's income comes from internal, and thus not covered by the CPS, sources. Therefore, one can actually expect a higher average income in this group when such sources are included. In reality, it is the income from these sources that permits the people "without income" to survive.

The intra-family sources might be very important for some age groups and individuals, but do not substantially change the overall distribution. The income obtained from external sources and estimated during the CPS is much larger than that redistributed inside family. Because the portion of population without income in smaller, the bias in mean income is larger in the youngest age group, and other age groups are characterized by more accurate income coverage. All in all, direct measurements of gross income and entire population represent a valid object for quantitative analysis and modelling.

Figure 1.6.5 displays dimensionless mean income as a function of work experience for some selected years between 1967 and 2002. All mean incomes are normalized to the largest values among all age groups in relevant calendar years. Therefore, the peak value of the normalized functions is always 1.0. The shape of the normalized curves evolves in time with a clear tendency of the critical age, $T_{cr}$, to grow. However, due to low age resolution it is difficult to estimate $T_{cr}$ accurately. Another feature should be mentioned. The curve for 1981 has an unexpected trough in the work experience interval between 30 and 40 years. We interpret this strong deviation as measurement error associated with the introduction of new questionnaire in 1980 and the new population controls after the 1980 decennial census. Such a prominent error in the estimation of mean income demonstrates the presence of problems in the CPS and puts the expected of uncertainty at a very high level.

Figure 1.6.6 presents the evolution of normalized mean income in each age group as a function of calendar year. Between 1967 and 2001, the peak mean income resides in two age groups: from 20 to 29 and from 30 to 39 years of work experience. The intercept occurred in the middle 1980s. Unexpectedly, the older group also had some years of superiority during the late 1960s and early 1970s. We interpret the intercept in 1974 as induced by substantial changes in the CPS. An important factor influencing the average income dependence on work experience consists in the evolution of the age-dependent portion of gross personal income estimate by the CB, in GDP estimated by the BEA, as discussed in §1.9.

From Figure 1.6.6, one can easily estimate where the peak mean income value was or will be in various age groups with time. Figures 1.6.7 through 1.6.10 present linear regressions of the normalized mean incomes. The obtained linear dependencies are extrapolated in the past or in the future before they intercept the unit line. In the youngest age group, the slope estimated by the linear regression is -0.004, as displayed in Figure 1.6.7. It gives the estimated intercept time around 1790. When we use the slope of -0.075, as obtained in other age groups, the intercept time moves to the beginning of the 20$^{th}$ century. Potentially, people between 15 and 24 years of age had a dominating income position in the 19$^{th}$ century. One should bear in mind that life expectancy a hundred years ago was very low compare to current one in the US.

In the age group from 25 to 34 years, the slop is -0.07, as shown in Figure 1.6.8. This group was at the top of personal income pyramid until the late 1940s. The group from 20 to 29 years had the peak mean income value until the middle 1980s. This was the only transition of peak mean income between adjacent age groups covered by the US Census Bureau. During the last 20 years, the peak mean income has had a tendency to move towards the group with 40 to 49 years of work experience. One can predict that this group will take the lead around 2015. It is not too far away and will be a good observation validating our concept – the growth in $T_{cr}$ is driven by the size of earning tools, $L$. Hopefully, one will be able to resolve the critical age with an appropriate accuracy by 2015.

As discussed in the Chapter 2, the value of critical work experience potentially defines the average rate of real economic development. During the last 60 years, the trend of real GDP growth or economic potential was exactly equal to the reciprocal value of $T_{cr}$. Effectively, if mean personal income grows during 50 years from zero to its peak value, one can suppose that average annual GDP growth is 1/50 or 2%. The current value of $T_{cr}$ in the USA is approximately 40 years. Thus, the current trend in real GDP growth is 2.3%. During the 1950s, when $T_{cr}$ was approximately 25 years, the trend was at the level of 4%.

Because the averaging intervals are relatively wide (10 years), it is difficult to determine the exact value of critical work experience value from a single distribution. The full set of curves, however, allows revealing some changes in the critical work experience value. Figure 1.6.11 demonstrates the evolution of $T_{cr}$ as a function of real GDP per capita according to relationship (1.3.5) for the years between 1950 and 2002. The predicted $T_{cr}$ was about 25 years in the late 1950s, reached the 30 years in the late 1970s, and is currently near the 40-year threshold. One has to take into account the difference between the theoretical $T_{cr}$ predicted for every single year of age, and the empirical $T_{cr}$ obtained from 10-year wide intervals. The latter has a several year lag relative to the former.

There is another defining parameter one should estimate from the average personal income – $\alpha_1$, which varies over time according to (1.3.6). The width of averaging window also causes substantial variations in the estimated values of the indices of exponential decay above critical work experience. Therefore, it would be difficult to exactly match all the observations of mean income in one set of model parameters. Figure 1.6.12 illustrates the difference in exponential decay obtained from 10-year and 5-year averaging intervals for the year of 2001. The latter intervals demonstrate much faster decay than the former ones. In order to fit the observed exponential decay beyond $T_{cr}$ for the period between 1967 and 2001 we fixed relative (i.e. normalized to the peak value) income at the age $A_r$=60 years to $M_r$=0.84. These values are slightly different from those in §1.3, but are inside the uncertainty associated with the fall of individual incomes beyond $T_{cr}$. Both sets provide very close trajectories of exponential fall and thus very close inputs to the mean income.

Total input of incomes above the Pareto threshold is completely defined by the factor 1.33, as discussed in §1.3. Technically, we calculate individual incomes for a given population, select those above the Pareto threshold, sum them up and multiply by 1.33. The result is the total income of all persons in the Pareto branch of PID. For the estimation of average income, the number of people and their total income is all we need.

Now we can predict the evolution of age dependent average income for the years between 1967 and 2001. Figure 1.6.13 displays some results of the mean income modelling. The observed curves are represented by mean incomes corrected for population without income, with the largest correction in the youngest group. Because the model calculates mean incomes in internal dimensionless units, we had to estimate the scaling factor to fit actual measurements of mean income. This factor has to be constant over years if the definition of personal income does not vary over time. Any change in the definition potentially results in relevant change in the portion of gross personal income (CB definition) in GDP (BEA definition) as well as in the redistribution of the GPI over age groups. In Figure 1.6.13 the scaling factor is 72. This factor has no physical sense and plays no role in the modelling itself. It is only used to compare predicted and observed mean income.

In 1967 and 2001, the observed and predicted curves practically coincide almost over the whole range of work experience between 0 and 60 years, except in the youngest age group, where the measured value in 1967 is smaller than the predicted one. The underestimation is likely induced by caveats in income definition. Supposedly, the overall fit between the 2001 curves is better because of improving precision of the CPS. A prominent and expected difference between the curves for 1967 and 2001 consists in the increase in $T_{cr}$. Therefore, the model predicts not only the shape of overall and age-dependent PIDs below the Pareto threshold, but also the observed evolution of mean income dependence on work experience. Again, we would like to stress that the prediction over the 35-year period was based on a few simple equations using the same set of empirically estimated defining parameters.

Figure 1.6.14 presents similar curves for 1974 and 1987, the years of considerable changes in CPS procedures. The Figure reveals fluctuations in the scaling factor from 72 in 1967 and 2001 to 79 in 1974 and 1987. Other years are also characterized by some variations in the scaling factor, but their amplitude is smaller than those observed in the original values of the mean income, as shown in Figures 1.6.15 and 1.6.16. The conversion factor for the original values varies in the range from 76 for 2001 to 95 for 1974. The conversion factor is 90 for 1967. There are two sources of these variations: 1) the fluctuation in the portion of GPI in GDP; 2) the change in the portion of people with income. The former will be analyzed in §1.8. Here we are more interested in the evolution of the shape of the dependence of mean income on work experience.

Unexpectedly for our modelling, the original mean incomes are modelled much better than the corrected ones. Neglecting the changes in the scaling factor in Figures 1.6.15 and 1.6.16, the predicted curves coincide with the measured ones over the whole income range, including the youngest age group. One can assume that the CPS provides a good cross section of actual personal income distribution, i.e. the shape of the curves, but a biased (under-) estimate of gross personal income, i.e. the level of the curve. The difference between the corrected and original mean incomes is the lowermost in the age groups where the portion of people without income is negligible, as will be discussed in §1.9.

Figures 1.6.17 and 1.6.18 illustrate some results for the years after 1994, when mean income readings are available in 5-year age intervals. Both, original and corrected mean income values are modelled. Due to higher resolution, the calibration of the decreasing branch is different from that used in the modelling of 10-year intervals: at the age of ($A_r$=) 80 years the level of mean income is only ($M_r$=) 0.45 of the peak value. Overall agreement between the curves is good, except two clear outliers in 1994: the observed mean income in the work experience groups between 30 and 34 years and between 40 and 44 years. Such outliers are not observed in other years between 1994 and 2001 and are likely associated with some changes in survey methodology in 1994. In the youngest age group, the predicted value of mean income in 2001 is slightly higher than the observed one. The original mean income values better fit the observations in the youngest group. A positive significant improvement upon the results for 10-year intervals consists in a much better fit beyond the critical work experience. All in all, the model provides an adequate description of actual mean incomes, including the effect of $T_{cr}$ increasing through time.

Direct modelling of the evolution of the $T_{cr}$ is hindered by the absence of measurements with appropriate resolution. True value of $T_{cr}$ usually resides somewhere between the end points of 10-year (5-year after 1994) intervals used in the CPS survey. As an alternative, one can model the evolution of (normalized) mean income in each work experience interval, as presented in Figure 1.6.6. The evolution should be defined by the growth in real GDP per capita, and thus, in $T_{cr}$. The evolution is not a mechanical and uniform increase of mean income in each of the age groups, but results from myriads of interactions between people of various ages and leads to the redistribution of income in favour of the age group with the largest mean income.

The evolution of mean income in a given age group over calendar time provides a representation equivalent to that used in Figure 1.6.13. Figures 1.6.19 through 1.6.23 display the results of modelling for five work experience groups: from the youngest group (from 0 to 9 years) to the oldest group (from 40 to 49 years). Two observed curves (original and corrected for population without income) in all the Figures are drawn in the whole range of available data and the predicted curves – between 1960 and 2001.

In the youngest age group (Figure 1.6.19), the corrected observed curve and the predicted mean income curve diverge considerably in line with the above discussion on the CPS questionnaire – the youngest group is the most problematic for measurements and thus for modelling. The original mean income values are much closer to the predicted ones. The same is valid for the other four groups: original mean income values are better fitted in relative terms. After 1980, the predicted and observed curves demonstrate similar downward trends. Furthermore, a plateau after 1995 is also a common feature.

The interval between 10 and 19 years of work experience, presented in Figure 1.6.20, is characterized by a better prediction. The theoretical curve has a slightly smaller slope than both empirical curves and no through near 1994, likely related to the changes in the CPS. Considering the full range of the change between ~0.65 and ~0.85 the prediction is accurate most of the time. As in the youngest age group, neither empirical nor theoretical mean income normalized to the peak value among all age groups can reach the unit line.

The first group reaching unit is between 20 and 29 years of work experience (Figure 1.6.21), where empirical $T_{cr}$ resided at least between 1967 and 1987. Theoretically, $T_{cr}$ left the group several years earlier – around 1980. As before, the empirical curves demonstrate fluctuations of higher amplitude. The predicted curve is smoother because it has no measurement errors. It is expected that the empirical curves, both having a small upward segment around 2000, will continue to evolve along the downward trend driven by increasing GDP per capita and $T_{cr}$.

The next group between 30 and 39 years, presented in Figure 1.6.22, shows an outstanding behavior – all curves stay at unit line almost all the time. It means that the peak mean income resides in this age group. The prediction is excellent, except the period between 1967 and 1974. Finally, the prediction in the eldest group, depicted in Figure 1.6.23, does not contradict the observations, if to take into account the range of overall increase from 0.83 to 0.88. Again, there is a trough in the empirical curves around 1994.

The above comparison of the predicted and observed curves shows an important overall agreement of the curves and a considerable divergence during some relatively short time intervals for some age groups. For example, there is an almost 10% deep trough between 1980 and 1990 in the observed mean income curve in the group between 30 and 39 years. The predicted curve does not show such behavior and retains its value close or equal to 1.0. There are also 2% to 4% amplitude variations in the observed curve from that predicted for the group between 20 and 29 years. This discrepancy can be partly related to CPS procedures and changes in the population estimates related to decennial censuses. The latter can reach several percent in some age groups. For example, one can compare two different mean income estimates obtained in 2000 and based on two different population estimates – postcensal and intercensal.

Thus, there is some concern about the accuracy of the mean income estimates. In fact, in order to derive exact mean incomes from the CPS one has to obtain true PID. Any error in the high-income end of measured PIDs leads to a large error in the mean income because of larger relative input of the high-income population in the net income. The low-income population does not add much to total income and usually is relatively better presented in the CPS surveys just because it is larger. The problem of people without income can also be resolved by adding people with zero incomes because their incomes are almost negligible in any case. There is an alternative to mean income.

A more accurate quantitative characteristic of income dependence on age is potentially median income, i.e. the income which divides personal income distribution into two equal parts. Median income is not sensitive to measurement errors at higher incomes – the number of people there is too small. Therefore, median income is closer to low incomes and represents a robust characteristic of PID and a good parameter to model.

The Census Bureau is a major source of data on median income in the U.S. It provides electronic tables since 1974. Figure 1.6.24 illustrates the difference between overall mean and median incomes, both expressed in 2001 CPI-U-RS adjusted dollars. The mean income grows much faster than the median one and the curves diverge over time. This divergence between the curves implies a faster growth of personal incomes at higher levels: the same portion of people above the Pareto threshold gets increasing portion of total personal income. A surprising feature is the presence of several quasi-flat segments in both curves; the longest is the period between 1974 and 1983 in the median income curve. The growth in real GDP per capita should be accompanied by proportional growth in GPI. Thus, the flat segments might be related to the decreasing portion of GPI in GDP.  Such effects can be incorporated in our model only in form of scaling factor, as shown for mean incomes.

In this paragraph, we are interested in age dependent measures of personal income. Figure 1.6.25 illustrates the effect of smaller growth rate of median income in two most important groups with work experience between 20 and 29 years and between 30 and 39 years. These are the groups where the critical work experience, $T_{cr}$, resides during the last 50 years. In the group between 20 and 29 years, the ratio decreases from 0.85 in 1974 to 0.75 in 2002. According to the model and following the observed trend, the ratio will continue to fall.

Having calculated all individual incomes for all ages one can find the median one without additional efforts. Figures 1.6.26 through 1.6.28 present results of the median income modelling. The observed and predicted median incomes are in a slightly better agreement than those for the mean income corrected for population without income. Since only aggregated income measures are available, one cannot correct median income for population without income as easily as mean income.

In the group between 10 and 19 years of work experience (see Figure 1.6.26), both the observed and predicted curves normalized to the peak value among all age groups demonstrate a downward trend. Therefore, this group is characterized by a diminishing relative income – people are getting poorer relative to the group with maximum median income. The model predicts this tendency with a good accuracy, if to treat the through near 1994 as an artificial one and associated with the CPS questionnaire.

The group between 20 and 29 years (Figure 1.6.27) contained $T_{cr}$ for a long time. The time when the critical work experience left the group is also well predicted. Fluctuations in the empirical curve are of higher amplitude, but otherwise are small – around 0.03.  One can expect that this group lost the largest median income forever, if the evolution of personal income distribution in the US will follow that predicted by our model.   The next age group (Figure 1.6.28) got the highest median income approximately in 1990. Since then it contains the $T_{cr}$ . Theoretically, the transition should happen around 1985. The difference between the predicted and observed curves between 1985 and 1990 is small and lays in the bounds of uncertainty associated with the CPS.

Our model meets a major problem associated with the accuracy of the mean and median income estimates. We model the observed values and obtain empirical estimates of the defining parameters, which correspond to the best fit model.  Any measurement errors in the observed values are directly transformed into equivalent errors in the defining parameters. On the other hand, the parameters obtained in the previous two paragraphs are the same as obtained from the modelling of the mean and median income measured during a relatively long period of time. This validates, to some extent, the model and the observations. Hence, one can conclude that our modelling is successful in spite of several problems remaining in the observational and modelling parts. Moreover, the model reveals weak points in the current procedure of personal income estimation and provides a good foundation for future corrections and improvements. This is a standard situation in the natural sciences, where the loop experiment-theory-experiment- … is infinite. To transform economics to a form appropriate for joining the club of hard sciences is our main goal.

The model provides an opportunity to extrapolate the observed behavior into the future and thus to test its predictive power. As shown in Chapter 2, the growth rate of real GDP has a trend associated with the reciprocal value of $T_{cr}$. The trend is near 1.6% between 2002 and 2022. We used this value to predict the evolution of the functional dependence of mean income on work experience. Figure 1.6.29 presents curves of real mean income (2001 dollars) with five year spacing. The evolution of population during these years is taken from the population projections also provided by the Census Bureau. By integrating the curves over work experience one obtains an estimate of total real GPI.

We failed to find similar personal income data sets in other developed countries. Equivalent income distributions would be of crucial importance for validation of the results obtained for the US. Fortunately, there exists an alternative income estimate. The UK Inland Revenue publishes distributions of taxable income, including mean income as a function of working experience for years 1999 through 2002. In order to compare PIDs in the UK and US one should scale them to the same currency. Also, for an adequate comparison the largest mean incomes over all age groups have to be equal because the critical age experience depends on real GDP (=GPI) per capita. For example, the level of real GDP per capita in the UK in 2002 was $26,500. The US reached the same level in 1986. Figure 1.6.30 compares relevant estimates of mean income normalized to the peak value among all age groups. The curves practically coincide most of the time, except the difference in trends at large work experience likely associated with shorter averaging intervals in the UK. The agreement between the curves is a quantitative evidence in favour of our model describing the distribution of personal income and its evolution with time. The model can be further validated using appropriate data from different developed countries and also the data from future surveys in the US. Both logical and historical inferences are useful.

This study uses GDP per capita as an external parameter. The distribution of personal income is proved to be a predetermined function of this parameter. One can interpret this relationship in opposite direction as well. Personal incomes, as a result of individual efforts to earn (or produce) money, represent the driving force of real economic growth expressed in monetary units. It is the sum of personal incomes that makes real GDP. So, the growth rate of real (and nominal) GDP per capita is unambiguously determined by current distribution of personal income, which in turn, depends on age distribution. As demonstrated in this Chapter, the shape of mean income dependence on time is controlled by values at two points - the starting point of the distribution and $T_{cr}$. The latter is defined by GDP per capita and the former is an external variable associated with the influx of new people in a given economy. One can expect that the number of newcomers somehow influences the growth rate of real GDP per capita. This intuition will be quantitatively tested in Chapter 2.

## §1.7. Modelling high incomes – the Pareto distribution

One of essential features of the PIDs measured in the United States and addressed in our microeconomic model is the existence of two inherently different regimes of money earning. Both regimes have one-to-one analogues in physics (Dragulesku & Yakovenko, 2001). The first regime, referred to as "subcritical", corresponds to the evolution of income proportional to the capacity of a person to earn money, i.e. to the product of the capability to earn money and the size of earning means. The capability is an inherent feature of each and every person in a given economy arising from numerous interactions between people as economic agents. The capability is evenly distributed among 29 ranks as well as the sizes of earning means. As a result, the model defines a discrete ranking of the capability to earn money, which accurately describes PIDs at low and middle incomes. The portion of population covered by the subcritical regime is ~90%.

The second regime is a "supercritical" one. It spans the range of higher incomes described by the Pareto law distribution. This regime starts at some relatively large income threshold, the Pareto threshold, and supposedly. In physical disciplines, the supercritical regime with power law distribution of sizes is an often phenomenon. The mechanisms leading to the Pareto law are not well understood or modelled (Lise & Paczuski, 2002) and usually are explained by a number of processes known as self-organized-criticality (SOC). We also do not understand the mechanisms behind the Pareto law for personal income distribution. Instead, we assume that the distribution is purely probabilistic any person in this income interval can reach any income with corresponding probability. Individual capability to earn money is not important above the Pareto threshold: all people reaching the threshold have equal probability to reach any feasible level of income. Thus, the only quantitative requirement for people dreaming to get rich is to reach the Pareto threshold.

A good (and close to our profession) example from hard sciences is the initiation of an earthquake. Share stresses in solid Earth should overcome some critical value in order to start fracturing. When fracturing is started, the crack can propagate any distance from the smallest to the largest possible. In other words, even such catastrophic earthquake as the one occurred on December 26, 2004 at Sumatra started as the smallest crack. Since the frequency distribution of earthquake sizes is also described by a power law any small crack obtains a non-zero probability to become the largest earthquake when stress overcomes the critical value or threshold. Same is valid for personal incomes: every person can reach any possible income level, but first s/he needs to reach the Pareto threshold.

As mentioned above, the mechanisms leading to scale free (power law) distributions of sizes are studied in more detail in the natural sciences. In economics, the nature of such mechanisms is still a big challenge, but one can conclude that the mechanisms work very fast. There is no delay between the moment, when some personal income reaches the Pareto threshold, and the moment, when the income leaps to its new position in the Pareto distribution. The observation behind this conclusion is simple: there is no deficiency in the observed PIDs at any income level, i.e. all vacancies arising from various processes are filled at no time. In solid Earth, final size of an earthquake is usually reached several seconds after fracturing starts.

Because the PIDs measured by the Census Bureau undoubtedly reveal the presence of Pareto distribution it is possible to directly incorporate this observation into the model with relevant empirical parameters. This is the simplest but not the best way to fully use available information. Despite the fact that we do not understand the mechanisms driving the Pareto law, there are several quantitative problems one can resolve in the framework of our microeconomic model. One important task is to accurately determine the Pareto threshold separating the subcritical and supercritical regimes. There is a relatively wide transition zone between the branches where the subcritical (exponential) and supercritical (power law) distributions practically coincide. The model distinguishes the zones by matching various characteristics of observed and predicted distributions.

The portion of people having incomes in the supercritical zone depends on work experience. Really, young people have no time to increase income to the Pareto threshold and elderly are losing income exponentially in average terms. One could expect the highest density of rich population in the mid-age group, i.e. in the work experience interval where $T_{cr}$ resides. Thus, the portion in the Pareto income zone should grow to some critical age and then fall. Such complex behavior should be quantitatively predicted by any model of personal income distribution. Moreover, such models should also predict the finer changes in the shape of personal income distribution evolving over time. This is a good quantitative test of predictive power.

Our model does not rely on any conventional economic theory or approach. It simulates a wide range of independent observations of personal income made in the United States. In part, these observations are carried out for economic purposes and are based on some definitions adapted from the field of theoretical economics. For example, gross domestic income is divided into personal and corporate portions, despite the fact that at the end of the day the latter also belongs to some selected people. In our framework, this partition is treated as an artificial one. Our model includes only variables related to economic system as a whole, because parts of the system are prone to random fluctuations. One can always introduce a formal model which links measured variables and take into account no external meaning of the data. Although our quantitative model has no roots in the mainstream economics, its merits have to be assessed by predictive power and resolution capabilities.

The overall PIDs from 1994 to 2001 with two branches are shown in Figure 1.4.6. The observed value of the power law exponent obtained by regression analysis is -3.97. The original annual distributions of the absolute number, as given by the U.S. Census Bureau, are normalized to the total population for corresponding calendar years. The obtained population density distributions are adjusted for the growth in nominal GDP per capita and the width of adjusted income bins according to the procedure described in §1.4. The normalized and adjusted curves demonstrate a high degree of similarity. This effect has been interpreted as the existence of a rigid PID structure.

Several PIDs in various open-end income intervals measured in 1994 are presented in Figure 1.7.1 as discrete functions of work experience with 5-year spacing. The curves are normalized to the peak value among all age groups in corresponding income intervals and illustrate the evolution of PID with increasing low-end threshold. The first interval starts with a zero income threshold. This interval also includes those people who are reported in the original Census Bureau table as having no income or losses. Second curve corresponds to the PID which includes all people with income above $10K, and so on with $10K increment. The last curve represents PID for the people with income above $100K.

The evolution of shape is remarkable. The first curve has its peak in the first work experience interval – between 0 and 9 years. As a rule, young people entering the US economy have very low income between $0 and $10K during the starting 10 years. Many of them stay in this interval forever. With increasing low-end threshold, the curves are gradually transforming into a bell-like (uni-modal) shape, with the peak value moving towards the work experience group between 30 and 39 years. As expected for a distribution governed by a power law, the curves with low-end threshold above $60K practically coincide, i.e. they are scale free. This observation also confirms the assumption that all high-income positions, even with the highest possible incomes, are filled momentarily. Otherwise, it would not be possible for the youngest age group to be characterized by a scale free distribution.

The portion of the observed PID described by Pareto distribution seems to be the simplest one. Really, the observed behavior of the PID at higher incomes obeys a simple law and should not change in time, at least theoretically. The total number of people with income in the zone controlled by the power law develops with time as a linear function of nominal GDP per capita and population growth. This theoretical conclusion is confirmed by observations. Figure 1.7.2 depicts the number of people with income above $100,000 normalized to relevant working

age population for the period between 1994 and 2001. The normalization is necessary to eliminate of the effect of growing working age population. The linear regression line demonstrates that our assumption is correct. In 2001, around 3% of working age population had incomes above $100K, and this portion will grow over time.

Figure 1.7.3 presents the number of people who reached $100K income as a function of work experience for selected years between 1994 and 2002. The numbers are normalized to total population in corresponding age group. There was no significant change in the shape of the curves over time, considering the accuracy of the CPS. Hence, the observed PIDs are fixed in relative terms at higher incomes confirming the existence of a rigid hierarchy of personal incomes independent on age. Is our microeconomic model capable to predict this observation?

The best-fit personal income distributions in the high income zone in the United States are obtained using the model with the following defining parameters: the start year is 1960, $T_{cr}(1960)$=26.5 years, $\alpha$=0.087, $\Lambda_{min}(1960)$=1.0 and $\Sigma_{min}(1960)$=1.0, and initial value of the Pareto threshold $M_P(1960)$=0.43. We also use varying index $\alpha_1$ with defining parameters $A_r$=80 years and $M_r$=0.45, as obtained in §1.6 for the period between 1994 and 2001.

Figure 1.7.4 compares predicted and observed number of people with income above the Pareto threshold in 1994 and 2002. Apparently, this number depends on work experience. In the youngest age group, one cannot expect too many rich people. Overall, the predicted values confirm this assumption. In the age groups well above the critical work experience, $T_{cr}$, the number of people with high incomes is also decreasing with age, in absolute and relative terms. The portion of people above the Pareto threshold has a peak near $T_{cr}$. It is worth noting again that the predicted curves better match the observed ones in age intervals where personal income has an adequate definition and the portion of people without income is the smallest.

In Figure 1.7.4, the Pareto threshold is evolving in time as real GDP per capita. Initial dimensionless Pareto threshold is 0.43 in 1960. In accordance with GDP growth, the threshold reaches the level of 0.829 in 1994 and 0.953 in 2002. The predicted curve is slightly lower than the observed one in 1994, but the curves for 2002 are in excellent agreement.

The discrepancy between the theoretical and observed curves might be induced in part by the uncertainty in income measurements. The resolution of income distribution is fixed at $2500, what also affects the amplitude of the discrepancy. One can only use discrete data with $2500 step. Effectively, one cannot distinguish between $50,000 and $52,499 – both values are inside the same income interval. The Pareto threshold increased in nominal terms from $43.5K in 1994 to $57.6K in 2002, i.e. only by ~$14000. For example, we used the income interval above $42.5K to present the actual distribution and obviously overestimated the total number of people in any work experience interval, because all the people with incomes between $42.5K

and $43.5K were counted in. In the model, the income and time resolution of income distributions is not limited. For practical purposes, we fixed the resolutions to $1K and 1 year.

All in all, the most important result of the comparison in Figure 1.7.4 is that the shape of the observed curves is accurately predicted over the whole range of work experience. This demonstrates the adequacy of our model in describing the underlying physical and social processes governing the principal features of the Pareto distribution.

Since the power law defines a scale free size distribution any threshold above the Pareto one should provide equivalent curves. The reason to increase the threshold far above the $M_P$ is to avoid the uncertainty in the estimation of the Pareto threshold. There is a transition zone between the sub- and supercritical regimes of income earning and the former might introduce a bias in the quantitative estimates of the latter near the boundary. Figure 1.7.5 presents the portion people who reached $100K (current dollars) as a function of work experience in 1994 and 2001. The predicted curves are in a good agreement with the observed ones in shape and level. Again, there is a slight difference in the initial parts of the theoretical and actual distributions. More people are counted in the very first work experience group in comparison with the predicted values. Despite a minor influence of this observation on the overall distribution in terms of the total income, one can argue that this difference is due to a wrong estimate of dissipation factor, α, used in the model. This factor defines the time constant in (1.3.3') and the rate of income growth at the initial part of individual income trajectory is the most sensitive to the factor.

We have no convincing quantitative explanation for the observed discrepancy, but should mention that actual start point of work experience for some people is well below 15 years of age and the accuracy of measurements at higher incomes for the youngest group is definitely low. Also, the resolution of income data is very low: the width of the first interval is 10 years. A data set with a finer resolution could help to reveal the reason for the discrepancy.

Figure 1.7.6 depicts theoretical curves of population density (income) distribution above the Pareto threshold as a function of work experience for selected years between 1980 and 2002. The curves are normalized to total population in the Pareto income zone for corresponding years and present a clear picture of the evolution during the modelled period. In the beginning of the 1980s, when effective dissipation factor, $\alpha/\Lambda_{min}$ was as large as 0.08, the work experience needed to reach the (normalized and dimensionless) Pareto threshold of 0.43 was lower and people with the highest capabilities, $S$, easily attained this level in the first 10 to 20 years of work. This corresponds to an almost linear growth in the number of people reaching the threshold in the first decade of work. With increasing $\Lambda_{min}$, effective dissipation factor was decreasing and the time needed to reach the threshold was growing. The start segment of the curves became smoother and the fastest growth migrated from the first to the second decade of work experience. Accordingly, the peak value has been shifting to larger work experiences. The usage of actual age pyramid introduced visible disturbances in the curves.

So, the model predicts the exact number of people reaching the Pareto threshold depending on work experience. Relevant population density distribution also evolves over time in a manner predicted by the model. The model can be also used to predict the future development if projections of population structure and GDP per capita are available. Figure 1.7.7 presents such a prediction for selected years between 2002 and 2023 based on the population projection published by the Census Bureau and the growth trend of real GDP per capita estimated as ~1.6% per year. This prediction is possible because there is no random or deterministic process which leads to observed features of power law distributions except the process of the personal income growth to the level of the Pareto threshold. In fact, if any other process does exist and adds (subtracts) sufficient number of people to that observed above the Pareto threshold, the agreement between the predicted and measured curves would be destroyed.

The model explicitly states that when all positions in the Pareto distribution are occupied there is no opportunity to create a new one with equivalent properties and occupy it. All positions are enumerated. Individuals may swap their positions, however. When a person has a low capability to earn money, there is no way to get rich because s/he cannot reach the threshold with that capability. If this person has high capability but small earning means, s/he is likely capable to change the means to a bigger one and reach the Pareto threshold. Theoretically, an exchange of capability between two people is not prohibited, but the ranking is rigid and it is difficult to imagine somebody to overcome the system of external evaluation, which put her/him to current rank.

When one's personal income reached ~$57,000 in 2005, it was a good start to obtain a higher income with the probability inversely proportional to the income cubed, as defined by the empirical exponent in actual Pareto distribution (see Figure 1.4.6). In §§1.8-1.9, we discuss the Pareto law index in detail, because its variation affects the estimation of Gini coefficient. In this paragraph, we are focused on the model, not on empirical findings.

Because the observed PIDs in the United States have demonstrated their rigidity over a long period one can conclude that just few people can ever reach the Pareto threshold and have a non-zero probability to become rich. The majority, about 90%, is below the Pareto threshold forever and it gets income exactly proportional to personal capability to earn money.

We developed the concept of personal income distribution and relevant quantitative model during the period between 2003 and 2004. The development included standard trial-and-error procedures with empirical assessment of defining parameters based on contemporaneous data. The period covered by the measurements of PID at high incomes was limited to 1994.

Later, the Census Bureau opened an access to PIDs for the whole period since 1947. These PIDs were given in form of scanned images of original CPS reports, what required some additional time and efforts to digitize them for usage in quantitative analysis. Therefore, these new PIDs served as an independent source of information, which allowed validating and testing the predictive power of the model. Really, all defining parameters of the model were estimated using data in a short time interval, and the future and past evolution was predicted beyond the interval. Figure 1.7.8 illustrates the predictive power. The observed and theoretical curves for the year of 1980 are in an excellent agreement over the whole range of work experience. The fluctuations in the measured curve are likely associated with measurement errors. In the following two paragraphs, the model is extended to a wider time interval between 1947 and 2006 and all characteristics of personal income distribution in the United Stated modelled in §§1.4-1.7 are also available. Hence, it is instructive to further test the predictive power of the model together with the estimate of income inequality as expressed by the Gini coefficient.

This is a good place to briefly speculate on the difference between capitalism and socialism as economic systems. Because this discussion is fully qualitative, it has no impact on our model and may be skipped by readers without any loss of consistency. In the first approximation, the socialist system is based on a theoretical assumption that personal income is proportional to the time necessary to produce some goods or service using some capability to produce varying among workers. This is the principle of socialism - to obtain income exactly proportional to the price of produced goods. The price is determined by (economic or political) authority according to some rules developed to balance inputs of time and productivity of population. As we have seen above, this assumption also works precisely in the capitalist economic system for the overwhelming majority of population. Ninety per cent of the population of 15 years of age and above gets personal incomes exactly proportional to their capability to produce income, as described by the microeconomic model. When extended to the whole population, this rule limits personal incomes of those ten per cent of the population, which has incomes above the Pareto threshold. In the socialist system, they would obtain incomes according to theoretical values, as determined by the model. Capitalism, however, has some extra feature. These ten per cent of the population have personal income not proportional to the capability. Fortunately for capitalism, their incomes are described by a power law distribution, which is extended to incomes of several million dollars and above. These people actually produce some additional income (and thus, GDP), exceeding the theoretical value by ~35%. In other words, they produce 45% of gross income, *GPI*. If they would produce in the subcritical regime, i.e. in the regime realized in the socialist economic system, their input would be only 34% of *GPI*. After simple calculations, one can conclude that capitalism has an advantage of personal income distributed by the Pareto law, which increases gross personal income or GDP by at least 20% compared to that in socialism. In the long run, this excess provides progressively increasing additional GDP. Hence, developed capitalist countries have been growing at a higher rate than socialist ones because of the presence of rich people. This might be a fundamental feature of capitalism.

**§1.8. Modelling Gini coefficient for personal incomes in the USA between 1947 and 2005**

The microeconomic model developed is previous paragraphs describes personal income distribution in the United States and its evolution through time. This model is based on the prediction of each and every individual income for the population of 15 years of age and over. It accurately predicts the overall PID, the average income dependence on work experience, the evolution of PIDs in narrow age groups, and the number of people and age dependence in the income zone described by the Pareto distribution. The model also provides quantitative predictions for these variables beyond the years where corresponding data are available. Having a complete and precise description of the US PID evolution one can compute the evolution of the Gini coefficient. This makes the Gini coefficient only of secondary importance.

The purpose of this paragraph is to accurately estimate the Gini coefficients associated with the personal income distributions provided by the US Census Bureau and to model the evolution of these coefficients between 1947 and 2005, i.e. during the period of continuous PID measurements. An extended analysis of the PIDs has been carried out and the discrepancy between observed and predicted Gini coefficients is interpreted in terms of the changing accuracy and methodology, including income definitions, used in the CPS during the studied period.

The Gini coefficient, *Gi*, is a standard measure of income inequality. By definition, *Gi* is the ratio of the area between the Lorenz curve related to a given PID and the uniform (perfect) distribution line, and the area under the uniform distribution line. The Lorenz curve, *Y=F(X)*, is defined as a function of the percentage *Y* of the total income obtained by the bottom *X* of people with income. Having measured values of individual incomes for all people with income and ranking them in increasing order one can precisely calculate corresponding Gini coefficient. It is also possible to include in the consideration those people who do not report nonzero income according to contemporary income definition. In reality, there are some difficulties potentially affecting the accuracy of the PID estimates and the uncertainty of associated Gini coefficients.

The US Census Bureau has been measuring personal income distribution in the USA since 1947 in annual current population surveys. Methodology of the measurements and sample size has been varying with time (US CB, 2002). Therefore, one has to bear in mind potential incompatibility of the CPS results obtained in different years. Changes in income definitions, sample coverage and routine processing influences the estimation of various derivatives of the PIDs, for example, measures of inequality. Moreover, such changes in procedures and definition are likely accompanied by some real changes in true PIDs - the latter changes are hardly distinguished from the former ones. The true PID is the distribution of incomes when all sources of personal income are included.

There are two principal effects of the changing income definitions on the measured PIDs. First, the number of people with income critically depends on definition of income near zero value. Due to a high concentration of people in the low-income range of the measured PIDs in the USA the number of people without income is prone to large variations dependent on introduction of new or exclusion of old sources of income in the CPS questionnaires. In addition, it is difficult to give accurate definitions to numerous potential sources of annual incomes near $1, and even more difficult to distinguish between $1 and $2 per annum. Due to high uncertainty and low resolution of the current CPS methodology in the low-income end it is practically impossible to measure true PIDs. Thus, the measured PIDs represent only a varying portion of the true PIDs, the latter being the actual object for our modelling. (Here we assume that the gross personal income is an exactly measured (true) variable and its distribution among people is fixed and can be also exactly measured. In this sense, true PID and Gini coefficient do exist and, theoretically, can be measured.) This variation creates some problems for the modelling and interpretation of results.

Figure 1.8.1 demonstrates the evolution of the ratio of the number of people with income to the working age population. There is a significant increase in this ratio: from the lowermost value of 0.64 in 1947 to the highest 0.93 in 1988. The ratio has been slightly decreasing since 1989 - to 0.89 in 2005. Such fluctuations should definitely introduce a significant bias in the estimates of Gini coefficient – people without income bring a large increase in the coefficient, if included. Therefore, when estimating the Gini coefficient one has to consider both cases – all population of working age and the portion with income. The true PID and Gini coefficient has to be somewhere between these two limiting cases. Considering the entire working age population, including persons without income according to contemporary definitions, one significantly overestimates the Gini coefficient. This effect is especially high in the beginning of the studied period. When only people with income are included, the Gini coefficient is obviously underestimated because zero income, quantitatively, is also income. With time, these two estimates have to converge as the portion of population without income decreases.

Second effect of the revisions to income definition and CPS procedures is related to the change in the portion of gross personal income in GDP. The introduction of new sources of income in the CPS questionnaires should result in an increase in the estimated GPI. Figure 1.8.1 depicts the evolution of the GPI portion in the US GDP: from 0.76 in 1951 to 0.86 in 2001. A significant and fast drop in the portion is observed between 2001 and 2005 – from 0.86 to 0.82.

The net change in the GPI portion between 1947 and 2005 is smaller than the change in the share of population with income.

A fundamental assumption of the model for the evolution of individual incomes presented in §1.3 is that all people older than 14 years have nonzero annual income and contribute to GPI, which is equivalent to gross domestic income and GDP under our framework. This assumption allows modelling the evolution of PID using real GDP per capita, which completely determines the time histories of the model defining parameters. The actually measured PIDs are associated with a changing portion of GDP.

In addition to the principal difficulties associated with definitions and procedures there are some technical problems for the estimation of Gini coefficient as created by the representation of relevant data and the resolution of the measured PIDs. The US Bureau of the Census has been publishing the numbers of people enumerated in income bins of varying width. There were only 14 bins, including the open-end one for the highest incomes, in 1947 and 48 bins in 2005.

In the absence of information on each and every individual income, the Gini coefficient can be calculated by some approximating relation. For example, if $(X_i, Y_i)$ are the values obtained from the CPS, with the $X_i$ indexed in increasing order $(X_{i-1} < X_i)$, where $X_i$ is the cumulated proportion of the population variable, and $Y_i$ is the cumulated proportion of the income variable, then the Lorenz curve can be approximated on each interval as a straight line between consecutive points, and

$$Gi_a = 1 - (X_i - X_{i-1})(Y_{i-1} + Y_i), \quad i=1, \ldots, n \qquad (1.8.1)$$

is the resulting approximation for *Gi*. One can also approximate the Lorenz curve using exponential function and a power law, where appropriate, for the interpolation of the underlying PID, as discussed in §1.4.

The choice of an appropriate function for the interpolation reveals an important pitfall of the CPS - the usage of the same income bins for representation of data counted during relatively long period of time. The growth rate of nominal GDP in the USA was high - more and more people obtained incomes above the upper limit in the CPS income reports and found themselves in the group " $MAX and over". So, the coverage of the populations below and above the Pareto threshold, also proportionally growing with time, was significantly different. This variation in the coverage might potentially result in a better or worse overall resolution and corresponding bias in the estimation of Gini coefficients.

The US Census Bureau provides several versions of the PIDs between 1947 and 2005. In some reports, there are presented the tables containing counts in year-specific income bins expressed in current dollars. Some reports give PIDs using CPI-U adjusted (constant) and/or current dollars but in the same income bins for all years staring from 1947 to the year of the report issuance. Figure 1.8.2 shows some selected original PIDs normalized to the total population (15 years of age and above) for corresponding years and additionally divided by the widths of corresponding income bins. These curves are population (or probability) density functions, *pdf*, and show the number of people in $1-wide bin for a given income level. Such a representation allows a direct comparison of the PIDs because they are independent on population size and reduced to the same income bins. As before, we associate the population density with mean income in given bins.

These "mean" densities obtained for bins of varying width might be a poor approximation for the densities at the edges of the bins. The wider is the bin the poorer is the approximation. It is worth noting that such a representation naturally excludes the open-end high-income bin because there is no width and mean income associated with this bin.

The PIDs between 1947 and 1987, shown in Figure 1.8.2a, are obtained using the same ten income bins as defined by the following boundaries expressed in current dollars: $0, $2000, $4000, $6000, $8000, $10000, $12500, $15000, $20000, $25000, and above $25000. The latter open-end bin is not shown in the Figure because it does not have finite width for normalization of the PID reading in this bin. Thus only nine bins describe the PIDs between 1947 and 1987.

Figure 1.8.2a illustrates the problems of resolution with constant income bins. The PID for 1947 (and also for the years between 1948 and 1950) does not contain any reading for incomes above $9000. This is due to the absence of people reporting such incomes in corresponding CPS population samples, but not because of the absence of such people at all. The best resolution (among the PIDs shown in the Figure) at high incomes, i.e. in the Pareto zone, is observed in 1957 – there were seven bins covering the zone. At the same time, there are only two bins covering the low-income zone in 1957. For the PID in 1987, the Pareto threshold is larger than $25000, and the PID contains only one reading in the Pareto zone corresponding to the open-end bin, and this reading is not shown in the Figure. Obviously, this PID provides the best resolution in the low-income portion of the distribution – nine bins. Therefore, the constant bins fail to provide a uniform description of the PIDs between 1947 and 1987 and the estimation of Gini coefficient can be severely biased.

The PIDs between 1947 and 2005 presented in Figure 1.8.2b are characterized by income bins which are better adjusted to the observed PIDs. These bins cover both low and high incomes better than in Figure 1.8.2a, with a varying resolution, however. As mentioned in §1.4, the years after 1994 are characterized by the highest resolution and the narrowest income bins of $2500 between $0 and $10000. Because of the increasing number of people with incomes over $100,000, three $50000-wide bins were introduced in 2000, covering incomes up to $250,000, extra to those provided by standard CPS reports. These wide bins allow a more accurate representation of the Pareto distribution and corresponding Gini coefficient.

As we have found in §1.4, the PIDs between 1994 and 2002 practically collapse to one curve, when normalized to the working age population and nominal GDP per capita. This observation demonstrates a fundamental property of the personal income distribution in the USA – it is characterized by a fixed hierarchy of incomes, which changes very slowly over time as induced by the evolution of age structure and real GDP per capita.

The PIDs measured for the years before 1994 allow to validate this property and to extend the presence of such a fixed hierarchy in the PIDs by 47 years back in the past (and 3 years ahead). There is a problem related to the normalization factor, however. The years between 1994 and 2002 are characterized by the constancy of the portion of the GPI in the GDP and the population with income in the working age population, as Figure 1.8.1 demonstrates.

This is not the case for the years before 1980, however. As a consequence, when normalizing to nominal GDP, one has to replace it with nominal GPI in order to accurately represent the evolution of the PIDs after 1947. Such a procedure should compensate the difference in the evolution of the GDP and GPI - less sources of personal income were considered in the earlier years and income scale was effectively biased down. Figure 1.8.3 displays the cumulative growth in the nominal GDP and GPI between 1947 and 2005 as reduced to the total working age population and the population with income. The curves diverge with time. The increasing deviation permits a more robust choice of an appropriate variable for a normalization, which we expect to be able to convert all PIDs for the years between 1947 and 2005 into one curve. Figure 1.8.4 depicts the PID for 2005 normalized to the four variables in Figure 1.8.3. One can clearly distinguish between the resulting normalized PIDs in the low- and high-income zones.

Figures 1.8.5 and 1.8.6 display some results of the normalization of the measured PIDs to the measured nominal GPI, as reduced to the people with income. For the period between 1947 and 1987 (Figure 1.8.5), where the PIDs were measured in the same bins, the normalized PIDs practically collapse to one curve with only minor deviations likely associated with measurement errors. For the period between 1947 and 2005, where a progressively higher resolution is available with the widths of income bins decreasing in relative terms, the normalized PIDs for population with income (Figure 1.8.6a) are also very close. Narrower bins result in higher fluctuations due to measurement errors, however. At the same time, the normalized PIDs for the entire working age population demonstrate a larger divergence with time because the normalization is associated with the nominal GPI reduced only to the population with income. So, the choice of normalization basis must correspond to the variable under consideration.

Overall, the normalized PIDs in Figures 1.8.5 and 1.8.6a are close. This observation extends the presence of a fixed hierarchy of personal incomes, as expressed by the portion of population having a given portion of gross personal income, to the years between 1947 and 1993, and beyond 2002. Therefore, one may expect only a slight variation in the Gini coefficient related to the PIDs measured since 1947. The presence of the hierarchy also represents a strong argument in favour of our model for the evolution of individual and aggregated income.

Having studied some principal properties of the PIDs for the years between 1947 and 2005, one can start a direct estimation of the Gini coefficient using (1.8.1). However, there are several technical problems related to the discrete representation to be first resolved. The PIDs provide only the estimates of total population but not the total income in given income bins. Only for the years after 2000, the mean income is determined for every bin allowing for an accurate estimate of cumulative income. No mean incomes are reported for the previous years, however.

When replacing true mean incomes with central points of corresponding bins, one introduces a slight bias in the estimate of Gini coefficient, as Figure 1.8.7 shows. Thus we need more reliable estimates of mean incomes. The best choice would be to approximate the observed PIDs in the low-income zone by exponential function, to determine corresponding exponent index for each year, and to calculate the Gini coefficient for in this approximation. This procedure might potentially provide a good estimate of the Gini coefficient if corresponding population estimates in given income bins are accurate. Unfortunately, the accuracy is inhomogeneous over the bins of varying width and the advantages of the exponential approximation might disappear, as Figure 1.8.8 demonstrates. Therefore, we use a different approach in the low-income zone.

The mean income estimates are available between 2000 and 2005 and it is easy find their average distance from central points of relevant bins. Figure 1.8.9 presents such deviations and corresponding regression line (mean deviation) for 2001 and 2005. The average dimensionless distance, i.e. the difference in $ divided by the bin width in $ ($2500 for the years between 2000 and 2005), is -0.12. Thus, in the following estimations of the Gini coefficient we use the mean income values corrected for this deviation from the centers of bins in the low-income zone.

In the high-income zone, a power low approximation is a natural choice for the PIDs, as demonstrated in Figures 1.8.5 and 1.8.6. Theoretically, the cumulative distribution function, CDF, of the Pareto distribution is defined by the following relationship:

$$CDF(x) = 1 - (x_m/x)^k$$

for all $x > x_m$, where $k$ is the Pareto index. Then, the probability density function, *pdf*, is defined as

$$pdf(x) = kx_m k/x^{k+1} \qquad (1.8.2)$$

The functional dependence of the probability density function on income allows an exact calculation of the population in any income bin, total and average income in this bin, and the input of the bin to relevant Gini coefficient because the *pdf* exactly defines the Lorenz curve. Thus, if populations are enumerated in a predefined set of income bins then relevant Lorenz curve can be easily retrieved using a known value of the Pareto index, $k$. Therefore, we use (1.8.2) in the following estimation of empirical Gini coefficients in the Pareto zone. As described in §1.3, the Pareto threshold (in current dollars) evolves proportionally to nominal GPI per capita. Such an evolution provides the rigid shape of the normalized PIDs because it retains unchanged the relative income level, where the transition from the low- to high-income zone occurs.

Now we are ready to estimate the Gini coefficient for the measured PIDs using the corrected mean incomes in the low-income zone and the power law approximation in the high-income zone. To begin with, we compare our estimates of *Gi* with those reported by the US Census Bureau, as shown in Figure 1.8.10. For the years between 1994 and 1997, the curves are very close. In 1998, a sudden drop by ~0.01 in the CB curve is not repeated by the estimated one. There is no clear reason for the drop – macroeconomic or that related to the CPS procedures. It is likely that there was some change in the Census Bureau's approach to the estimation of Gini coefficient in 1998. After 1998, the curves continue to slightly diverge, but move in sync otherwise. The difference between the curves reaches 0.01 in 2005. Overall, our estimates of *Gi* seem to be consistent with the CB's ones. This observation partly validates the Gini estimation procedure we have developed.

Figure 1.8.11 presents the estimates of *Gi* for the PIDs in current dollars, which do not include people without income. There are "crude" estimates of *Gi* obtained for the populations counted in the same income bins between 1947 and 1987. A "fine" PID is available from the year specific bins for the period between 1947 and 2005. Despite corresponding income bins in the second case also were used several years in a raw, the overall resolution of the PIDs is higher and *Gi* estimates are of a lower uncertainty. Two curves in Figure 1.8.11b present the evolution of *Gi* for the two sets of PIDs – crude and fine ones. In 1947, the curves are spaced by 0.1. When approaching 1970, they slowly converge. Between 1974 and 1984, the curves are hardly distinguished. In 1985, a new period of divergence starts. The observed discrepancy between the curves is related to the coverage of the PIDs by corresponding sets of income bins.

Figure 1.8.11a illustrates the difference between two Lorenz curves for 1947. The crude set of bins does not resolve the Lorenz curve well and relevant Gini coefficient is highly underestimated as compared to that estimated from the fine set. For the years between 1974 and 1984, both sets provide a compatible resolution (the number of bins is 10 and 18, respectively) and the estimates converge.

In fact, the Gini curve associated with the fine PIDs hovers around 0.51 between 1960 and 2005 despite the increase in the GPI/GDP ratio and the portion of people with income during this period (see Figure 1.8.1). This is a crucial observation because of the active discussion on the increasing inequality in the USA as presented by the Gini coefficient for households. Supposedly, the increasing Gini for households reflects some changes in their composition, i.e. social but not economic processes defined by the distribution of personal incomes.

Between 1947 and 1960, the fine *Gi*-curve monotonically grows from 0.45 to 0.50. This growth may be associated with the increasing resolution in corresponding PIDs. One can expect a further increase in the estimates of Gini coefficient when a finer grid is used. The possibility of a slight increase in the estimates of *Gi* associated with the inclusion of new (and true) income sources is also not excluded. All in all, the Gini coefficient for the true PIDs is likely to be higher than that predicted using the fine PIDs for population with income.

In the absence of the true PIDs, it is possible to carry out an estimate for the limit case – to include all people without income in the first income bin with zero width, i.e. from \$0 to \$0. It is difficult to believe that a person without income might potentially survive. However, current income definitions do not cover the sources, which bring actual personal incomes of people "without income". In any case, the inclusion of these people in the PIDs creates a problem for the Gini coefficient estimation. Figure 1.8.12b presents two time series of Gini estimates for the crude and fine bin sets. In 1947, the difference is 0.05 what can be explained by the properties of corresponding Lorenz curves, as Figure 1.8.12a depicts. Then, the curves converge and intersect in 1971. Between 1971 and 1984, the curves are very close and diverge again since 1985. These observations are similar to those associated with the PIDs for the population with income. The only difference is that the curves for the PIDs with total working age population undergo an expected decrease with time according to the decreasing portion of population without income. Therefore, the Gini curves associated with the total working age population and with its portion having nonzero income should converge over time. When the portion with income reaches 1.0, i.e. everybody has a nonzero income, the curves will become identical. So, where is our prediction of *Gi(t)* relative to the empirical curves?

In the model, the evolution of personal incomes is defined by a number of parameters, which we have determined empirically in previous paragraphs. For the estimation of Gini coefficient a crucial parameter is the Pareto law index, *k*, which defines how "thick" is the PID tail in the high-income zone. There are two independent techniques for the estimation of *k*.

First, for a Pareto distribution with index *k* and minimum value, $x_m$, the mean value is

$$x_{av} = (k+1)x_m/k.$$

Therefore, the measured average incomes for the open-end income bins provide valuable information on corresponding Pareto indices: $k=x_m/(x_{av}-x_m)$. Figure 1.8.13 presents the estimates of index *k* for the years between 2000 and 2005. These values allow multiple estimates using the increasing number of people with incomes above $250,000, $200,000, $150,000, and $100,000 – all in the Pareto zone. For example, the average income for people with incomes above $100,000 in 2005 is $176,068 and $x_{av}$ for people with incomes over $250,000 is $470,616. Corresponding Pareto indices are 1.31 and 1.13, respectively. The latter estimate is obtained using the average incomes for male and female separately, as presented by the Census Bureau. In 2005, there were 10,896,000 people with income above $100,000 and only 1,334,000 above $250,000. Bearing in mind that the population estimates are also obtained using only a relatively small population sample (~80,000 households), one can consider the Pareto index for the population with income over $250,000 as less reliable than the former value. Also, the average income in the open-end bin may be slightly shifted up due to the effect of few super-rich people, who do not obey the Pareto distribution. Such a deviation from the power law distribution is also often in the natural sciences and usually considered as statistical fluctuation related to the under-representation in any finite subset of infinite distribution. For example, catastrophic earthquakes may not obey the Gutenberg-Richter frequency-magnitude relation. According to Figure 1.8.13, *k*=1.3 is our best choice.

Second method is a direct estimation of *k* using a linear regression technique in the Pareto income zone. For such a regression, we represent the probability density functions in the log-log coordinates, as shown in Figure 1.8.14. The slope of the regression line is -3.36. Therefore, the Pareto index is *k*=3.36-2=1.36, i.e. consistent with the results obtained by the first method.

Figure 1.8.15 demonstrates the effect of *k* on the Gini coefficient predicted by our model. Obviously, lower *k* values create "thicker" tails in corresponding PIDs, i.e. more people with higher incomes, and larger *Gi* values. The effect of *k* on Gini is a nonlinear one and the difference of 0.3 units in the index results in the Gini coefficient difference of 0.01 to 0.015. One should not neglect such a difference when comparing predicted and measured Gini coefficients.

Another parameter of the model, which critically depends on the Pareto index, is the effective increase in income production in the model relative to that in the sub-Pareto income zone (see §1.3 for details). Figure 1.8.16 depicts the dependence of the corresponding ratio on $k$. As obtained previously, the empirical value of 1.33 exactly corresponds to $k$=1.35. This ratio is very sensitive to $k$, and the effect is also slightly nonlinear.

Having estimated the empirical parameters defining the model and the age structure of the US population between 1947 and 2005 one can predict the evolution of Gini coefficient (for personal incomes) during the studied period. Figure 1.8.17 compares the measured and predicted Gini coefficients. The predicted curve is in a good agreement with that obtained using the PIDs for the persons with income. The latter curve lies below the former one during the entire period. The empirical Gini coefficient for the PIDs including all working age population is always above the predicted curve. Hence, the predicted curve always takes the place just between the empirical ones and the latter two likely will converge to the predicted curve in the future, when accurate definitions of income are introduced.

This is an expected result of the modelling – neither of income definitions given by the Census Bureau can provide an adequate description of the true personal income distribution and, thus, all of them fail to predict the true Gini coefficient. The usage of biased Gini values may lead (and leads!) to economic misinterpretation and social confusion. The Gini coefficient for personal incomes in the USA underwent a slight increase between 1947 (0.5346) and 1962 (0.5378), and then has been monotonically decreasing to the current value of 0.524. There was no significant increase in income inequality in the USA during the last 60 years, as expressed by the Gini coefficient for personal incomes predicted by our model.

There are several simple, but meaningful findings related to the estimation of the empirical Gini coefficients. First and most important consists in the fact that the estimates of Gini coefficient critically depend on definition of income. The inclusion of new income sources in the CPS has resulted in a large change in the number of people with income and also in the ratio of GPI and GDP. The current set of definitions is far from the true PIDs.

Second, the Gini coefficient associated with the entire population of 15 years of age and over and that associated with people with nonzero income converge with time as the portion of people without income decreases. The true Gini coefficient has to be somewhere between these two estimates. Thus, the empirical estimates can not be considered as reliable for the purposes of economics as a theory.

Third, the resolution of the empirical PIDs directly influences the estimates of Gini coefficient. A higher resolution guarantees a smaller variation in Gini coefficient over time. Poor resolution leads to a negative bias in the Gini estimates.

Fourth, the empirical PIDs collapse to one curve when normalized to the cumulative growth in nominal GPI for the studied period between 1947 and 2005. The remaining differences in the PIDs are well reflected in the changes of Gini coefficient obtained using the population with income.

The model predicts the unchanged (normalized) PIDs and Gini coefficient between 1947 and 2005. Some weak changes in the PIDs and Gini are related to economic growth and the changes in the age structure of American population. The decreasing portion of young and thus relatively low-paid people in the working age population effectively leads to a decrease in the Gini coefficient. The increasing portion of the population older than the critical age $T_{cr}$ (55 years in 2005) results in an increase in the portion of relatively poor people because of the exponential decrease of personal income (including average one) with age. As a net result of these effects, the empirical Gini coefficient has a minimum of 0.5238 in 1990 and then starts to grow again, reaching 0.5266 in 2005.

Such defining model parameters as the Pareto law index (1.35) and the ratio of the efficiency of money earning in the Pareto zone relative to that predicted by the model (1.33) are well calibrated by the empirical PIDs and Gini coefficient. Our microeconomic model is very sensitive to these parameters.

The empirical Gini curves converge to the predicted one. Asymptotically, the empirical curves should collapse to the theoretical one when all the working age population will obtained an appropriate definition of their incomes. This convergence should be seen more clearly in the age dependent PIDs, where the portion of population without income decreases with work experience. For example, in the age group between 45 and 54 years this portion increased from 0.78 in 1960 to 0.94 in 2005. Hence, the portion was consistently larger and changed less than that for the working age population. One can expect a lower difference between the two empirical estimates and a better prediction.

The Gini coefficient is a crude and secondary measure of inequality for economics as a science. It could be useful for social and political discussions as a relative and operational measure without any specific meaning of its absolute value. What is important and has a primary significance for scientific models are the PIDs, which demonstrate a fixed hierarchy during a very long period between 1947 and 2005. (It is very unlikely that this hierarchy will be destroyed in the near future.) The shape and the evolution of the measured PIDs are well predicted for the whole period between 1947 and 2005.

The Census Bureau focuses its attention on the Gini coefficients related to the measurements of income inequality at a family and household level. Corresponding coefficients change over time and are presented as evidence in favour of the increasing economic inequality in the United States. Our estimates of the Gini coefficient for the PIDs, both empirical and theoretical, demonstrate that the inequality is not changing so dramatically. Therefore the Gini coefficient associated with households should be affected primarily by some changes in their structure.

### §1.9. Modelling the evolution of age-dependent Gini coefficient between 1965 and 2006

Understanding and modelling of the age-dependent personal income distribution deserves special attention. Dramatic changes in the shape of PID are observed with age. In §1.5, we successfully modelled the age-dependent PIDs in the United States for the period between 1994 and 2002. Our microeconomic model quantitatively describes the evolution (with age and over time) of each and every personal income as a function of individual capacity to earn money and real economic growth. The sum of all personal incomes predicted by the model builds a macroeconomic model. The modelling of the age-dependent PIDs was not accompanied by an explicit estimation of the level of income inequality.

Slight changes observed in the overall PIDs and the evolution of the *Gini* is related to economic growth and changes in the age structure. We have also demonstrated that empirical estimates of Gini coefficient converge to theoretical ones when all working age population has income. We have suggested that such convergence might be clearly observed in age-dependent PIDs since the portion of population without income decreases with age.

The age-dependent PID in the youngest group is characterized by large differences from the overall PIDs. Obviously, all individuals start with zero income and the initial part of personal income time history is close to exponential growth. In the mid-age groups, PIDs are similar to the overall PID. In the eldest age group, PID is also different and is closer to that in the youngest group. Accordingly, the Gini coefficient undergoes a substantial change from the youngest to the oldest age groups.

The purpose of this paragraph is to present accurate estimates of Gini coefficients associated with the age-dependent PIDs published by the US Census Bureau. We also model the evolution of *Gini* in various age groups between 1967 and 2005, i.e. during the period where the estimates of total personal income in each of these age groups are available.

The portion of population with income varies over age and time. These variations might affect the estimation of Gini coefficient associated with personal incomes, as demonstrated in §1.8. In the youngest age group between 15 (14 before 1987) and 24 years of age, only from 65% to 80% reported some income, as presented in Figure 1.9.1. Corresponding curve has a peak in 1979. Since then, the portion of people with income in this age group has been decreasing. This effect, obviously, needs a thorough examination and might be induced by the appearance of some new actual sources of income, which were not included in the contemporary CPS questionnaires. An increasing level of intra-family income redistribution is a potential mechanism to consider.

Figure 1.9.1 demonstrates that the portion of people with income increases with age before reaching its peak and then falls again. In 2005, the largest portion of around 98% was measured in the group between 55 and 64 years of age. This observation is consistent with the fact that the critical work experience, $T_{cr}$, in our model was moved in this age group. Between 1967 and 1977, the curves for all age groups, except the youngest one, were converging; and after 1979 the scatter was almost constant. An important observation is the presence of a step in all time histories (except the youngest one) between 1977 and 1979. According to the Census Bureau, this step is related to the introduction of new income definitions and significant changes to the CPS methodology. In average, this step is of 10 percentage points. For example, the portion of population with income in the working age population as a whole jumped from 83% in 1977 to 92% in 1979. One can expect that further elaboration of income definition will finally result in the 100% participation in income distribution. There should be no persons without income.

For people of 44 years of age and above, the portion with income is more than 95%. Therefore, in corresponding age groups, the difference between the Gini coefficient associated with people having income and that associated with the working age population as a whole has to be the smallest among all age groups. These age groups provide the best opportunity to test our model because almost everybody has some reported income, which might be biased by inaccurate definition, however.

The procedure of Gini coefficient estimation is detailed in §1.8. The main amendment consists in the dependence of index $k$ on age, as Figure 1.9.2 demonstrates. The evolution of the Pareto law index (slope) with age is as follows: $k$=-1.91 for the age group between 25 and 34 years; $k$=-1.48 between 35 and 44; $k$=-1.38 between 45 and 54; $k$=-1.14 in the age group between 55 and 64. It is clear that $k$ declines with age. Obviously, smaller index $k$ corresponds to a larger population density at higher incomes and a larger Gini coefficient. The decrease in $k$ deserves a special study because it should be inherently linked to some age-dependent dynamic processes above the Pareto threshold. We limit our analysis by the empirical findings, however.

The declining $k$ is a specific feature of the age-dependent PIDs which should be incorporated in our model. In §1.8, we found that $k$=-1.35 for the population of 15 years of age and over, i.e. within the range of its change with age. It is not excluded, however, that the age-dependent and overall $k$ might also undergo some changes over time. The latter index may vary just because of the changing age pyramid, i.e. varying input of various ages to the net $k$. For the empirical estimates of the Gini coefficients carried out below, the observed variation in this index plays insignificant role because we use actual income distributions. For theoretical estimates, the Gini coefficient might be overestimated for the youngest age group and underestimated in the oldest age group when one uses $k=-1.35$ everywhere.

As mentioned before, the Census Bureau presents two versions of PID – for total working age population and for that with reported income. We have calculated empirical *Gini* in several (fixed) age groups between 1967 and 2005. Figure 1.9.3 displays its evolution in all groups except in the youngest one. The latter group is characterized by severe variations in methodology and definition of income. This makes it impossible to distinguish actual and artificial features in the evolution of *Gi*. The curves associated with all people aged in given ranges are marked "all", and those including only people with incomes – "w/income". The major revision to income definition between 1977 and 1979, which dramatically increased the portion of people with income, induced sharp decrease in the curves named "all", and opposite changes in the curves "w/income". For obvious reasons, the Gini coefficients for people with income are systematically lower than those for the entire population. The curves in Figure 1.9.3 have to be predicted by our model.

Before 1977, the portion of population without income was big enough to introduce a significant bias in the estimates of Gini coefficient. It was overestimated for the entire population and was underestimated for the population with income. Same effect is observed for the age-dependent Gini. Before 1977, one can observe large changes over time. After 1977, all curves are approximately horizontal, with only a slight decline. Hence, one can expect large deviation between these empirical curves and theoretical ones before 1977.

The accuracy of theoretical estimates of Gini coefficient is related to the quality of PIDs' prediction. Figure 1.9.3 demonstrates that the Gini coefficients for the age groups over 34 years vary in a narrow range. This observation presumes that underlying PIDs are very similar. We have already demonstrated that the PIDs for the entire working age population (with income) for the years between 1967 and 2005 collapse practically to one curve when normalized to populations and nominal GPI (instead of GDP). Real GDP drives two key parameters in our model: critical work experience, $T_{cr}$, and the size of earning tools, $Λ(t)$. However, when GPI is not equal to GDP (the equality is assumed in the model) one should use the former variable for the normalization of the PIDs. The GPI/GDP ratio has been varying through time since the start of the CPS.

Figure 1.9.4 (similar to Figure 1.8.3) presents the evolution of various measures of mean income (i.e. GPI per capita) using: GDP; GPI reported by the BEA; and GPI reported by the Census Bureau, as estimated in the annual CPS. Two population estimates are used for calculations of these mean values – total working age population (all) and people reporting income (with income). According to current income definitions, the GPI reported by the BEA is larger than that estimated by the CB because the former includes additional sources of income.

In the case of age-dependent PIDs, one should separately estimate total personal income in each age group. Accordingly, Figure 1.9.5 presents the evolution of mean personal income in various age groups. There are two cases shown: for all people of given age (including those with no income) and for people who reported income. (By definition, mean personal income is the ratio of total personal income and relevant population.) One can observe significant differences between the youngest people and those in the groups with the largest mean income. These curves, obtained from the estimates provided by the Census Bureau, are used to normalize relevant age dependent PIDs.

Before normalizing the age-dependent PIDs to total income and population one needs to reduce them to the same units of measurements; originally, the PIDs are obtained in income bins of varying width. For example, Figure 1.9.6a displays the PIDs for the age group between 35 and 44 years in 1967 and 2005. Income bins are not uniform in 1967 creating local troughs and peaks. In 2005, income bins are uniform between $0 and $100,000. Obviously, the number of people in a given bin depends on its width and position in the distribution. As discussed early in this Chapter, a reasonable way to reduce these inhomogeneous distributions to the same units is to divide the number of people in a given bin by its width. This mathematical operation defines population density, i.e. the number of people per $1 at a given income level. Figure 1.9.6b depicts the PIDs (shown in Figure 1.9.6a) normalized to the width of relevant income bins. The troughs and peaks are essentially smoothed in the density curves. It is likely that the true population density distribution can be represented by an exponent undergoing a smooth transformation into a power law function near the Pareto threshold.

Finally, we have population density curves, which are defined in the same units. To reduce the curves to one dimensionless scale, we normalize them to the total population and to the increase in total personal income over years in relevant age intervals, as defined in Figure 1.9.5. We expect that the normalized curves should collapse to one within the bounds of uncertainty related to measurement errors. Figure 1.9.7 displays the normalized PIDs in various age groups for years 1967, 1993, and 2005. There is no significant difference between the curves except that in the age group between 15 and 24 years of age, where the data are available only since 1974. Unfortunately, we have to exclude the latter age group from the modelling due to very high uncertainty in income measurements. It is likely that any conclusion drawn from this group would be severely biased. The overall PIDs are also presented and match the expectation.

Apparently, the similarity between the normalized PIDs results in practically constant Gini coefficients in all age groups between 1967 and 2005. On the other hand, this similarity supports our basic assumption that relative distribution of personal income has not been changing over time not only overall, as shown in §1.4, but also for any given age above 15 years. One can conclude that there exist internal (economic, social, etc.) forces, which always return personal income distribution to its fixed shape. In other words, PID is an invariant in the US economy. This is an observation, not an assumption.

*Comparison of observed and predicted Gini coefficients*

To obtain theoretical estimates of the age-deponent Gini coefficients we start with the modelling of corresponding PIDs. Following our analysis of the observed age-dependent PIDs in Section 2 we have predicted PIDs in the same age groups. The model is characterized by a resolution of 1 year of age and 1 year in calendar time. Therefore, we have to aggregate all personal incomes in the age groups predefined by the Census Bureau. The start year of the model is 1967 with the following defining parameters: $\alpha_0=0.071$; $T_{cr}(1967)=32.0$ years; $M_P(1967)=0.43$. Index $k$ is taken for given age groups from the empirical estimates in Figure 1.9.2. Other parameters are the same as in §1.8. The age distribution was reported by the Census Bureau, and thus, is prone to future revisions. For selected age groups, such revisions may reach several percentage points. This might result in slight deviations in the predicted Gini coefficients.

Figure 1.9.8 depicts predicted and observed PIDs for the age groups between 24 and 35 years of age, between 45 and 54 years of age, and for the entire population over 15 years of age. For the narrow age groups, the PIDs measured in 1993 were chosen, and for the whole working age population the year of 2005 was modelled. Corresponding indices are those estimated empirically and are as follows: $k=-1.91$; $k=-1.38$, and $k=-1.35$. These values precisely fit the slope in relevant PIDs' above the Pareto threshold. In the low-income zone, the best fit is observed for the whole population. This is likely the result of a better resolution in the entire population curve at lower incomes in 2005. In 1993, the resolution at low incomes was poor. This was one of the reasons for new questionnaire and methodology introduced in 1994. The number of income bins underwent a dramatic increase from 23 (including the open-end one for the highest incomes) to 42. All in all, the observed and modelled PIDs demonstrate very good similarity.

The microeconomic model predicts the evolution of each and every personal income. Therefore, it allows the prediction of an exact Gini coefficient for a given set of defining parameters because the construction of an exact Lorenz curve is possible. The empirical Gini coefficients were obtained separately for the whole working age population and for people with income. These empirical coefficients provide only some estimates of the range, where the true age-dependent Gini coefficients are likely to reside.

Figure 1.9.9 presents the evolution of the observed and predicted Gini coefficient in four age groups and for the whole population over 15 years of age. For the sake of simplicity, we have predicted $Gi$ using the same index $k=-1.35$ for all ages. In the age group between 25 and 34 years of age, the predicted curve is close to that obtained for the entire population in this group. Because actual index is $k=-1.91$, there is a slight overestimation of the predicted coefficient, but it still resides between the empirical curves. Since 1994, the predicted curve has been deviating from the curve for the whole population and approaching that for population with income. This might be an effect of a higher resolution related to the introduction of new income bins.

In the age group between 35 and 44 years of age, the empirical curves are closer to each other. The predicted curve stays between them, but much close to the curve for population with income. In the age group between 45 and 54, where theoretical index $k$ is close to actual one, the predicted curve reproduces the decline in both empirical curves observed after 1983 and lays much close to the empirical curve for population with income. This is likely that the true Gini coefficient in this age group is consistent with the predicted one. In the age group between 55 and 64, the predicted curve is also close to that for the working age population, but still between the empirical curves. As expected, the level of income inequality in this age group is larger than in any other age group.

It is worth noting that the gap between the empirical Gini curves is between 0.02 and 0.03. The gap between the predicted and the closest empirical curve is usually less than 0.01. This is less than the uncertainty of the estimation of Gini coefficient as related to the discrete representation of the observed PIDs.

In all age groups, the level of personal income inequality, as expressed by Gini coefficient, has been decreasing (with small local peaks) since 1967. This empirical and theoretical observation is especially important for the age groups above 45 years, where the portion of population with income is close to 100%.

The predicted and measured curves demonstrate that the true Gini coefficient is definitely age dependent. Figure 1.9.10 displays two empirical curves of Gini dependence on age obtained in this study for 1967 and 2005, two theoretical curves predicted by our model for the same years, and a curve reported by the Census Bureau. Due to low resolution and large measurement errors in the youngest and oldest age groups we limit our illustration to the age between 25 and 65. In this range, all curves are very close. However, the CB's curve goes beyond the limits and demonstrates that there is a turning point at the age between 65 and 70. Our model supports this observation and the average income for people above the critical age, $T_{cr}$, falls exponentially with age. This fast decay is also reflected in a severe drop in the number of people with income above the Pareto threshold and corresponding decrease in Gini coefficient.

This study was primarily carried out for validation of the microeconomic model defining the evolution of personal incomes in the United States. In previous paragraphs, we have revealed some problems with income definition, which did not allow a comprehensive description of the overall PID. The most important problem was that a large portion of population did not report any income. Another problem is a poor resolution before 1977.

In the model, everyone is assigned a non-zero income. This discrepancy may results in a significant deviation between observed and predicted Gini coefficients. The age-dependent PIDs allow overcoming this discrepancy because the portion of population without income is very low (~2%) for ages over 45 years. Therefore, one could assume a more precise prediction of the Gini coefficient in these age groups. This paragraph confirms the assumption: the evolution of Gini coefficient for the years with a good PID resolution was accurately (<0.005) predicted.

## §1.10. Inequality estimates: Census Bureau vs. Internal Revenue Service

In previous paragraphs, we have found that PID in the United States has not been significantly changing in relative terms since 1947. As a consequence, the Gini coefficient has been varying in a very narrow range around 0.51. Hence, the inequality in personal incomes has not been growing, as many economists report using income estimates from the Internal Revenue Service. We suppose that the estimates made by the Census Bureau are valid because they are consistent through time and well described by our microeconomic model which also demonstrates its high predictive power in geomechanics.

Then a natural question arises. What is the problem with the IRS based measures of income, which result in changing inequality as expressed by the Gini coefficient? This paragraph develops a simple answer - these inequality measures are based on income definitions allowing floating low-end income threshold. In other words, the portion of population used by the IRS for the estimation of income inequality fluctuates randomly or according to some predetermined rule.

The effect of changing population basis due to numerous revisions to income definition is also observed in the CB's income data. The portion of people with income severely changes over time, as discussed in §1.2. It was increasing in the 1960s and 1970s due to a strong growth in women's participation rate. It has been falling since 1990, however. When people without income are included in calculations of income inequality, the Gini coefficient actually has been intensively falling since 1947 due to a strong growth of the portion of people with income. So, one can conclude that the driving force behind the increasing personal income inequality, as reported by the IRS, likely consist in biased measurements and inconsistent definitions.

It is of principle importance for the current study that despite the changing income definition and corresponding population basis the estimates of income inequality were not changing in the group with non-zero income. This observation contradicts the changing inequality as obtained from the IRS data. Only quantitative analysis can resolve this conflict. The resolution of the conflict is the purpose of this paper. Because the results showing increasing inequality are quantitative, it is feasible to exactly show the reasons for the observed contradiction and indicate caveats in the Krugman's approach.

Original (real gross) income distributions are reported by the IRS. Table 1.10.1 provides the numbers of people in predefined income bins (in chained 1990 dollars) for 1990 and 2004. Also listed are widths of income bins, which are used below for calculations of population densities, and centers of income intervals (see §1.4 for details of the normalization procedure).

These income bins are fixed over time and not adjusted for the growth of real economy and the increase in working age population. The lowermost income bin contains zero income and net losses. The highest income bin includes those income reports, which exceed $10,000,000. This is an open-end income bin without the estimate of average income. Fortunately, only several thousand people have incomes above $10,000,000. First thousands is not the number which could influence the overall income inequality estimate. Moreover, these richest people also distributed according to the Pareto law, i.e. measured and estimated total income in this bin should not differ much.

The IRS income tables provide a basis for current estimates of economic inequality in the USA. Conventional conclusion about income inequality is very consistent among economists – the inequality has been rising during the last 20 years. At first glance, this conclusion is quantitatively correct, but we will argue that it is wrong due to potential inaccuracy in methodology and unacceptable misinterpretation of quantitative results.

Figure 1.10.1 compares income distributions for 1990 and 2004 listed in Table 1.10.1. Since the income bins presented in the Table are of increasing width one can observed some spikes in the distributions. These spikes are, obviously, related to those income bins, which are wider than their predecessor. For example, the bin between $25,000 and $30,000 ($5000-wide) is followed by the bin between $30,000 and $40,000, ($10000-wide). Therefore, one can expect a larger number of people in the latter bin than in the former one. This effect is clearly observed in Figure 1.10.1, where the enumerated populations are assigned to the centers of corresponding income bins. Here and below, we prefer to use the log-log coordinates in order to present highly changing population (and population density) distributions in the range of income spanning seven orders of magnitude. The lowest income bin, corresponding to zero and negative (loss) reported incomes, is artificially associated with $1 income. The bin with incomes above $10,000,000 is not shown because of the absence of mean income estimate in this bin.

One can easily derive an obvious conclusion from Figure 1.10.1- there are more people with lower, middle and high incomes in 2004 than in 1990. This is a mechanical result of increasing population – more and more people get income as working age population is growing.

One should normalize the curves to total population (with reported income) in given years in order to obtain population independent results. In addition to this normalization one can use population density instead of original population estimates in width changing bins. Income bin width would not be a problem for constant widths. Therefore, when the measured populations are normalized to corresponding income bin widths one obtains density of population as a function of income, i.e. the number of people per $1 bin. As before, we assign the obtained population densities to the centers of corresponding bins. The assignment of the

density readings to the centers can potentially to disturb the observed curves when income bins are very wide and income distribution is described by a power law.

Figure 1.10.2 depicts the population density curves obtained after the normalization of the curves in Figure 1.10.1 to the number of people with (IRS reported) income, which includes also the people without income and those with incomes above $10,000,000. The normalized valued are divided by widths of corresponding income bins. At higher incomes, both curves accurately follow the Pareto law distributions, which are represented by straight lines in the log-log coordinates. The most prominent features of the obtained curves are the increasing deviation between them staring from $62,500 (1990 dollars) and the fact that they are practically indistinguishable below this income threshold. As a rule, modern studies of increasing income inequality find their conclusions in these population density curves. Seemingly, the curves demonstrate that the portion of population with higher incomes has been growing since 1990 and as a consequence the inequality has been increasing.

This is not the end of the story, however. There is one question left. What is the effect of increasing gross personal income on the observed population density distribution? Actually, total personal income grew from $3.41E+12 to $4.70E+12 (1990 dollars) between 1990 and 2004. So, larger gross personal income is a possible reason for the increased number of people with higher incomes. Then the same level of population density at lower incomes might be an artifact associated with inaccurate measurements at very low incomes or exclusion of some categories of income from the IRS definition. This may be a big problem for the compatibility of estimates over time, as the Census Bureau discusses in methodological documents.

What does really happen when dimensionless income distribution is used instead of that obtained in absolute income values? Two curves in Figure 1.10.3 represent those in Figure 1.10.1, which are additionally normalized to the gross personal income reported by the IRS. Income scales in 1990 and 2004 are also normalized to these incomes and represent now dimensionless portions of total income. As a result, the widths of the income bins in 1990 and 2004 also became different since relevant income scales were compressed by different factors. Also, the centers of the original income bins which were the same in 1990 and 2004 are now shifted relative to each other.

The curves in Figure 1.10.3 represent population density as a function of dimensionless income and practically coincide at higher incomes and diverge at low incomes. (There is a deviation at very high incomes, but it is much smaller than in Figure 1.10.2 and hardly can affect the estimate of Gini coefficient.) Therefore, density of population at higher incomes, as measured in dimensionless portions of total income, is practically the same in 1990 and 2004, considering the efficiency of the IRS work and possible measurement errors. All in all, rich people have the same (within the uncertainty of income measurements) portion of gross personal income. In relative terms, these high-income people in 2004 are not richer than in 1990.

In the low income zone, the distributions are diverging with time. There are several explanations of this observation. First, this is the results of some real (objective) processes of income redistribution between rich, middle class and poor people in the US. This is a common opinion in economic literature and media. Because of the changes in the measured personal income distributions one needs some driving force explaining the process. Second reason for the changing distribution is not related to increasing income inequality but is associated with lower (and varying) accuracy of income measurements at smaller incomes (possibly driven by definitions).

In the case of actual income redistribution process, one can expect some consistency between measures of income inequality provided by different agencies. For example, the inequality estimates provided by the US Census Bureau, which include many taxable income sources and some extra sources as well, would be expected to confirm the IRS results. This is not the case, however, as shown in §1.8.

So, there is a conflict between quantitative estimates based on the IRS and Census Bureau data sets. Which measure is a more reliable one? Let's consider two aspects of relevant income distributions – population basis and total personal income reported by the IRS and Census Bureau. It is likely that larger portions of working age population and real GDP potentially provide more reliable estimates of inequality.

Figure 1.10.4 presents the evolution of the portion of working age population with income as reported by the IRS and Census Bureau between 1990 and 2004. The number of people with IRS reported income is about 113,000,000 in 1990 and 132,000,000 in 2004. The Census Bureau reported ~181,000,000 in 1990 and 205,000,000 in 2004 from total working age population of ~194,000,000 in 1990 and ~230,000,000 in 2004. Corresponding portions are 0.93 and 0.58 in 1990, and 0.89 and 0.57 in 2004, respectively. Therefore, the CB surveys cover a larger portion of population with income measurements.

Moreover, the surveys include taxable incomes as a subset of all measured incomes. Figure 1.10.5 illustrates the differences in income definitions between the IRS and CB. Gross personal income measured by the CB is ~70% of real GDP – falling from 73% in 1990 to 67% in 2004. At the same time, the IRS reports only from 58% of real GDP in 1990 to 54% in 2004. It is also important that the IRS curve is of a higher volatility. This observation is potentially related to changes in (taxable) income definition.

Apparently, the IRS covers a smaller portion of population and gross personal income than the Census Bureau. Basically, the IRS reports some income subset relative to the Census

Bureau. Therefore, the observed difference between economic inequality estimates based on IRS and Census Bureau data is likely results from lower reliability of the IRS estimate. IRS income reports cannot provide a consistent measure of personal income.

## §1.11. Conclusion

This Chapter is very important for understanding the behavior of a developed economy as a physical system. In classical mechanics, there exist several well-known invariants, i.e. a property of a closed system, which does not change under certain transformations. The invariants facilitate quantitative description of very sophisticated mechanical systems. One of famous invariants is associated with the fundamental law of energy conservation. No process in a closed physical system can violate this law and all individual components of the total energy must sum up to its constant value, whatever happens.

No economy can be considered as a completely closed system because it is usually exposed to such external forces as weather, export-import, and immigration, and internal processes like demographic booms and declines. Nevertheless, there is a quantitative parameter in the US economy, which could be treated as an invariant characterizing the economy as a whole. This is the personal income distribution, as it measured by the US Census Bureau in the Annual Social and Economic Supplement to the Current Population Survey.

This observation goes beyond the representation of the distribution in econophysics, where standard methods and models of statistical physics are used (Dragulesku and Yakovenko, 2001). It is exciting to interpret economic variables in terms of "temperature" and "energy". However, it does not add much to the understanding of internal and external forces driving the process of personal income distribution and evolution. Econophysics provides rather a convenient interpolation of quantitative observations than a "first principle" model.

On the contrary, we have developed a model which is based on "first principles", as presented in §1.3. It does not approximate or interpolate observed data at each time step, but accurately predicts the dependence of all defining parameters on real GDP per capita and age structure. An essential feature of the model is its simplicity: there is only one first order ordinary differential equation completely defining individual, but irrelevant to personalities, income trajectories. Also, the model has deep roots in natural sciences that suggest that economic activity is just a natural process governed by laws inherently following physical laws. Economics often considers human behavior as unpredictable and even stochastic. A good analogue of such a system in physics is an ensemble of gas particles in a box. Nobody can predict the trajectory for a given particle. One can predict, however, the most probable number of particles in a given energy range and such macro parameters as temperature and pressure. Similarly, our model can exactly predict the number of people in a given income range, but does not predict their names.

The orthodox economics treats economic systems as chaotic heaps of unshaped bricks: external or internal perturbations induce unexpected and random reactions such as avalanches in stock market, inflation creep, swells of economic bubbles, and troughs of recession. There is a term invented to express the nature of these unpredictable phenomena – "exogenous stochastic shocks". This is the core component of economic development – stochastic but inevitable fluctuations. As a part of the mainstream economics, the economics of personal incomes follows this tradition and presumes the existence of some stochastic exogenous forces driving the changes in the distribution of income and related inequality. As a consequence, the trends observed in the distribution and inequality are treated as intrinsically stochastic ones.

Our concept denies that the evolution of personal income distribution, in particular, and developed economies as a whole is stochastic. We have found that the observed variations in the shape of PID in the US are small and entirely related to the changes in age structure and real GDP growth. In other words, the PID has an almost infinite rigidity to any other external of internal disturbances, at least to those occurred since 1947. Otherwise, the model would not fit the observations.

As a thought experiment, imagine a rigid construction like a concrete building. How does it react to various external forces like wind, changing atmospheric pressure, elastic waves from earthquakes, etc.? It needs no technical knowledge to conclude that such rigid systems always react in a consolidated way, i.e. internal bonds cause various parts of the system to interact and create counterforce balancing the disturbance and preserving the integrity. Otherwise, the system would collapse, as often happens when the amplitude of external forces exceeds relevant strength.

How does react the rigid system of personal incomes to internal and external forces? It is observed a cooperative reaction retaining the hierarchy of incomes! In other words, both real economic growth and inflation do not disturb the PID, in relative terms. A counter question can be raised then. Since people are the only source of economic growth, how the variations in macroeconomic variables are possible? One could explain a constant growth rate in per capita values as a result of some permanent onward motion or translation in technical terms. Really, a rigidly structured society where relative changes in the number of income positions are prohibited should produce a constant flow of economic efforts through relevant economy and should advance at a constant pace.

If the internal human structure of a developed economy can not be the source of the observed fluctuations in economic growth, what is the driving force? The answer is obvious - the changes in boundary conditions, i.e. primarily the changes in the number of young people joining the economy. Chapters 2 through 6 present quantitative evidences in favour of this assumption.

Chapter 1 evidences that personal income distribution in the US can be exactly predicted by our model together with the time history of the PID, as observed since 1947. Step by step, we modelled various aggregate and fine quantitative features from the overall distribution to the age-dependent Gini coefficient. At each step, we consistently extended the data set of personal income or/and the length of the studied period. A special attention has been paid to the Pareto law at the highest incomes.

In §1.4, we have revealed the rigid hierarchy of personal incomes as a whole. In order to prove the existence of the fixed income structure, we had to normalize the PIDs measured by the Census Bureau to corresponding gross personal income and total working age population. The normalized overall PIDs between 1994 and 2002 are shown to collapse to one curve unveiling the underlying hierarchy. This fixed hierarchy defines the first ever economic invariant. Taking the classical mechanics as an example, one might construct an economic theory around such an invariant, with intrinsic links expressed in equations similar to homonymic relationships of mechanics.

First and most important conclusion is that the existence of the invariant PID evidences that the US economy is a *physical* system and the evolution of economic state, as defined by some measured variables like GPI or GDP, obeys some strict laws. Second, there exist economic variables linked by deterministic relations which can be expressed mathematically. This would characterize economics as a hard science with all the arsenal of ideas and methods developed in physics to be potentially applicable. Third, it builds a bridge between micro- and macro levels, the latter being a simple aggregation of the former, as in thermodynamics. It would mean that there is no macro-variable, whose behavior is not completely determined by the quantitative properties of micro-objects. And those macro-variables which cannot be reduced to micro-level are void.

In §1.5, we have validated the microeconomic model by demonstrating accurate predictions of the shape and evolution of PIDs in various age groups. The change in the shape over age is so dynamic that provides a very good resolution of relevant time history: from practically exponential fall in the youngest and eldest age groups to a piecewise function in the mid-age groups. All these features have been successfully modelled during the relatively short period between 1994 and 2002. For the age-dependent modelling we have used the same defining parameters as those for the overall PIDs. This is the best quantitative test and validation of the model – one can accurately predict beyond the set of data formally used for the estimation of empirical parameters and coefficients.

Therefore, the microeconomic model quantitatively describes the evolution (with age and over time) of each and every personal income as a function of three measurable parameters: the capacity to earn money, the size of earning means, and real GDP per capita.

In §1.6, we have modelled two aggregate variables – the mean and median income and their dependence of on work experience. For these variables, quantitative estimates are available from 1967. So, we extended the modelled period by 27 years back in the past compared to the previous paragraphs. The entire period since 1967 was described by a model with the defining parameters obtained for a shorter period and different data set. In this paragraph, we have also introduced a new feature to model - the PID at higher incomes is approximated by the Pareto law.

Disappointedly, a trouble has come from the side of income definition and methodology of the CPS. Our modelling has met significant difficulties related to the changes in the portion of gross personal income in GDP. Significant revisions to the CPS and the population estimates after decennial censuses create artificial steps in the GPI and, thus, in mean and median income. The latter income represents a more robust variable due to lower sensitivity to higher incomes. As a result, its dynamics is relatively better predicted by the model. Overall, the dependence of mean and median on work experience and the evolution of the dependence with time both validate the model in logical and historical sense.

Paragraph 1.7 addresses a different type of problem, which is a consequence of the self-organized similarity (SOC) reining at higher incomes. The Pareto distribution of incomes, being a manifestation of the SOC, starts at a relatively high level of income and controls only ~10 percent of working age population. Internal mechanisms of the Pareto distribution are beyond the scope of the microeconomic model and we use its properties as measured. This might be a challenge for both economics and physics. Our model has successfully predicted the time history of the portion of people in the Pareto distribution as a function of work experience. Basically, the age groups with low mean income are characterized by lower portions than those with high mean income. And again, the microeconomic model has shown its capability to accurately predict the sought variables, and thus the overall input of high incomes.

This is a good place for a short speculation. What does the Pareto law says us about rich people? Is there any quantitative characteristic or property, which is definitely necessary for a person to become rich? Under our framework the answer for the second question is "no". One really needs to have a job with a large size of earning means and a high capability to earn money. Supposedly, the latter is more important because it is more difficult to change personality than job. But even these two factors do not guarantee the highest income in the top 1 per cent. To get the highest income one needs a property, which is not trained – good luck, as described by the Pareto laws as a probability function. Therefore, there is no individual quality making people super-rich. As a result, the choice is random and there is no fairness in the process. This conclusion does not deny the possibility of many rich people emerging from some tight group. For example, many Russian oligarchs have similar backgrounds. Such backgrounds were not associated with the desire of super- high incomes, however.

By definition, income inequality is a derivative from PID. Therefore, having an accurate PID one should not meet any difficulty to quantitatively predict income inequality. In §1.8, we have modelled the evolution of Gini coefficient for the overall PID since 1947. So, we have extended the modelled period by another twenty years. No further extension into the past is possible because of the absence of measurements before 1947 and the model has finally covered the entire period with CPS income reports.

The modelling has unveiled severe problems with the resolution of the CPS – there were too wide income bins for the estimation of the Gini coefficient. However, relevant overall PIDs evidence that the fixed income hierarchy has been observed from the very beginning of the CPS. There is no reason to assume that the hierarchy will fail in the near future because it has its roots in the ranking of people's capability to earn money. This feature is a solid one since it has come from centuries of economic interactions between human beings as economic agents. It should not fade away as a modern pop melody.

There are several important findings related to the estimation of empirical Gini coefficients associated with the US income distribution:

1. These estimates of Gini coefficient critically depend on definition of income.
2. The Gini coefficient associated with the whole population 15 years of age and over and that associated only with people with income *converge* with time as the portion of people without income decreases.
3. Resolution of the measured PIDs (i.e. a proportional coverage of population with income bins) and interpolation of the PIDs inside these bins influences the estimates of Gini coefficient.
4. The empirical PIDs practically collapse to one curve when normalized to the cumulative growth in nominal Gross Personal Income (GPI) for the studied period between 1967 and 2005.

Slight changes observed in these PIDs and the evolution of *Gi*-values are related to real economic growth and changes in the age structure. We have also demonstrated that the empirical estimates of Gini coefficient converge to theoretical ones when all individuals in working age population have income. Such convergence might be clearly observed in age-dependent PIDs, since the portion of population without income decreases with age.

Paragraph 1.9 has supported general findings of the previous paragraph and also has extended them to the data in various age groups. Both PIDs and Gini coefficients are successfully modelled for the years after 1965. The lower limit comes from the availability of the age-dependent gross personal income, which is necessary for the normalization of the PIDs.

As expected, the gap between the Gini coefficient associated with the entire working population in a given age interval and that associated with people reporting income converge with the decreasing portion of people without income. The true Gini coefficient had to be somewhere between these two estimates. In the group between 45 and 54 years of age, this portion is approximately 3% and the gap (in Gini coefficient) is less than 0.02. As the overall PIDs, the empirical age-dependent (density) PIDs collapse to practically one curve when normalized to cumulative growth in personal income and total population for the period between 1967 and 2005.

In all age groups, the model predicts slightly decreasing Gini coefficients between 1967 and 2005. The overall $Gi$ is approximately constant, however and minor changes are related to real economic growth and the changes in age structure.

The Pareto law index, $k$, undergoes significant changes over age: increases from the youngest age to approximately 67 years of age, and then drops. Such an evolution could be expected but its actual behavior deserves a deeper study.

The age-dependent PIDs demonstrate a fixed hierarchy during a very long period between 1967 and 2005. The shape and the evolution of the measured PIDs are well predicted for the whole period. This allows exact prediction of Gini coefficient and other measures of inequality, which are defined by personal income distribution.

In §1.10, we have discussed the observation related to the increase in economic inequality in the United States, which contradicts our model. We have demonstrated quantitatively that the estimates of income inequality associated with the Internal Revenue Service are not reliable. The principal problem of the IRS estimates is an intrinsic one to almost all income studies, which base their approaches on varying portions of a system as a whole. From physics, it is well-known that parts of a closed system are always characterized by high volatility of measured parameters and no relationship revealed for a part works for the system as a whole. In the case of the IRS, the problem is the loose boundary associated with highly volatile incomes of people in the low-end of income distribution. The loose definition of boundary condition and the volatility related to measurement errors, changes in definitions and/or improper reporting *must* result in the observed changes in income inequality. When the entire working population is considered no changes in income inequality are observed, as the case for the age group between 45 and 54 years of age evidences.

There is also a professional discussion and active area of economic research around the idea of the influence of income inequality on real economic growth. Overall, relevant results are controversial. Under our framework, the answer is obvious – income inequality is a secondary effect of the ranking in the capability to earn money. In turn, the ranking has come from the history of economic, social, psychological, etc. links between people. These links is the force shaping current income distribution. On the other hand, variations in the growth rate of real economic growth are inherently related to demographic variables through the rigid PID. In that sense, the US economic system is a self-consistent one and is driven internally by the PID, and externally by demography.

What our model can propose for other developed and many developing countries? This is a crucial question for our model to be further validated by empirical data. The only obstacle on this road is the absence of reliable data. When and where such data will become available, we will continue the modelling.

Summarizing all findings and discussions in one sentence we would like to conclude that the developed microeconomic model accurately describes the shape of the US PID and predicts its evolution during the past sixty years.

# References

Dragulesky, A., and Yakovenko, V. (2001). Exponential and power-law probability distributions of wealth and income in the United Kingdom and the United States. Physica A: Statistical Mechanics and its Applications, 299, 1-2 , 213-221

Galbraight, J. (1998). Created Unequal. A crisis in American Pay. The Free Press, New-York.

Lise, S., Paczuski, M., (2002). A Nonconservative "Earthquake Model of Self-Organized Criticality on a Random Graph", cond-mat/0204491, v1, 23 Apr 2002, pp. 1-4.

Neal, D., Rosen, S., (2000). Theories of the distribution of earnings. In: Handbook of Income Distribution, (Eds.) Atkinson, A. and Bourguinon, F., 379-427, Elsevier 2000.

Rodionov, V.N., V.M.Tsvetkov, I.A.Sizov. Principles of Geomechanics. Moscow. Nedra, 1982, pp.272. (in Russian)

West, K., Robinson, J., (1999). What Do We Know About The Undercount of Children? , Population Division, U. S. Census Bureau Washington, DC 20233-8800 August 1999 Population Division Working Paper No. 39

Yakovenko, V., (2003). Research in Econophysics. Cond-mat/0302270. Retrieved April 8, 2004 from http://www.physics.umd.edu/news/photon/isss24/Yakovenko_article.pdf

# Tables

Table 1.4.1. GDP growth rates from 1950 to 2002

| year | current dollar GDP growth rate | real GDP growth rate | per capita current dollar GDP growth rate | per capita real GDP growth rate | per capita current dollar GDP growth rate. age above 15 | per capita real GDP growth rate. age above 15 |
|---|---|---|---|---|---|---|
| 1950 | 1.099 | 1.087 | 1.077 | 1.065 | 1.085 | 1.073 |
| 1951 | 1.155 | 1.077 | 1.135 | 1.059 | 1.144 | 1.067 |
| 1952 | 1.056 | 1.038 | 1.038 | 1.021 | 1.046 | 1.028 |
| 1953 | 1.059 | 1.046 | 1.041 | 1.029 | 1.049 | 1.036 |
| 1954 | 1.003 | 0.993 | 0.985 | 0.976 | 0.992 | 0.983 |
| 1955 | 1.090 | 1.071 | 1.071 | 1.053 | 1.079 | 1.060 |
| 1956 | 1.055 | 1.019 | 1.036 | 1.001 | 1.043 | 1.008 |
| 1957 | 1.054 | 1.020 | 1.035 | 1.002 | 1.041 | 1.007 |
| 1958 | 1.013 | 0.990 | 0.996 | 0.974 | 0.999 | 0.977 |
| 1959 | 1.084 | 1.071 | 1.066 | 1.053 | 1.070 | 1.057 |
| 1960 | 1.039 | 1.025 | 1.023 | 1.009 | 1.027 | 1.012 |
| 1961 | 1.035 | 1.023 | 1.018 | 1.006 | 1.022 | 1.011 |
| 1962 | 1.075 | 1.061 | 1.059 | 1.044 | 1.054 | 1.040 |
| 1963 | 1.055 | 1.044 | 1.040 | 1.029 | 1.038 | 1.027 |
| 1964 | 1.074 | 1.058 | 1.059 | 1.043 | 1.057 | 1.041 |
| 1965 | 1.084 | 1.064 | 1.070 | 1.051 | 1.067 | 1.048 |
| 1966 | 1.096 | 1.065 | 1.083 | 1.053 | 1.078 | 1.048 |
| 1967 | 1.057 | 1.025 | 1.045 | 1.014 | 1.039 | 1.008 |
| 1968 | 1.093 | 1.048 | 1.082 | 1.038 | 1.075 | 1.031 |
| 1969 | 1.082 | 1.031 | 1.071 | 1.021 | 1.064 | 1.014 |
| 1970 | 1.055 | 1.002 | 1.043 | 0.990 | 1.035 | 0.983 |
| 1971 | 1.085 | 1.034 | 1.072 | 1.021 | 1.065 | 1.014 |
| 1972 | 1.099 | 1.053 | 1.087 | 1.042 | 1.078 | 1.033 |
| 1973 | 1.117 | 1.058 | 1.106 | 1.048 | 1.096 | 1.038 |

| Year | | | | | | |
|---|---|---|---|---|---|---|
| 1974 | 1.085 | 0.995 | 1.075 | 0.986 | 1.065 | 0.977 |
| 1975 | 1.092 | 0.998 | 1.081 | 0.988 | 1.072 | 0.980 |
| 1976 | 1.114 | 1.053 | 1.104 | 1.043 | 1.094 | 1.034 |
| 1977 | 1.113 | 1.046 | 1.102 | 1.036 | 1.093 | 1.028 |
| 1978 | 1.130 | 1.056 | 1.118 | 1.045 | 1.110 | 1.037 |
| 1979 | 1.117 | 1.032 | 1.105 | 1.020 | 1.098 | 1.014 |
| 1980 | 1.088 | 0.998 | 1.081 | 0.991 | 1.078 | 0.988 |
| 1981 | 1.121 | 1.025 | 1.112 | 1.017 | 1.109 | 1.014 |
| 1982 | 1.040 | 0.981 | 1.031 | 0.971 | 1.028 | 0.969 |
| 1983 | 1.087 | 1.045 | 1.076 | 1.036 | 1.074 | 1.033 |
| 1984 | 1.112 | 1.072 | 1.102 | 1.062 | 1.100 | 1.060 |
| 1985 | 1.073 | 1.041 | 1.064 | 1.032 | 1.061 | 1.030 |
| 1986 | 1.057 | 1.035 | 1.048 | 1.025 | 1.045 | 1.023 |
| 1987 | 1.062 | 1.034 | 1.053 | 1.025 | 1.051 | 1.023 |
| 1988 | 1.077 | 1.041 | 1.067 | 1.032 | 1.067 | 1.032 |
| 1989 | 1.075 | 1.035 | 1.065 | 1.026 | 1.066 | 1.027 |
| 1990 | 1.058 | 1.019 | 1.048 | 1.009 | 1.049 | 1.010 |
| 1991 | 1.033 | 0.998 | 1.023 | 0.988 | 1.025 | 0.990 |
| 1992 | 1.057 | 1.033 | 1.045 | 1.021 | 1.046 | 1.023 |
| 1993 | 1.050 | 1.027 | 1.039 | 1.015 | 1.039 | 1.016 |
| 1994 | 1.062 | 1.040 | 1.052 | 1.030 | 1.052 | 1.030 |
| 1995 | 1.046 | 1.025 | 1.036 | 1.015 | 1.036 | 1.015 |
| 1996 | 1.057 | 1.037 | 1.047 | 1.027 | 1.045 | 1.026 |
| 1997 | 1.062 | 1.045 | 1.052 | 1.035 | 1.051 | 1.034 |
| 1998 | 1.053 | 1.042 | 1.043 | 1.032 | 1.041 | 1.030 |
| 1999 | 1.060 | 1.044 | 1.050 | 1.035 | 1.049 | 1.034 |
| 2000 | 1.059 | 1.037 | 1.048 | 1.026 | 1.046 | 1.024 |
| 2001 | 1.032 | 1.008 | 0.998 | 0.974 | 0.997 | 0.974 |
| 2002 | 1.035 | 1.019 | 1.024 | 1.008 | 1.022 | 1.006 |
| Total increase | 35.69 | 5.67 | 18.94 | 3.01 | 17.55 | 2.79 |

Table 1.10.1. Personal income distribution according to the IRS

| Income bin | Width | Center | 1990 | 2004 |
|---|---|---|---|---|
| No adjusted gross income | | [1] | 904876 | 1854886 |
| $1 under $5,000 | 5000 | 2500 | 16478272 | 17039057 |
| $5,000 under $10,000 | 5000 | 7500 | 14952855 | 17211889 |
| $10,000 under $15,000 | 5000 | 12500 | 13922750 | 15889660 |
| $15,000 under $20,000 | 5000 | 17500 | 11543228 | 13056490 |
| $20,000 under $25,000 | 5000 | 22500 | 9572317 | 10990767 |
| $25,000 under $30,000 | 5000 | 27500 | 7838225 | 8567162 |
| $30,000 under $40,000 | 10000 | 32500 | 12282786 | 13309262 |
| $40,000 under $50,000 | 10000 | 35000 | 8837067 | 9928723 |
| $50,000 under $75,000 | 25000 | 62500 | 10944102 | 13635393 |
| $75,000 under $100,000 | 25000 | 87500 | 3276142 | 4934480 |
| $100,000 under $200,000 | 100000 | 150000 | 2329562 | 4213077 |
| $200,000 under $500,000 | 300000 | 350000 | 644027 | 1211221 |
| $500,000 under $1,000,000 | 500000 | 750000 | 130252 | 240876 |
| $1,000,000 under $1,500,000 | 500000 | 1250000 | 29060 | 61800 |
| $1,500,000 under $2,000,000 | 500000 | 1750000 | 11581 | 26977 |
| $2,000,000 under $5,000,000 | 3000000 | 3500000 | 15331 | 39047 |
| $5,000,000 under $10,000,000 | 5000000 | 7500000 | 3184 | 9625 |
| $10,000,000 or more | >10000000 | | 1522 | 5651 |

**Figures**

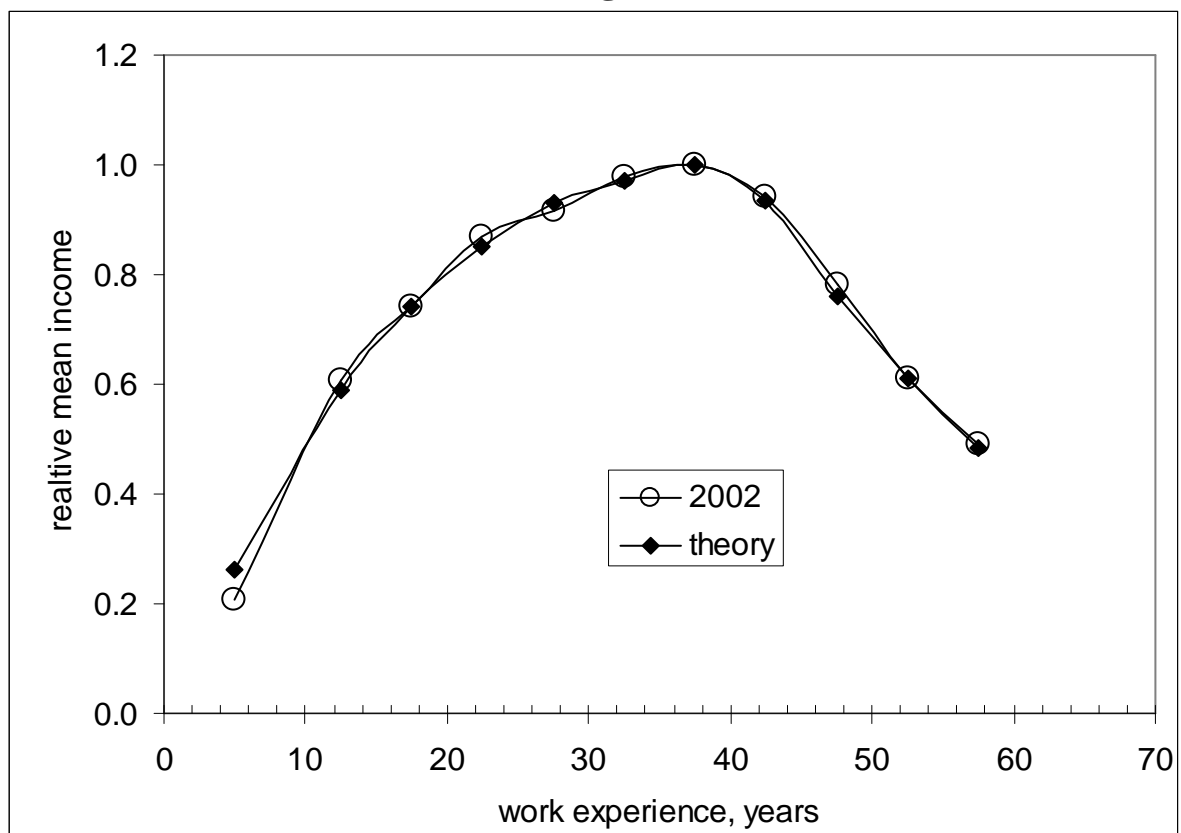

Figure 1.3.1. Approximation by two exponential functions of the normalized mean income dependence on work experience in the USA in 2002.

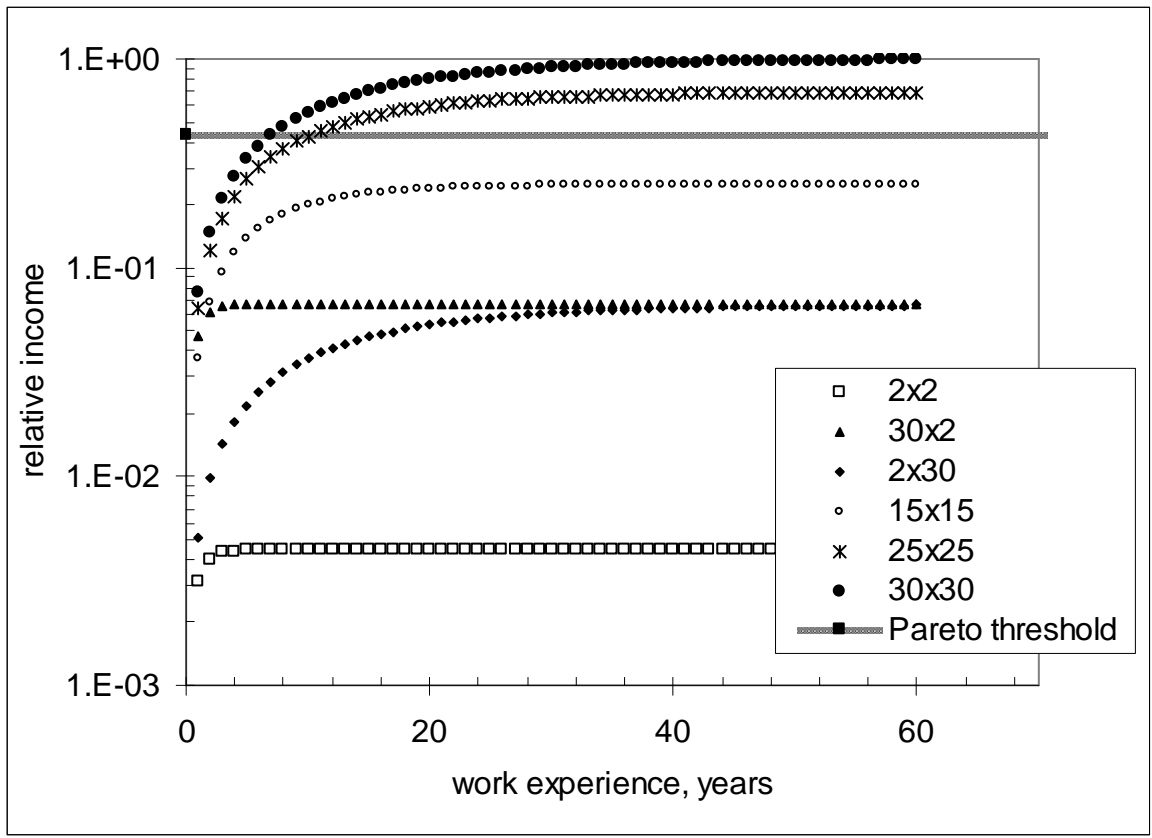

Figure 1.3.2. Evolution of personal income for various combinations of earning means size, *L*, and personal capability to earn money, *S*. The (dimensionless) Pareto distribution threshold is 0.43. Only people with high *S* and *L* can eventually reach the threshold. The duration of period needed to reach the maximum potential income depends on *L*. Compare cases 30x2 and 2x30. Because of smaller *L*=2, the first person reaches maximum much faster than the second person with the means of size 30.

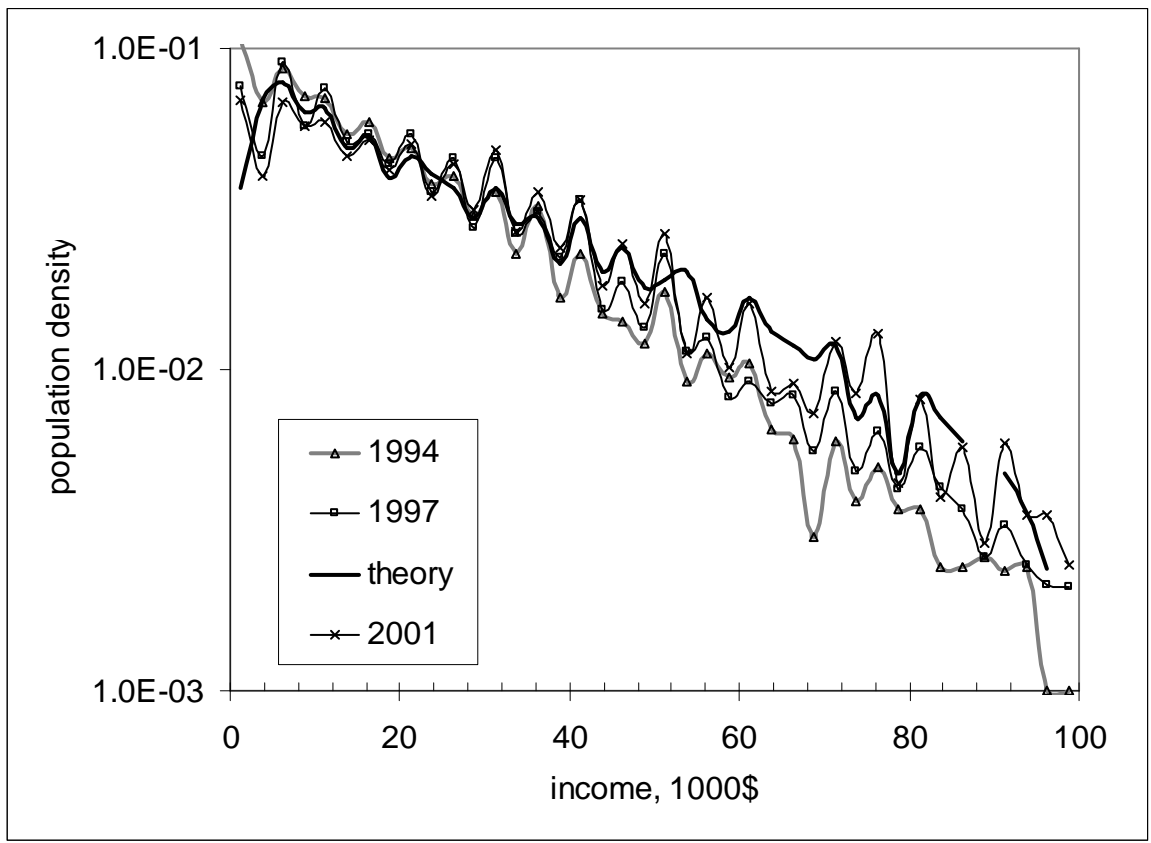

Figure 1.3.3. Comparison of the theoretical PID with integer $S$ and $L$ distributed evenly between 2 and 30 and those observed for the years 1994, 1997, and 2001.

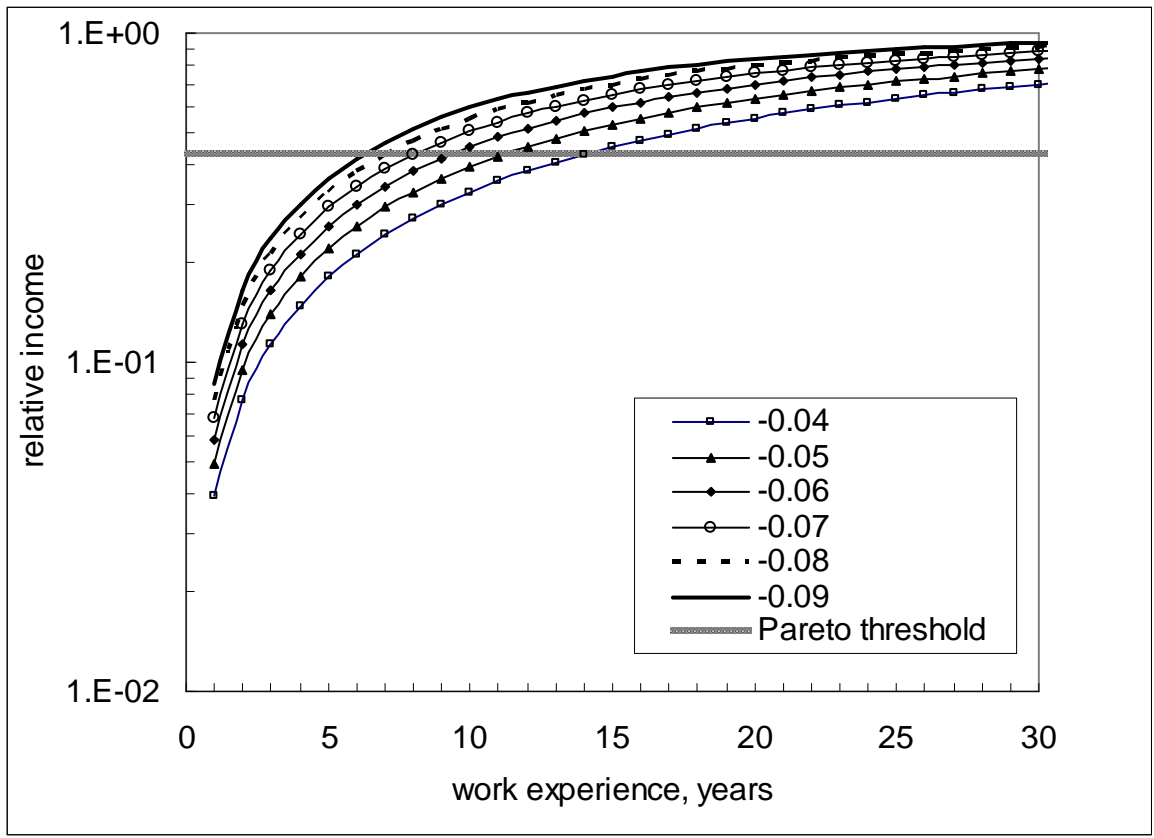

Figure 1.3.4. Evolution of personal income for various dissipation factors ☐. The earning means size $L_{29}$=30/30, and personal capability to earn money $S_{29}$=30/30. The Pareto distribution threshold is 0.43. The duration of period needed to reach maximum income depends on ☐: larger ☐ corresponds to shorter time needed to reach the threshold.

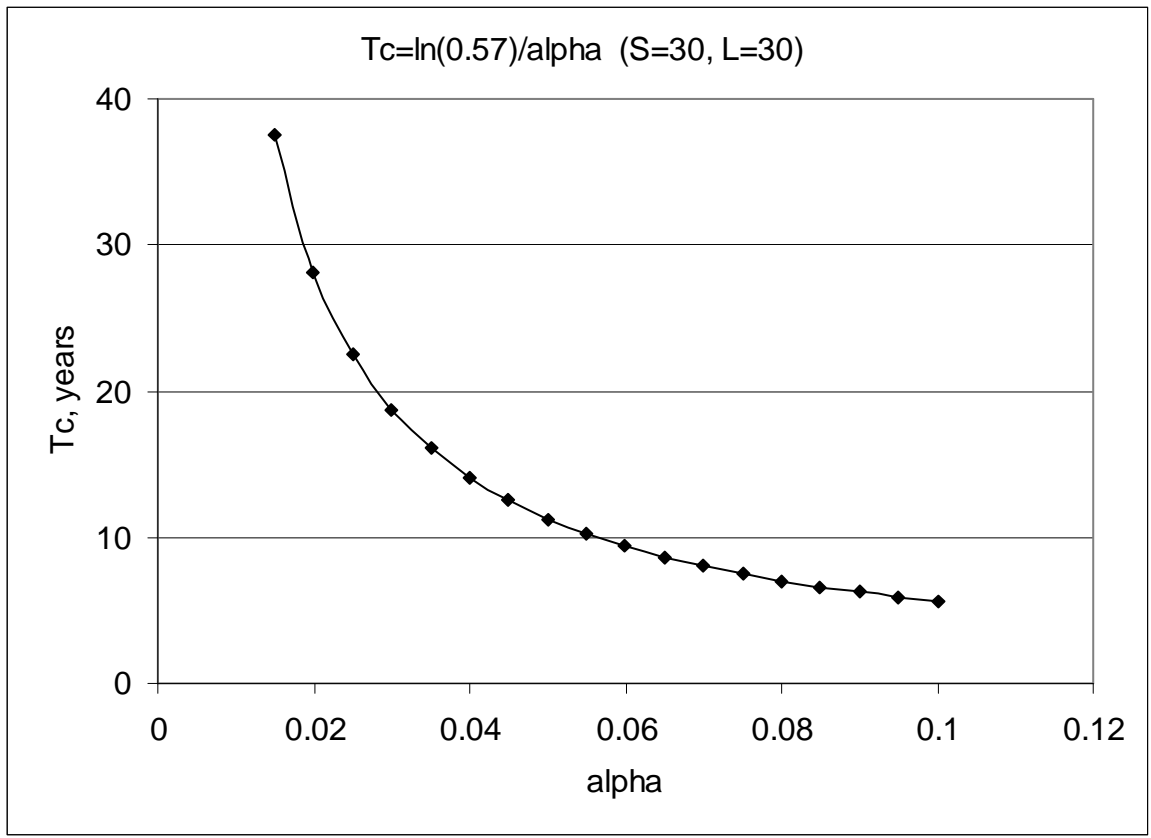

Figure 1.3.5. Time needed to reach the Pareto threshold as a function of effective dissipation factor $\alpha$. The earning means size $L_{29}=30/30$, and personal capability to earn money $S_{29}=30/30$.

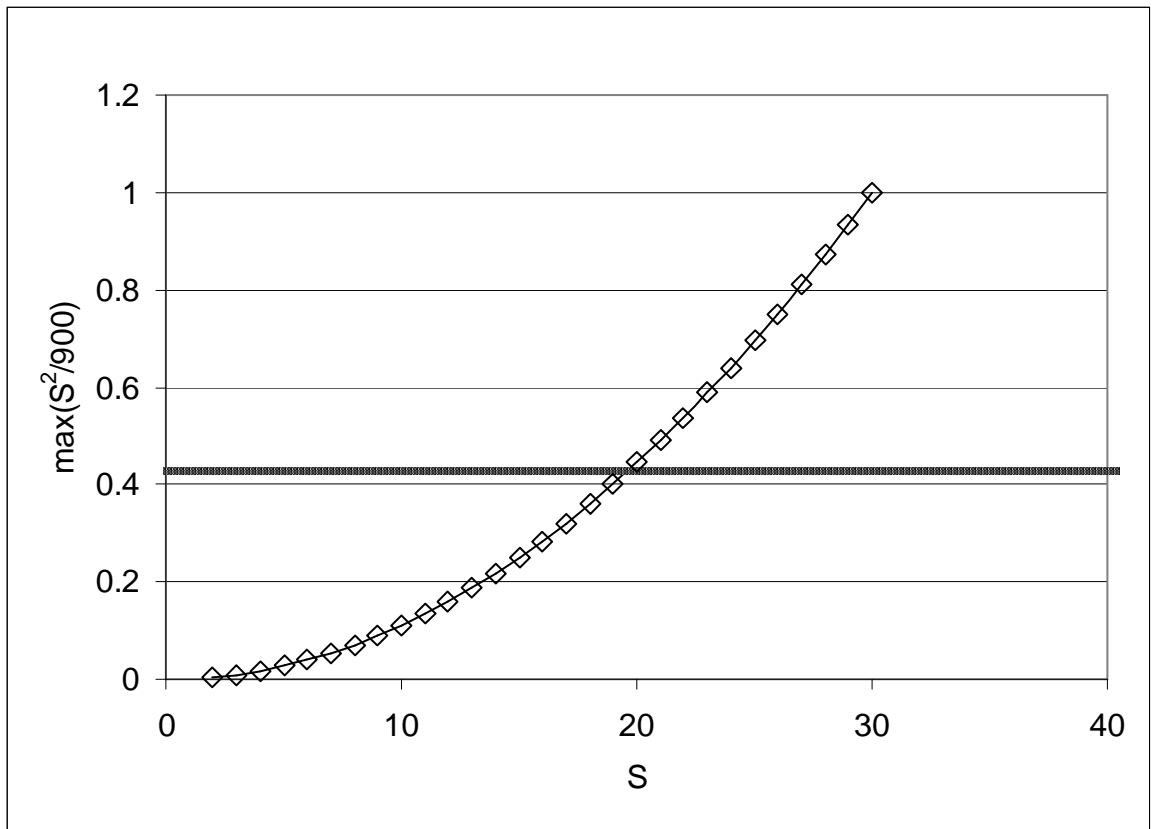

Figure 1.3.6. Peak income value ($S_i^2/900$) reached by a person with a capability to earn money $S_i$. The size of earning means is constrained to be the same $L_i=S_i$. The Pareto distribution threshold is 0.43. Nobody with both $S_i$ and $L_i$ below 20 can reach the threshold. Approximately 10 per cent of total population can eventually reach the Pareto threshold.

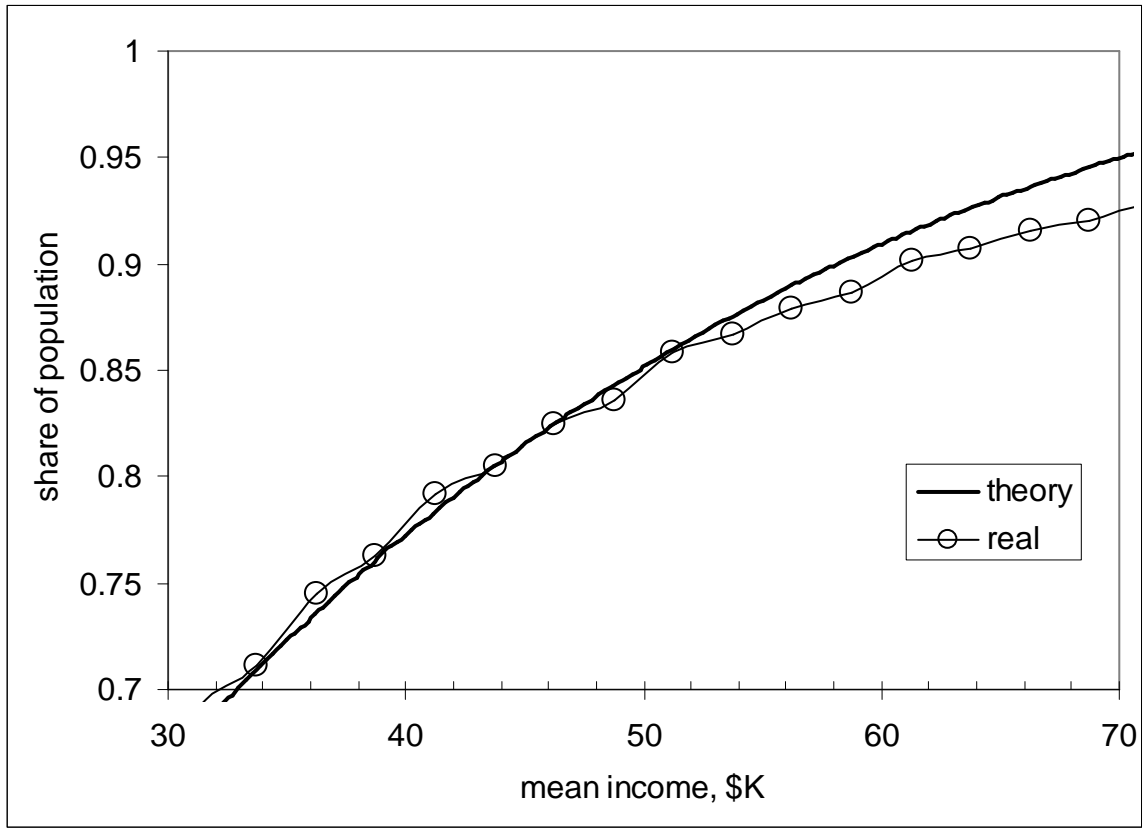

Figure 1.3.7. Comparison of the observed and predicted personal income distributions for the year 1999 – a portion of population with income below a given value. The curves diverge at income of $54K – the Pareto threshold.

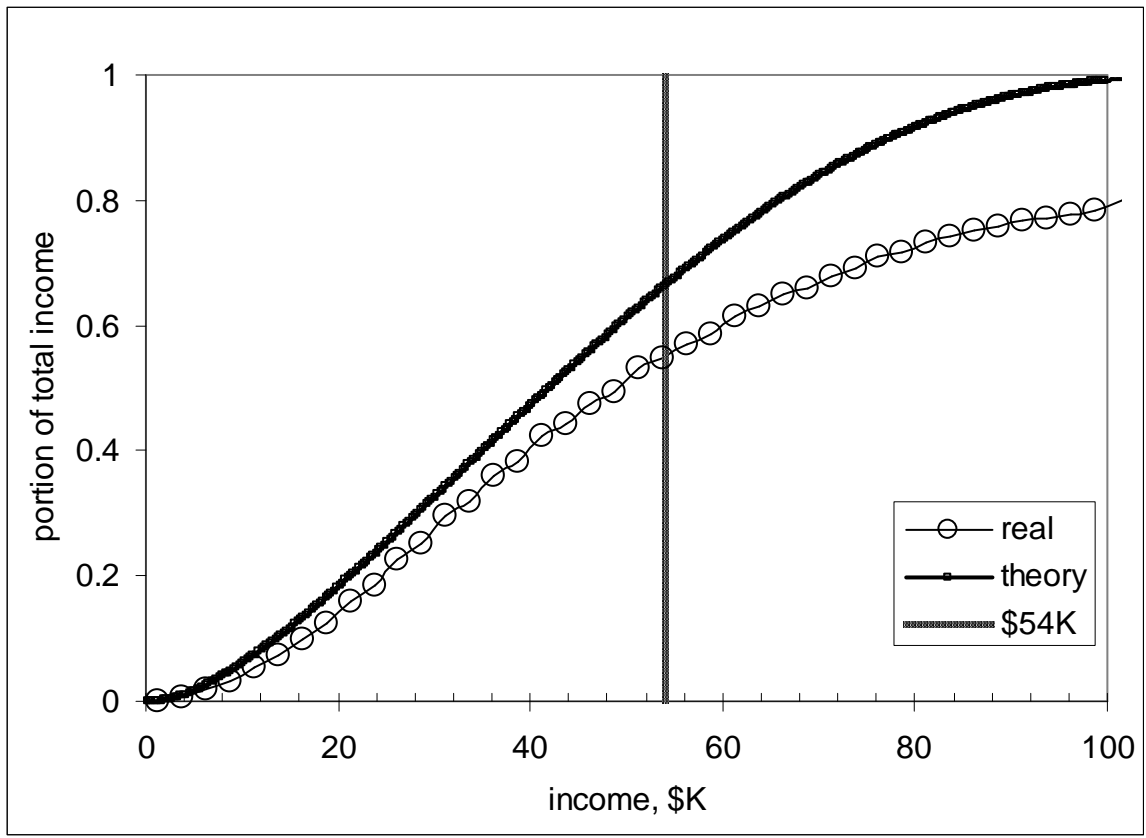

Figure 1.3.8. Comparison of the observed and predicted cumulative personal income distributions for 1999 – a portion of total income received by population with income below a given value. The ratio of the observed cumulative income of the population above the Pareto threshold (0.450 – intercept of the vertical line and the solid curve) and the corresponding theoretical value (0.333 – intercept of the vertical line) and is equal to 1.35. This value is considered as an effective increase of the average capacity to earn money for people above the Pareto threshold.

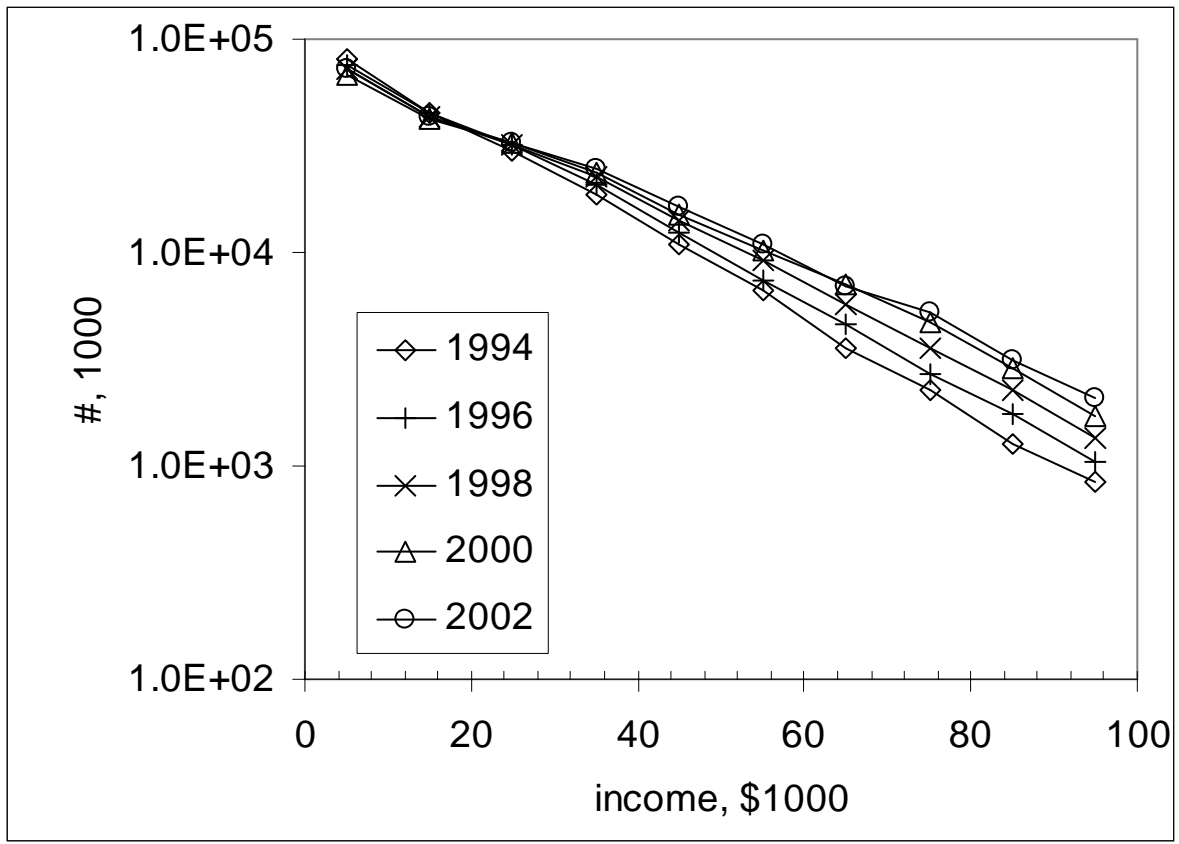

Figure 1.4.1. Personal income (current dollars) distributions in the USA from 1994 to 2002. Odd years are skipped for the sake of clarity. Absolute number (thousands) of people with income in $10K bins are shown. Notice the lin-log coordinates.

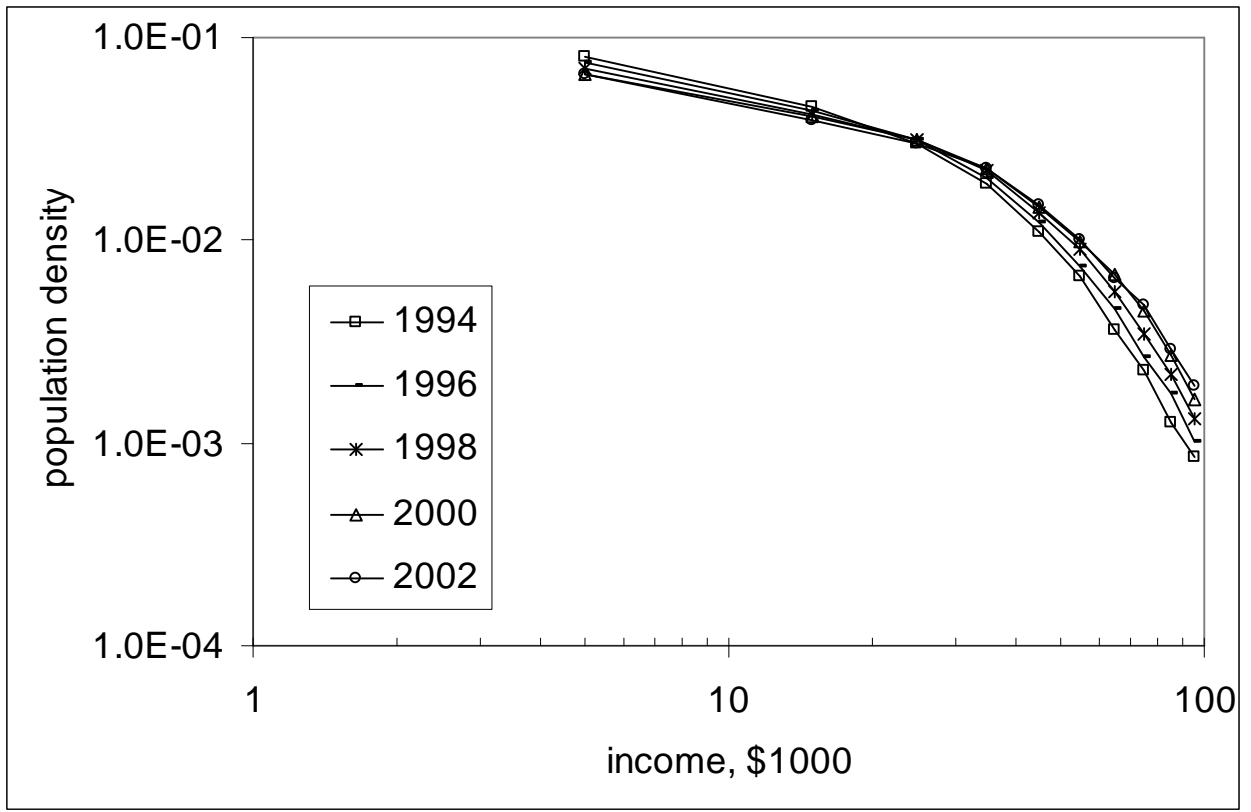

Figure 1.4.2. Selected personal income distributions normalized to relevant midyear populations. Notice the log-log coordinates.

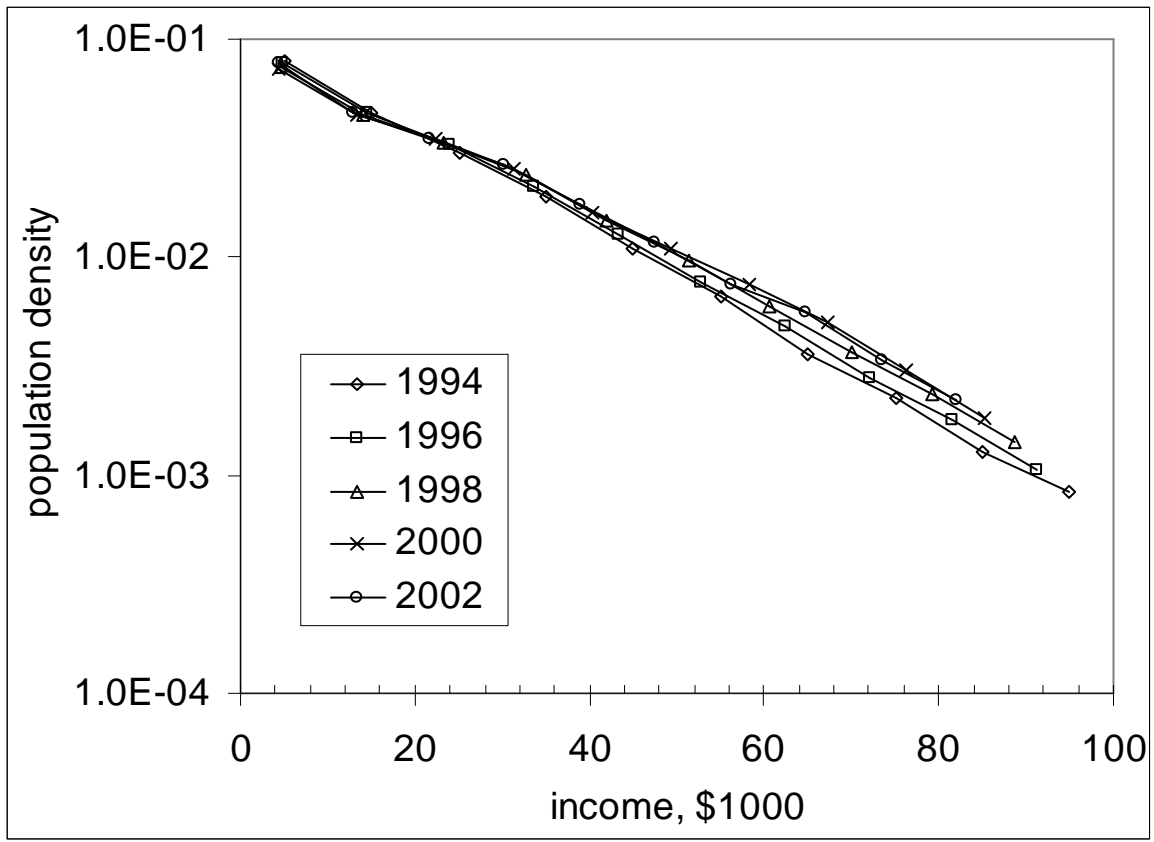

Figure 1.4.3. Selected population density distributions corrected for GDP deflator measured by the BEA.

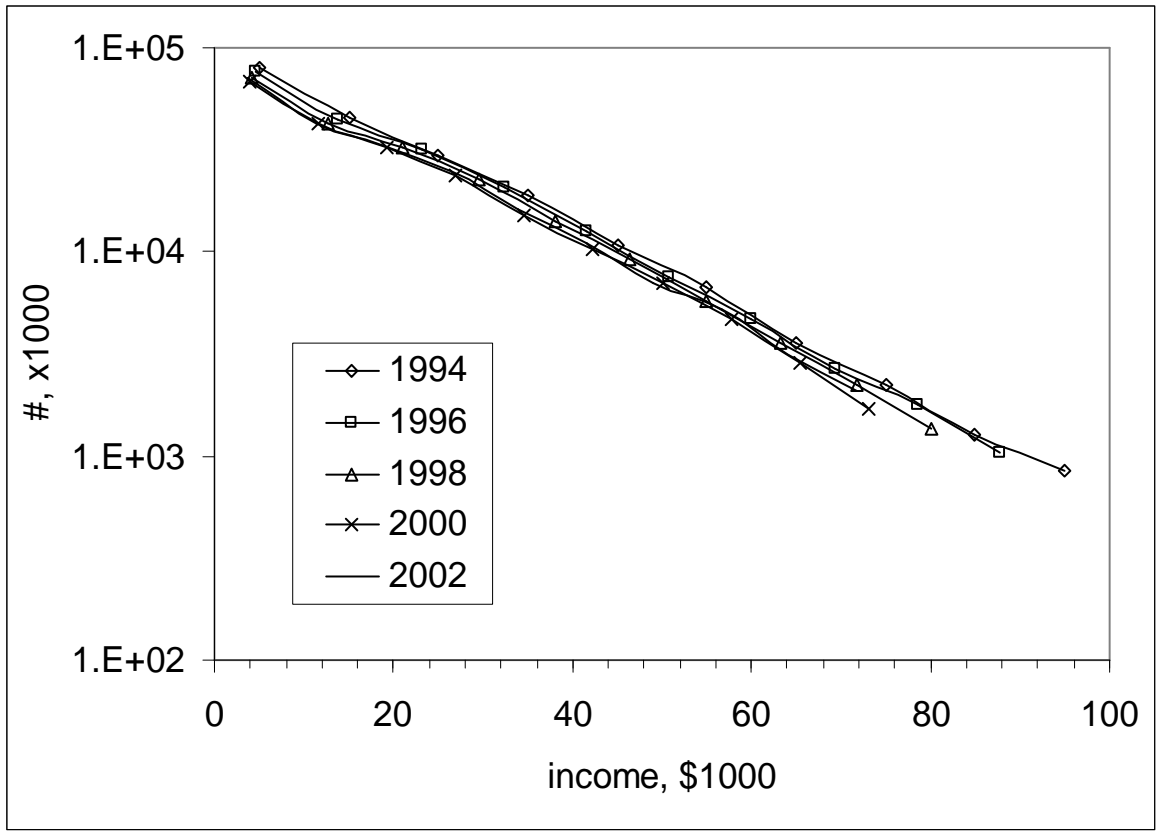

Figure 1.4.4. Personal income distribution corrected for the growth in nominal GDP.

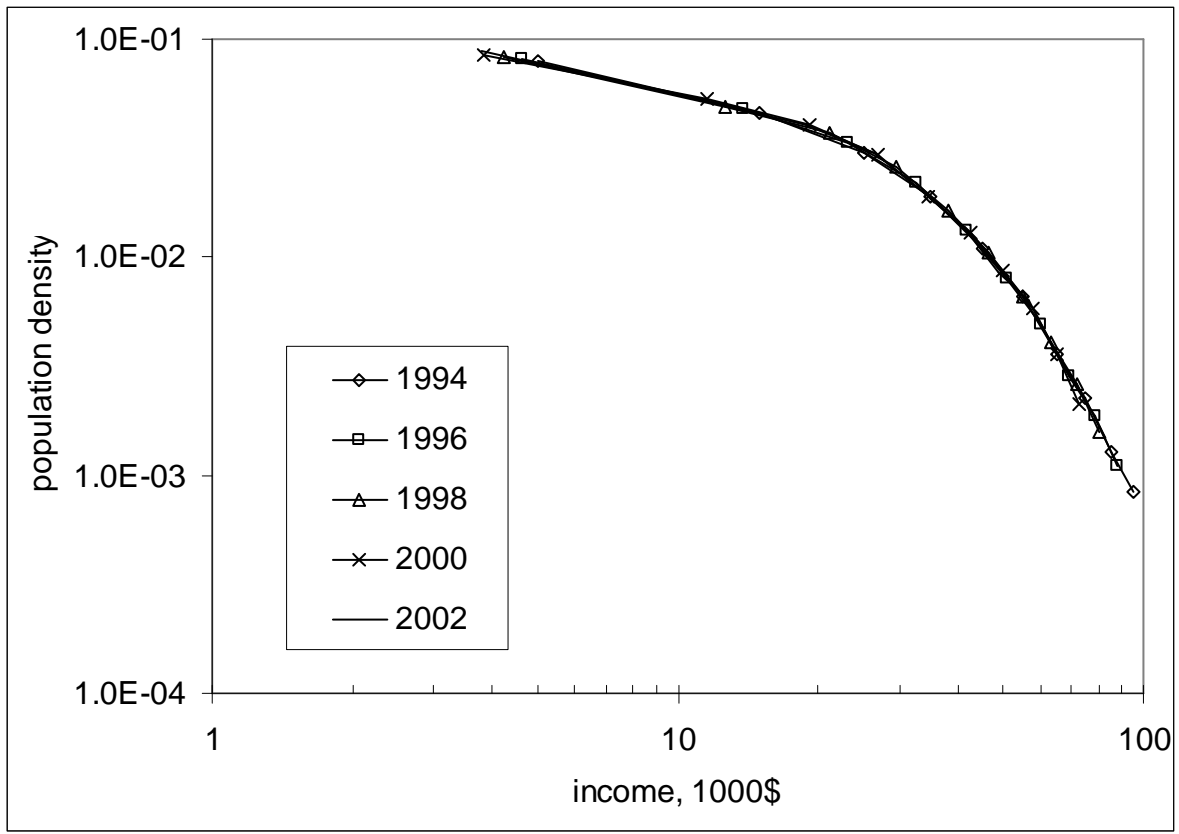

Figure 1.4.5. PDDs corrected for the growth in nominal GDP per capita. Notice the log-log coordinates.

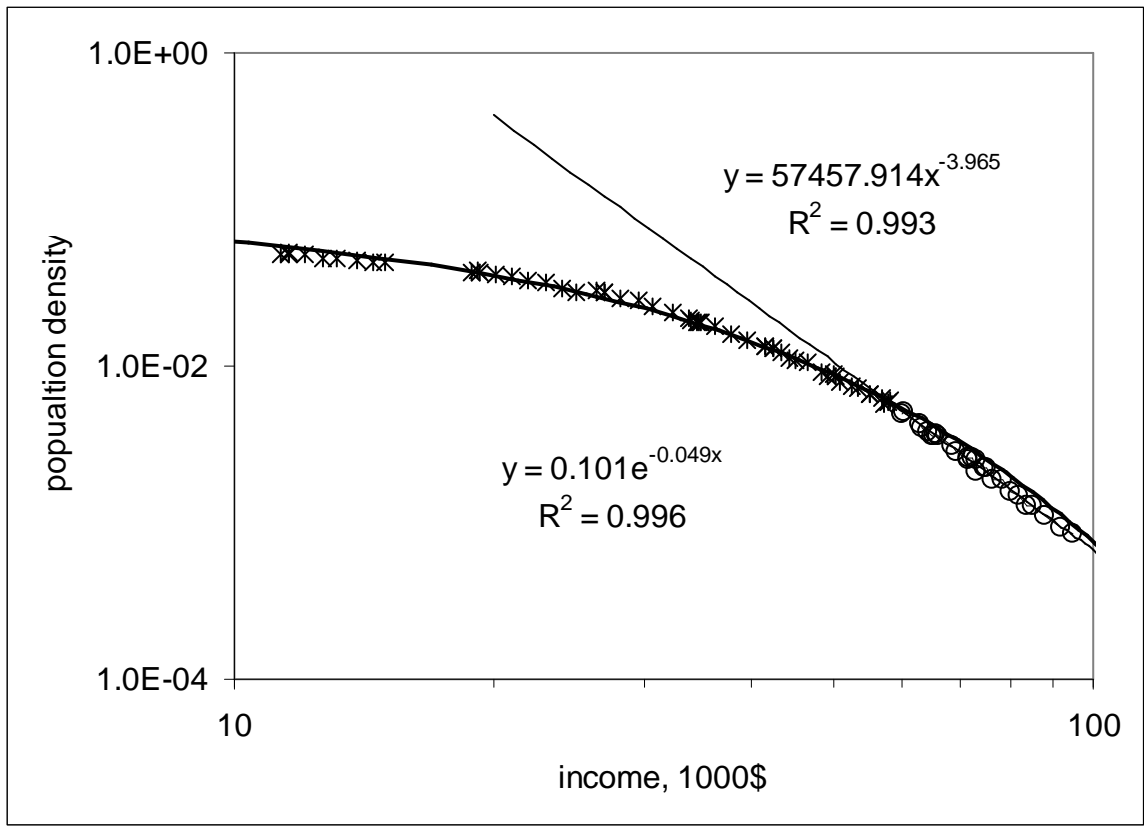

Figure 1.4.6. Personal income density distributions below and above the Pareto threshold for years between 1994 and 2001. The distributions are adjusted for the nominal per capita GDP growth. A power law regression demonstrates that the adjusted distributions practically coincide.

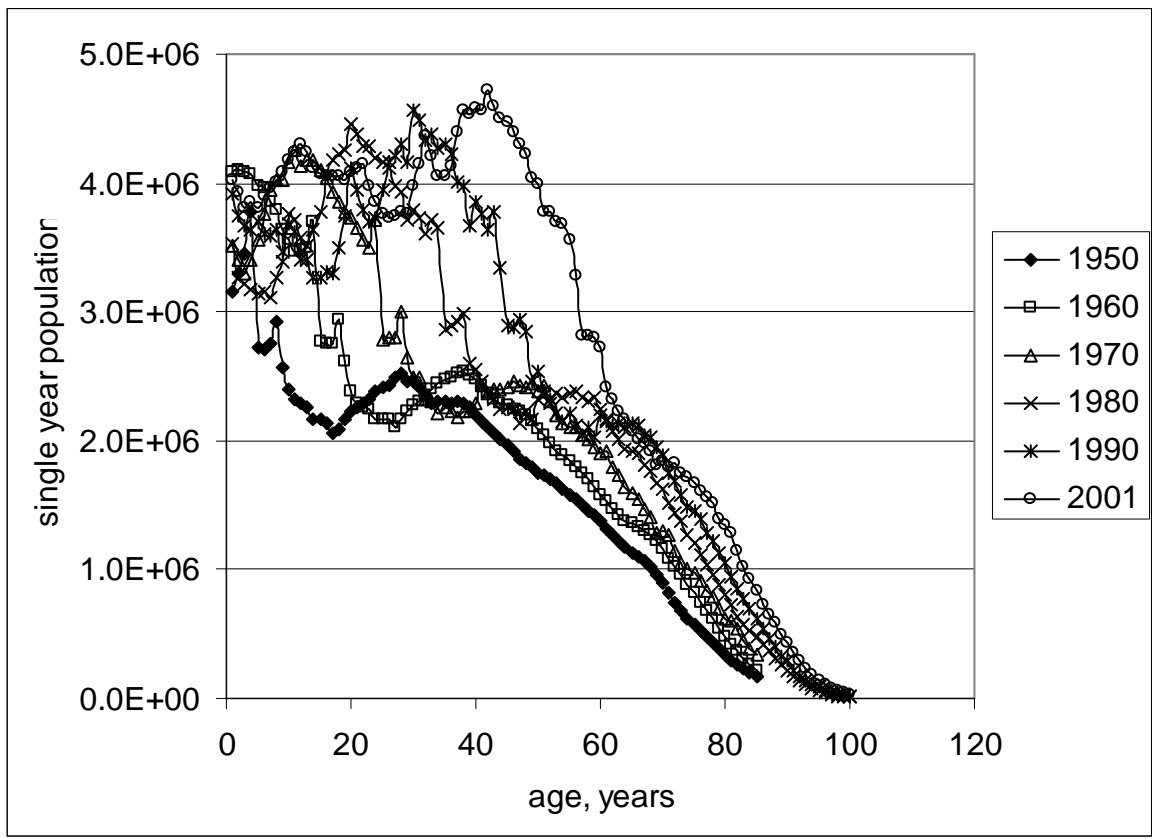

Figure 1.4.7. Population estimates for the calendar years of 1950, 1960, 1970, 1980, 1990, and 2001 (The US Census Bureau (2004b)).

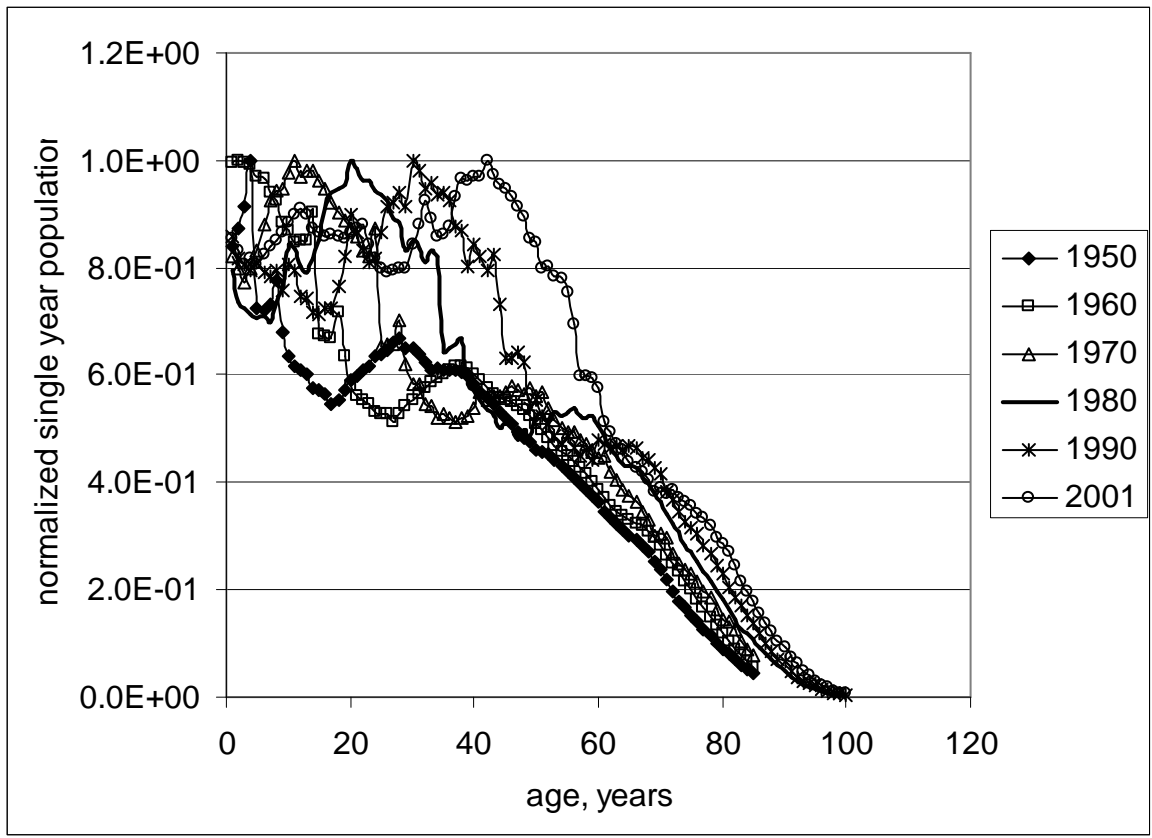

Figure 1.4.8. Normalized population estimates for the calendar years of 1950, 1960, 1970, 1980, 1990, and 2001.

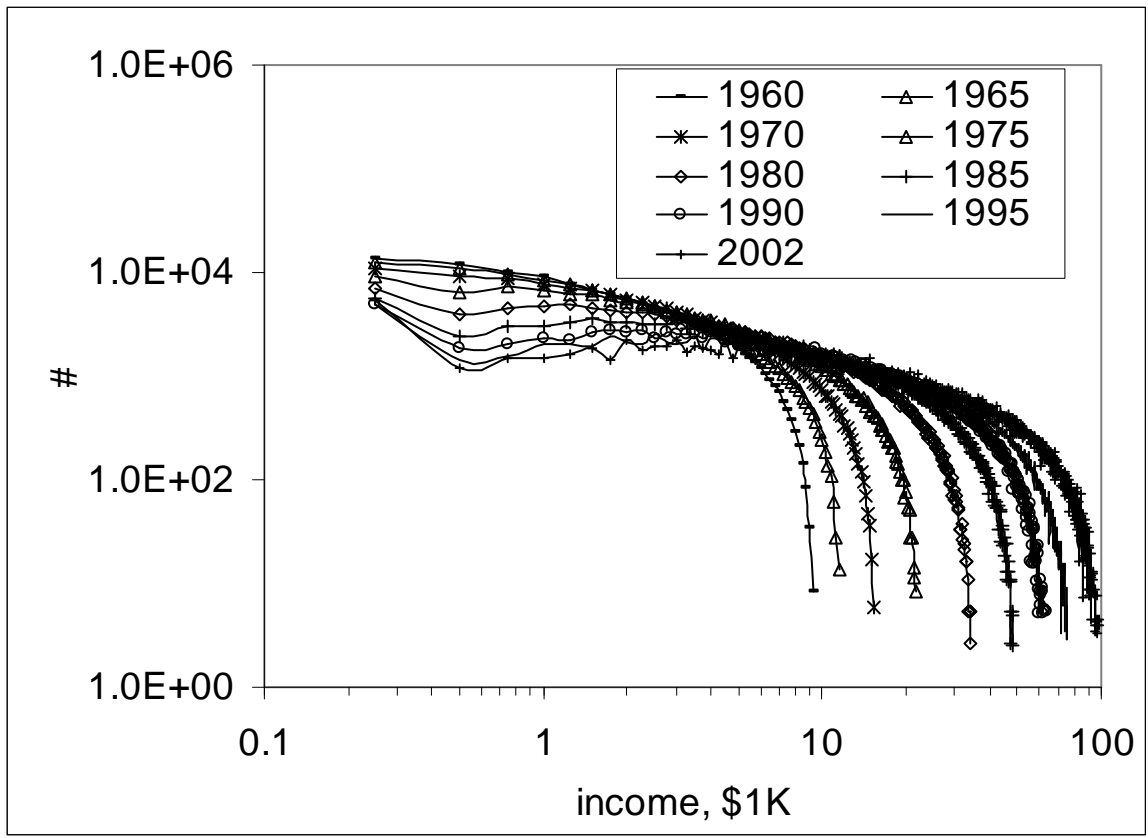

Figure 1.4.9. The evolution of predicted personal income distribution between 1960 and 2002.

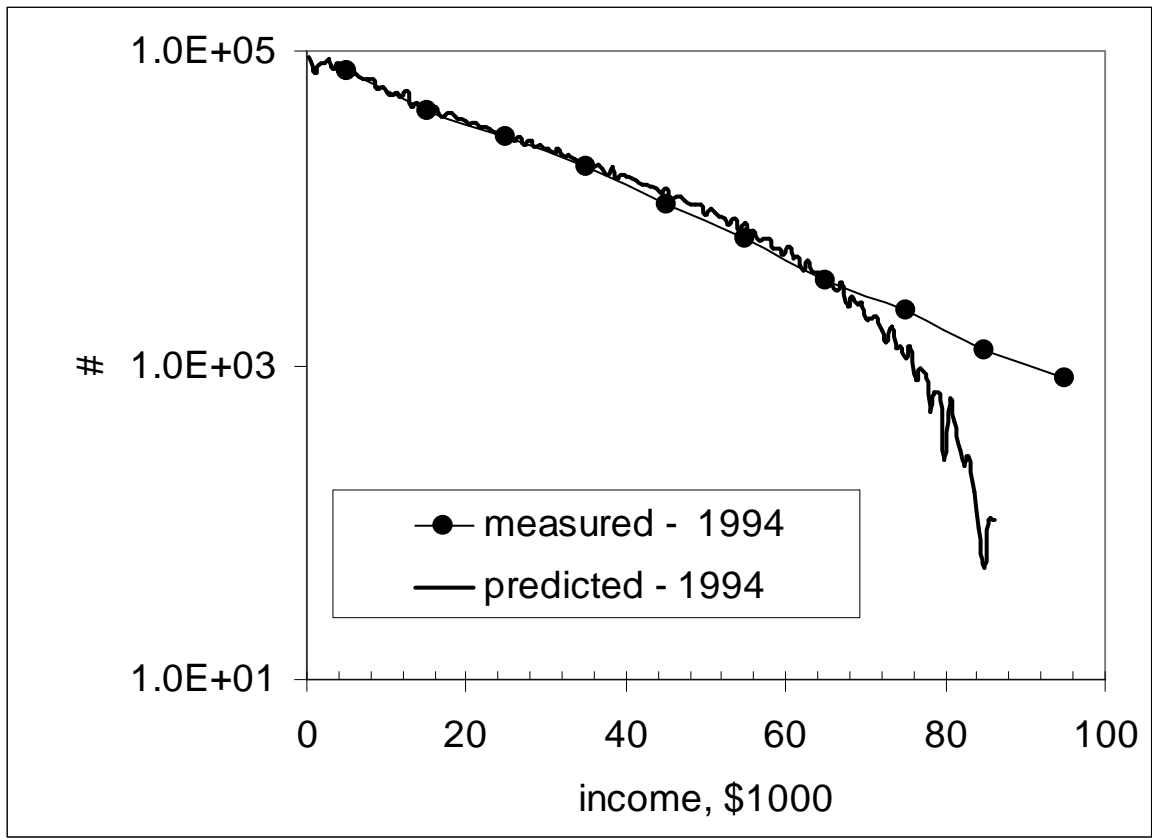

Figure 1.4.10. Comparison of predicted and observed personal income distributions for 1994. The Pareto threshold is between $35K and $45K.

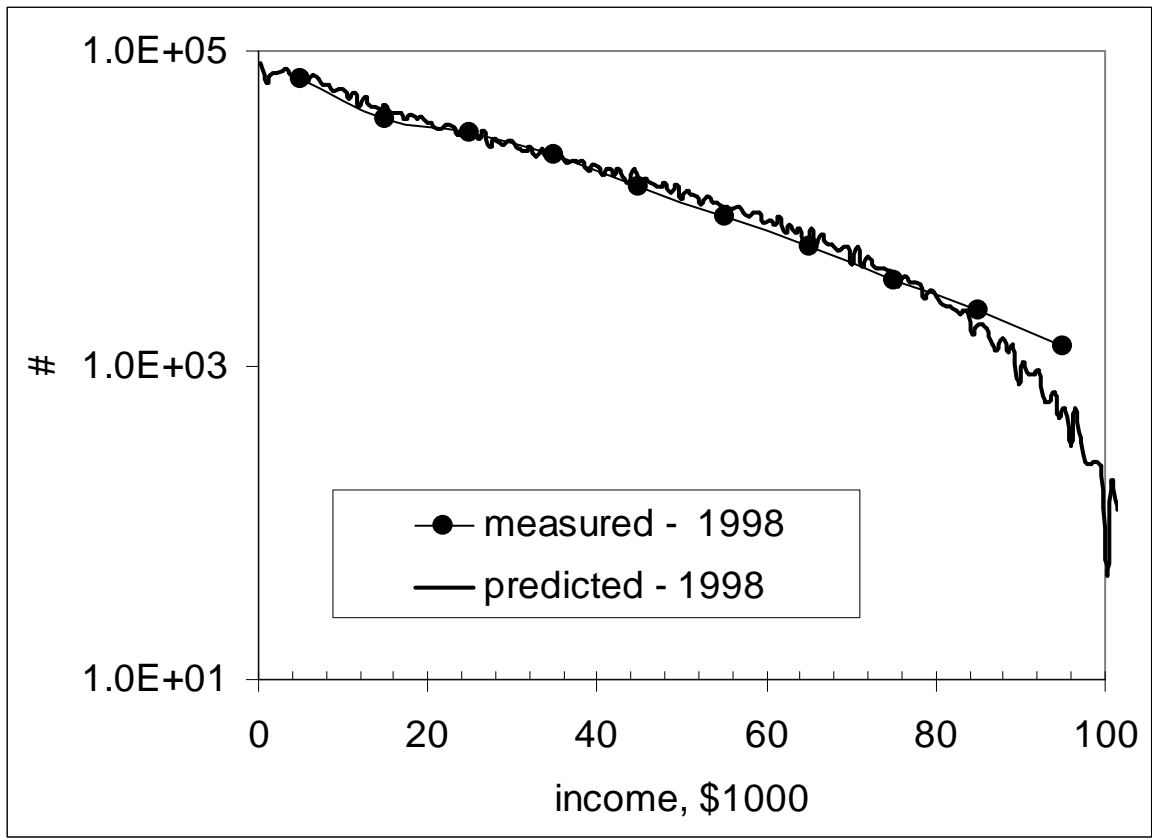

Figure 1.4.11. Comparison of predicted and observed personal income distributions for the year of 1998. The Pareto threshold is between $45K and $55K.

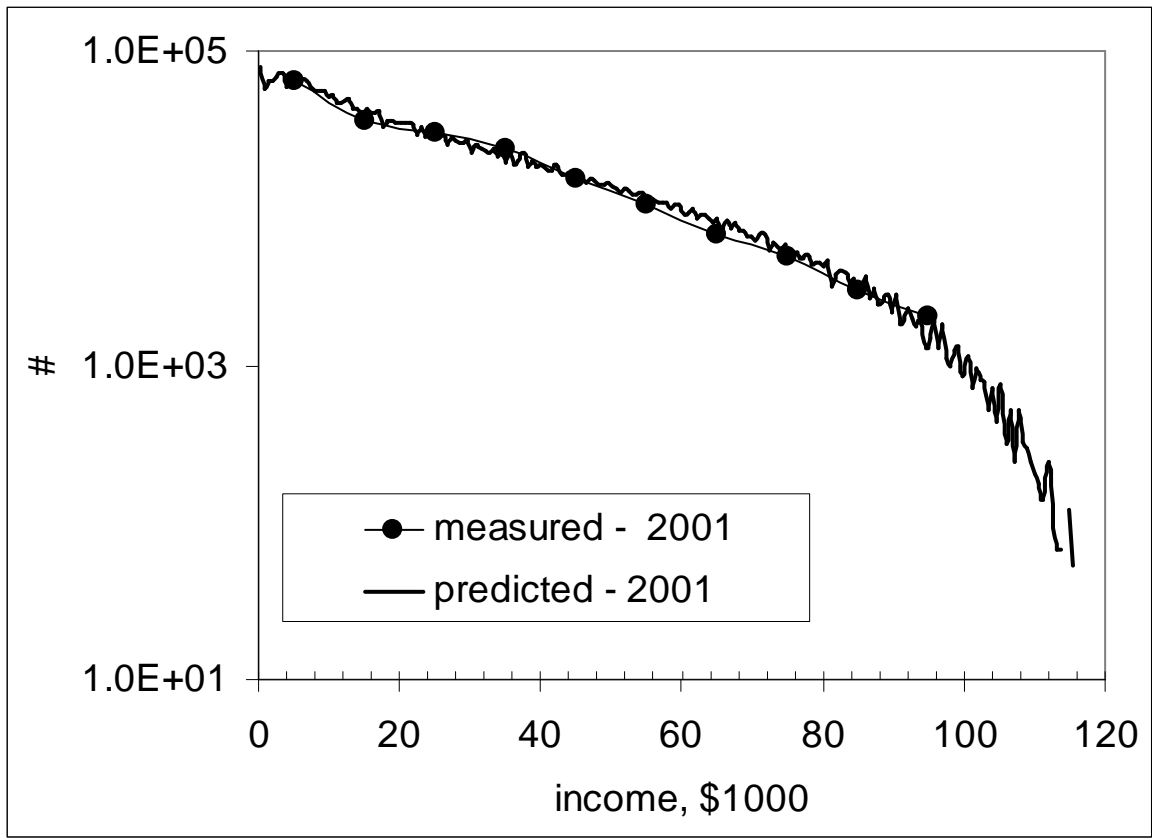

Figure 1.4.12. Comparison of predicted and observed personal income distributions for the year of 2001. The Pareto threshold is between $55K and $65K.

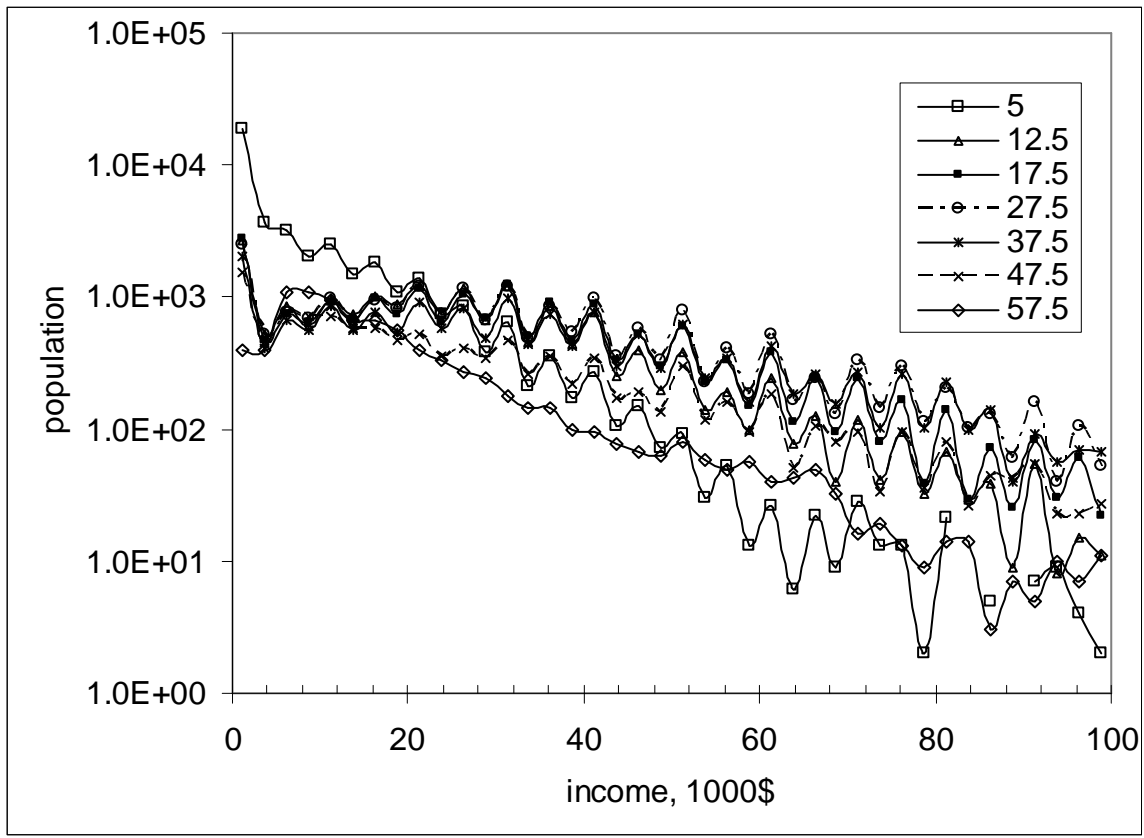

Figure 1.5.1. Personal income distribution in various age groups: from 15 to 24 years (5 – central point of corresponding work experience interval from 0 to 9 years), from 25 to 29 years (12.5), …, from 70 to 74 years (57.5). In the first age group (5) - from 0 to 9 years of work experience, an exponential decrease in observed.

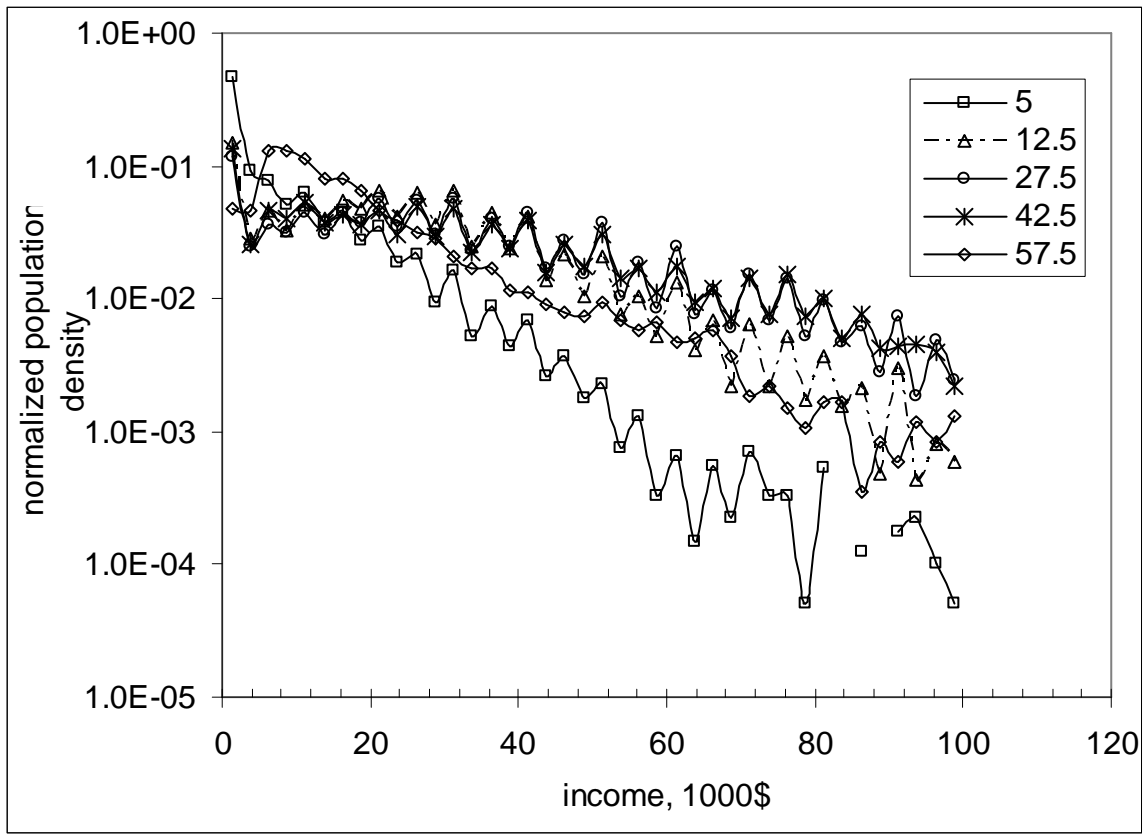

Figure 1.5.2. Population density vs. personal income in age groups: from 15 to 24 years (5 – central point of corresponding work experience interval from 0 to 9 years), from 25 to 29 years (12.5), …, from 70 to 74 years (57.5). In the first age group an almost exponential decrease in observed.

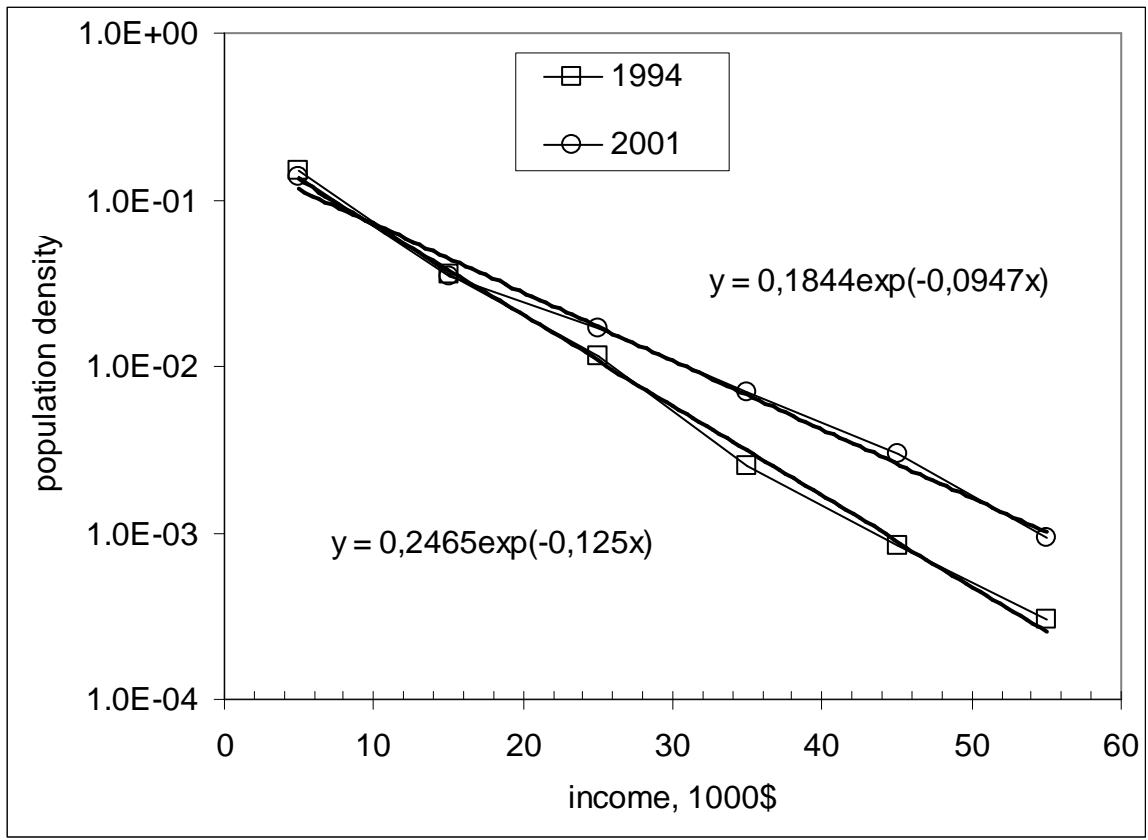

Figure 1.5.3. Population density vs. personal income in the age group from 15 to 24 years. The index of exponential regression function increases with time from -0.125 to -0.095.

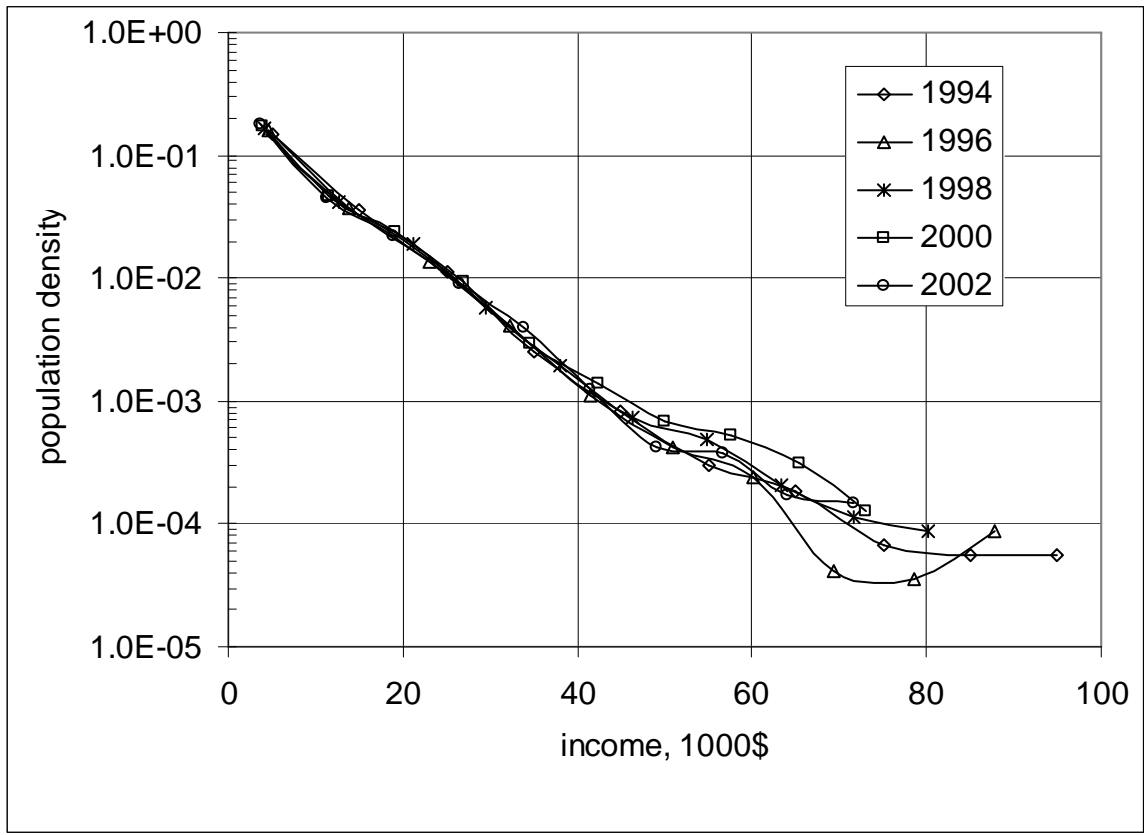

Figure 1.5.4. Population density distribution in the age group from 15 to 24 years adjusted for the per capita nominal GDP growth. A strong scattering at the highest incomes is induced by lack or resolution power of the current ASEC due to undercoverage of the population. Population density drops by three orders of magnitude with income increase from $5,000 to $45,000.

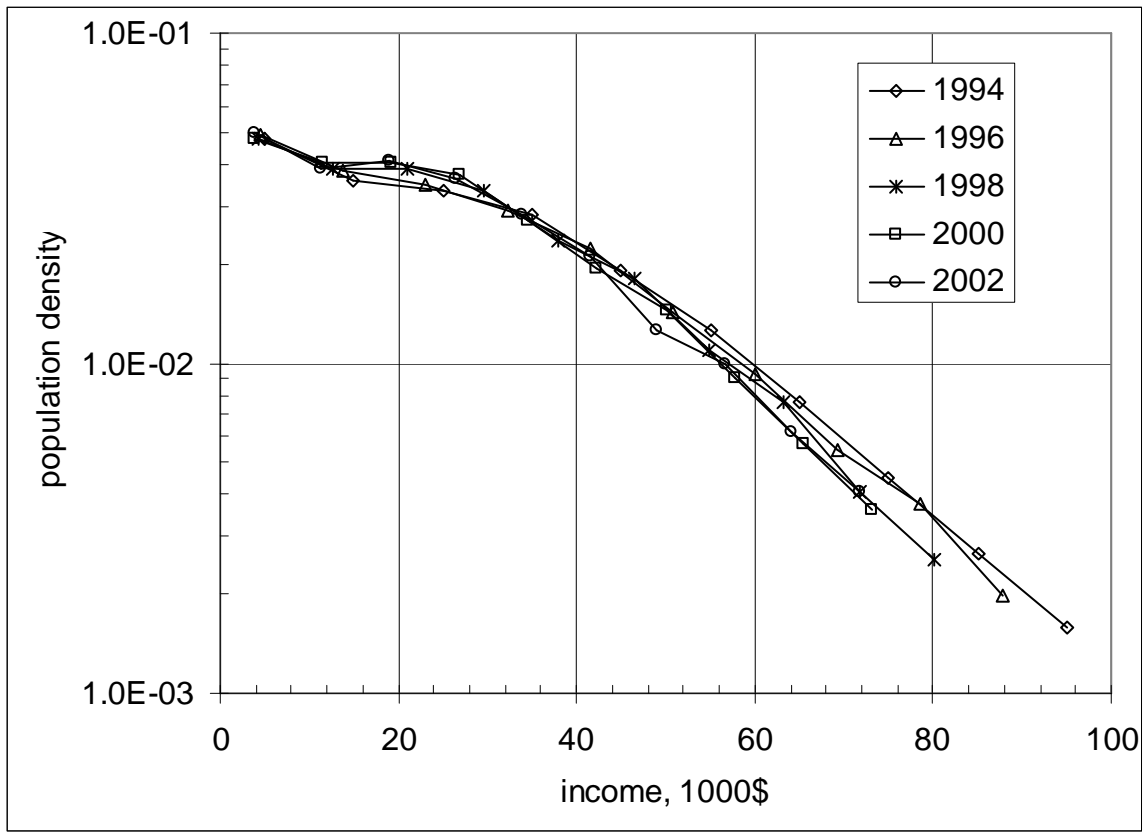

Figure 1.5.5. Population density distribution in the age group from 45 to 49 years adjusted for the per capita nominal GDP growth.

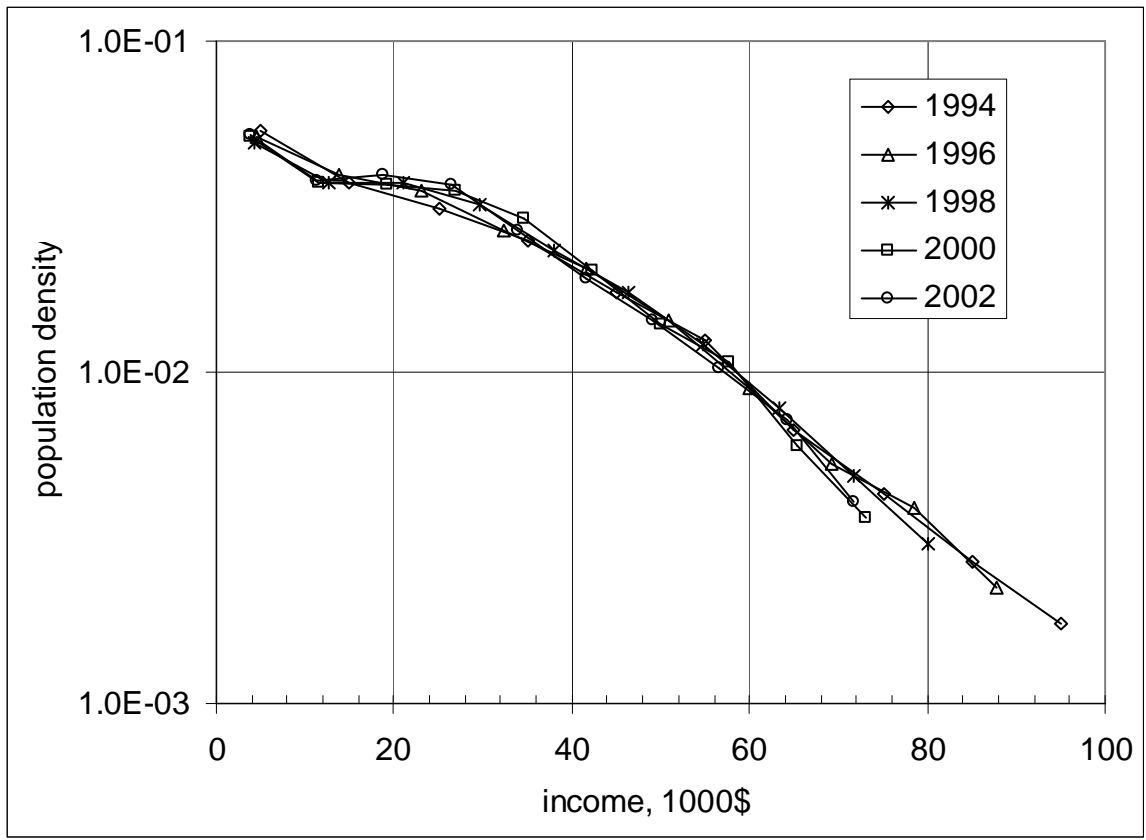

Figure 1.5.6. Population density distribution in the age group from 50 to 54 years adjusted for the per capita nominal GDP growth.

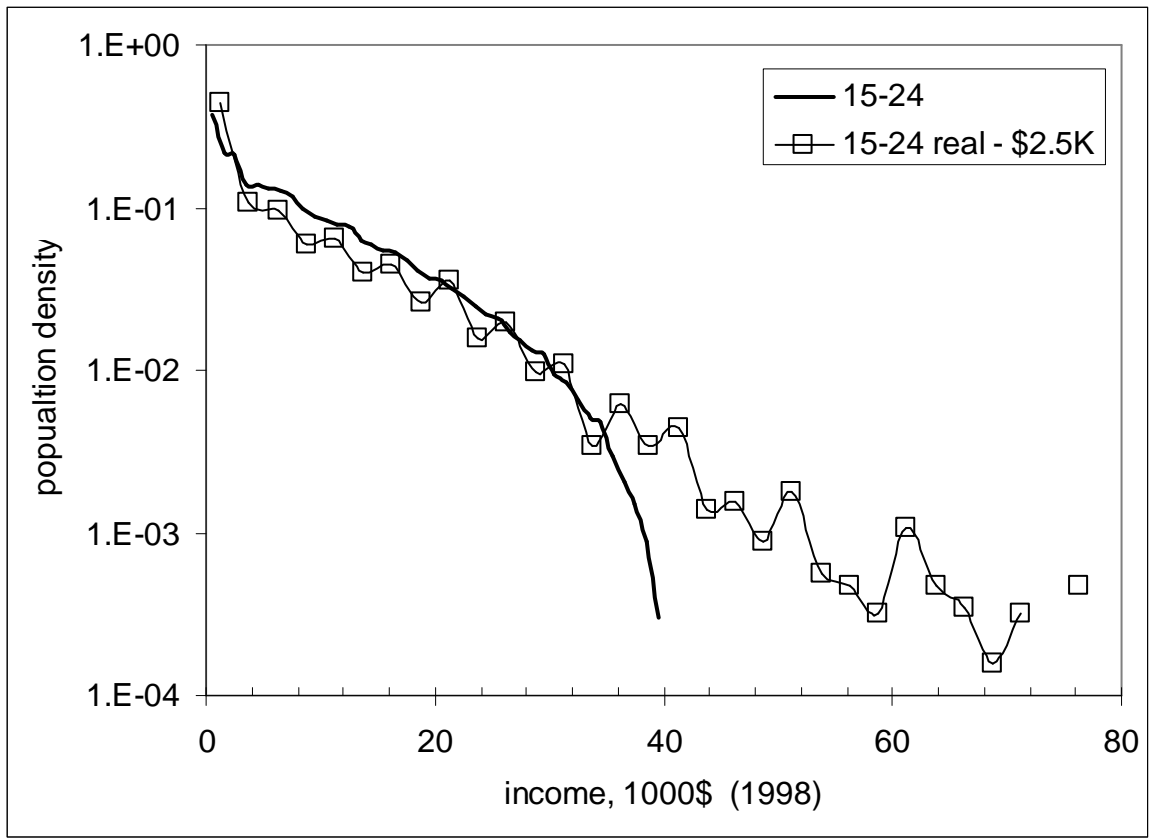

Figure 1.5.7. Comparison of predicted and measured personal income distributions in the age group from 15 to 24 years. The actual distribution is characterized by an emergent Pareto part and lies almost everywhere below the predicted curve. The difference is associated with a poor procedure of the income survey which does not take into account money redistribution.

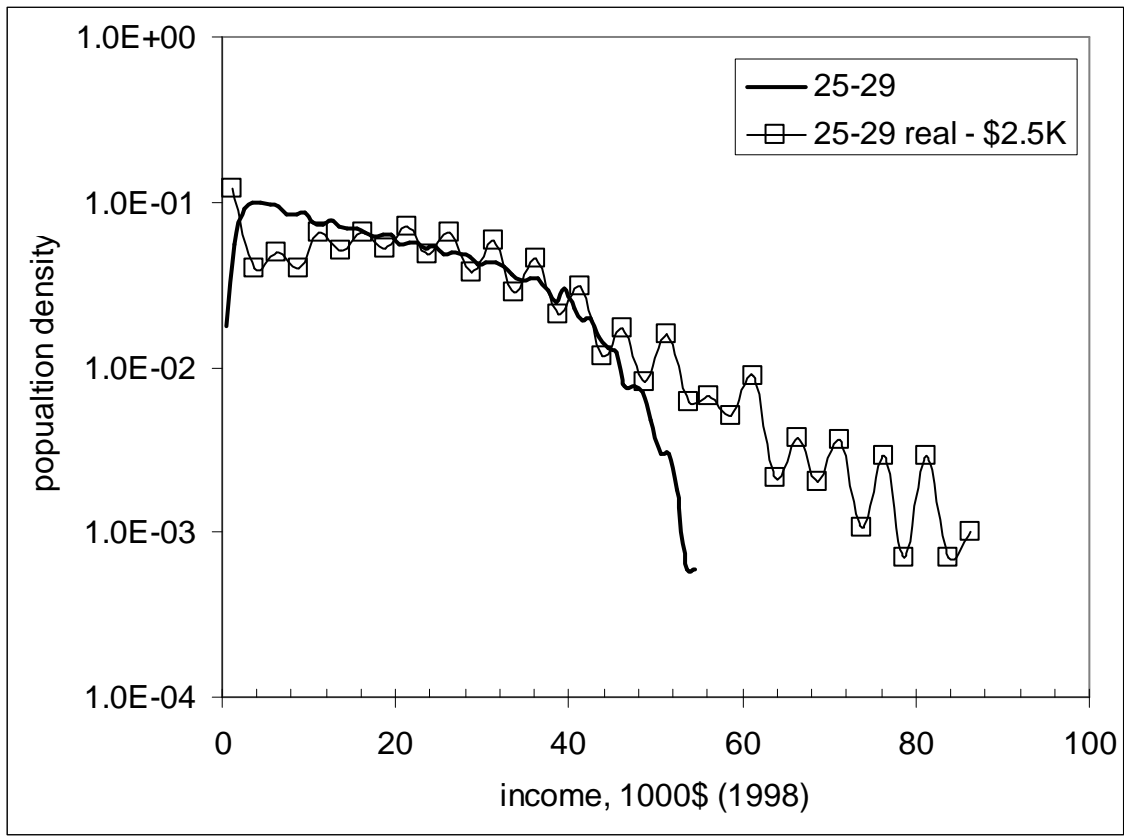

Figure 1.5.8. Comparison of predicted and measured personal income distributions in the age group from 25 to 29 years. The actual distribution is characterized by the presence of a part with well established Pareto distribution. The difference between the curves at low incomes is associated with a poor procedure of the income survey.

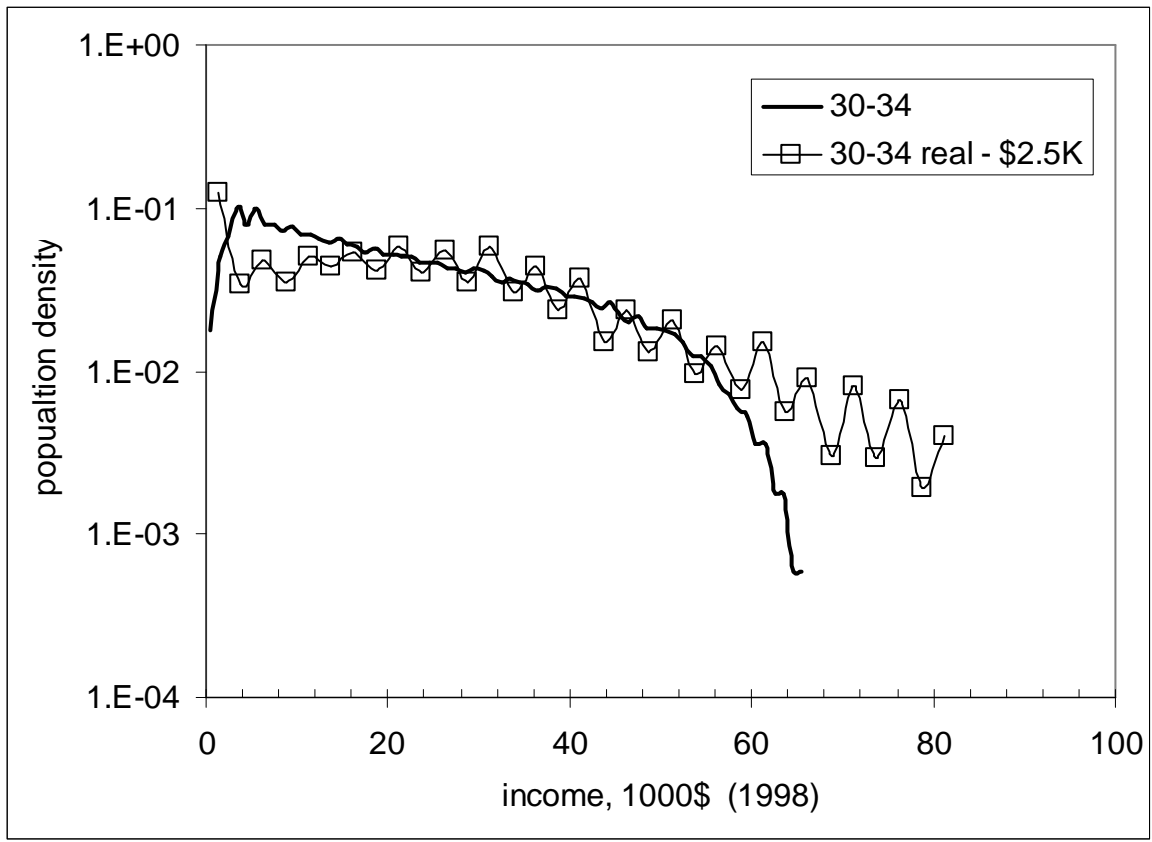

Figure 1.5.9. Comparison of predicted and measured personal income distributions in the age group from 30 to 34 years. The difference between the curves at low incomes is associated with poor procedure of the income survey.

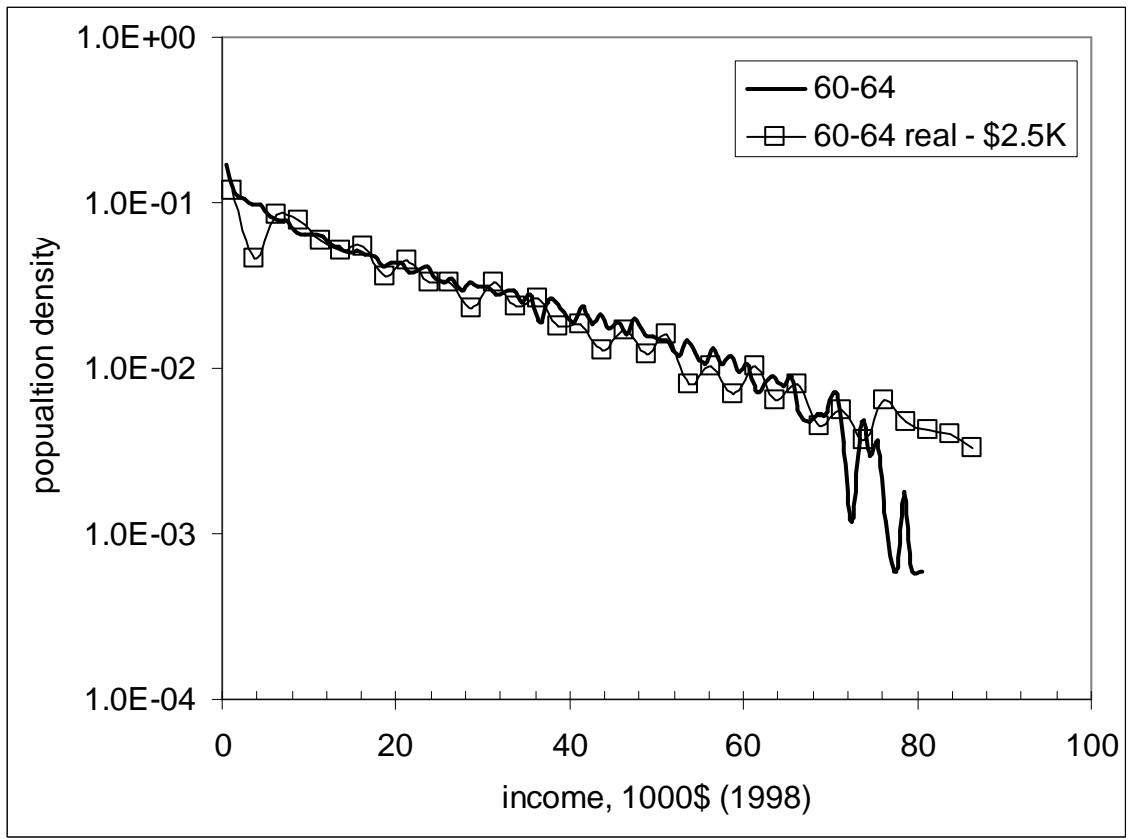

Figure 1.5.10. Comparison of predicted and measured personal income distributions in the age group from 60 to 64 years. Both distributions are quasi-exponential.

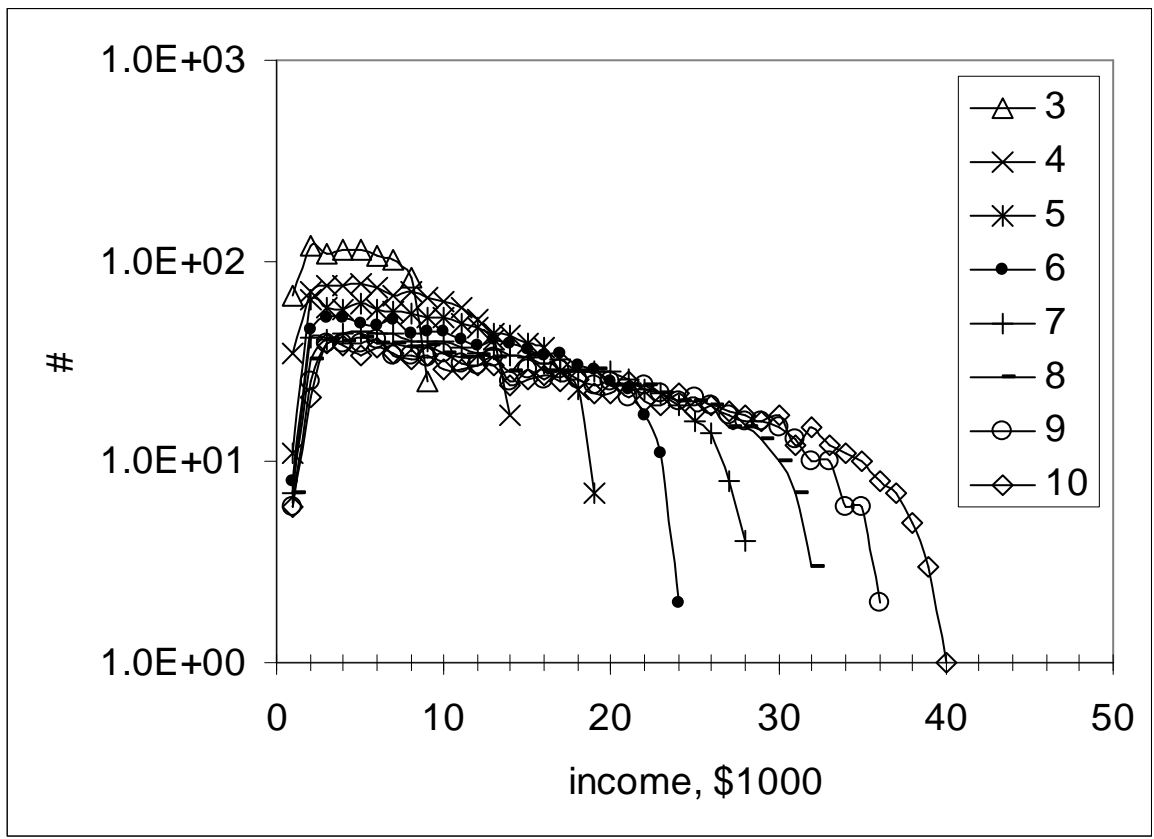

Figure 1.5.11. Evolution of the predicted personal income distribution (absolute number of people) in single-year-of-age intervals between 3 and 10 years of work experience.

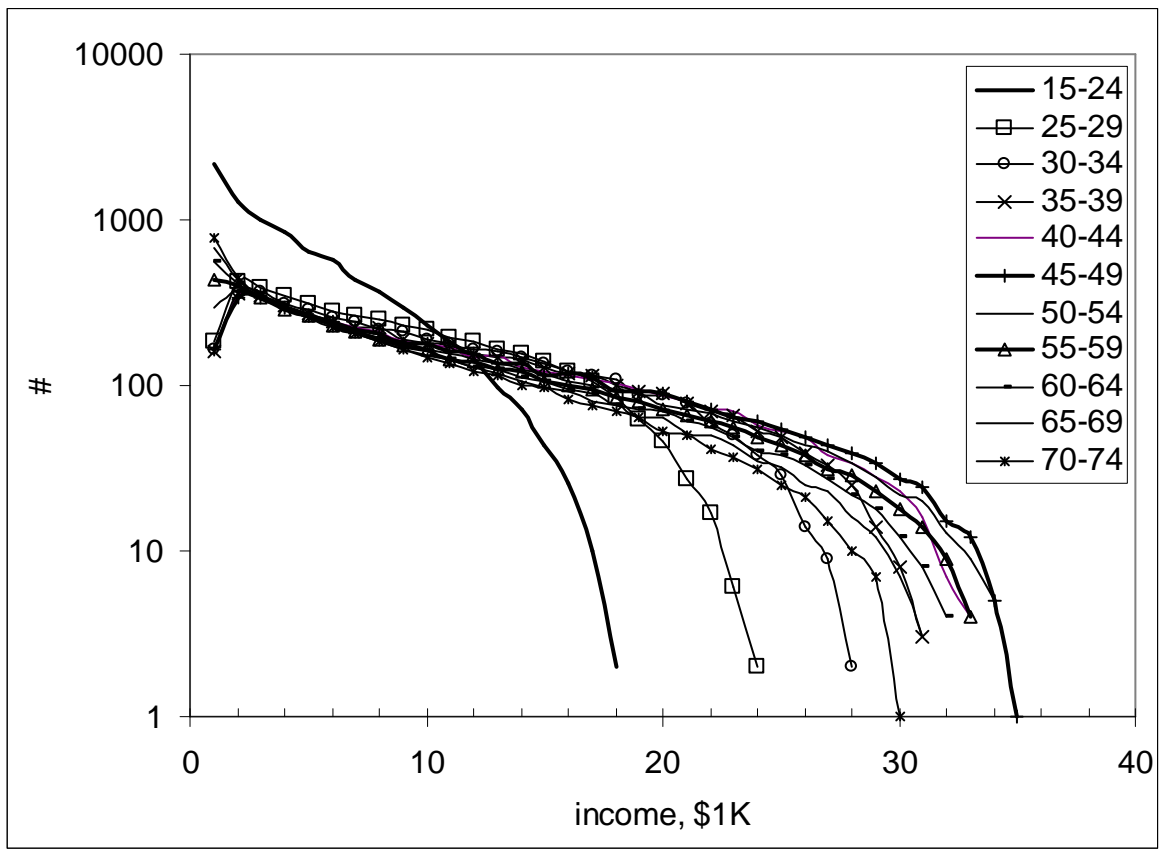

Figure 1.5.12 Evolution of the predicted personal income distribution (absolute number of people) in 1980.

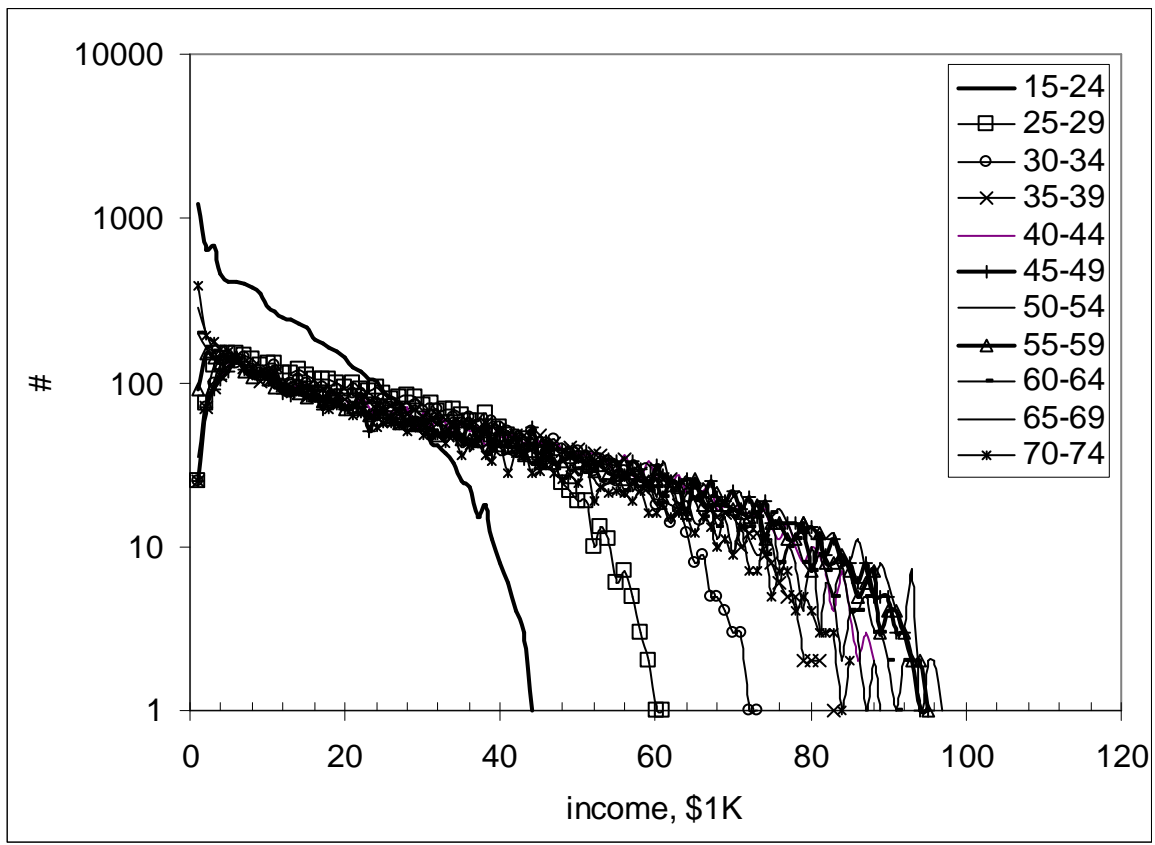

Figure 1.5.13. Evolution of the predicted personal income distribution (absolute number of people) in 2002.

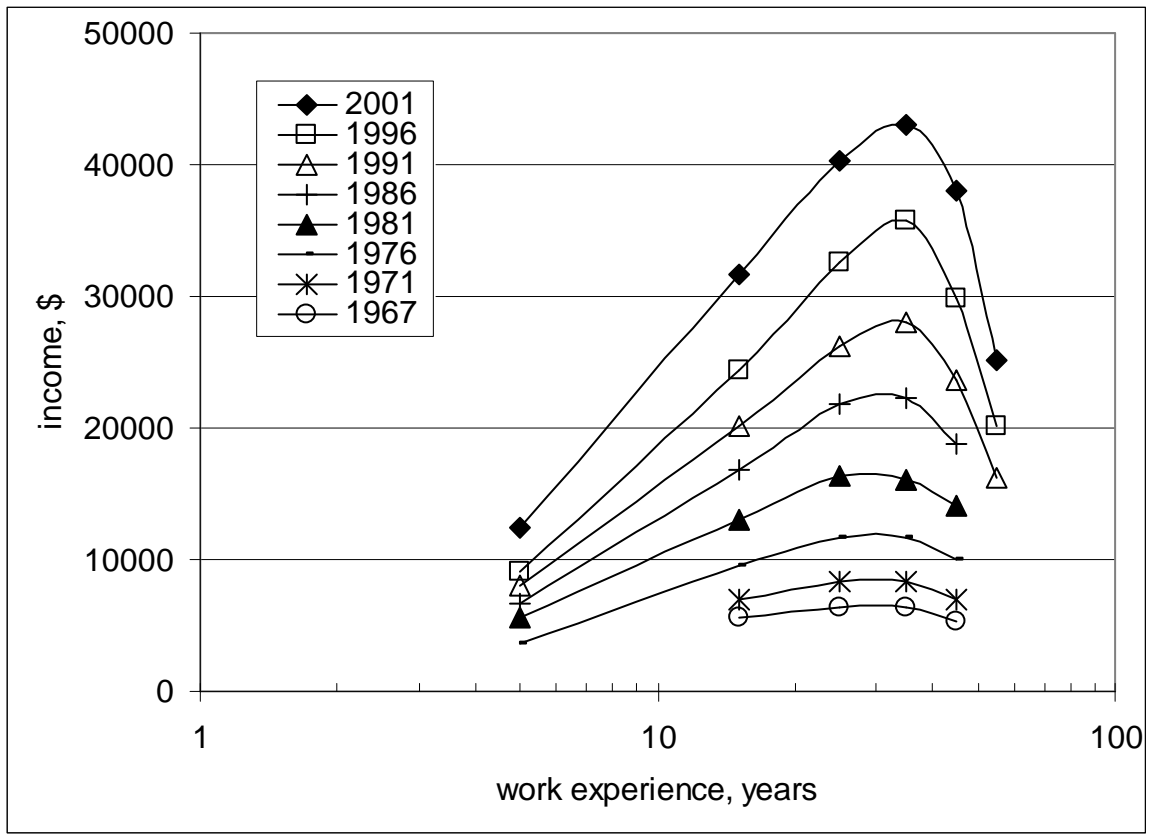

Figure 1.6.1. Mean personal income (current dollars) as a function of work experience for selected years between 1967 and 2001.

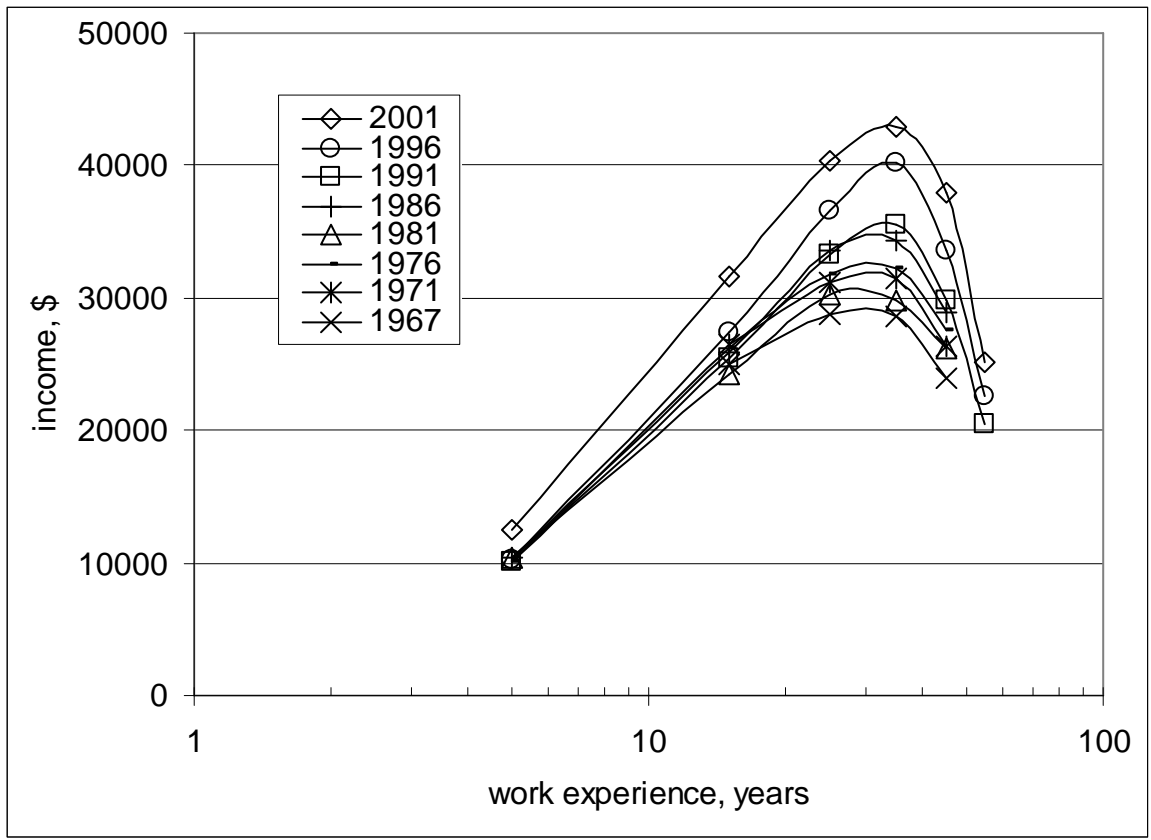

Figure 1.6.2. Mean personal income (chained 2001 dollars) as a function of work experience for selected years from 1967 to 2001.

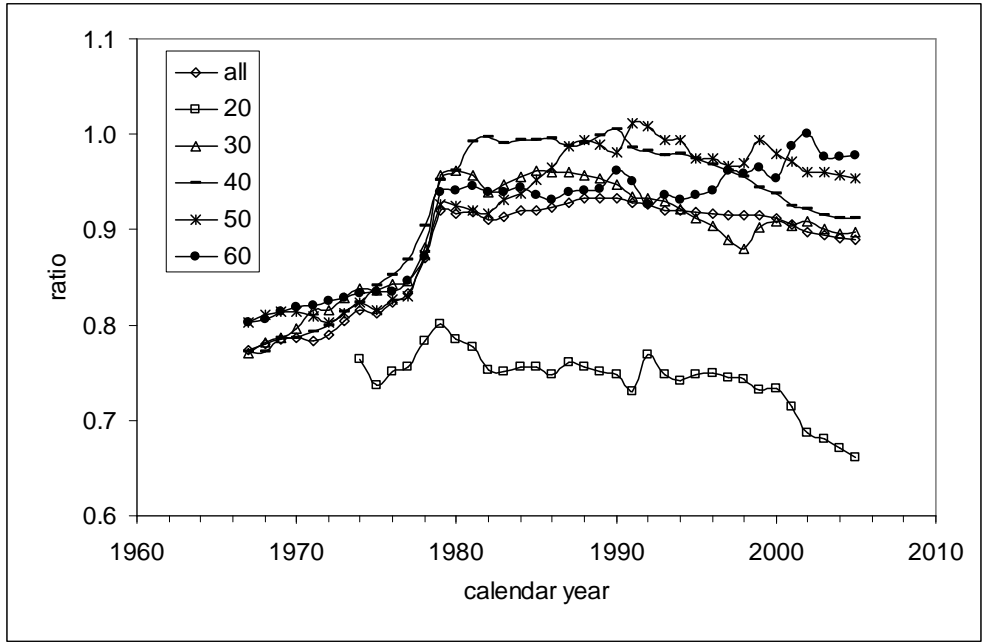

Figure 1.6.3. Evolution of the portion of population with income in various age groups: all – above 14 years of age, 20 – from 15 to 24 years of age, 30 – from 25 to 34 years, 40 – from 35 to 44 years, 50 – from 45 to 54 years, 60 – from 55 to 64 years. In the group between 16 and 24 years of age, the portion has been falling since 1979. Notice the break in the distributions between 1977 and 1979 induced by large revisions implemented in 1980 – "Questionnaire expanded to show 27 possible values from 51 possible sources of income." The participation factor in other age groups increased from 0.82-0.85 to 0.92-0.99.

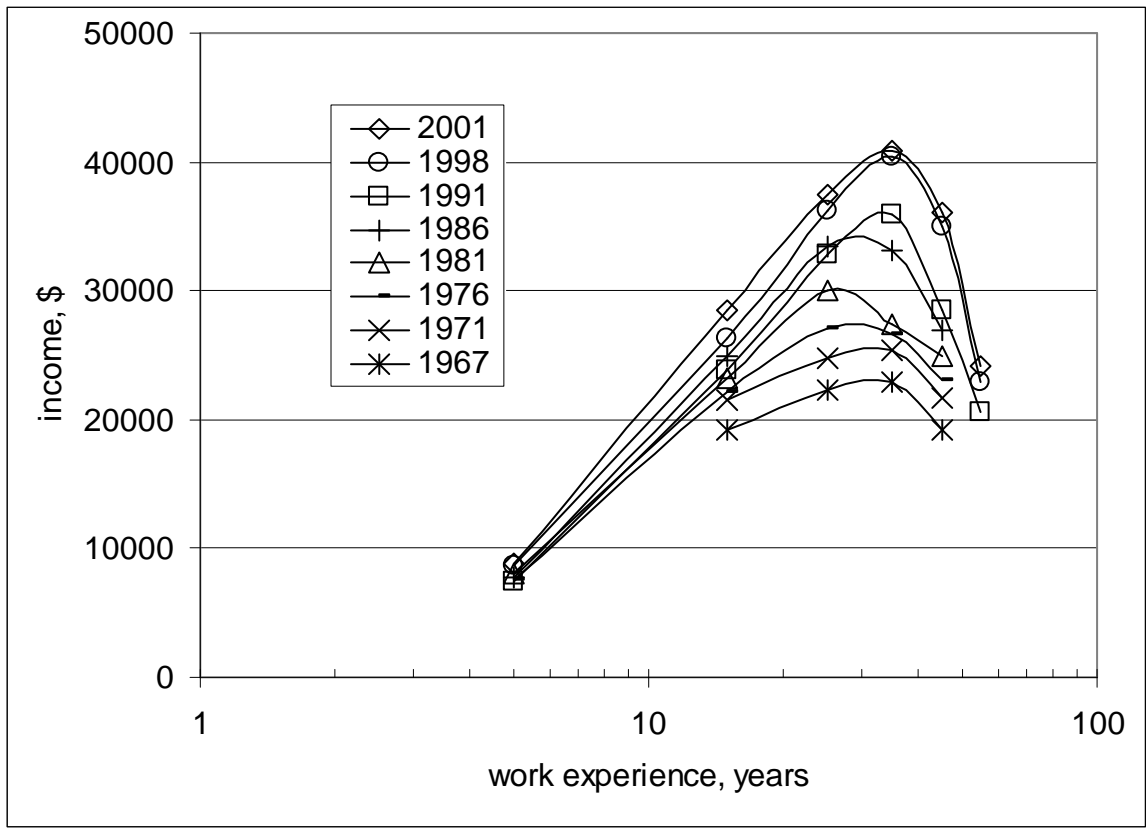

Figure 1.6.4. Mean personal income (chained 2001 dollars) as a function of work experience for years from 1967 to 2001. The mean income readings are corrected for the number of people without income.

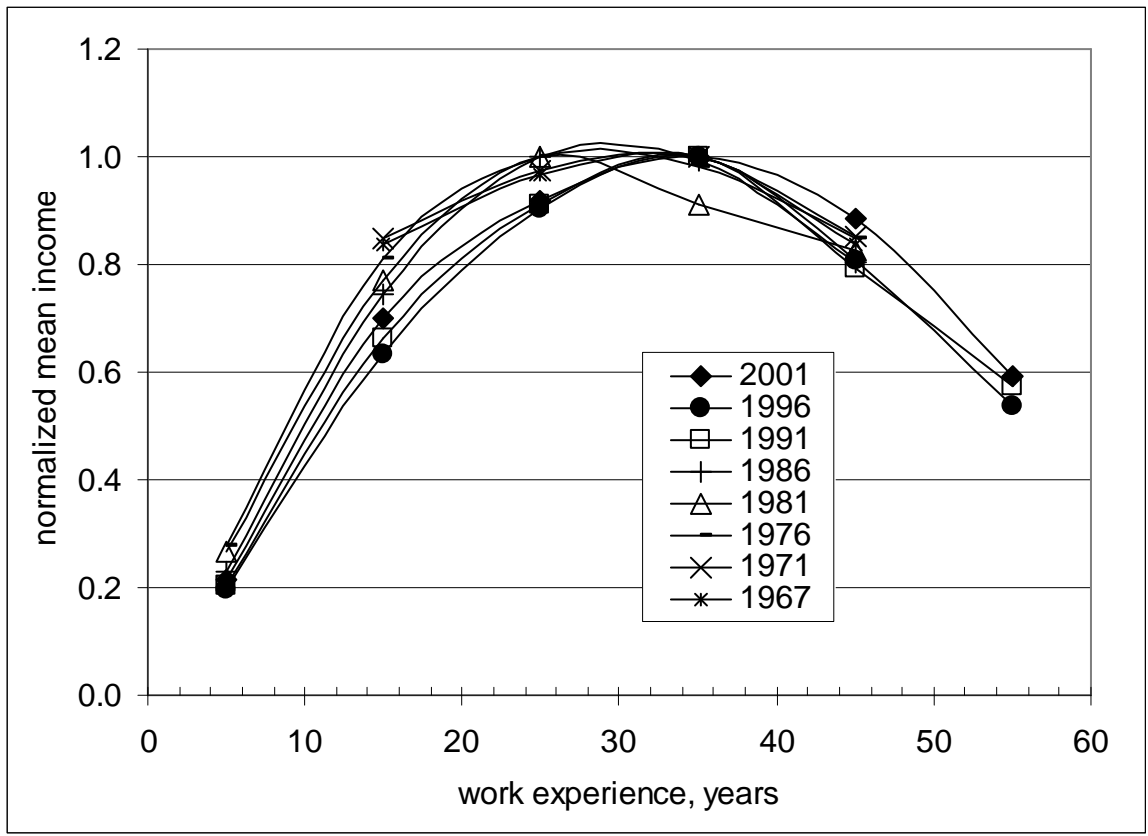

Figure 1.6.5. Normalized mean personal income as a function of work experience for years from 1967 to 2001.

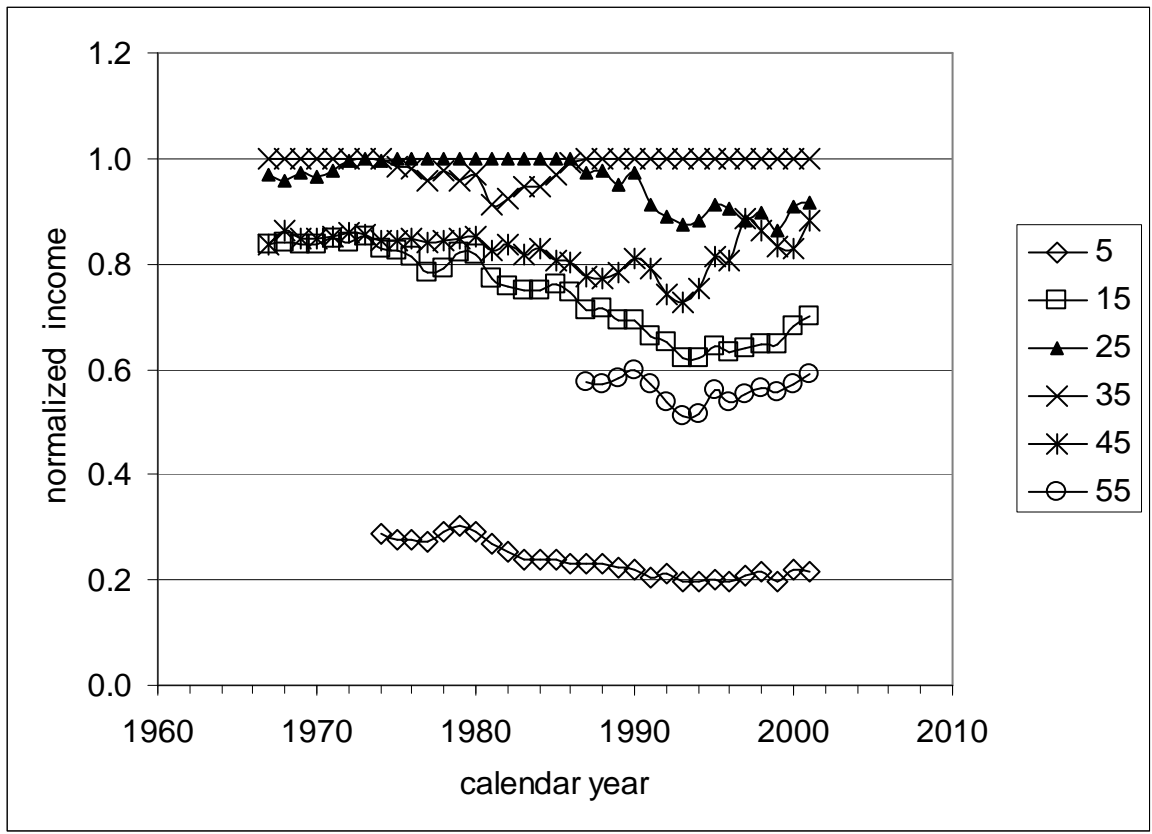

Figure 1.6.6. Mean income in various age groups (5 – from 0 to 9 years, 15 – from 10 to 19 years, …, 55 – from 50 to 59 years of work experience) normalized to the peak value in corresponding years.

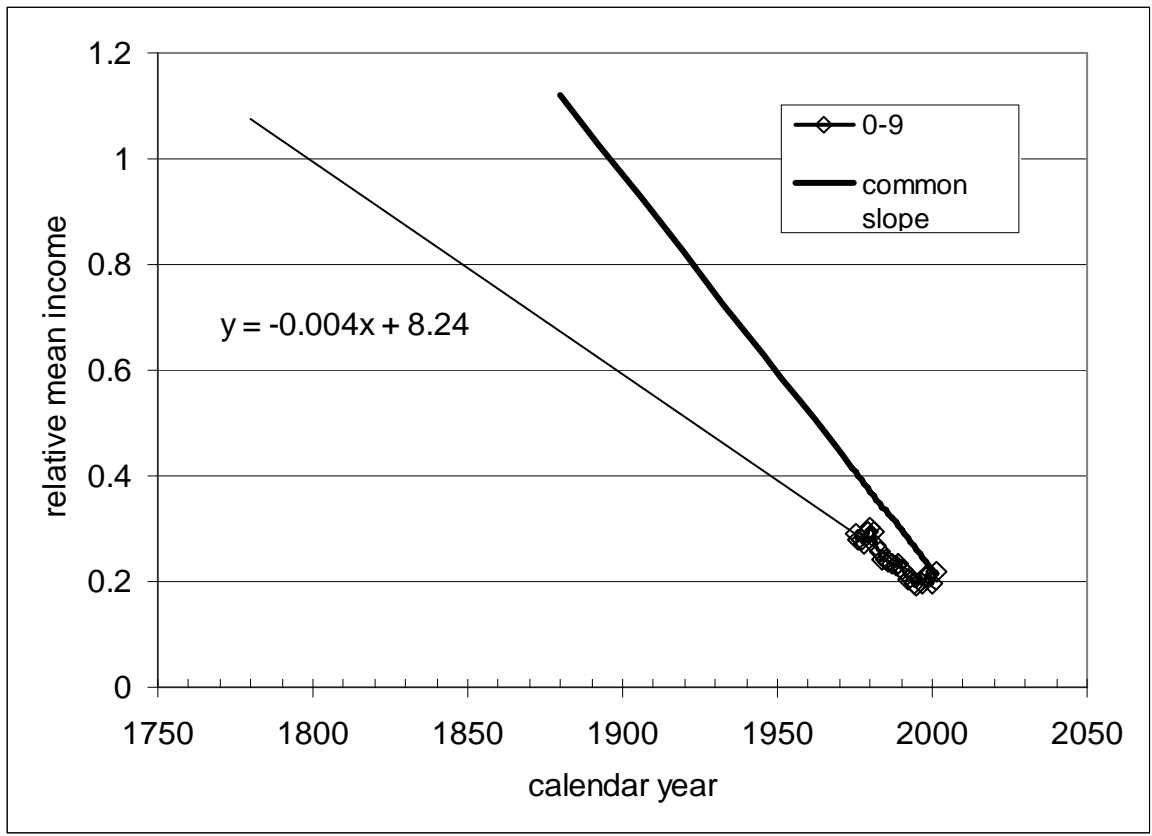

Figure 1.6.7. Mean income in the age group from 15 to 24 years normalized to the peak value in corresponding years as a function of calendar year from 1967 to 2001. Linear regression indicates that this age group could contain the peak value in the last quarter of 18[th] century. If the slope of -0.075, as obtained for the older age groups, is used the group could have the peak value near the end of the 19[th] century.

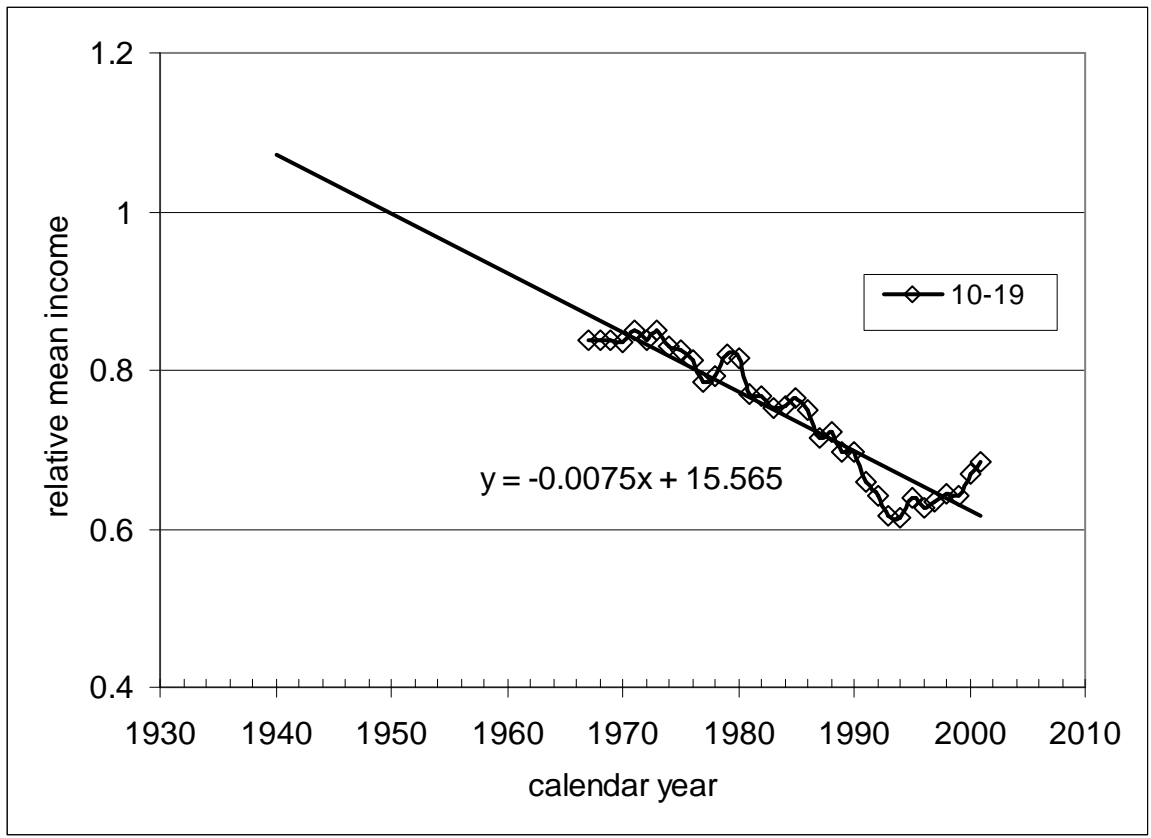

Figure 1.6.8. Evolution of mean income in the age group from 25 to 34 years normalized to the peak value in corresponding years. The slope is -0.075. This age group reached the peak value between 1940 and 1950.

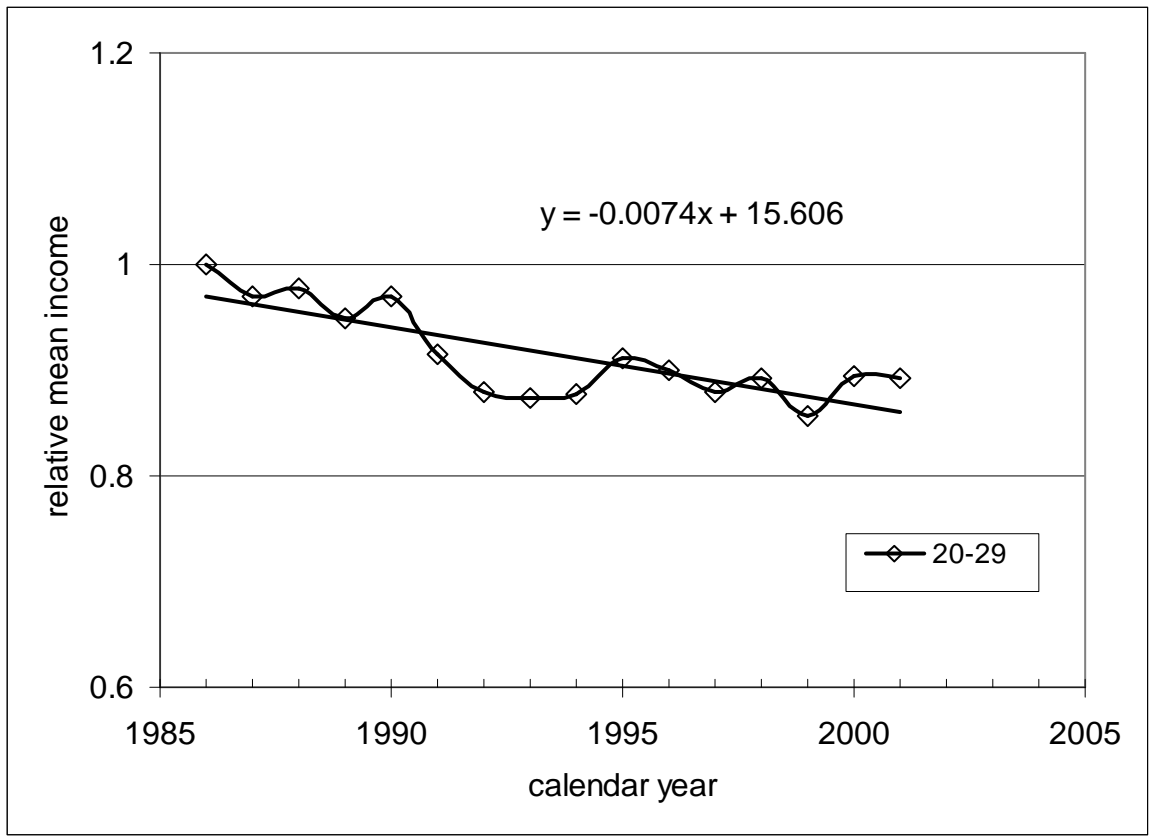

Figure 1.6.9. evolution of mean income in the age group from 35 to 44 years normalized to the peak value in corresponding years. The slope is -0.074. This age group reached the peak value in 1986.

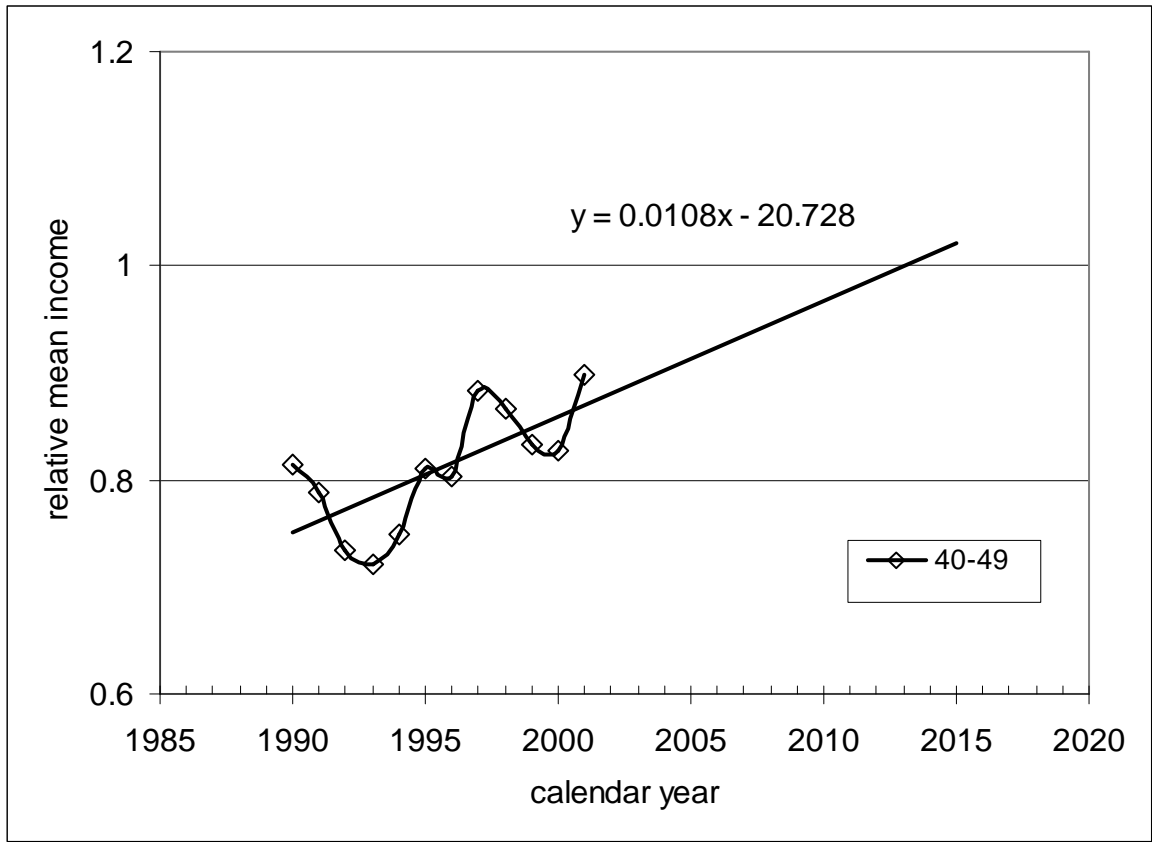

Figure 1.6.10. Evolution of mean income in the age group from 55 to 64 years normalized to the peak value in corresponding years. The slope is +0.011. This age group will reach the peak value in around 2015.

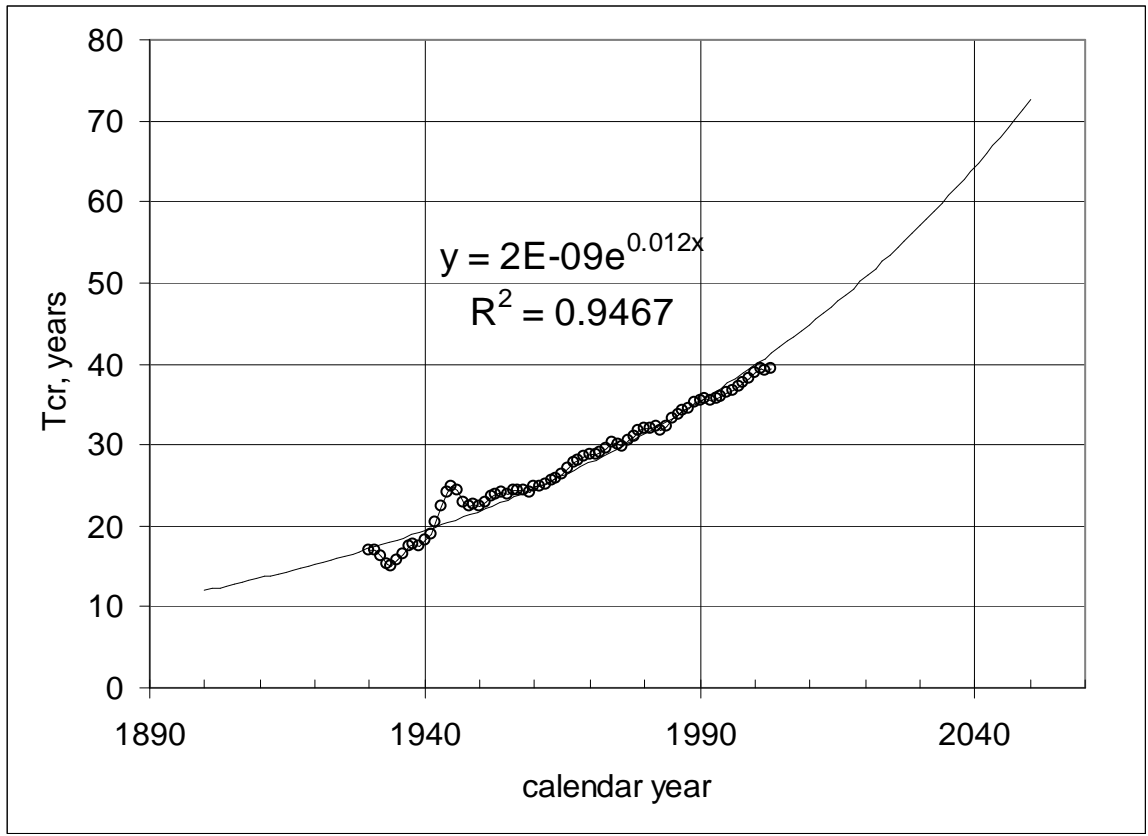

Figure 1.6.11. Critical work experience as a function of calendar year as calculated from the growth rate of real GDP per capita using relationship 1.3.5. $T_{cr}$ was about 25 in the late 1950s, reached the 30 year level in the late 1970s, and is currently near the 40-year threshold.

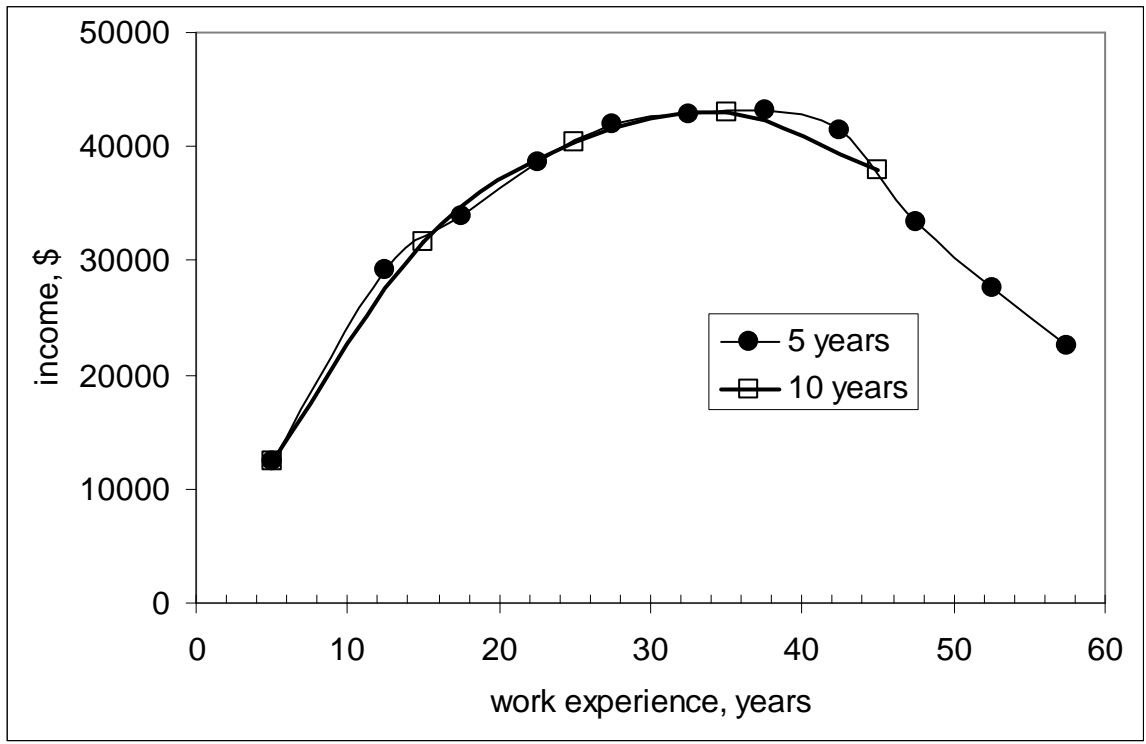

Figure 1.6.12. Comparison of the mean income dependence on work experience as obtained in 10-year and 5-year wide intervals for the year of 2001. The distributions have quite different asymptotic decay above $T_{cr}$.

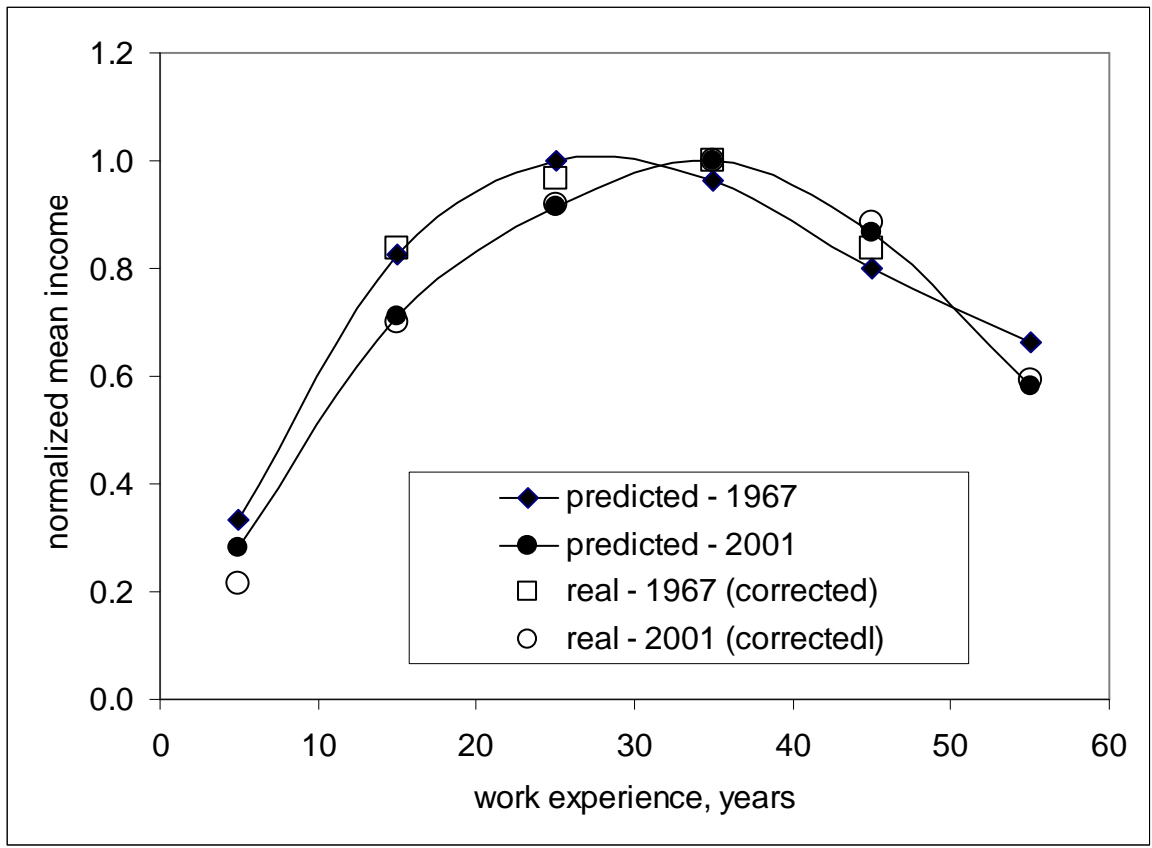

Figure 1.6.13. Comparison of observed and predicted mean income dependence on work experience in 1967 and 2001. The observed mean incomes are corrected for the population without income. Averaging is accomplished in 10-year intervals of work experience. The factor used to scale the predicted values is 72 in 1967 and 2001.`

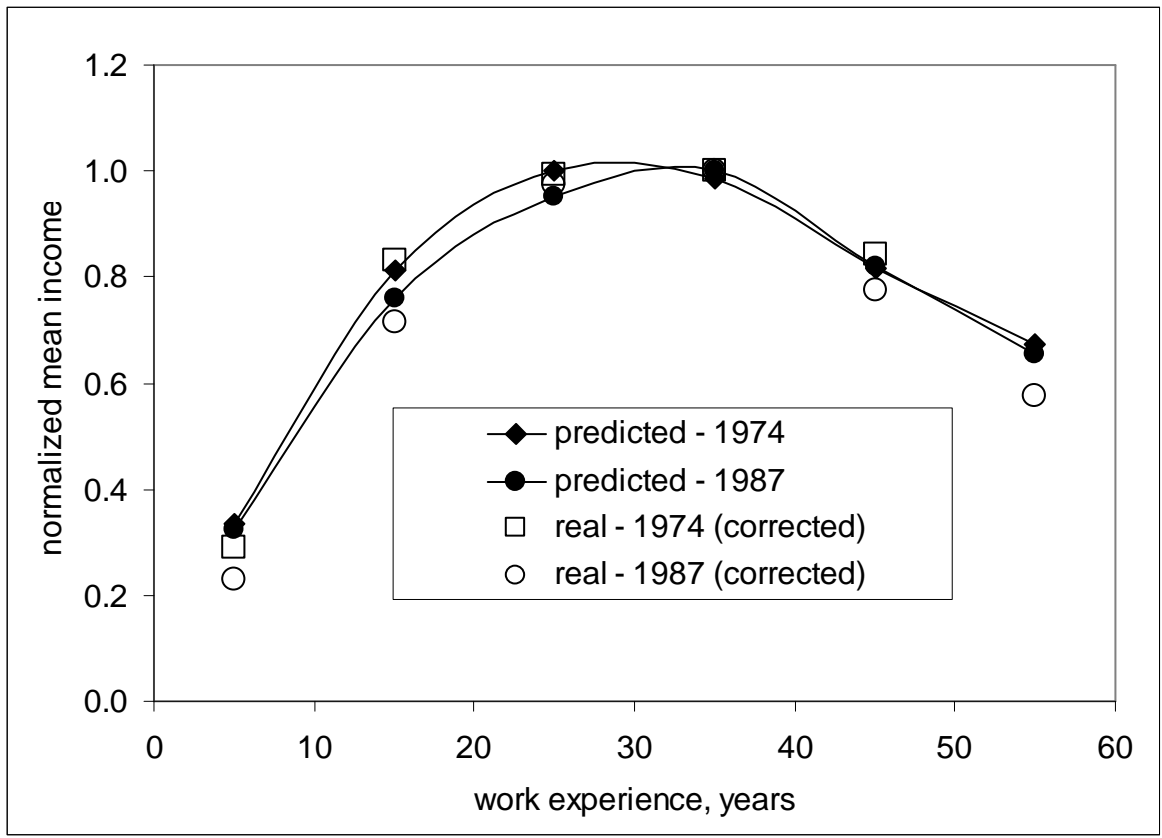

Figure 1.6.14. Comparison of observed and predicted mean personal income dependence on work experience in 1974 and 1987 (the years of major changes in CPS procedures). The observed mean incomes are corrected for the population without income. Averaging is accomplished in 10-year intervals of work experience. The factor used to covert the predicted values is the same in 1987 and 1974 - 79. This value is different from that used in 2001 and 1967.

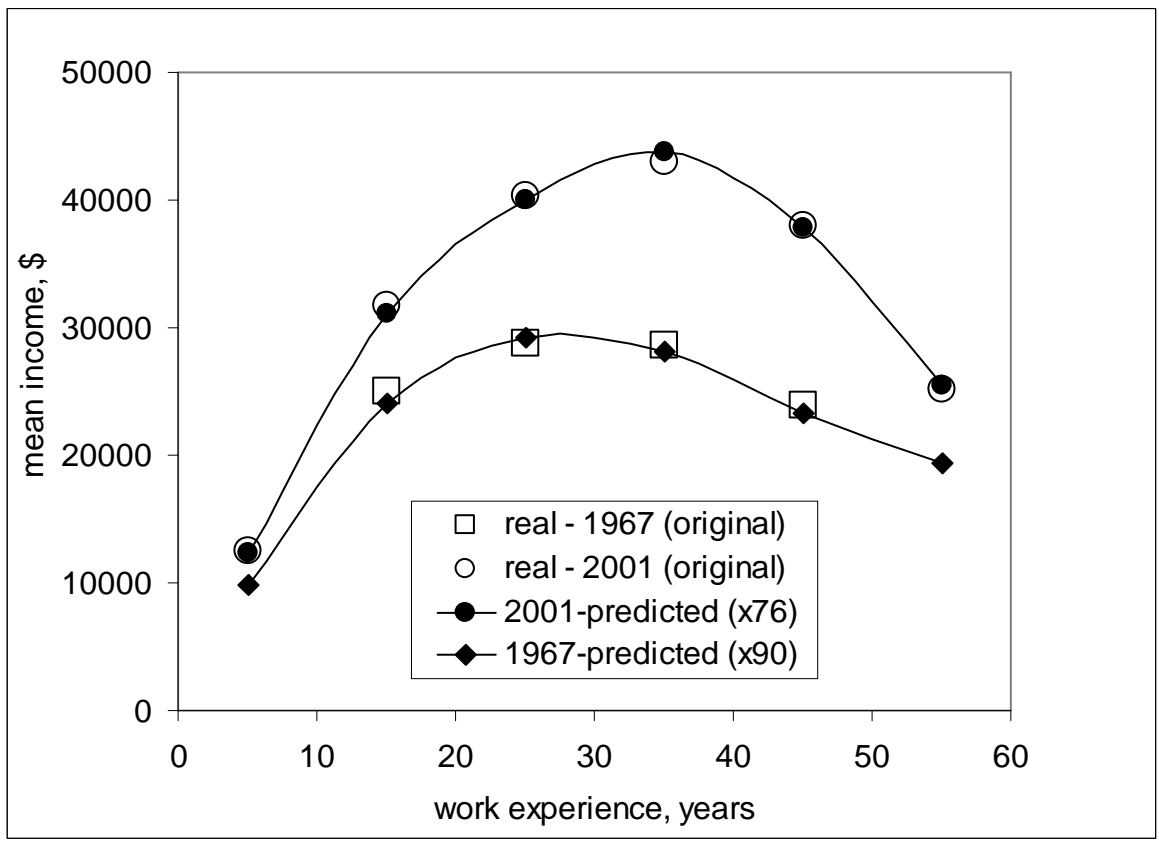

Figure 1.6.15. Comparison of observed and predicted mean personal income dependence on work experience in 1967 and 2001. The observed mean incomes are as in the original tables. Averaging is accomplished in 10-year intervals of work experience. The factor used to covert the predicted values is 76 in 2001 and 90 in 1967.

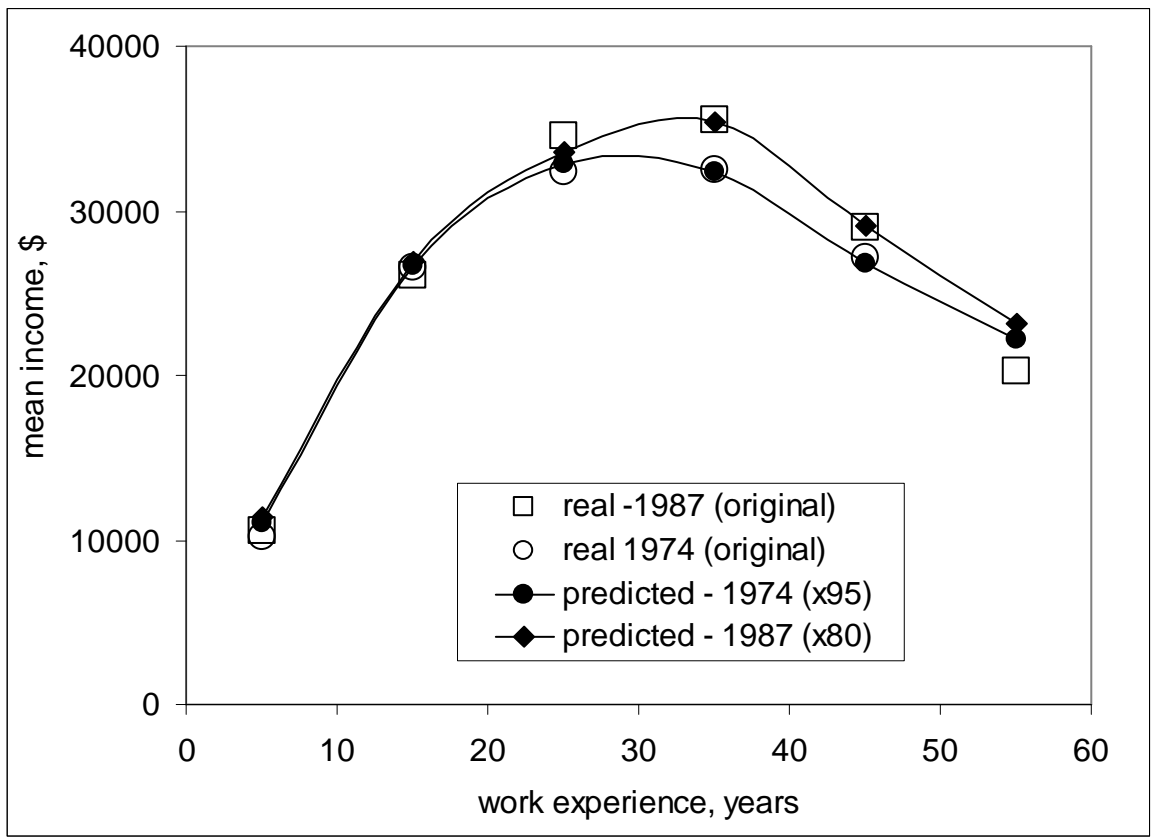

Figure 1.6.16. Comparison of observed and predicted mean personal income dependence on work experience in 1974 and 1987. The factor used to covert the predicted values is 80 in 1987 and 95 in 1974.

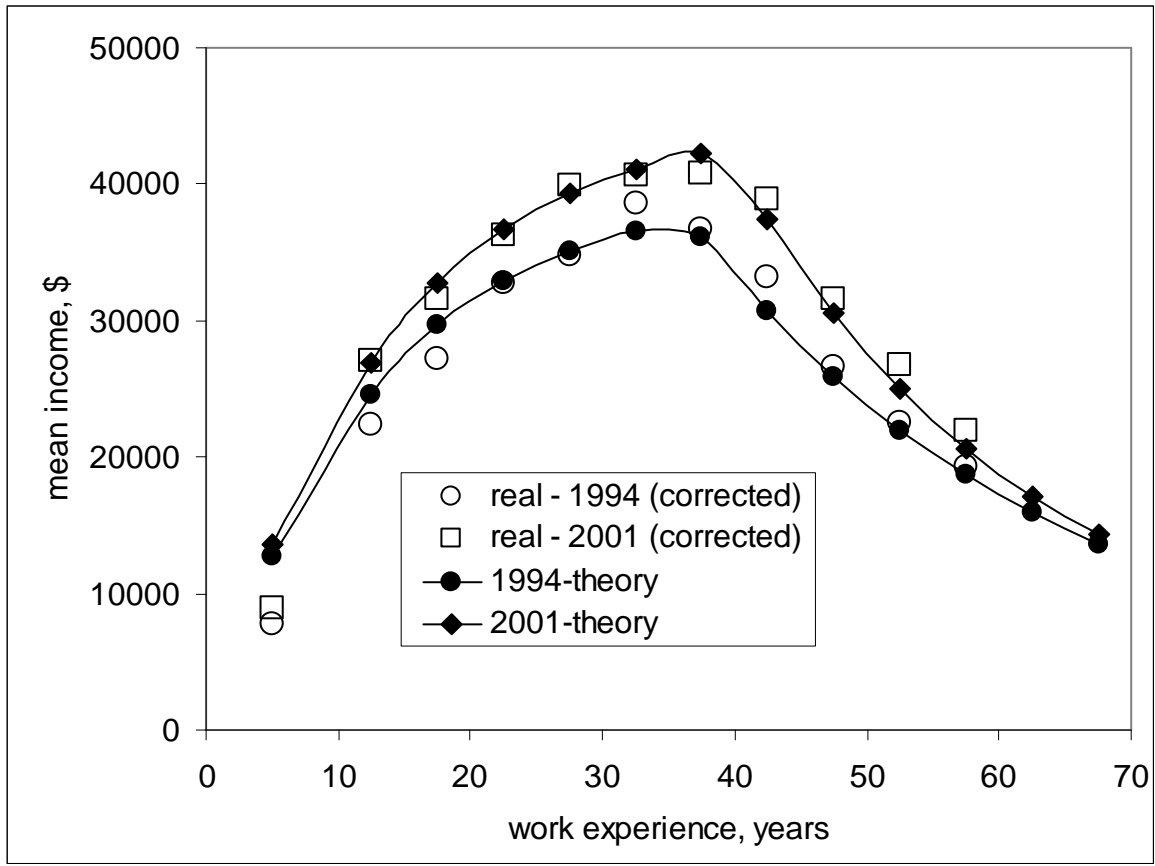

Figure 1.6.17. Comparison of observed and predicted mean personal income dependence on work experience in 1994 and 2001. The observed mean incomes are corrected for the population without income. Averaging is accomplished in 5-year intervals (except the first, which is 10-year interval) of work experience. The boundary condition for the mean income at the age of 67 years is 0.45.

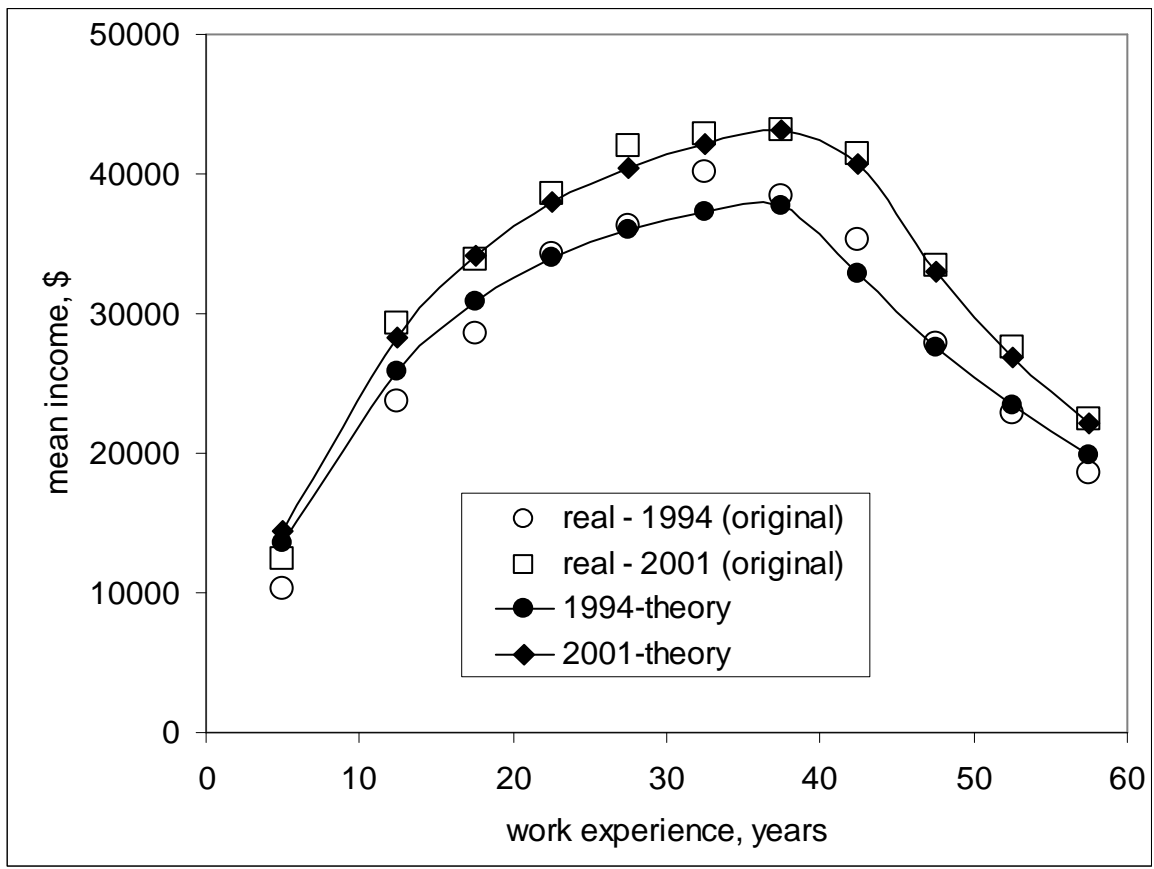

Figure 1.6.18. Comparison of actual and predicted mean personal income dependence on work experience in 1994 and 2001. The observed mean incomes are as in the original. Averaging is accomplished in 5-year intervals (except the first one) of work experience.

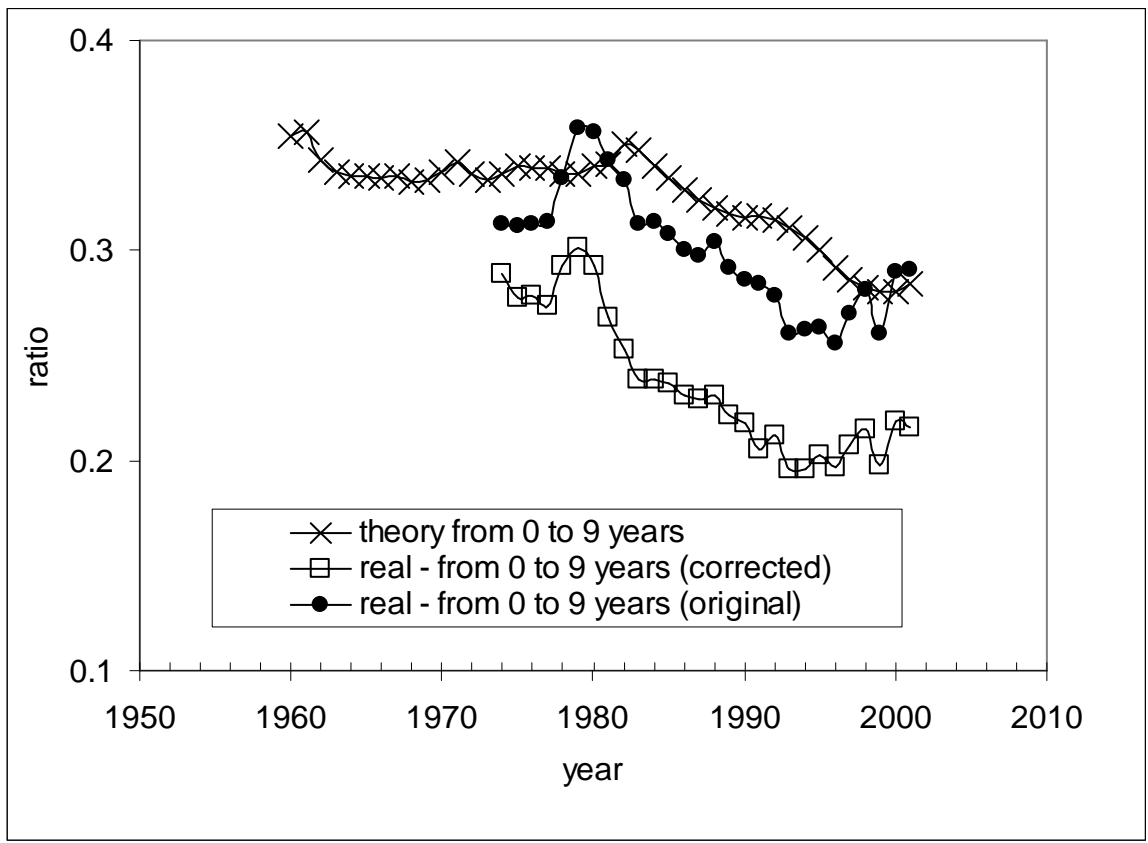

Figure 1.6.19. Evolution of observed (original and corrected for the population without income) and predicted average income value in the work experience group from 0 to 9 years normalized to the peak average income over all work experience groups.

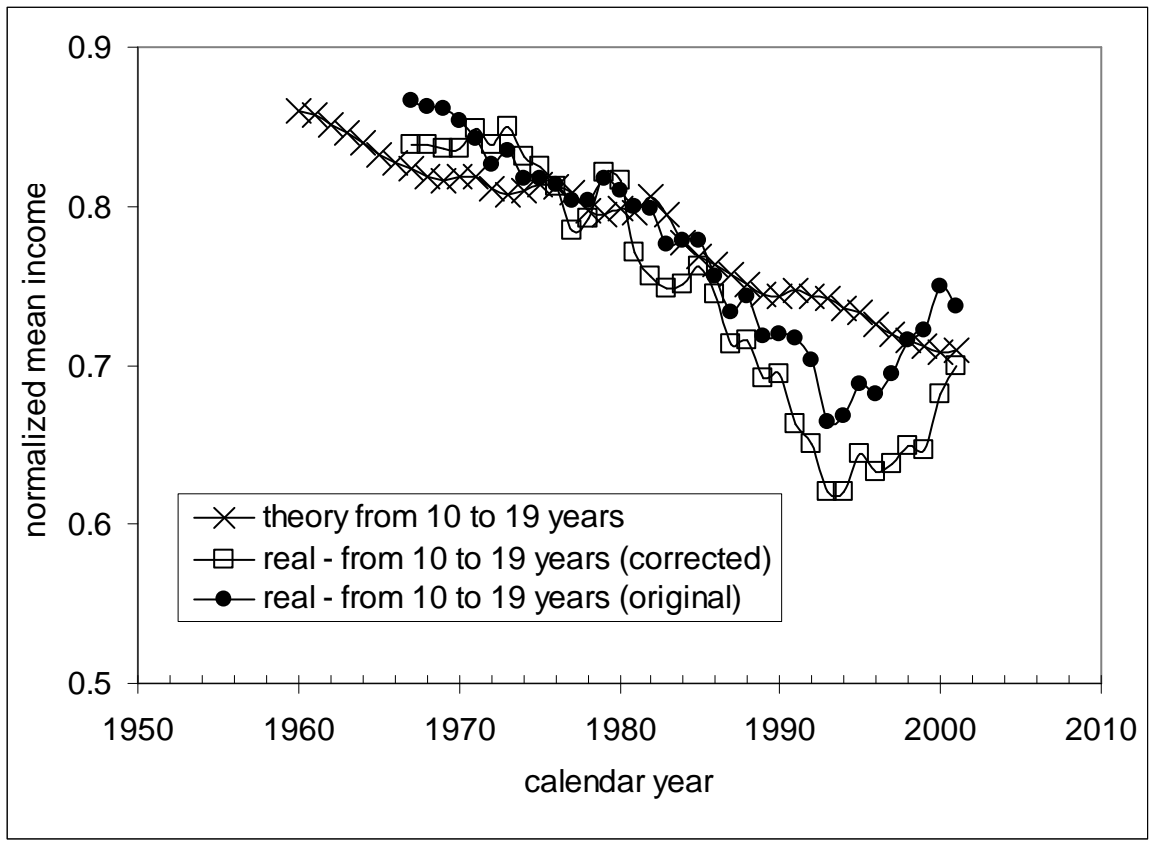

Figure 1.6.20. Evolution of observed and predicted average income value in the work experience group from 10 to 19 years normalized to the peak average income over all work experience groups.

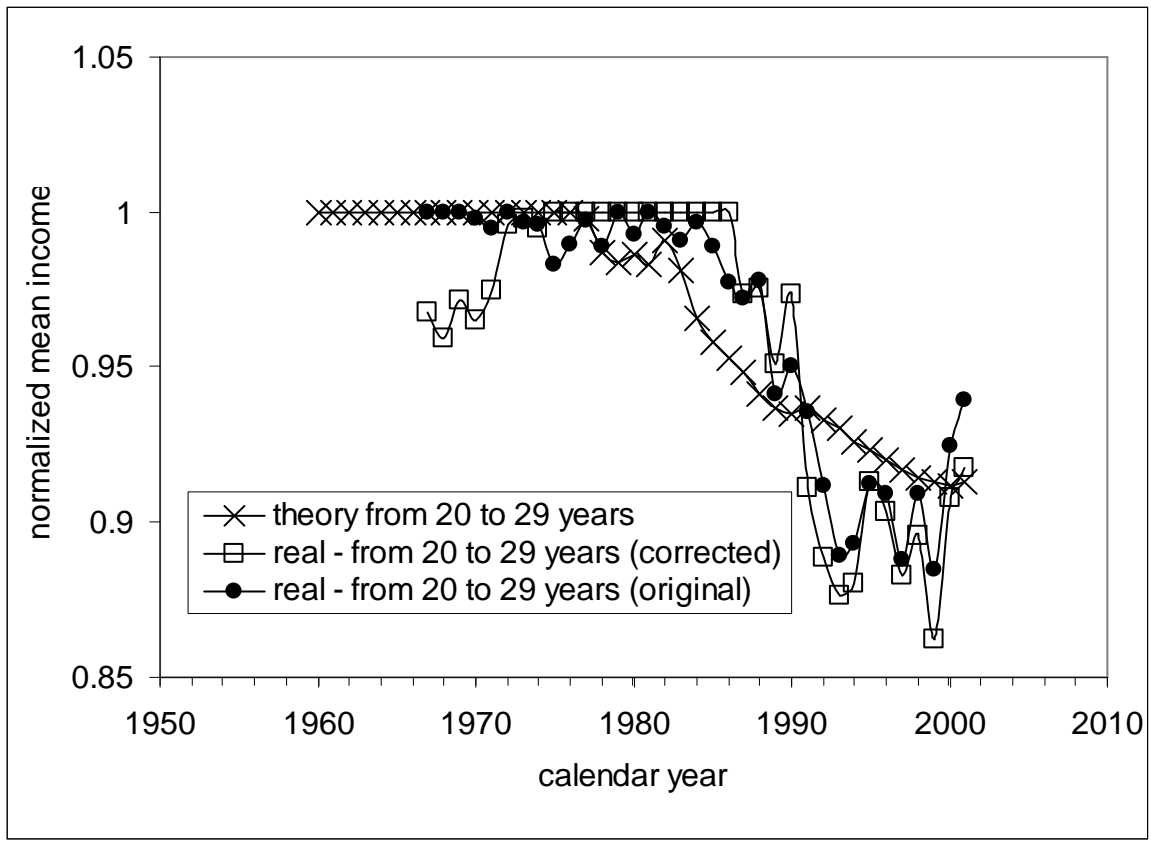

Figure 1.6.21. Evolution of observed and predicted average income value in the work experience group from 20 to 29 years normalized to the peak average income over all work experience groups.

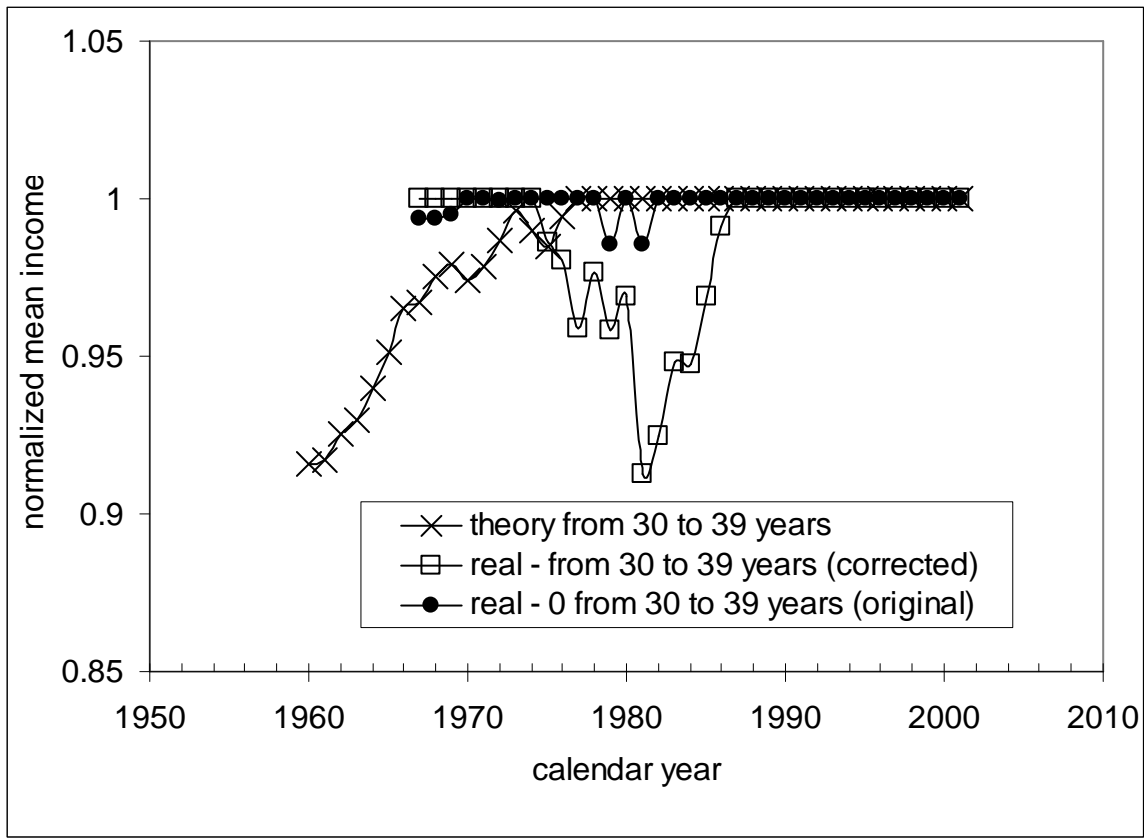

Figure 1.6.22. Evolution of observed and predicted average income value in the work experience group from 30 to 39 years normalized to the peak average income over all work experience groups.

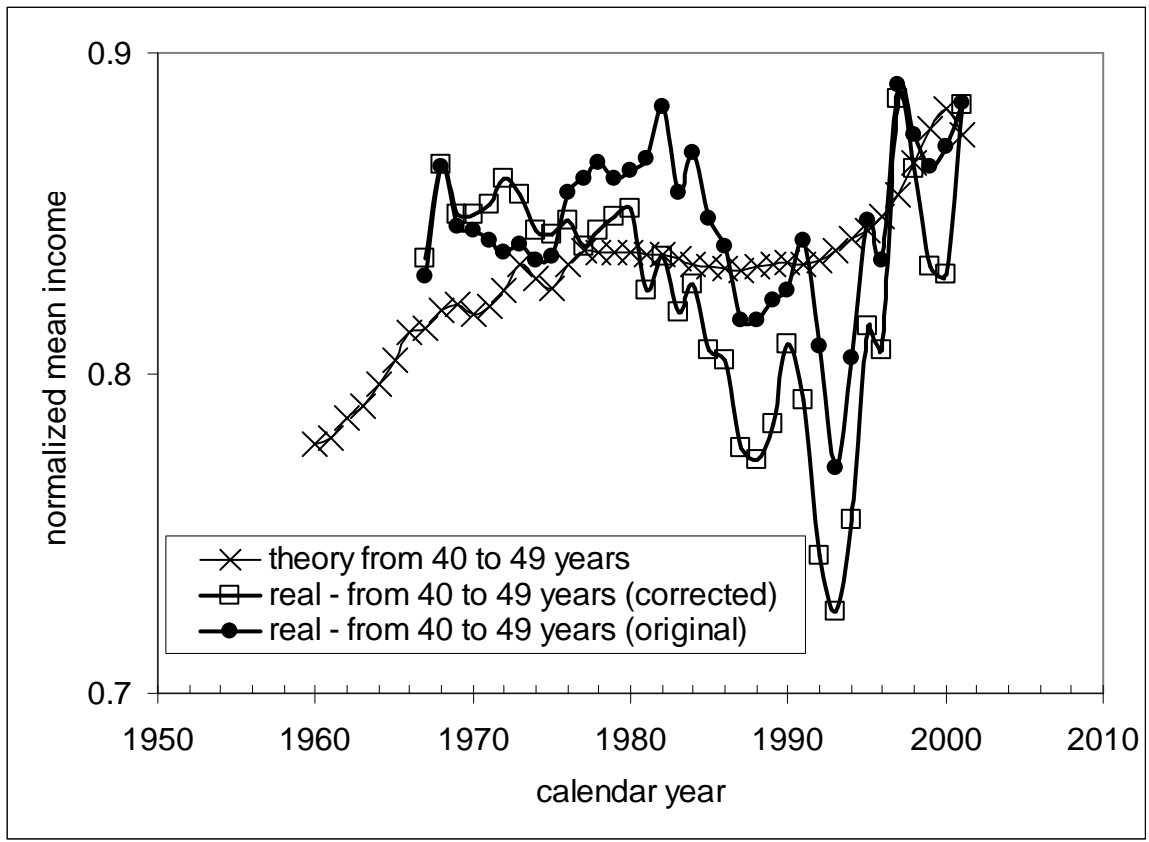

Figure 1.6.23. Evolution of observed and predicted average income value in the work experience group from 40 to 49 years, normalized to the peak average income over all work experience groups.

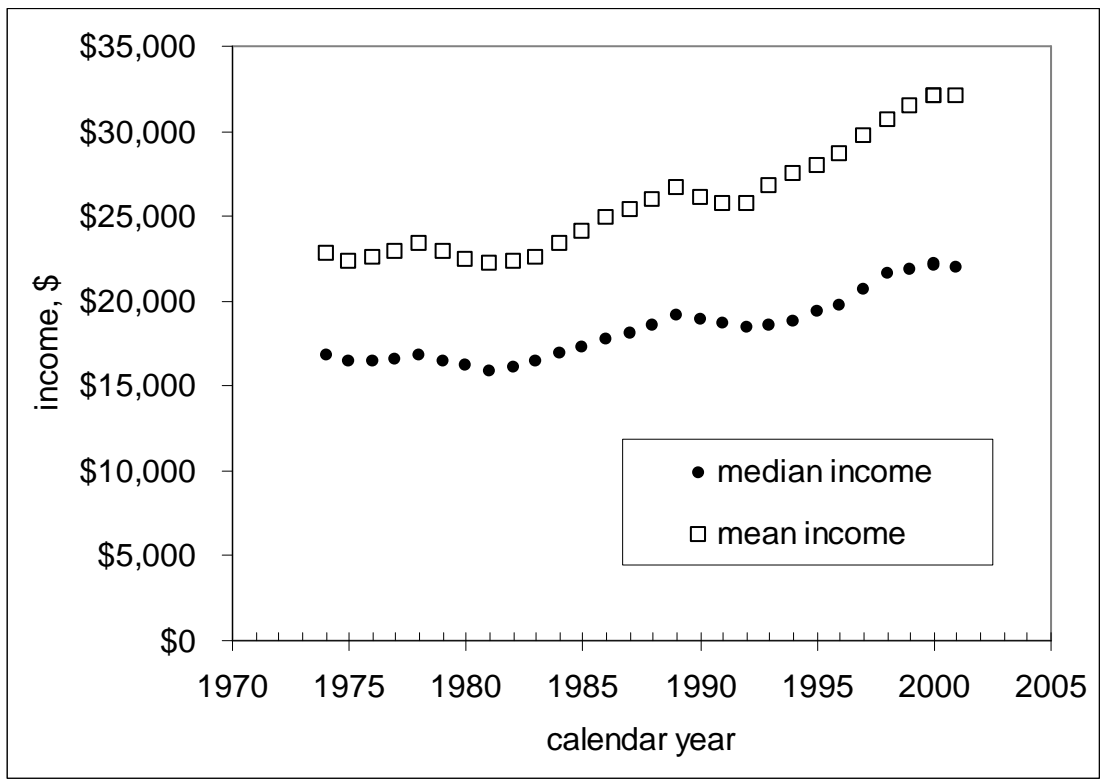

Figure 1.6.24. Evolution of observed overall mean and median income (in 2001 CPI-U-RS adjusted dollars) in the USA between 1974 and 2001.

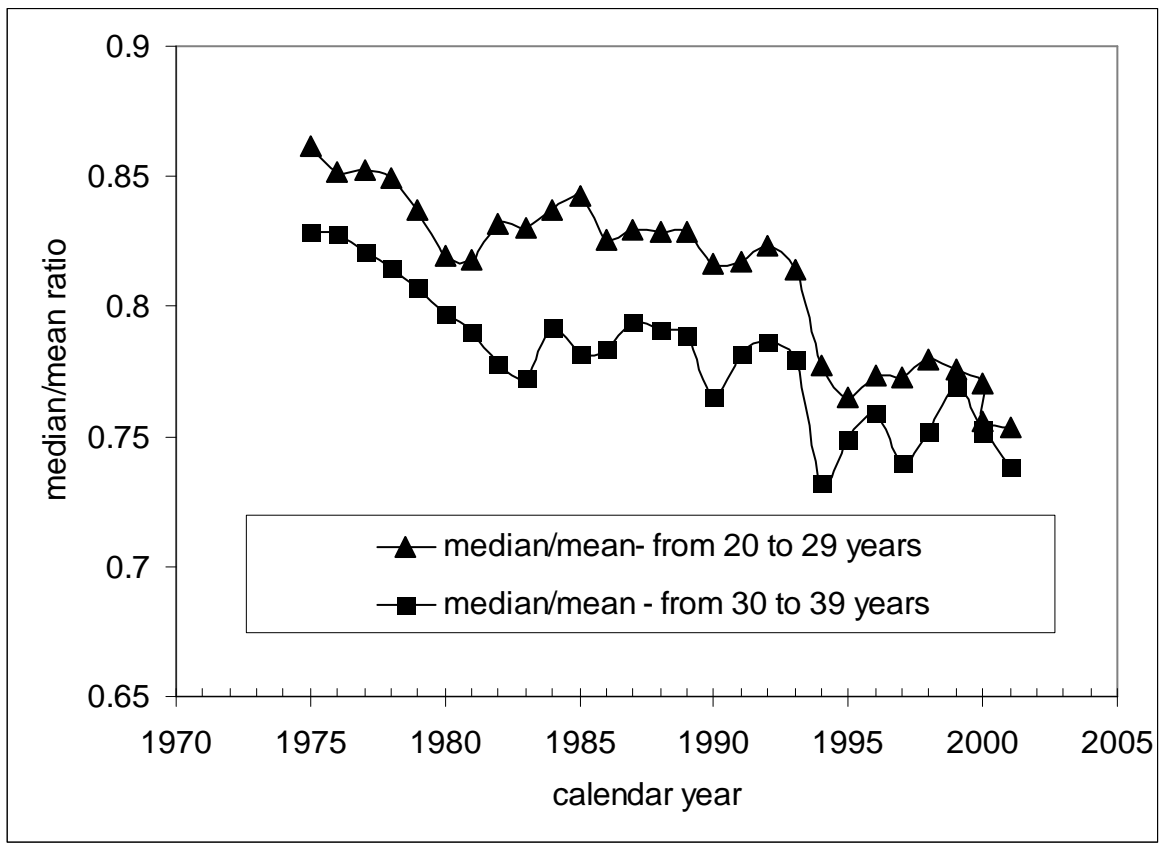

Figure 1.6.25. Ratio of median and mean income in work experience groups from 20 to 29 years and from 30 to 39 years. A decrease in the ratio in both groups is observed from 1974 to 2002.

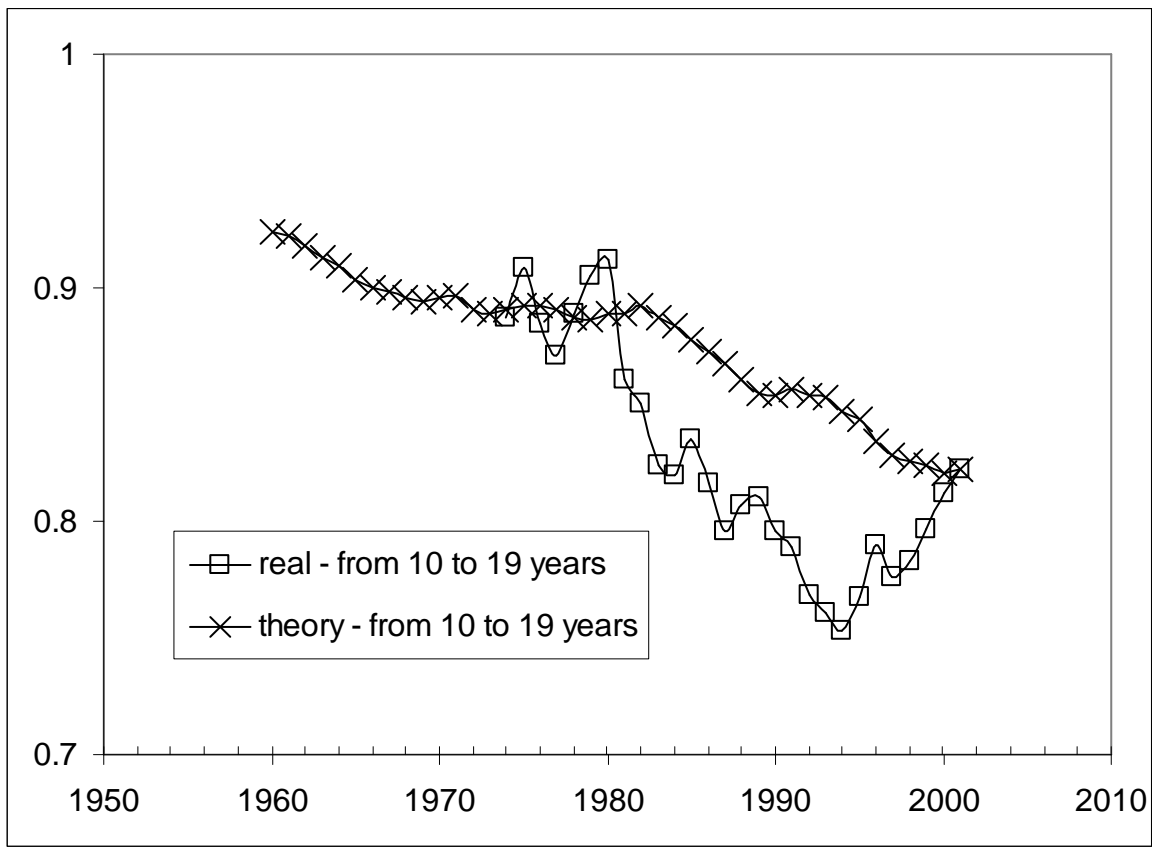

Figure 1.6.26. Evolution of observed and predicted median income in the work experience group from 10 to 19 years, normalized to the peak average income over all groups.

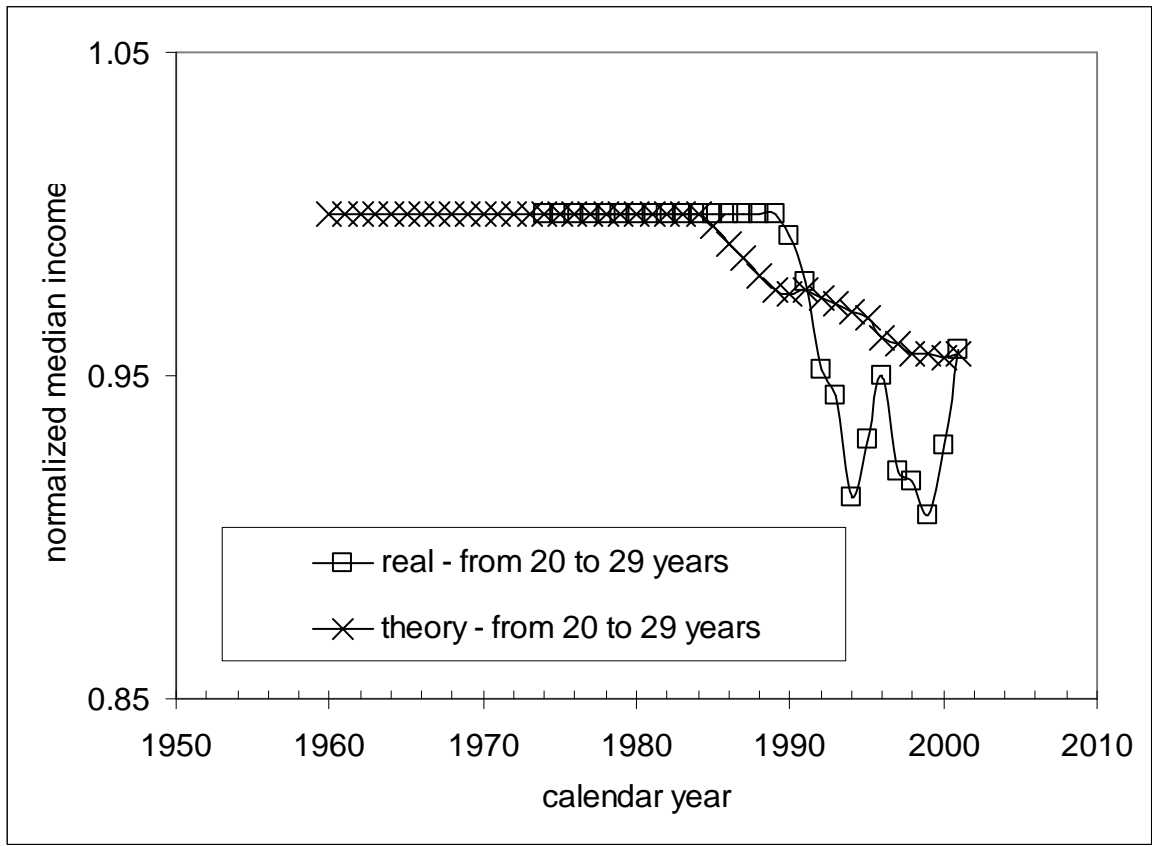

Figure 1.6.27. Evolution of observed and predicted median income in the group from 20 to 29 years, normalized to the peak average income over all groups.

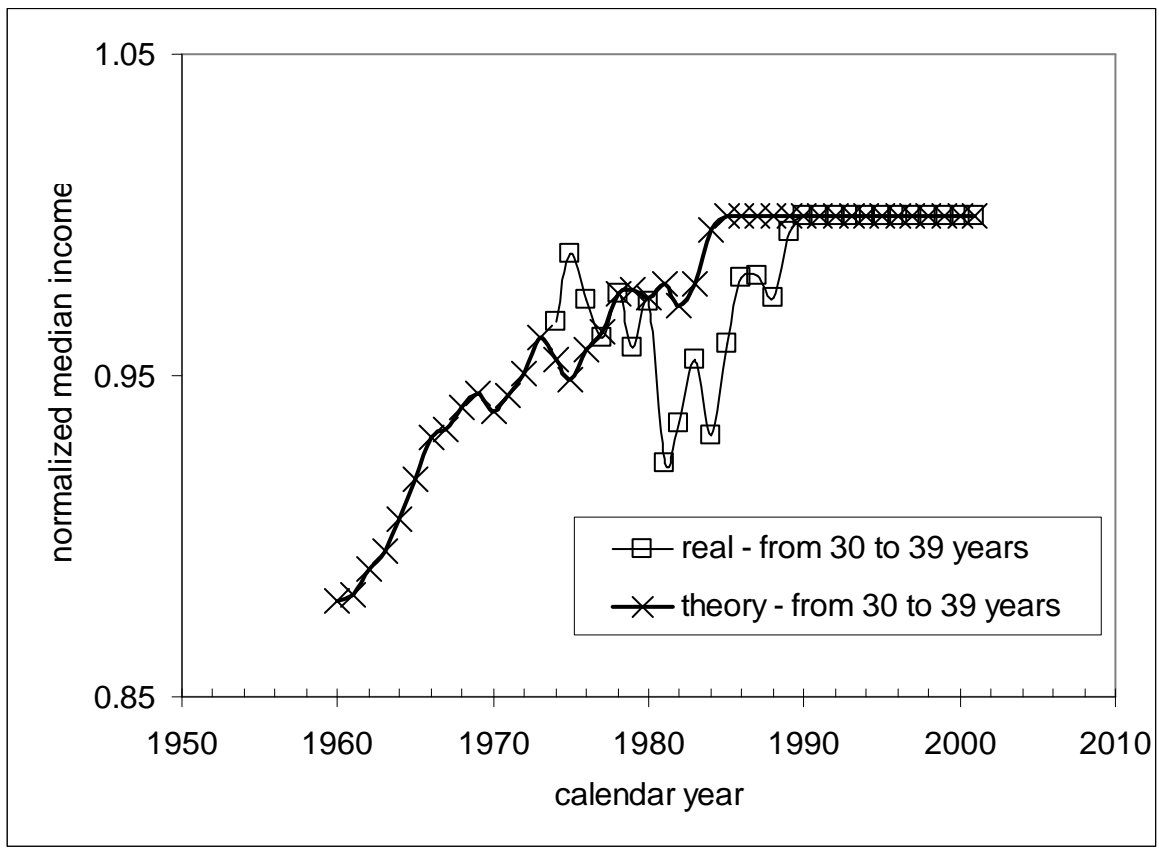

Figure 1.6.28. Evolution of observed and predicted median income in the group from 30 to 39 years, normalized to the peak average income over all the work experience groups.

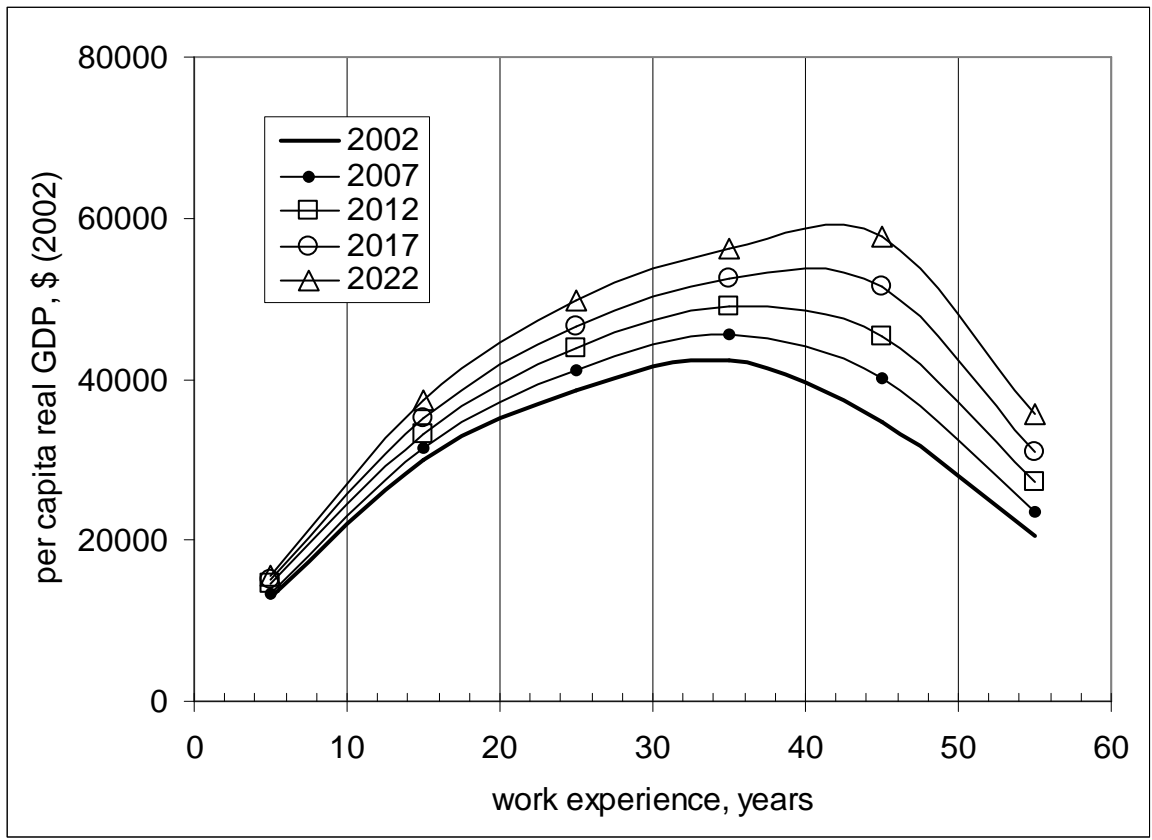

Figure 1.6.29. Evolution of the mean income distribution. The growth rate of per capita real GDP is 1.6% per year. Population projections provided by the US Census Bureau are used.

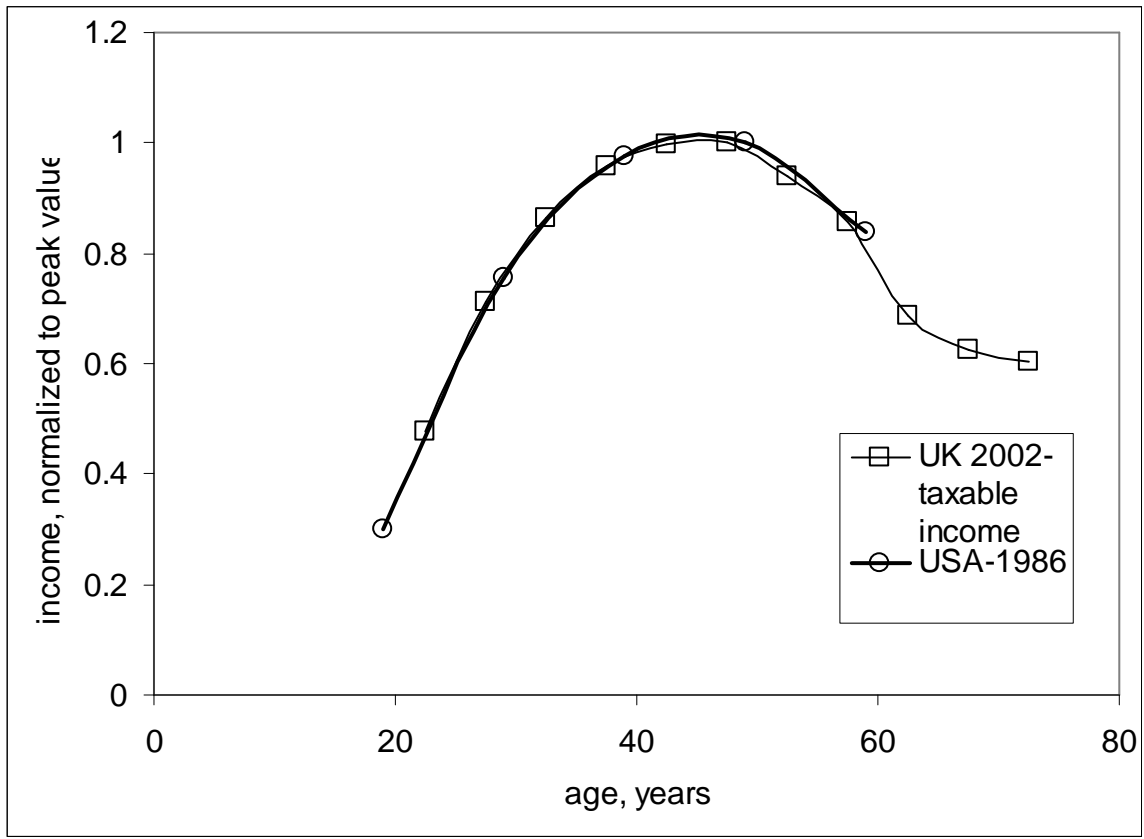

Figure 1.6.30. Mean individual taxable income in the UK (current prices) in 2002 and average income distribution in the US in 1986

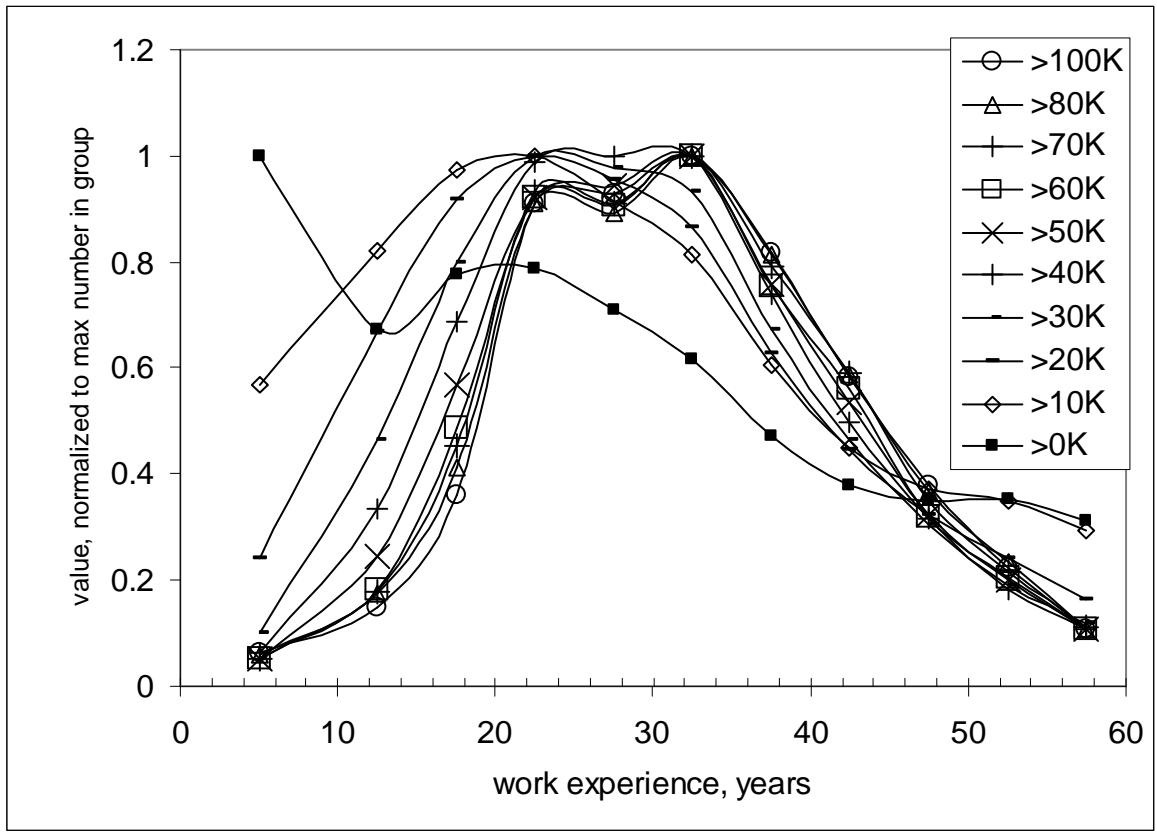

Figure 1.7.1. Evolution personal income distribution in 1994 for incomes above a number of thresholds: >$0K (all personal incomes), >$10K, …, >$100K, each normalized to relevant peak value among all age groups. The normalized distributions for income above $60K are similar. Thus the underlying size distribution is scale free.

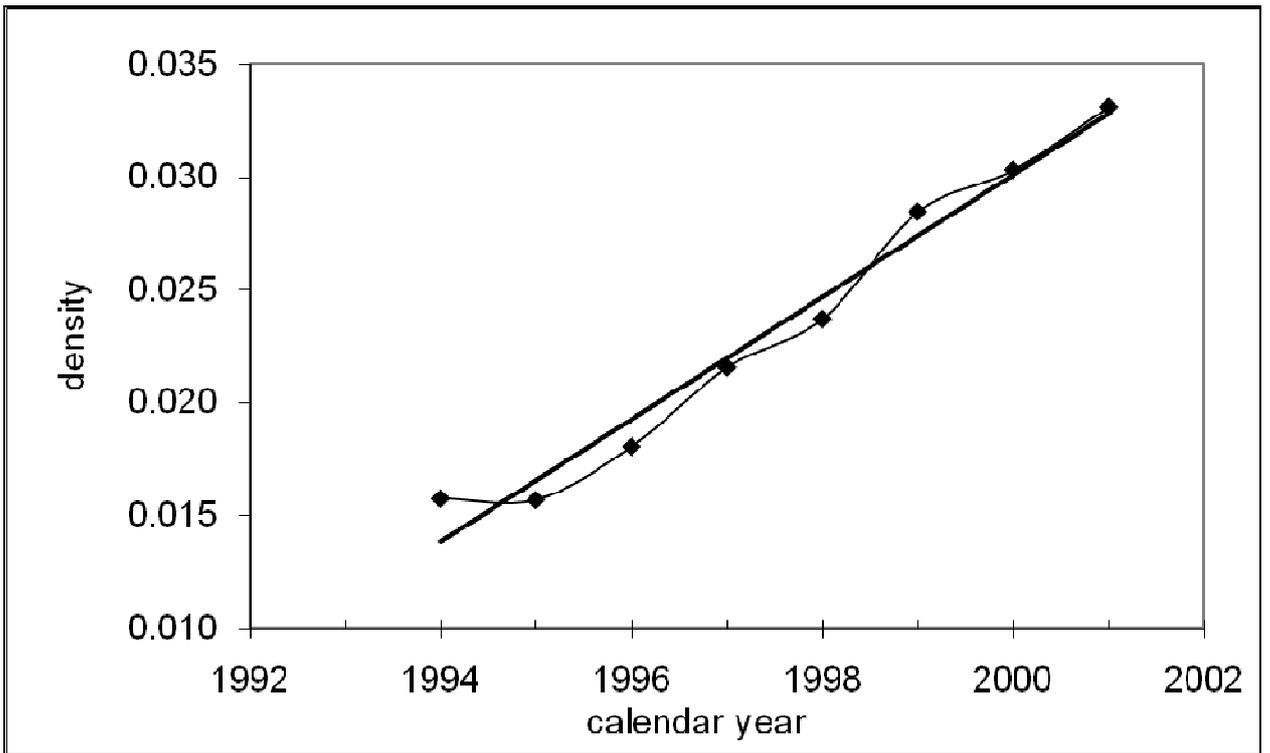

Figure 1.7.2. The evolution of the number of people with income above $100K normalized to relevant working age population. A linear increase with nominal GDP growth is expected.

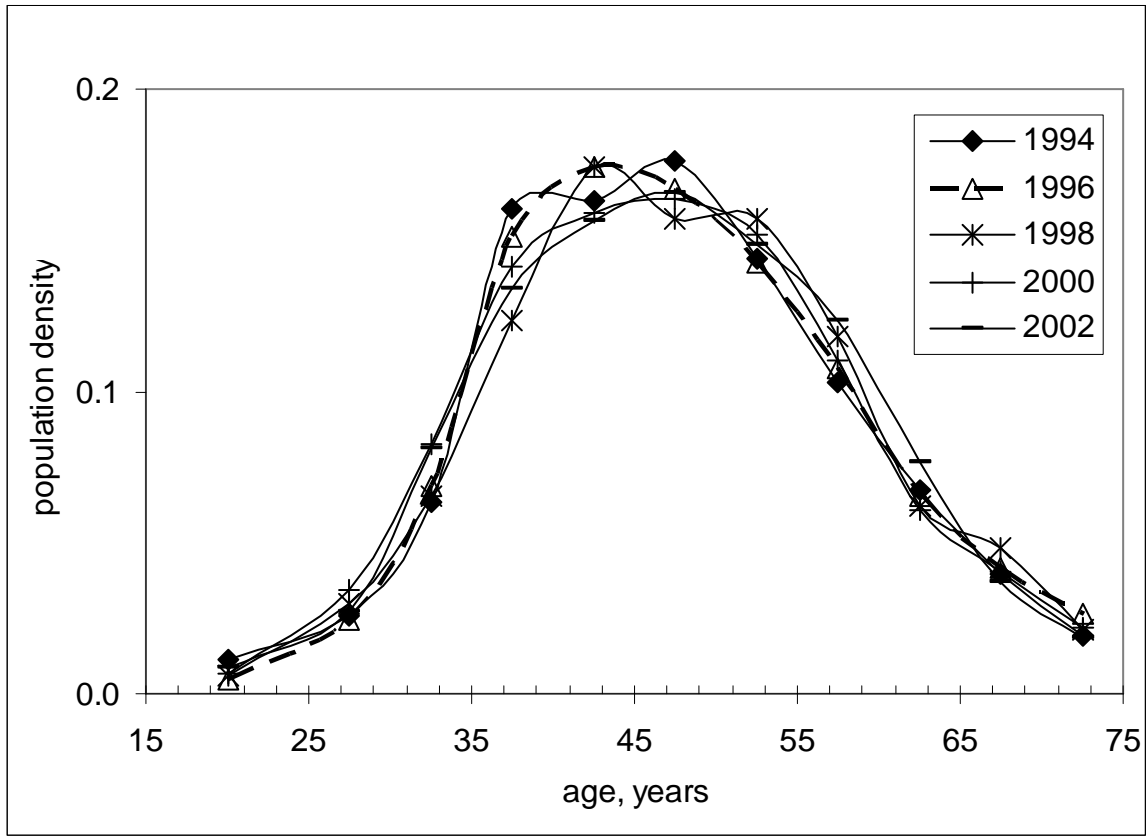

Figure 1.7.3. Normalized personal income distribution for incomes above $100K as a function of work experience. The curves are shown for even years from 1994 to 2002.

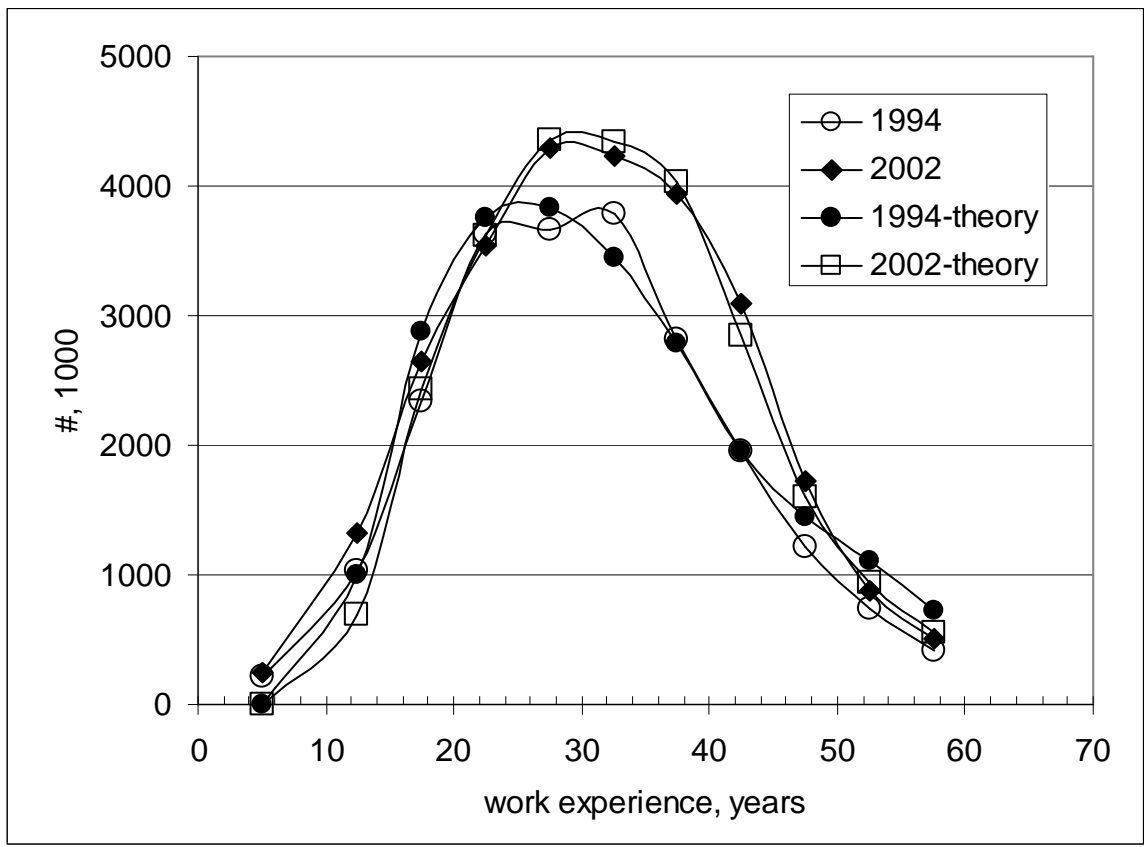

Figure 1.7.4. Comparison of observed and predicted dependence on work experience of the number of people with income above the Pareto threshold.

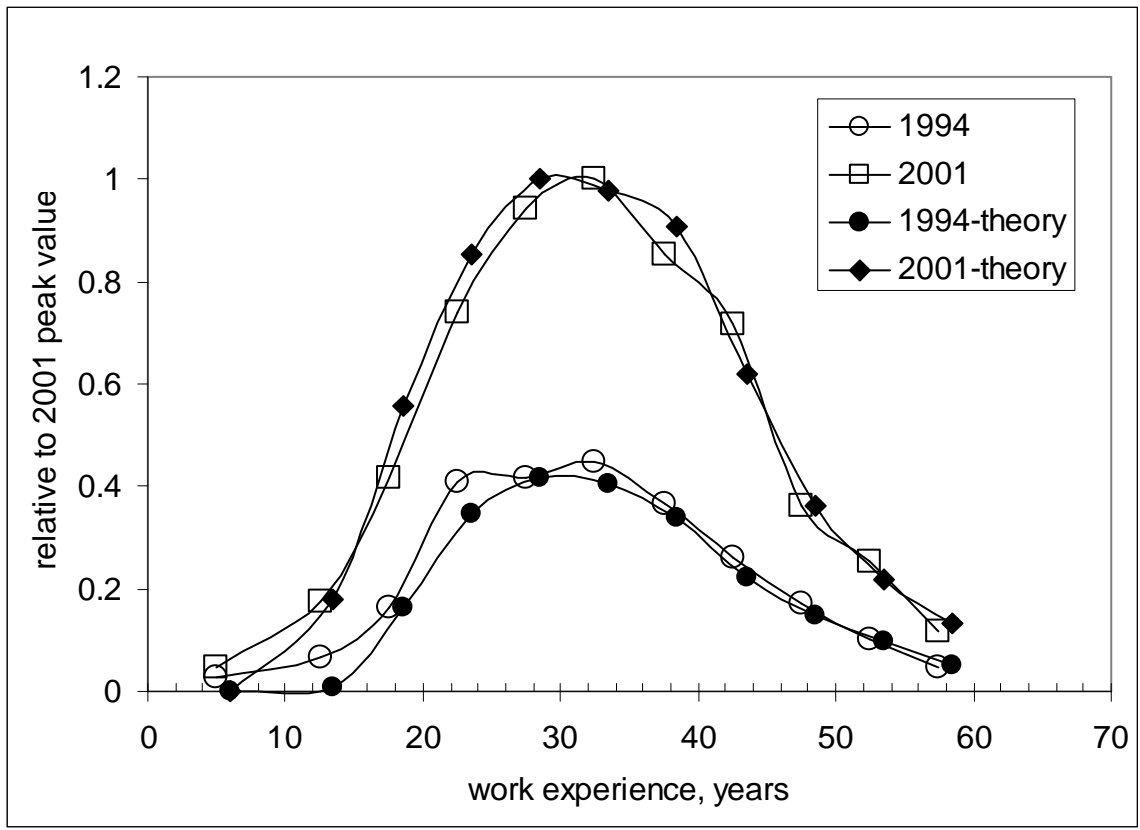

Figure 1.7.5. Comparison of observed and predicted dependence on work experience of the number of people with income above $100,000 (current dollars). All curves are normalized to the peak value in 2001.

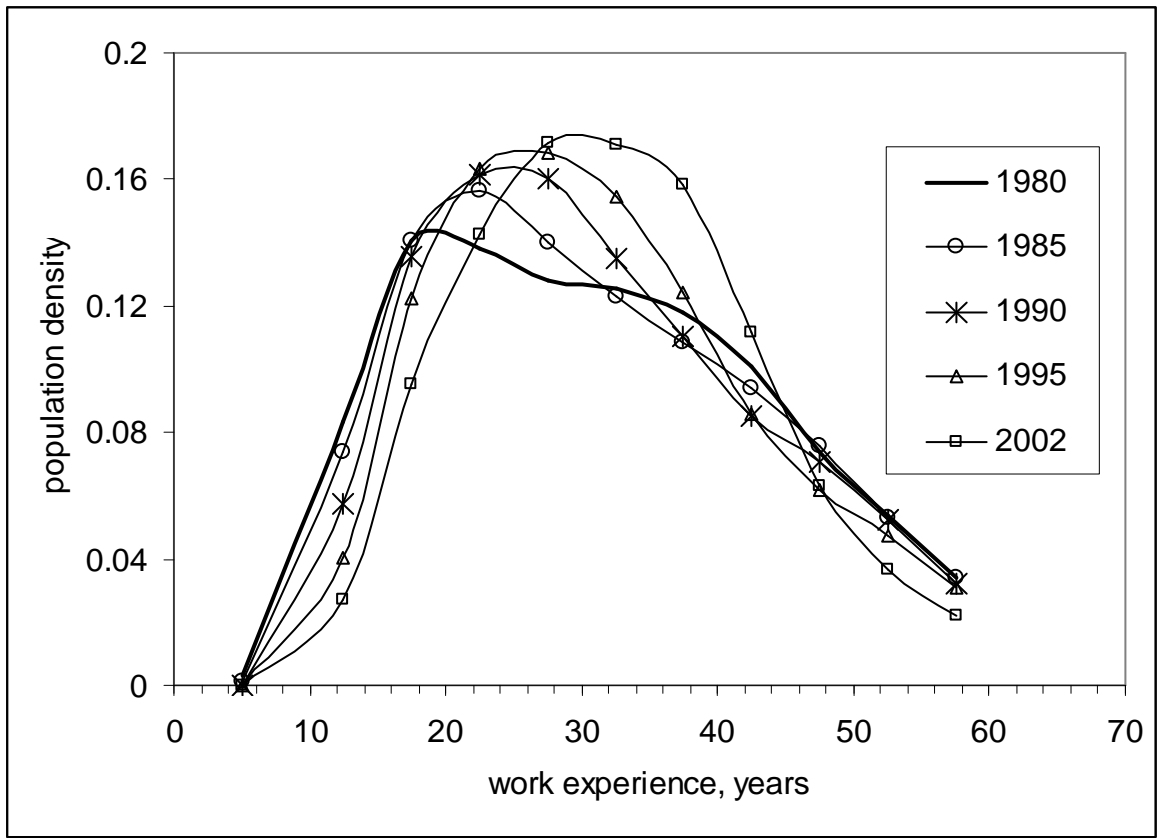

Figure 1.7.6. Evolution of population density distribution for people with income above the Pareto threshold for calendar years 1980 through 2002. Notice the increase in time needed to reach the Pareto threshold from 1980 to 2002. This is the result of the decrease in effective dissipation factor, $\alpha/L$, with increasing size of earning means, $L$.

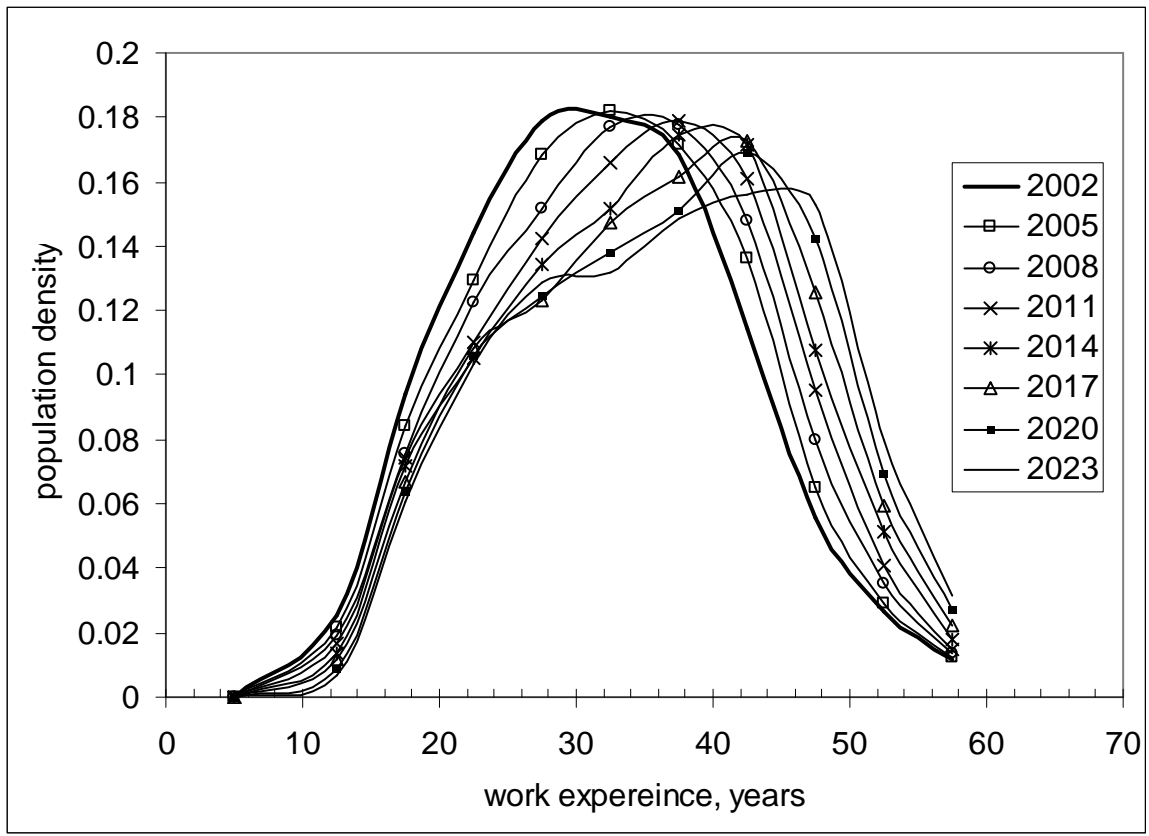

Figure 1.7.7. Evolution of normalized population density distribution for people with income above the Pareto threshold for selected calendar years between 2002 and 2023. The growth rate of real GDP per capita is set to 0.016 per year. Population projections are obtained from U.S. Census Bureau.

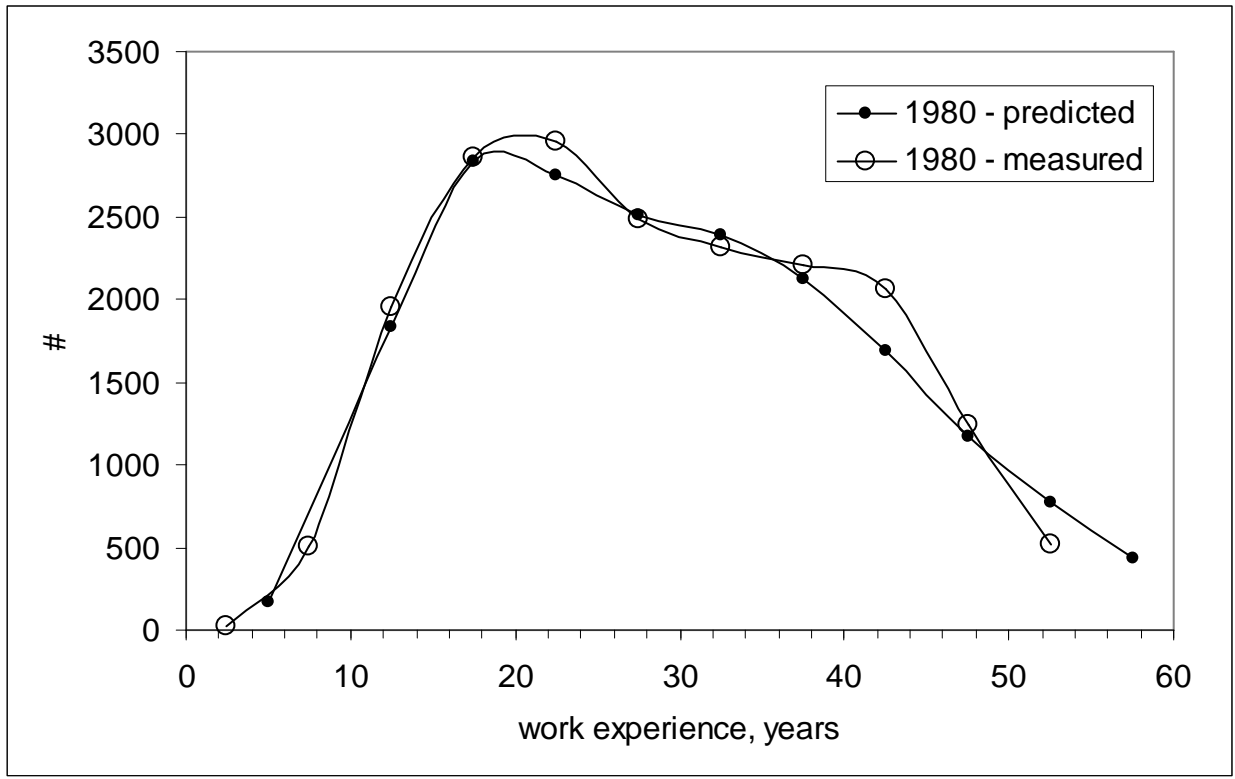

Figure 1.7.8. Observed and predicted number for people with income above the Pareto threshold in 1980.

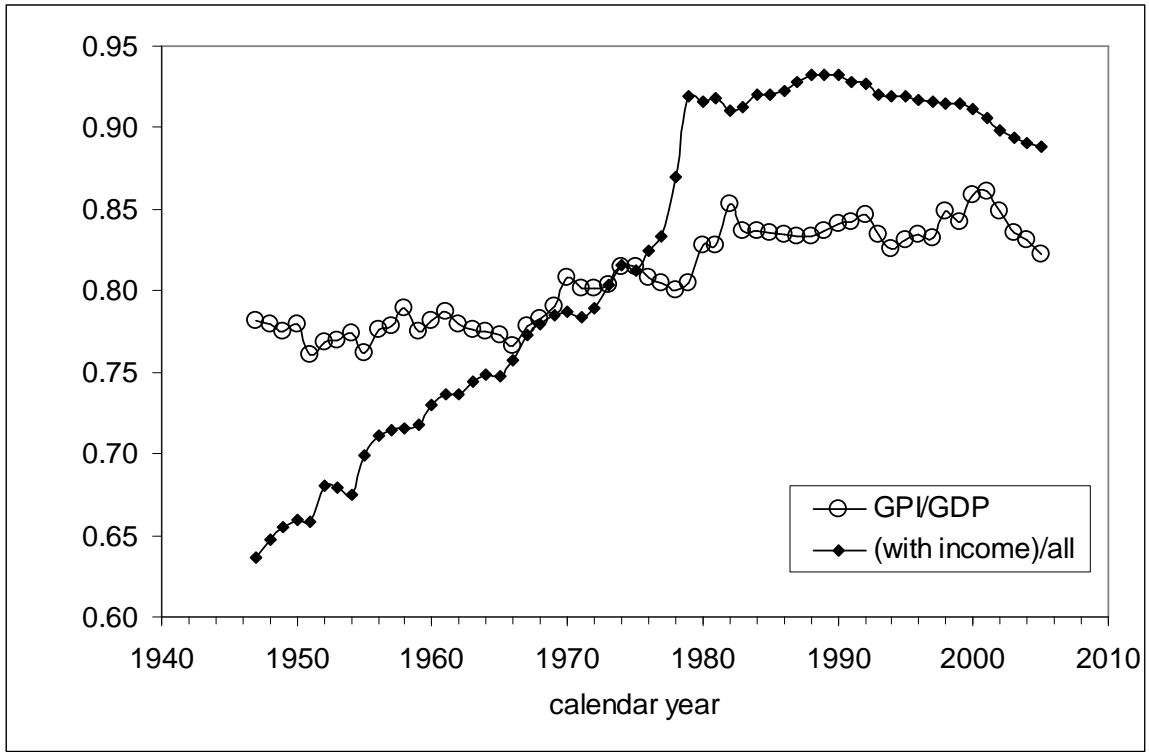

Figure 1.8.1. Ratio of nominal gross personal income, GPI, and nominal GDP for the same year, and the portion of the US working age population with income, as reported by the US Census Bureau.

a)

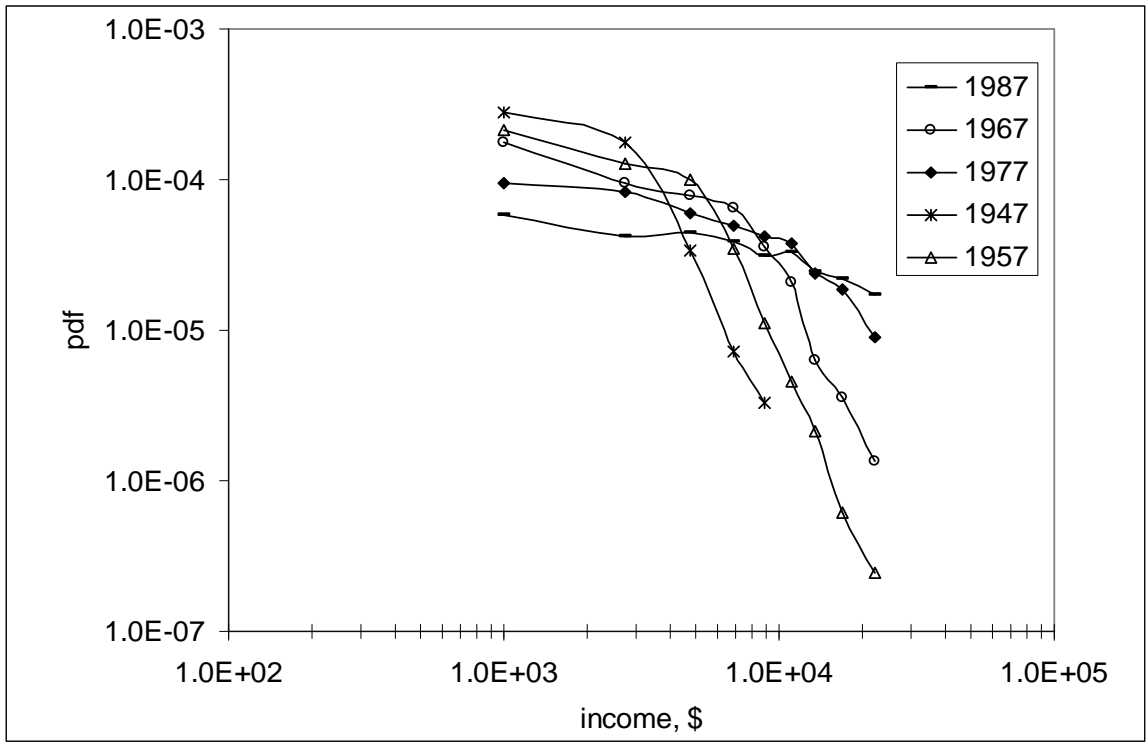

b)

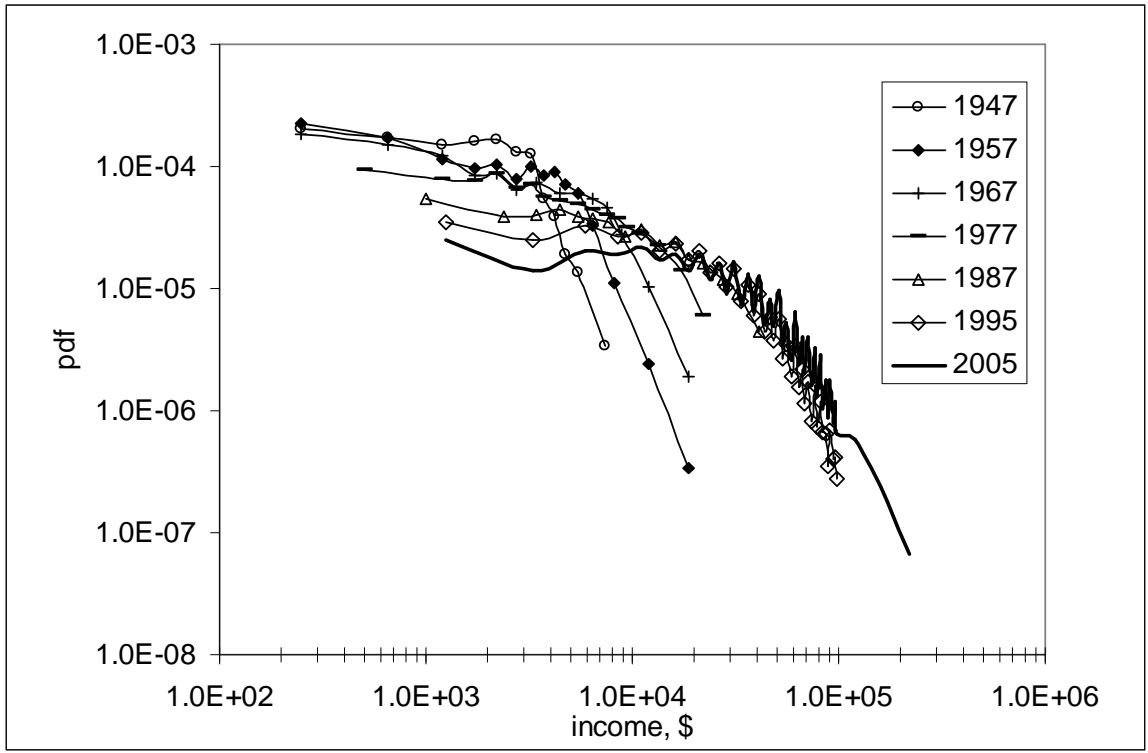

Figure 1.8.2. PIDs for selected years measured in current dollars: a) – for the years between 1947 and 1987 in constant income bins; b) – for the years between 1947 to 2005 in varying income bins. The PIDs are normalized to the total population and reduced to width of corresponding income bins.

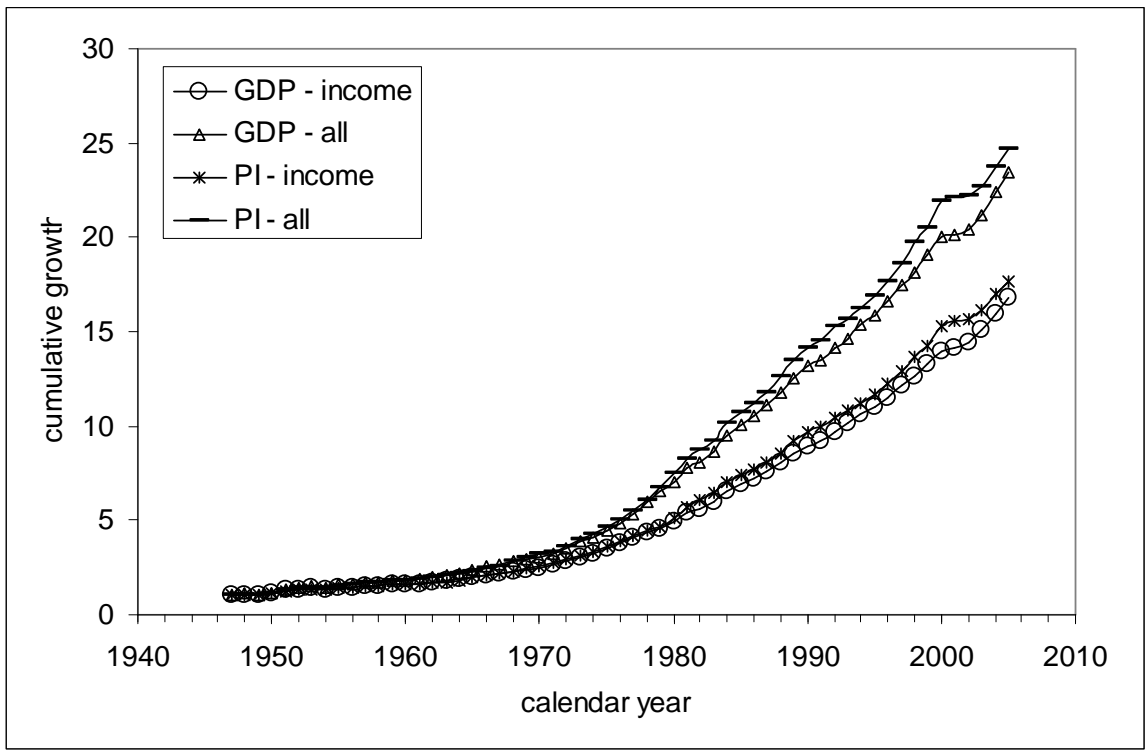

Figure 1.8.3. Cumulative growth of the nominal GDP per capita and nominal GPI per capita reduced to different population groups.

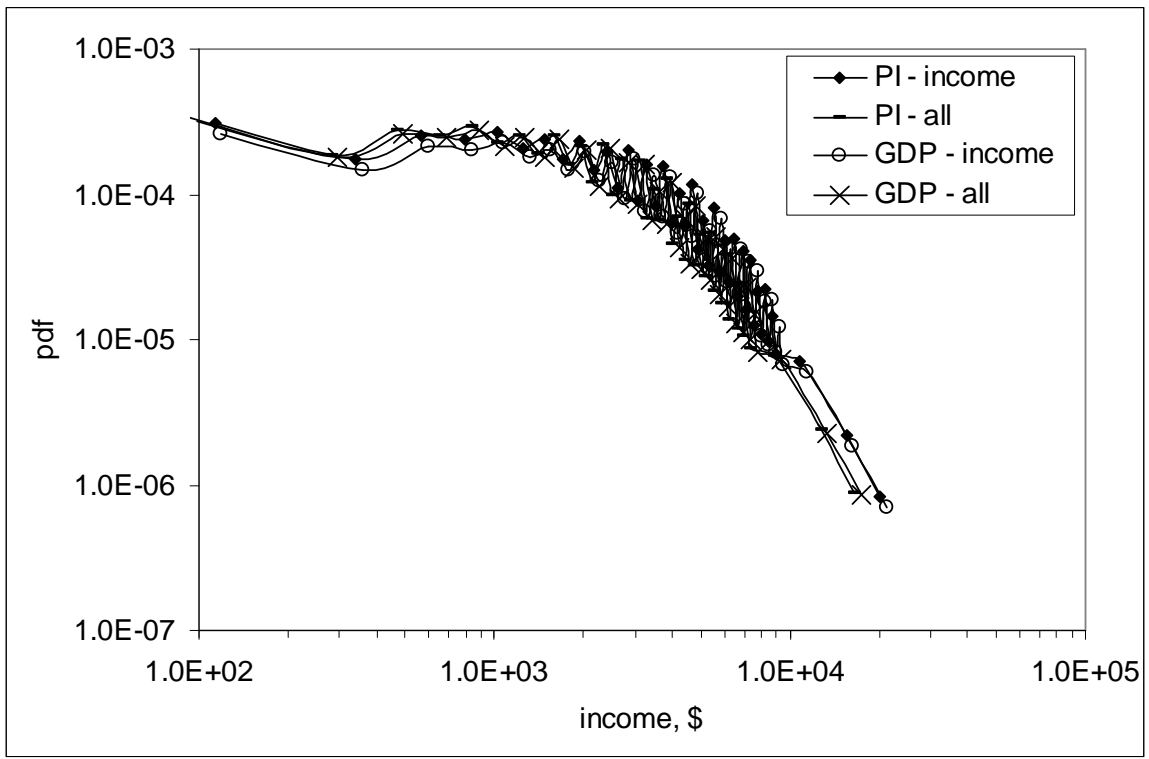

Figure 1.8.4. The PID for 2005 reduced to the cumulative growth between 1947 and 2005 of the four variables presented in Figure 1.8.3.

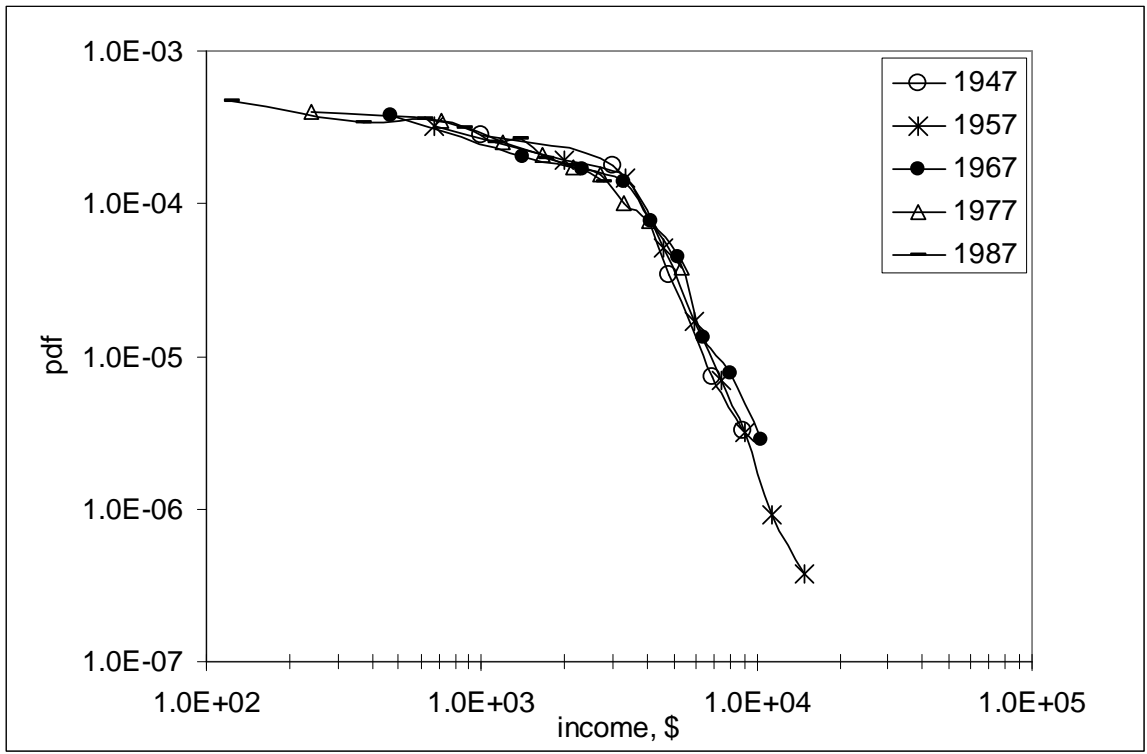

Figure 1.8.5. PIDs for some selected years between 1947 and 1987. The income scale is reduced by the cumulative growth of the nominal GPI per capita since 1947, as obtained for people with income. Notice consistent behavior of the PIDs between 1947 and 1987. One can expect an approximately constant true Gini coefficient for the years before 1987.

a)

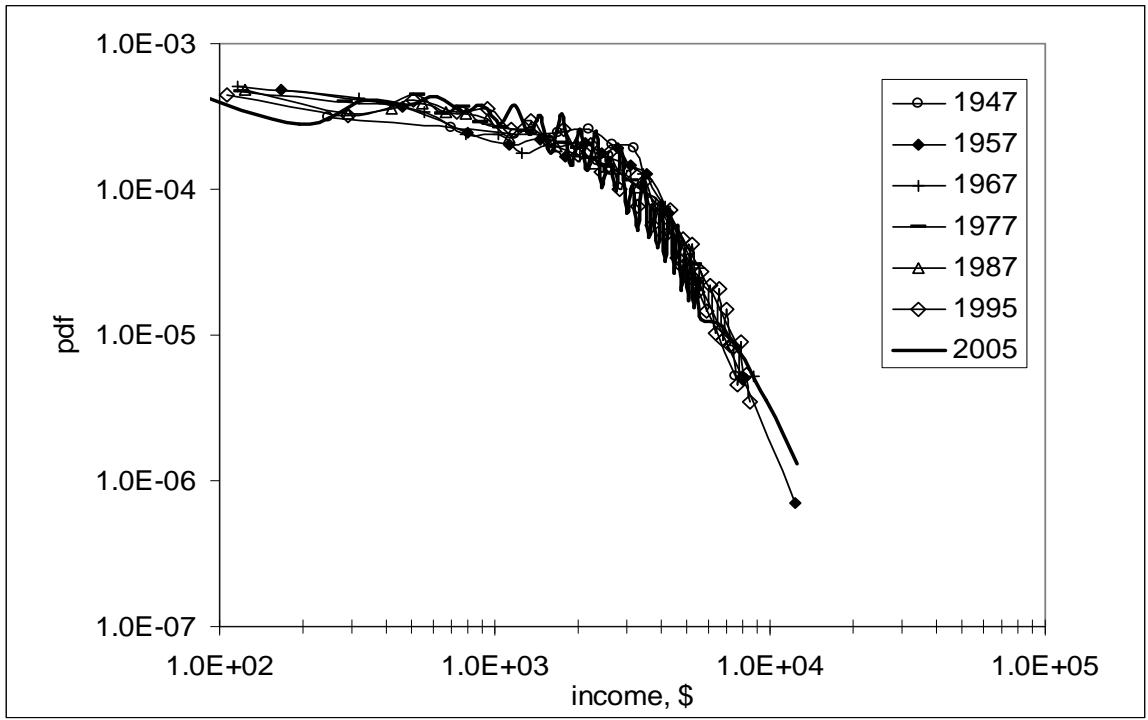

b)

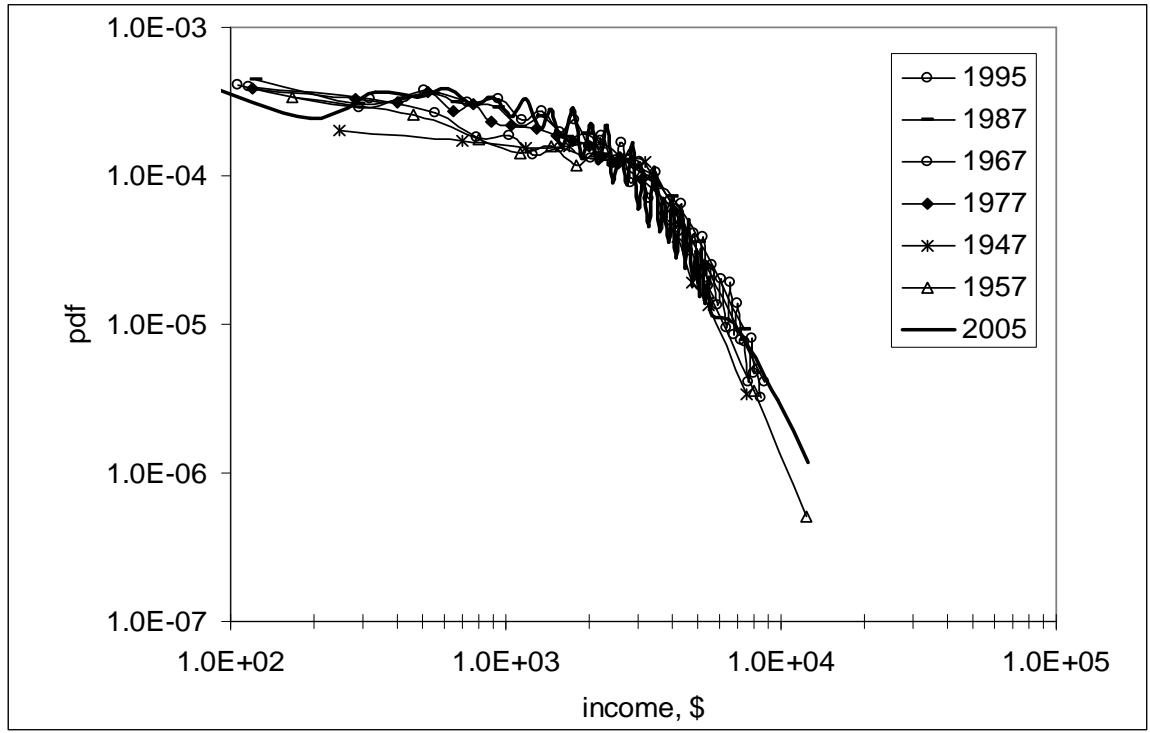

Figure 1.8.6. PIDs for some selected years between 1947 and 2005: a) – for people with income and b) – for the working age population as a whole. The income axis is reduced by the cumulative growth in the nominal GPI per capita since 1947, as obtained for people with income. Notice the deviation between the curves for 1957 and 2005 at higher incomes, which likely manifests measurement errors induced by income definition.

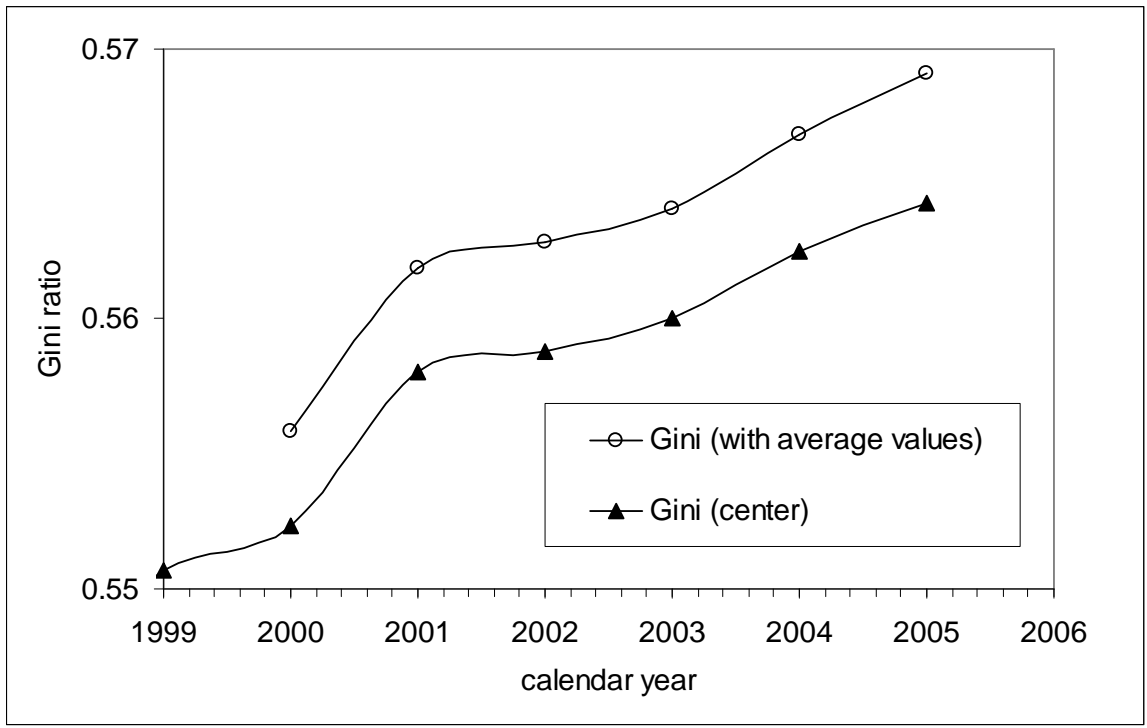

Figure 1.8.7. Comparison of Gini coefficients with average income values and centers of income bins. The curve with the average values also has three additional income bins between $100,000 and $250,000.

a)

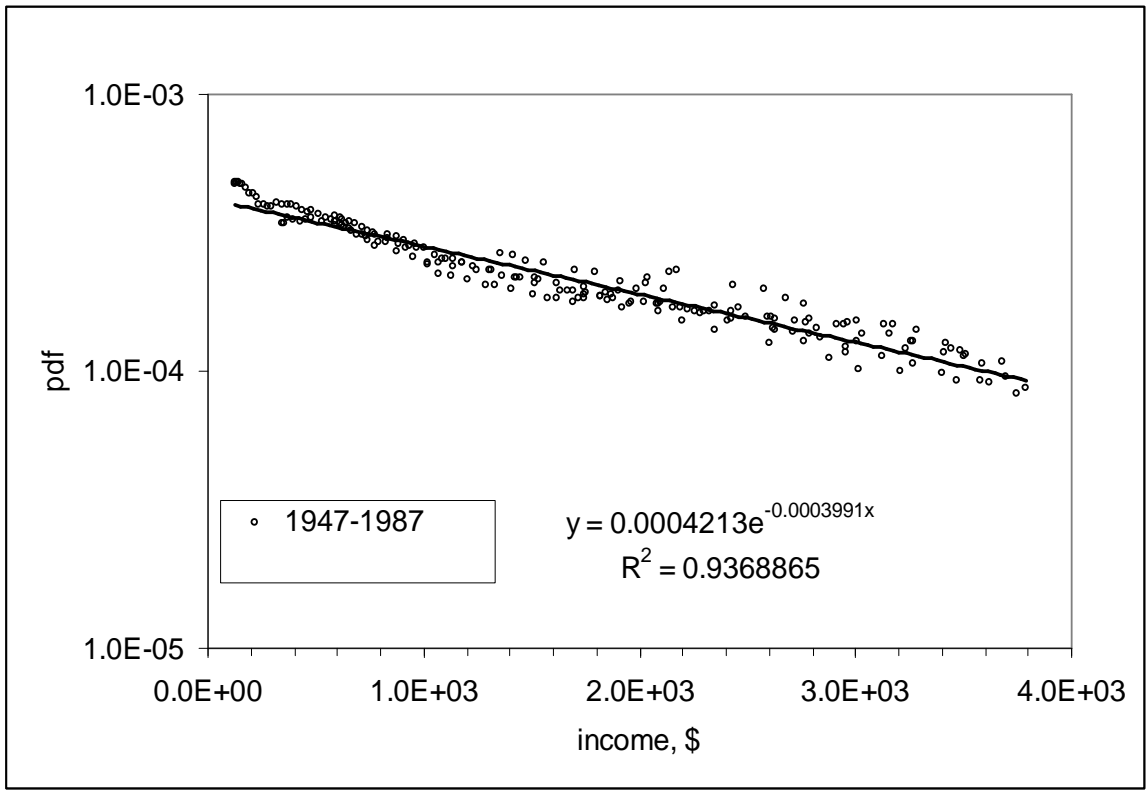

b)

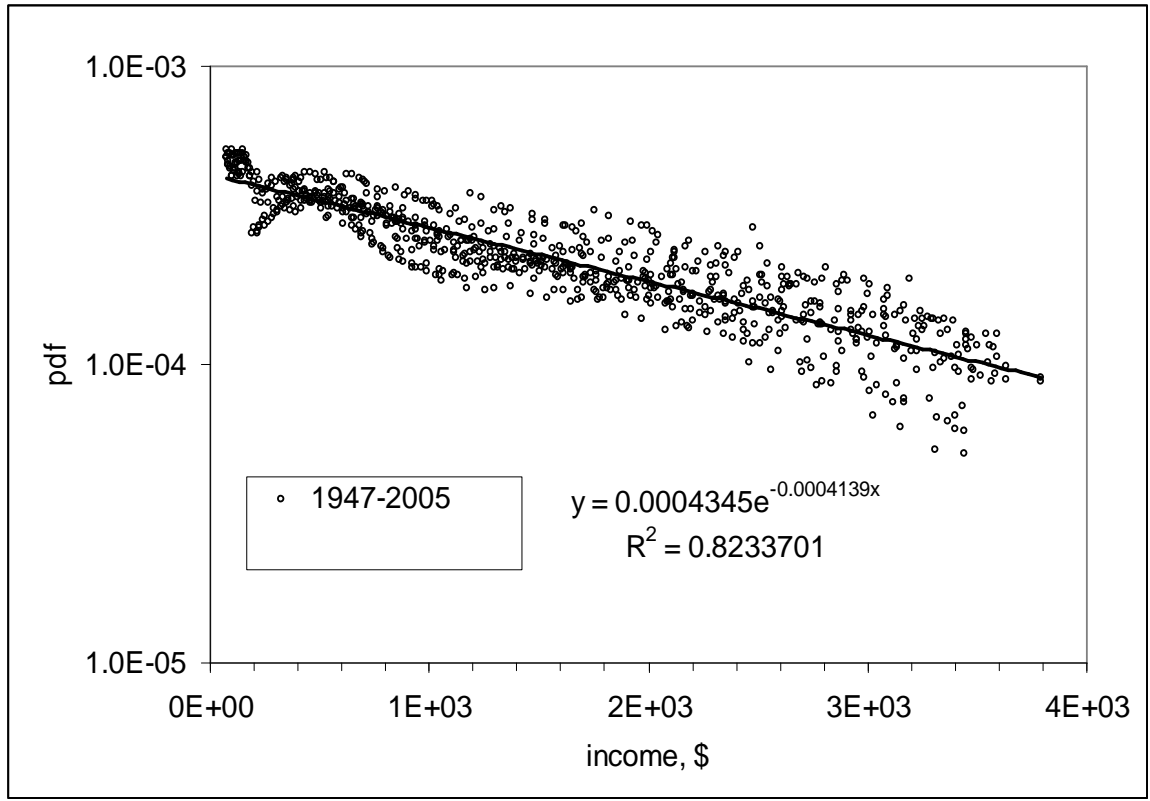

Figure 1.8.8. Approximation of the PIDs by exponential functions – a) between 1947 and 1987; and b) between 1947 and 2005. Obtained indices are very close, but scattering is larger in the second case, which might be of a higher resolution.

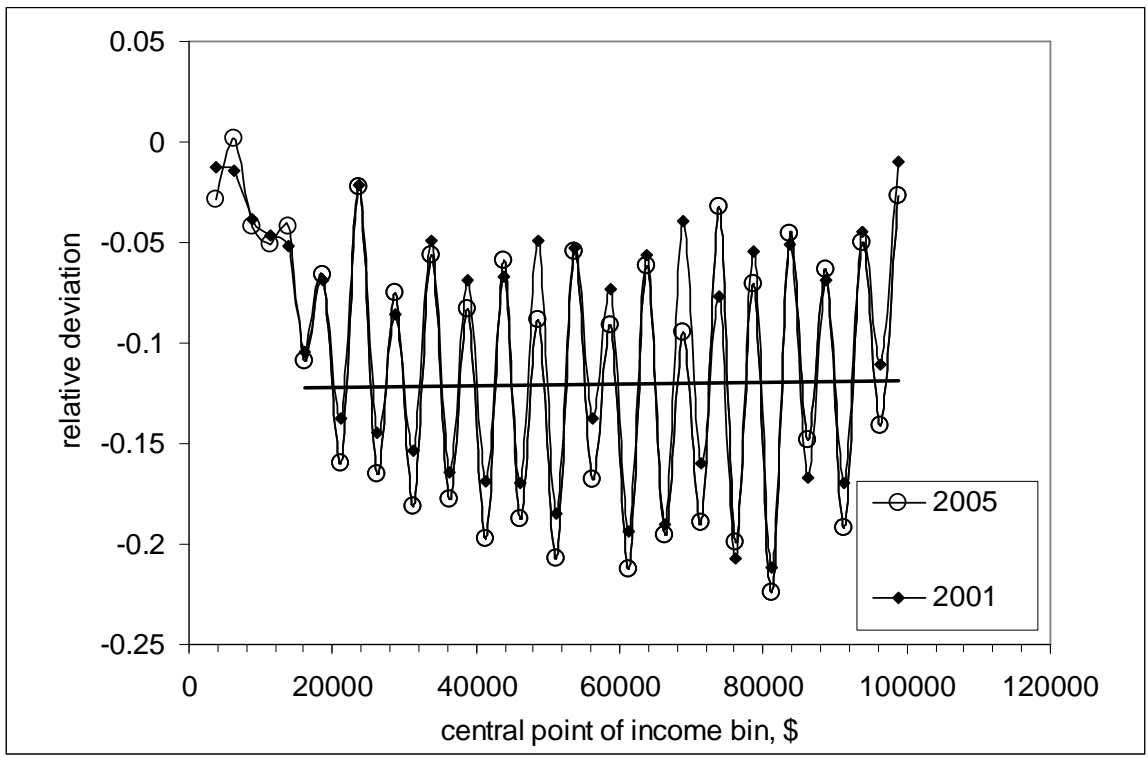

Figure 1.8.9. Relative deviation of the average value in income bins, $X_{av}$, from the central point of the bin, $X_c$ : $(X_{av}-X_c)/dX$, where $dX$ is the width of corresponding bin. The CPS reports for 2001 and 2005 are compared.

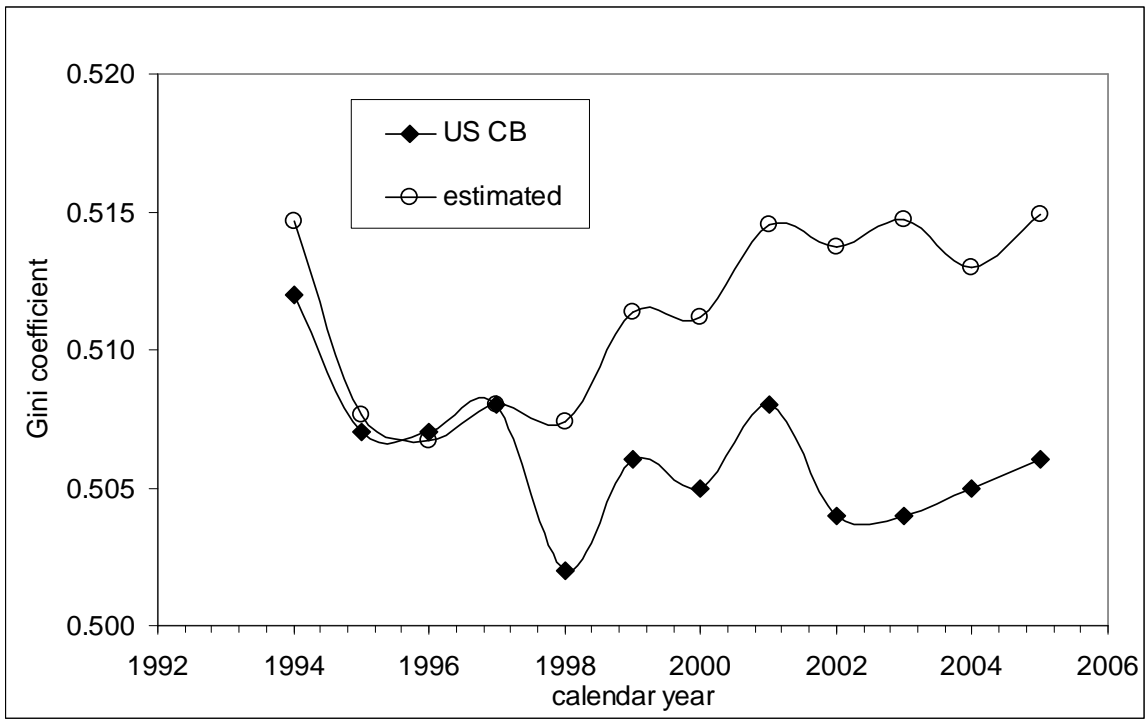

Figure 1.8.10. Comparison of the Gini coefficient reported by the US Census Bureau for the years between 1994 and 2005 with that estimated in this study. Obviously, the US CB changed the procedure for the estimation of Gini coefficient in 1998.

a)

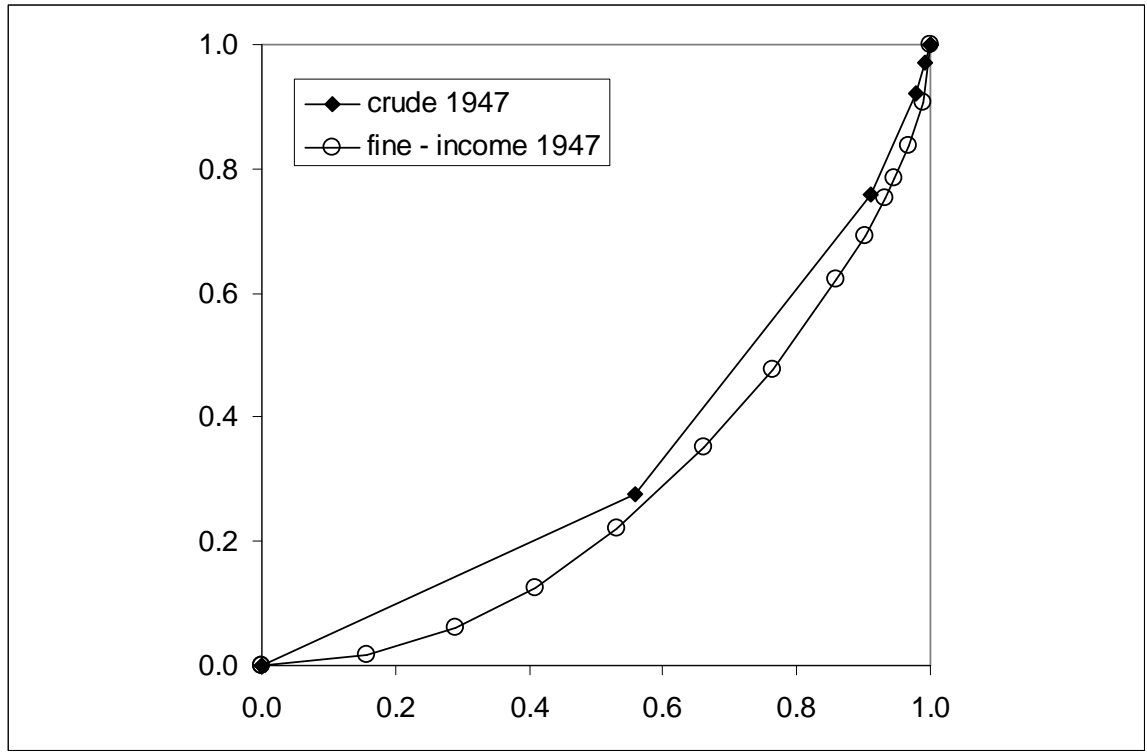

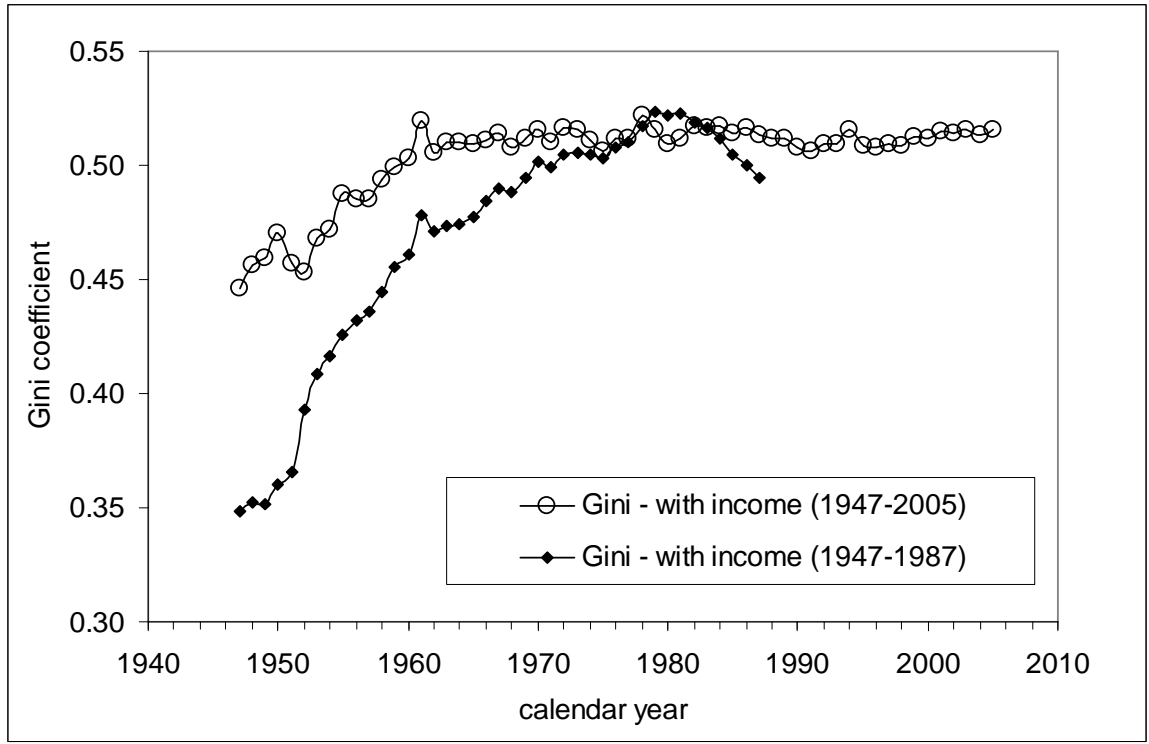

Figure 1.8.11. a) Comparison of two Lorenz curves for 1947 associated with the crude and fine PIDs. b) Comparison of two estimates of Gini coefficient between 1947 and 2005 using the crude and fine PIDs. Both coefficients are obtained for population with income.

a)

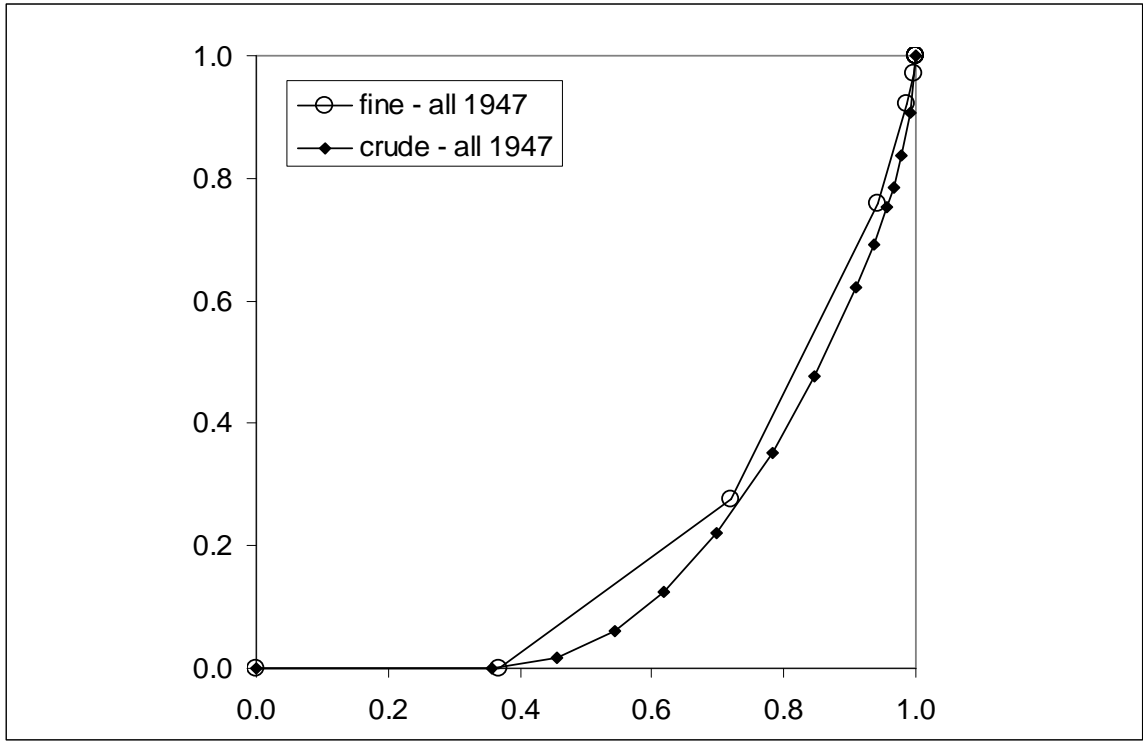

b)

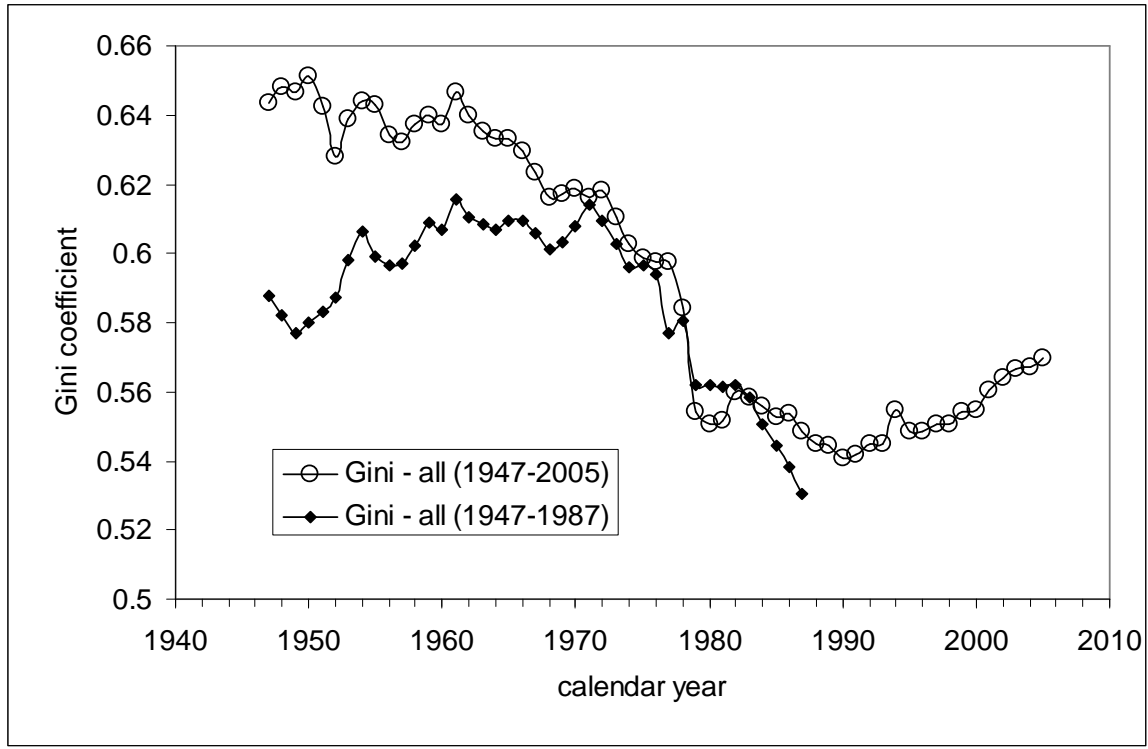

Figure 1.8.12. a) Comparison of two Lorenz curves for 1947 associated with the crude and fine PIDs. b) Comparison of two estimates of the Gini coefficient between 1947 and 2005 using the crude and fine PIDs. Both coefficients are obtained for the working age population as a whole. The observed change in the actual PIDs is not well described by the fixed income bins. Nevertheless, the years between 1970 and 1983 are characterized by a good agreement between the curves.

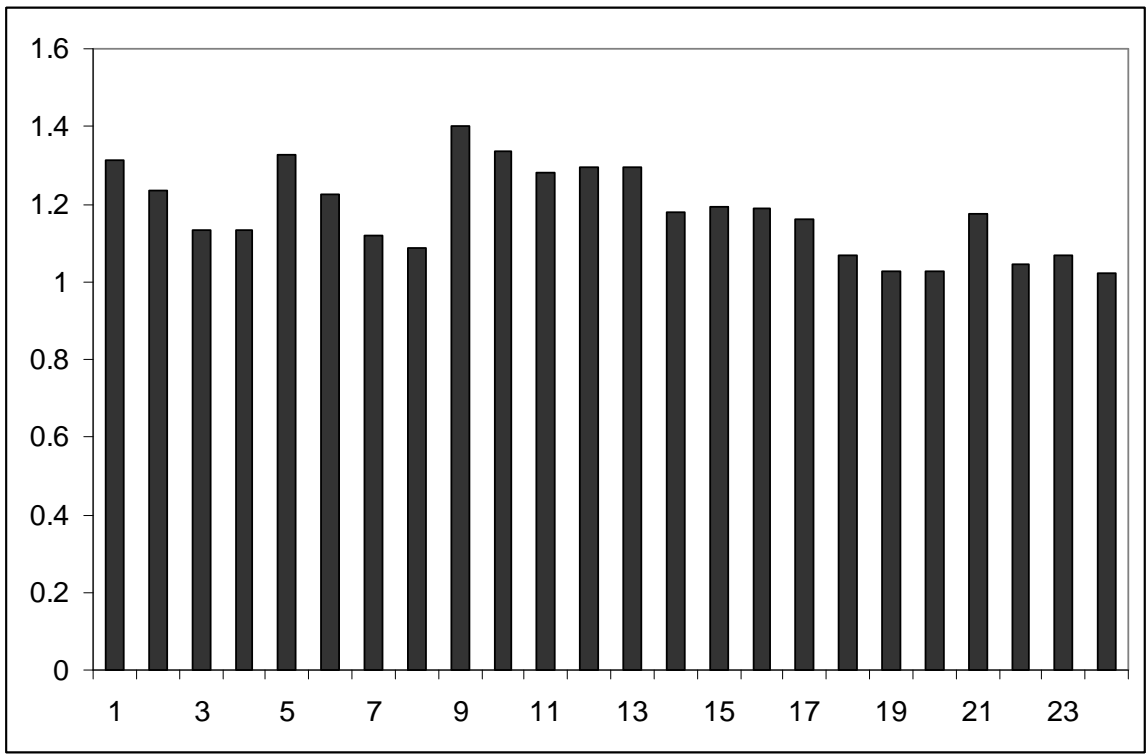
Figure 1.8.13. Estimates of index *k* obtained from the average values of income in the Pareto income zone – above $100000, $150000, $200000, and $250000 for the years between 2000 and 2005.

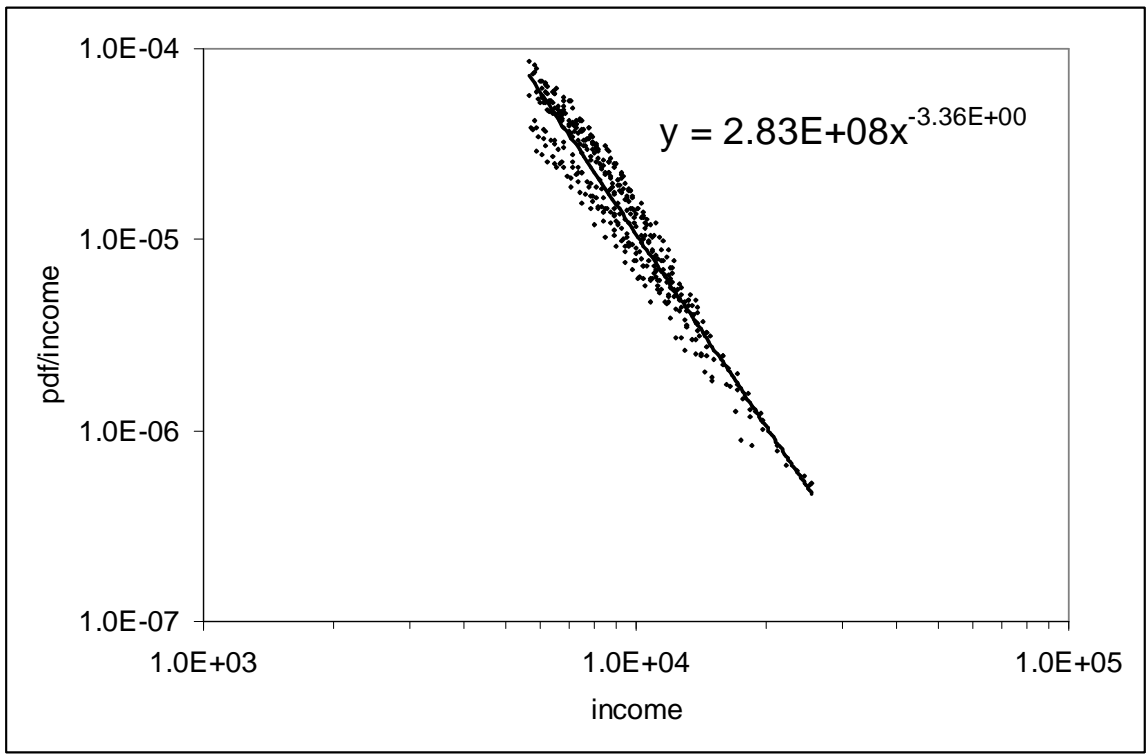

Figure 1.8.14. Linear regression of the probability density functions in the Pareto zone (the log-log coordinates). The Pareto index is ($k=$) 1.36. This estimate is consistent with that obtained using the average values above $100000 in Figure 1.8.13.

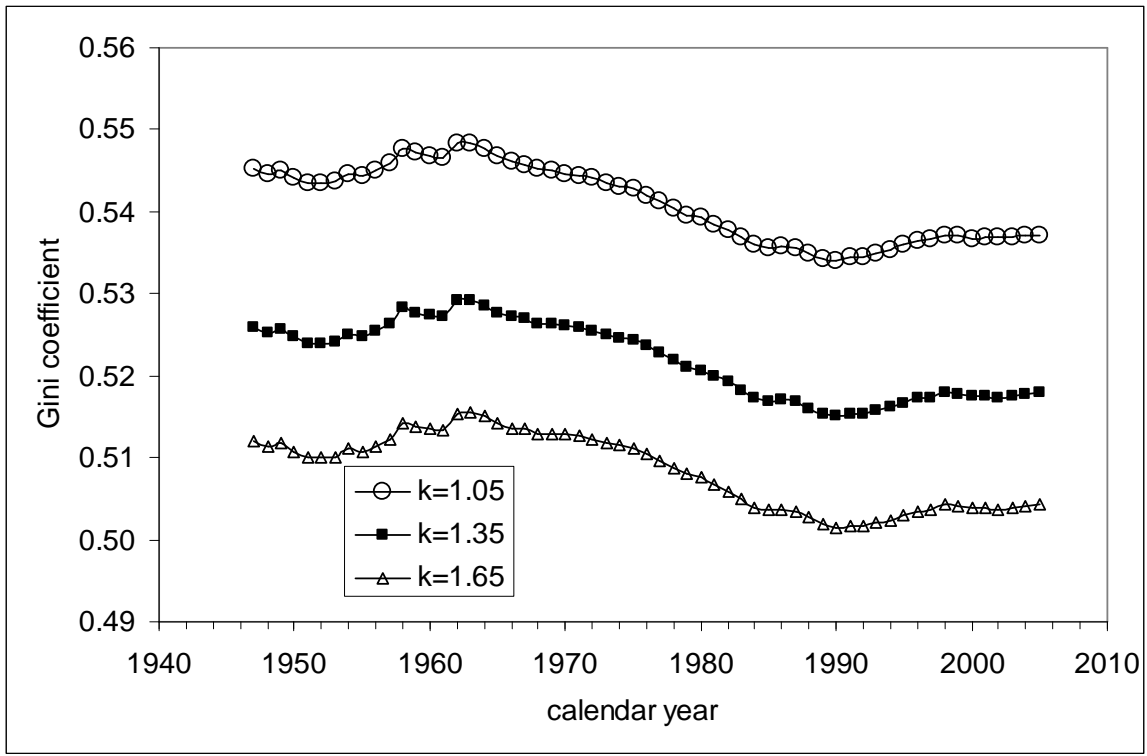

Figure 1.8.15. Dependence of the predicted Gini coefficient on the Pareto index, $k$. Lower $k$ values correspond to "thicker" tails in the PIDs and larger *Gini* values. The effect of $k$ on *Gini* is nonlinear.

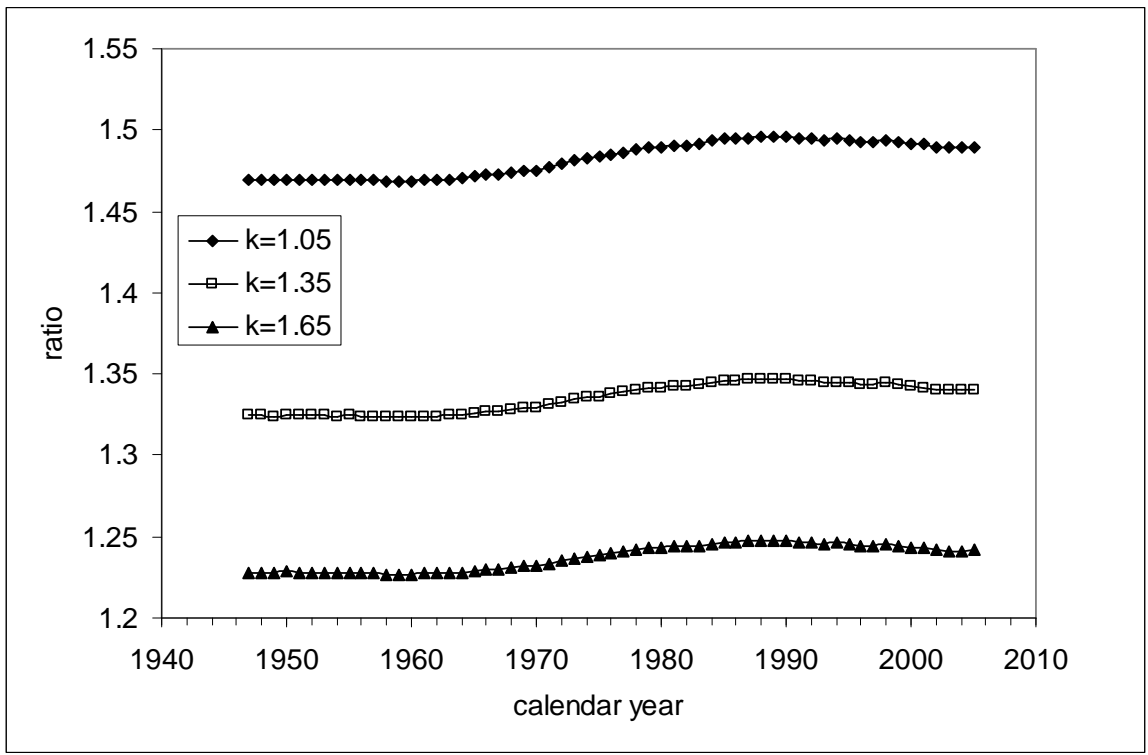

Figure 1.8.16. Dependence of the effective increase in income production (extra income) in the model relative to that in the sub-Pareto income zone. Theoretical value is 1.33 and corresponds to $k$=1.35. The effect of $k$ on the ratio is nonlinear.

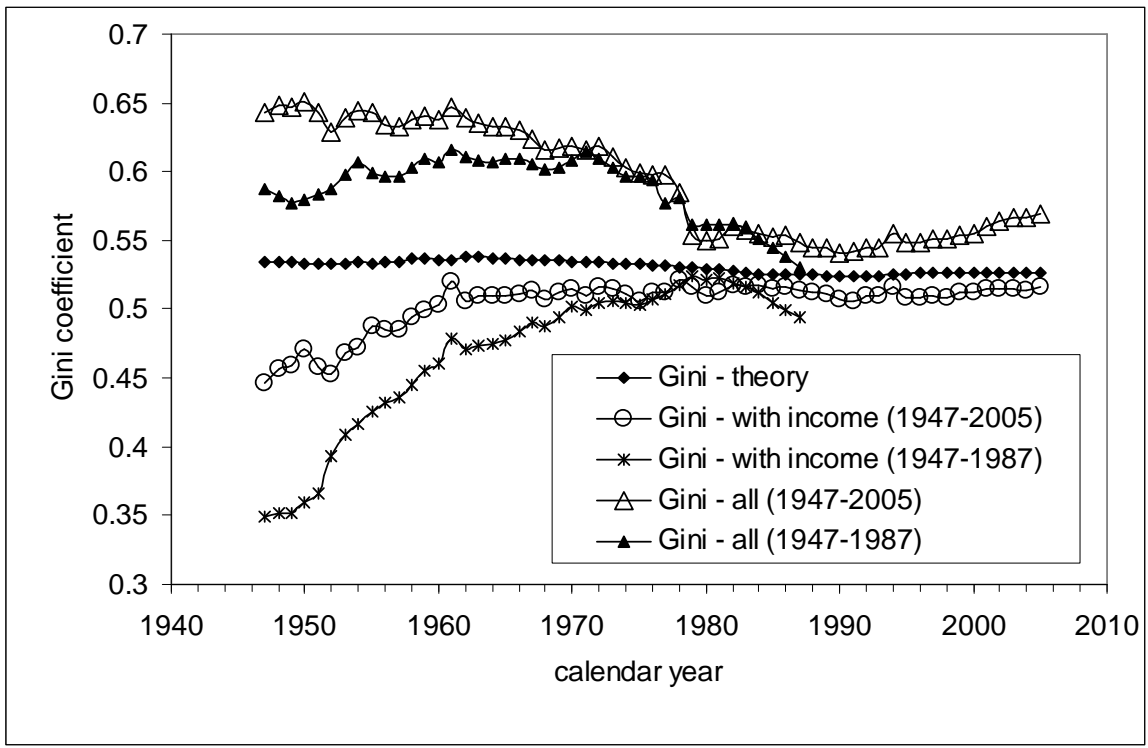

Figure 1.8.17. Comparison of the estimated and predicted Gini coefficients. The predicted curve lies between the two estimated curves, which converge as the portion of population without income drops. One can consider the predicted curve as representing the true Gini coefficient for the period between 1947 and 2005.

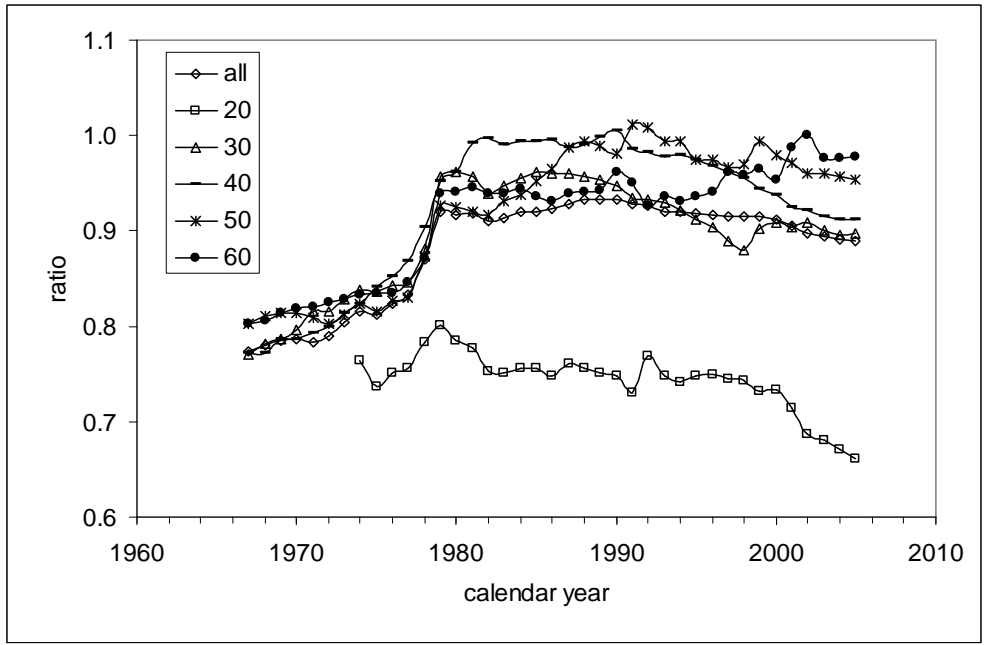

Figure 1.9.1. Evolution of the portion of population with income in various age groups: all – above 15 years of age, 20 – from 16 to 24 years of age, 30 – from 25 to 34 years, 40 – from 35 to 44 years, 50 – from 45 to 54 years, 60 – from 55 to 64 years. In the group between 16 and 24 years of age, the portion has been falling since 1979. Notice the break in the distributions between 1977 and 1979 induced by large revisions implemented in 1980 – "Questionnaire expanded to show 27 possible values from 51 possible sources of income."

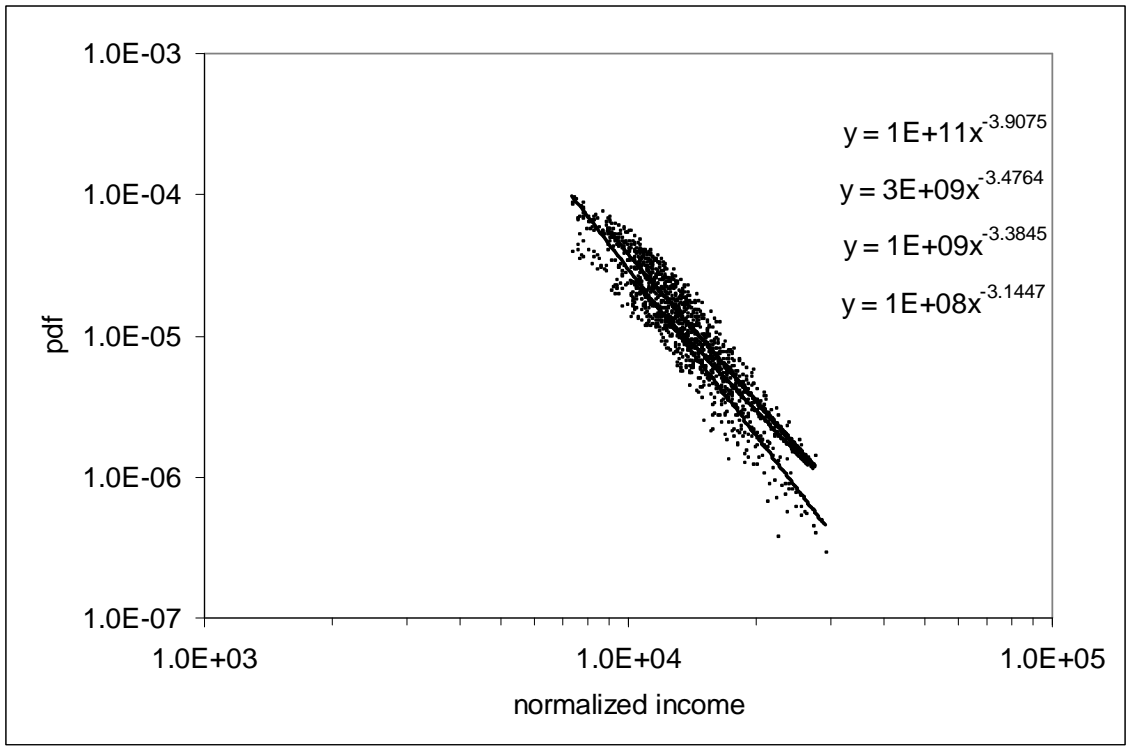

Figure 1.9.2. Evolution of the Pareto law index (slope) with age: $k$=-1.91 for the age group between 25 and 34 years, $k$=-1.48 between 35 and 44, $k$=-1.38 between 45 and 54, and $k$=-1.14 in the age group between 55 and 64.

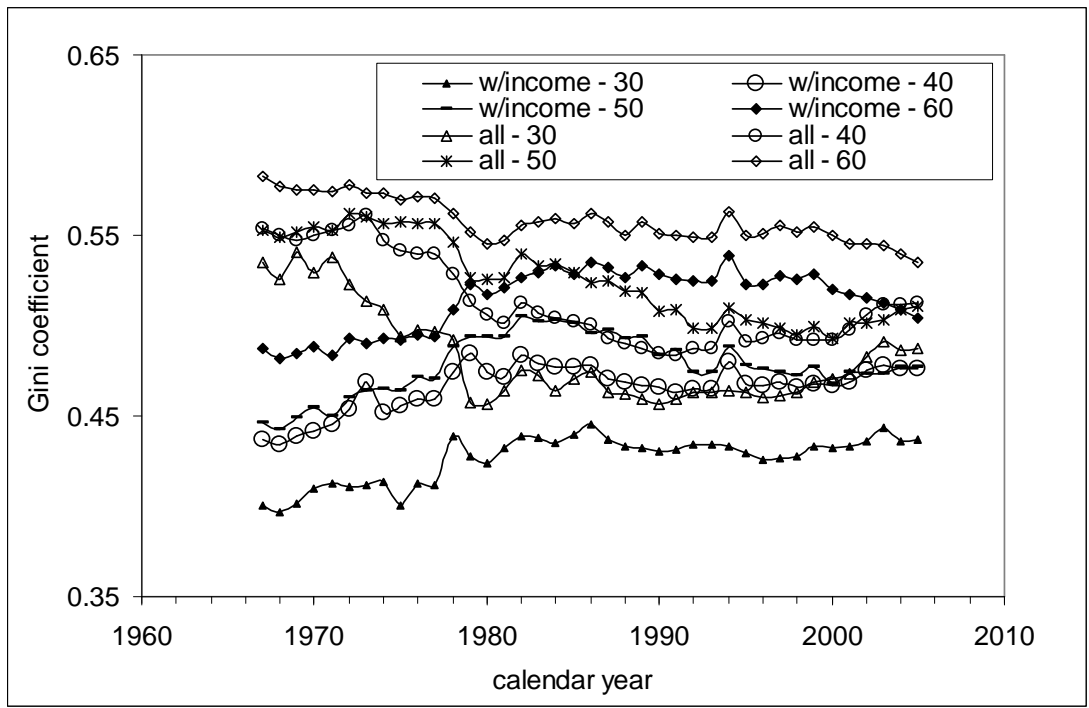

Figure 1.9.3. Evolution of the estimated Gini coefficient for personal incomes in various age groups between 1967 and 2005. There are two versions in each age group - first includes all people aged in given range (all), and second includes only those with nonzero income (w/income). Obviously, the Gini coefficients for people with income are systematically lower than those including all population with given ages.

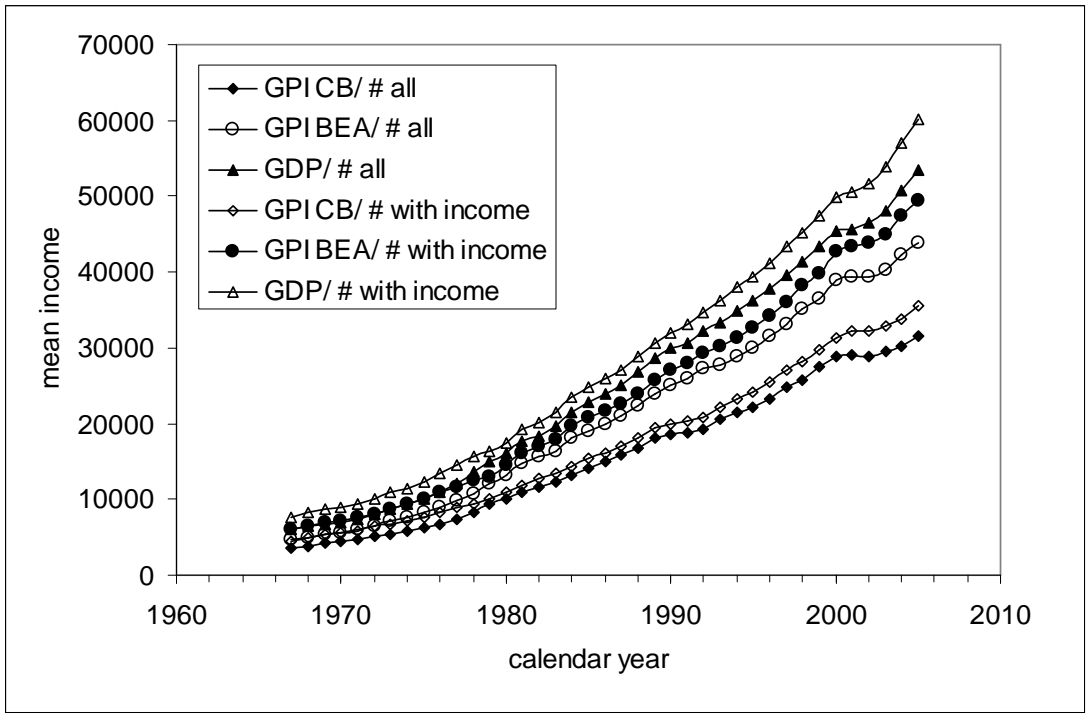

Figure 1.9.4. Evolution of various measures of the overall mean income: using GDP; GPI reported by the BEA; and GPI reported by the Census Bureau as estimated in annual CPS. Two population estimates are used for calculations of the mean values – total working age population (all) and people reporting income (with income). According to current income definitions the GPI-BEA is larger than the GPI-CB because the former includes additional sources of income.

a)

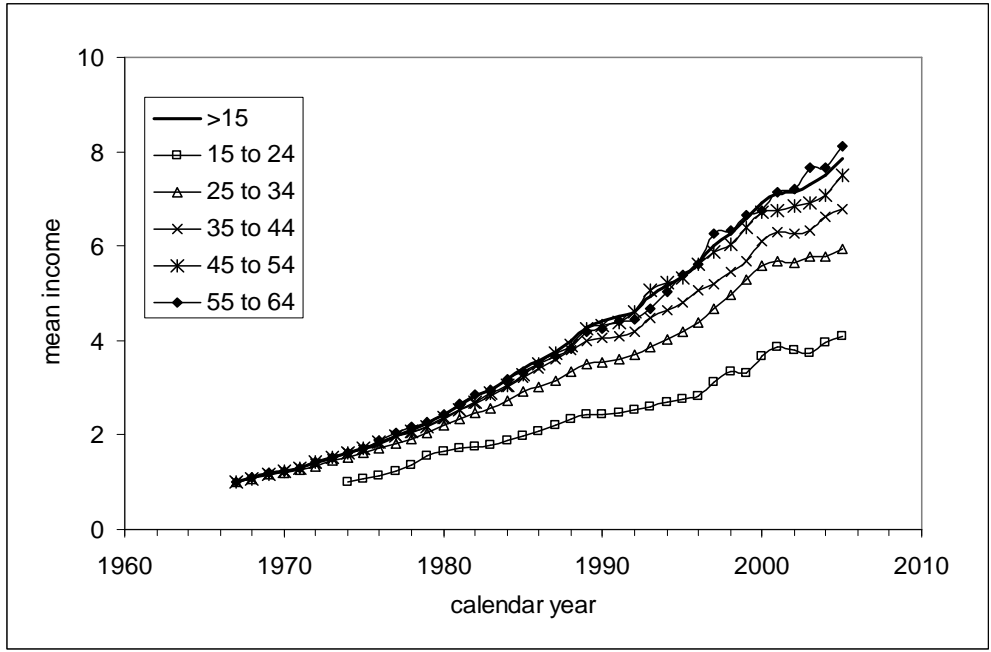

b)

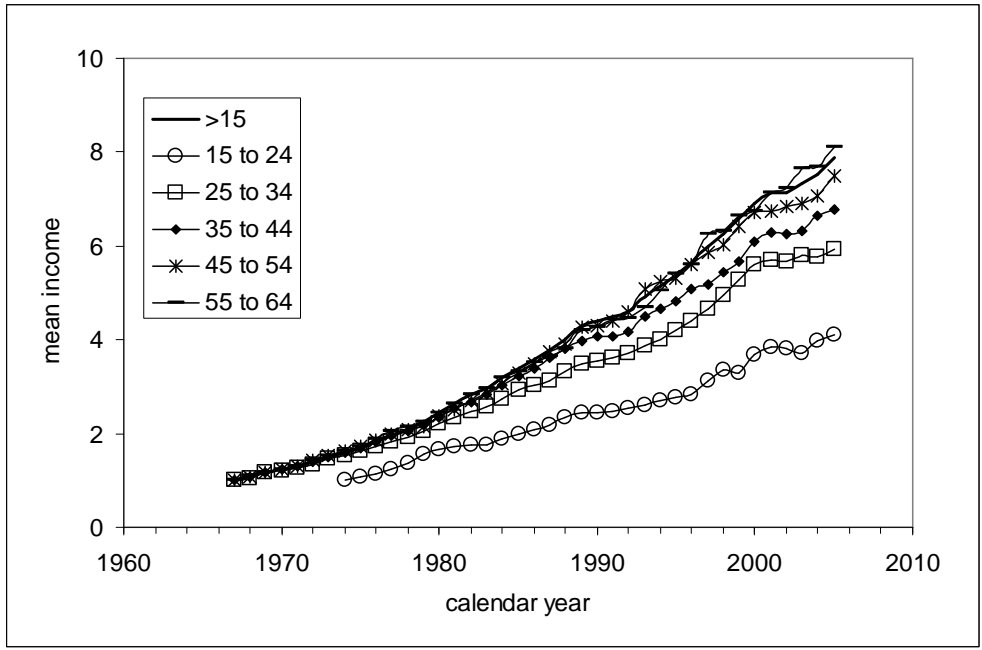

Figure 1.9.5. Evolution of mean income (normalized to that in 1967) in various age groups as estimated using: a) total working age population; b) only people with income. The curves are used to normalize corresponding PIDs.

a)

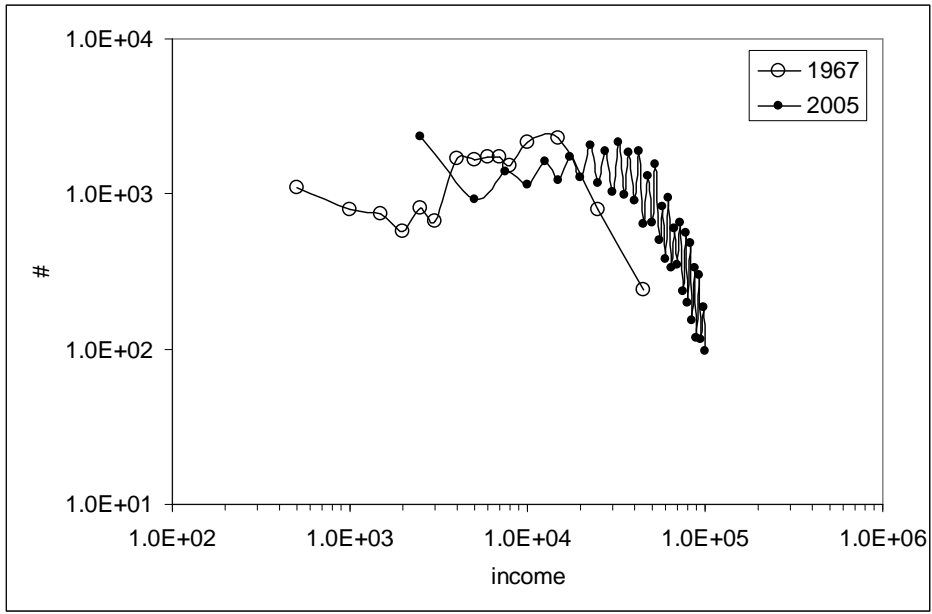

b)

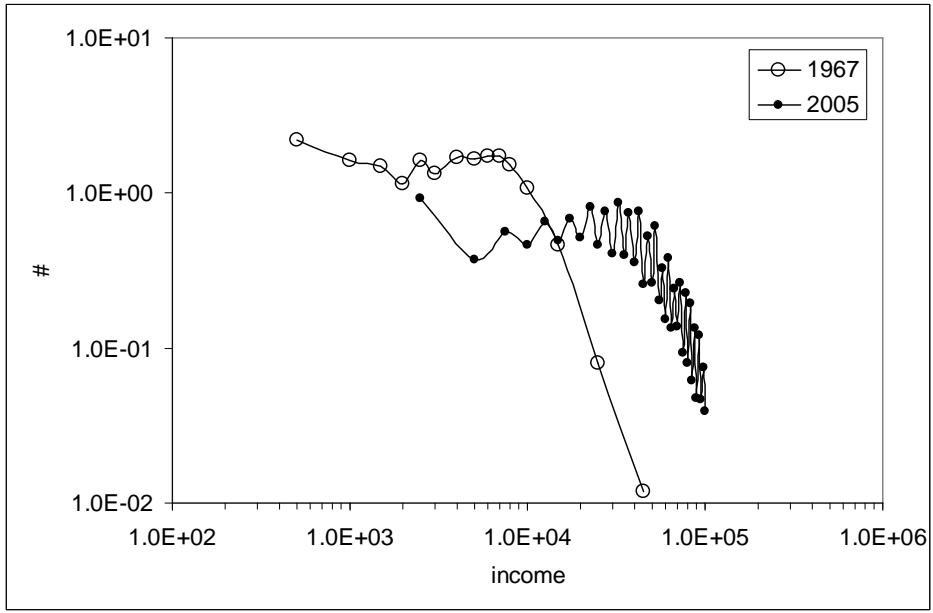

Figure 1.9.6. Comparison of a) original; and b) normalized personal income distribution (in current dollars) in the age group between 35 and 44 years in 1967 and 2005. Original distributions published by the Census Bureau are normalized to the width of relevant income bins in order to obtain population density distribution. Income bins are not uniform in 1967 creating local troughs and peaks. In 2005, income bins are uniform between $0 and $100,000. Three $50,000-wide bins above $100,000 are not shown. More people and larger GPI in 2005 are observed.

a) 15 to 24

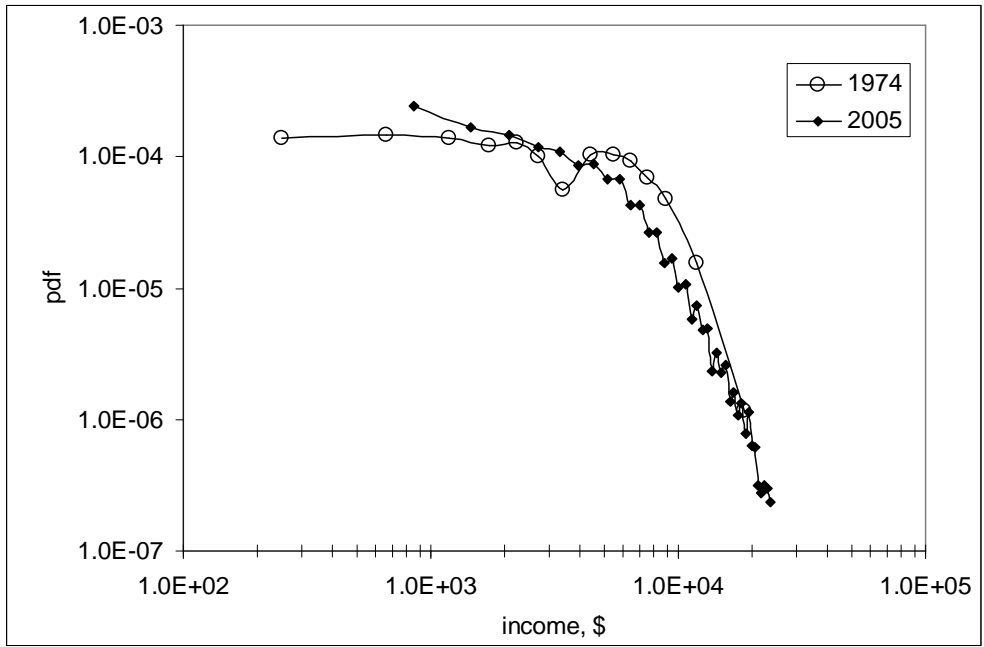

b) 25 to 34

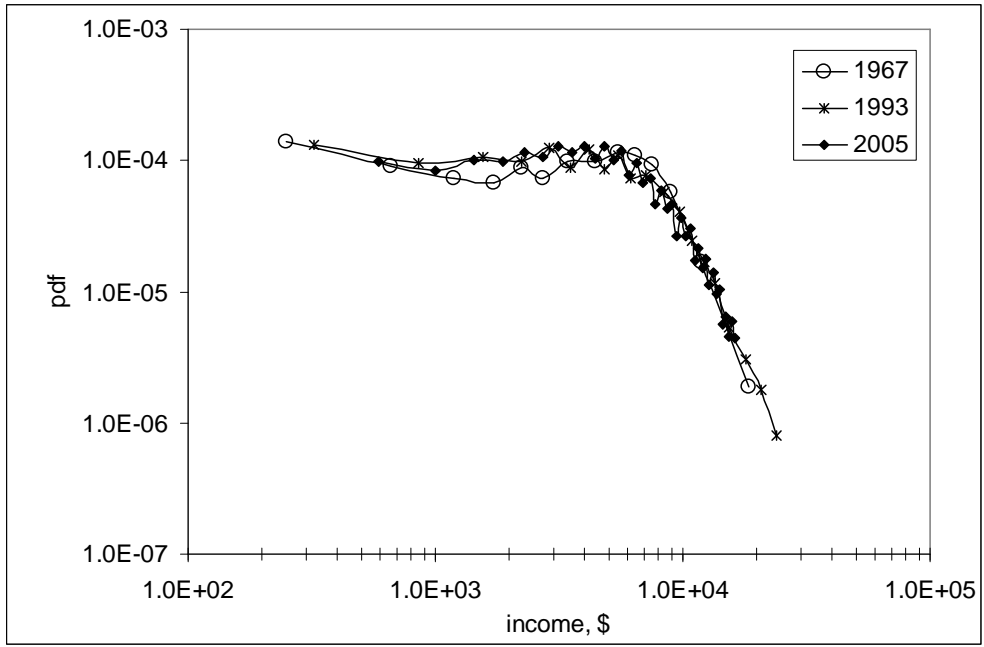

c) 45 to 54

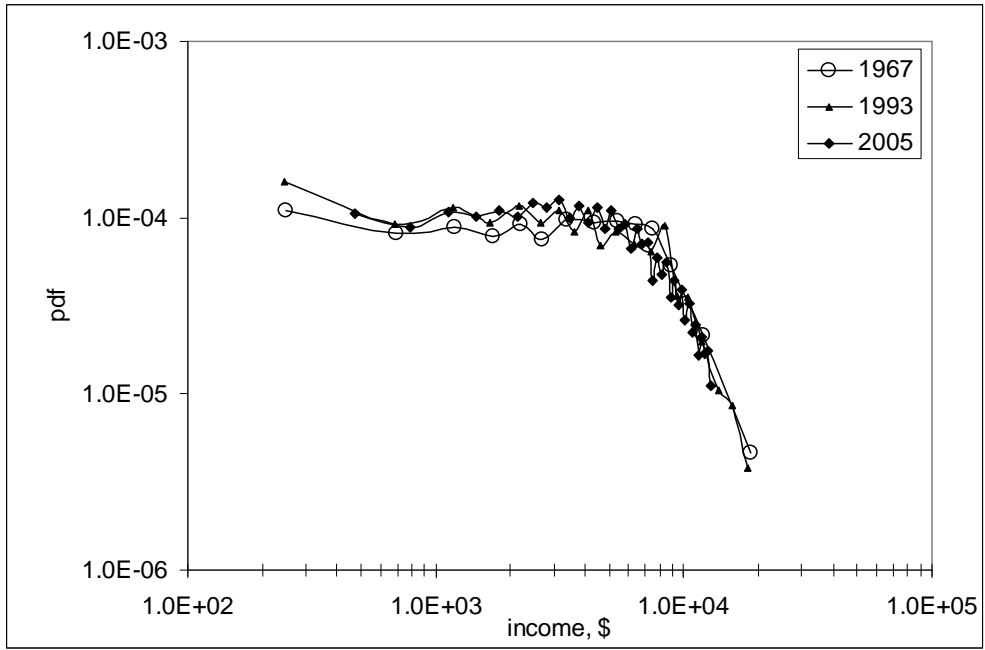

d) 55 to 64

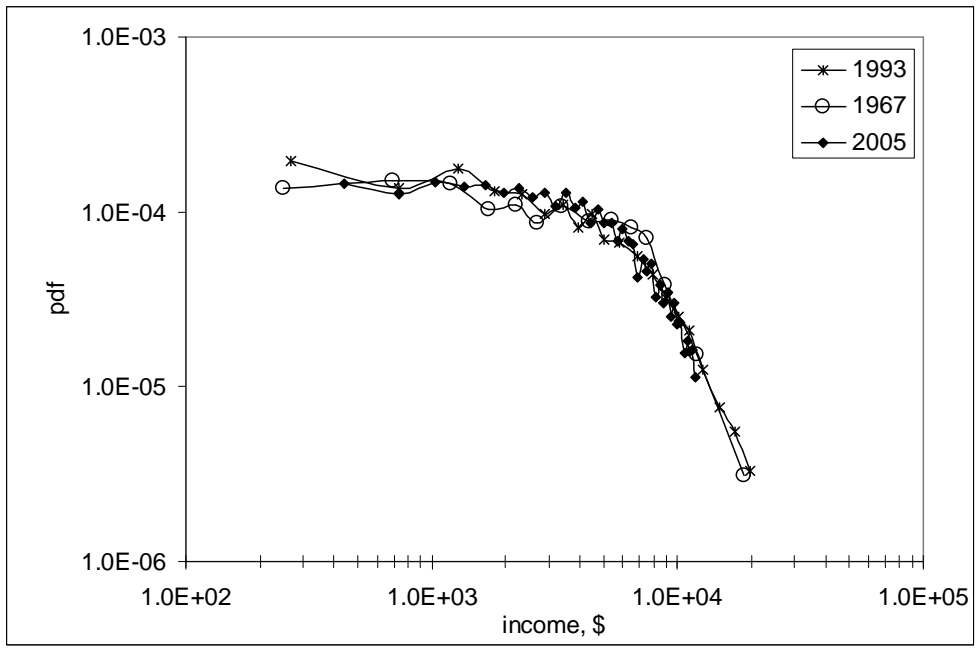

e) 15 years of age and over

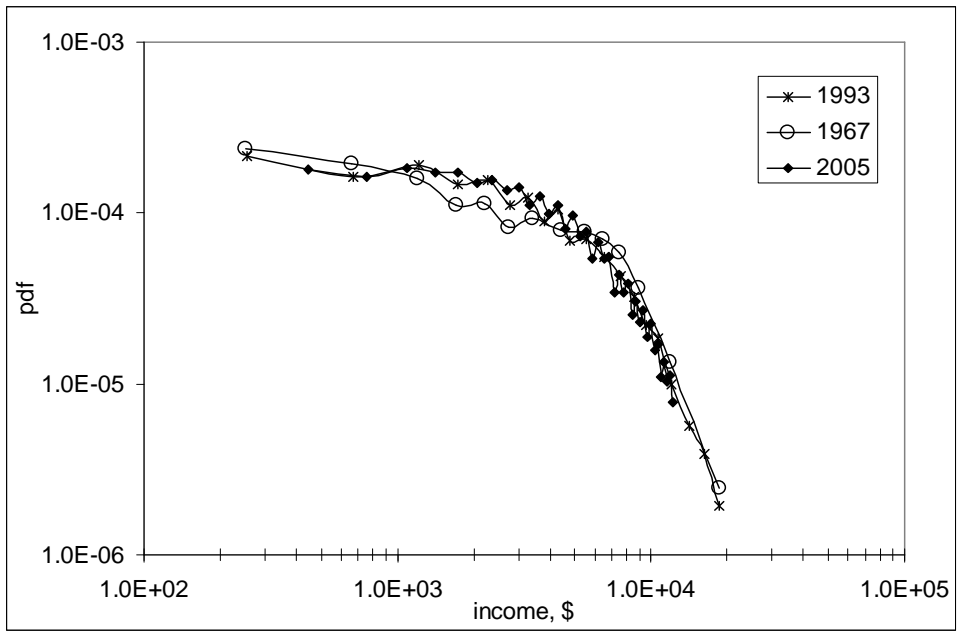

Figure 1.9.7. PIDs in various age groups for people with income normalized to the increase in total income and total population in given group. Years 1967, 1993, and 2005 are presented. There is no significant difference between the curves except in the age group between (14) 15 and 24 years of age.

a) 25 to 34 years of age

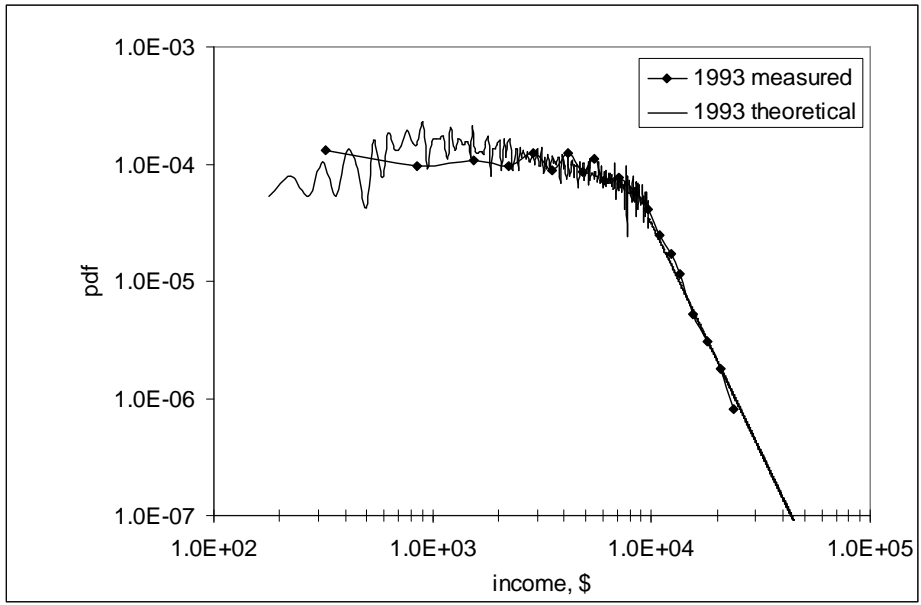

b) from 45 to 54 years of age

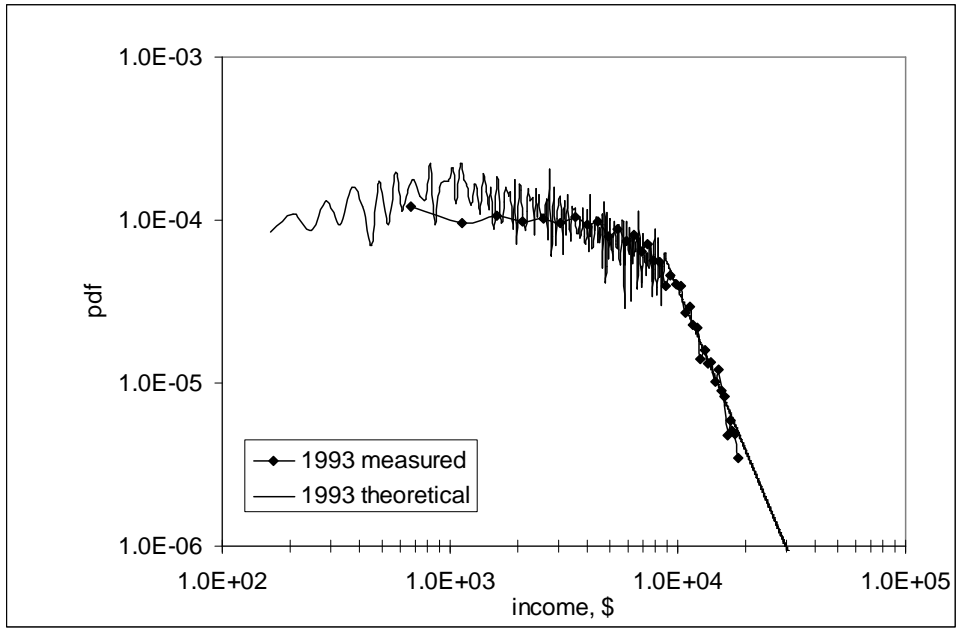

c) over 15 years of age

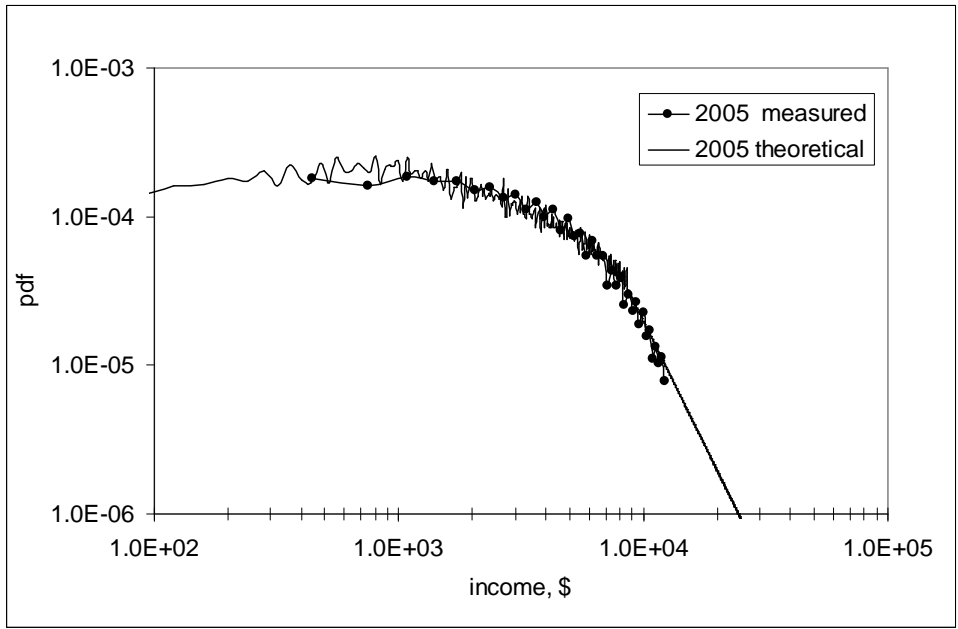

Figure 1.9.8. Comparison of measured and predicted PIDs in some age groups. High incomes are describes by a power law with $k=-1.91$; $k=-1.38$; and $k=-1.35$, respectively.

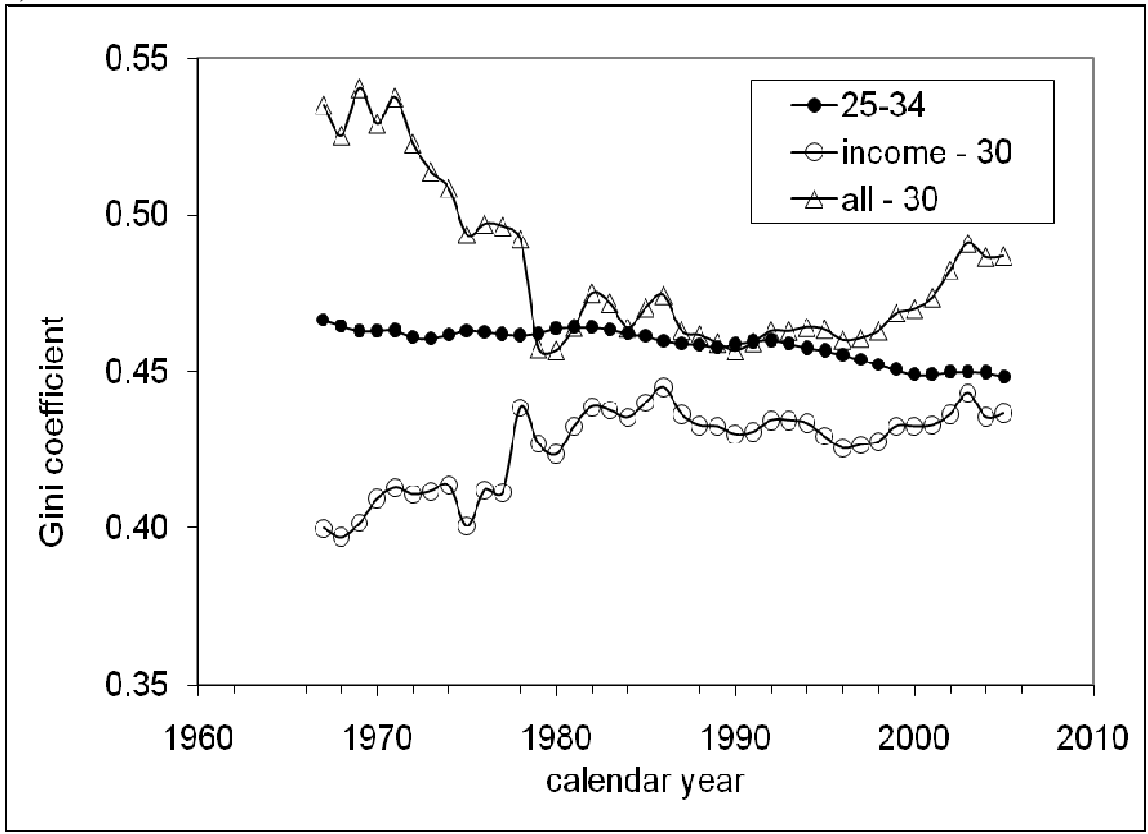

a)

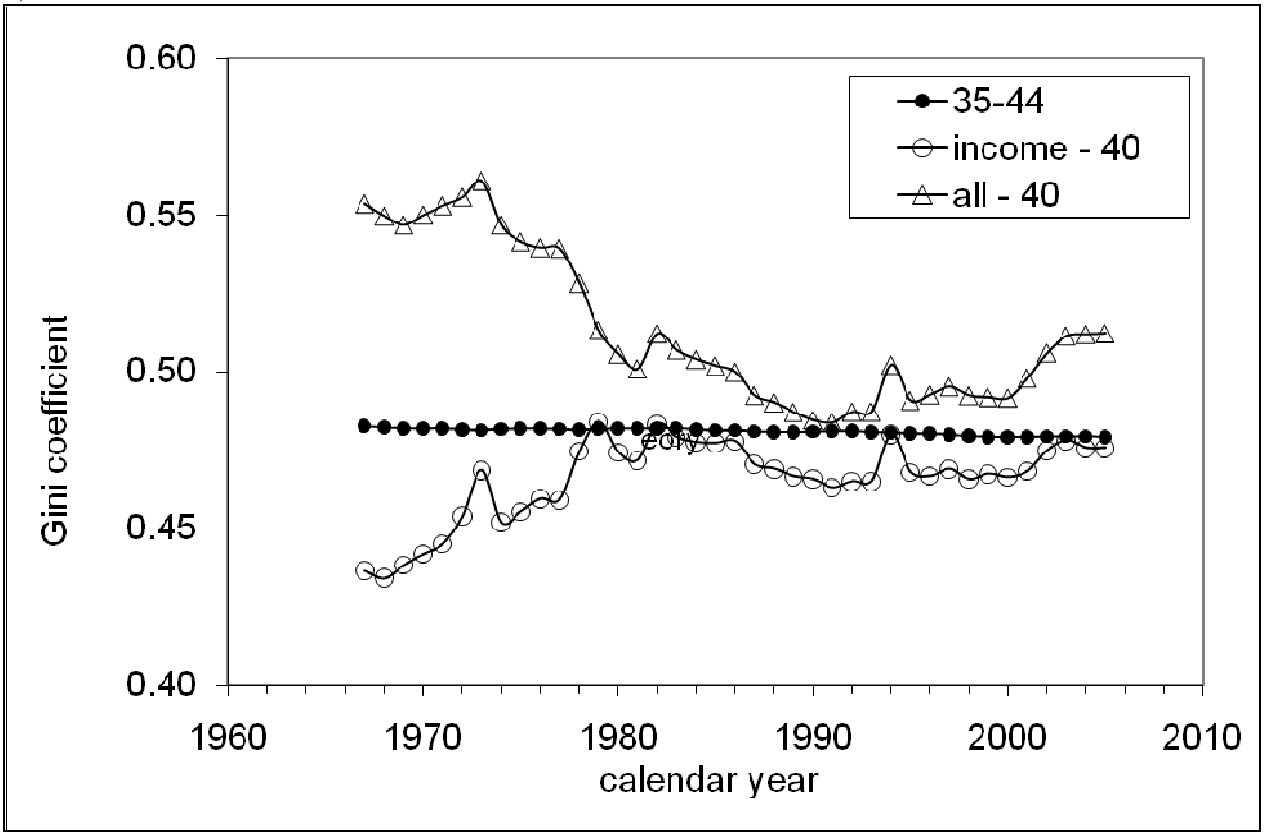

c)

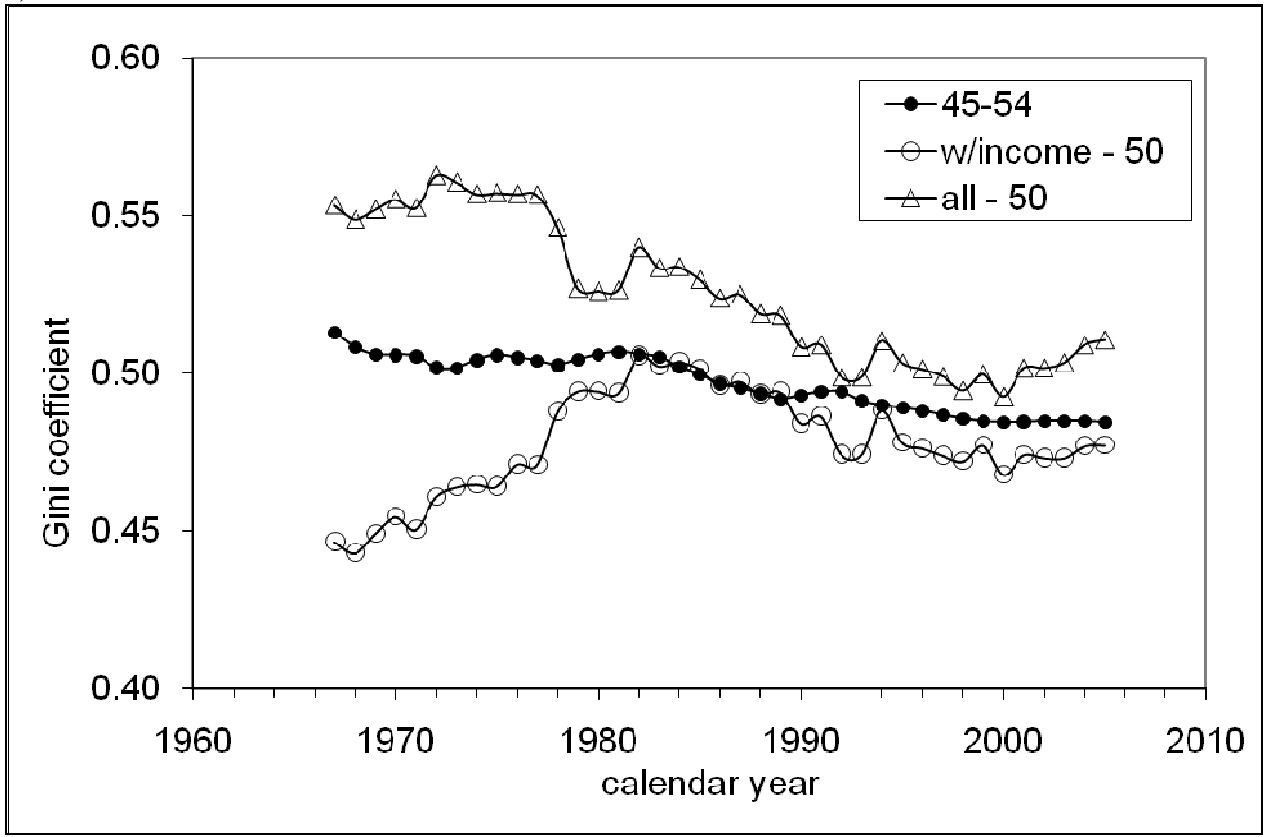

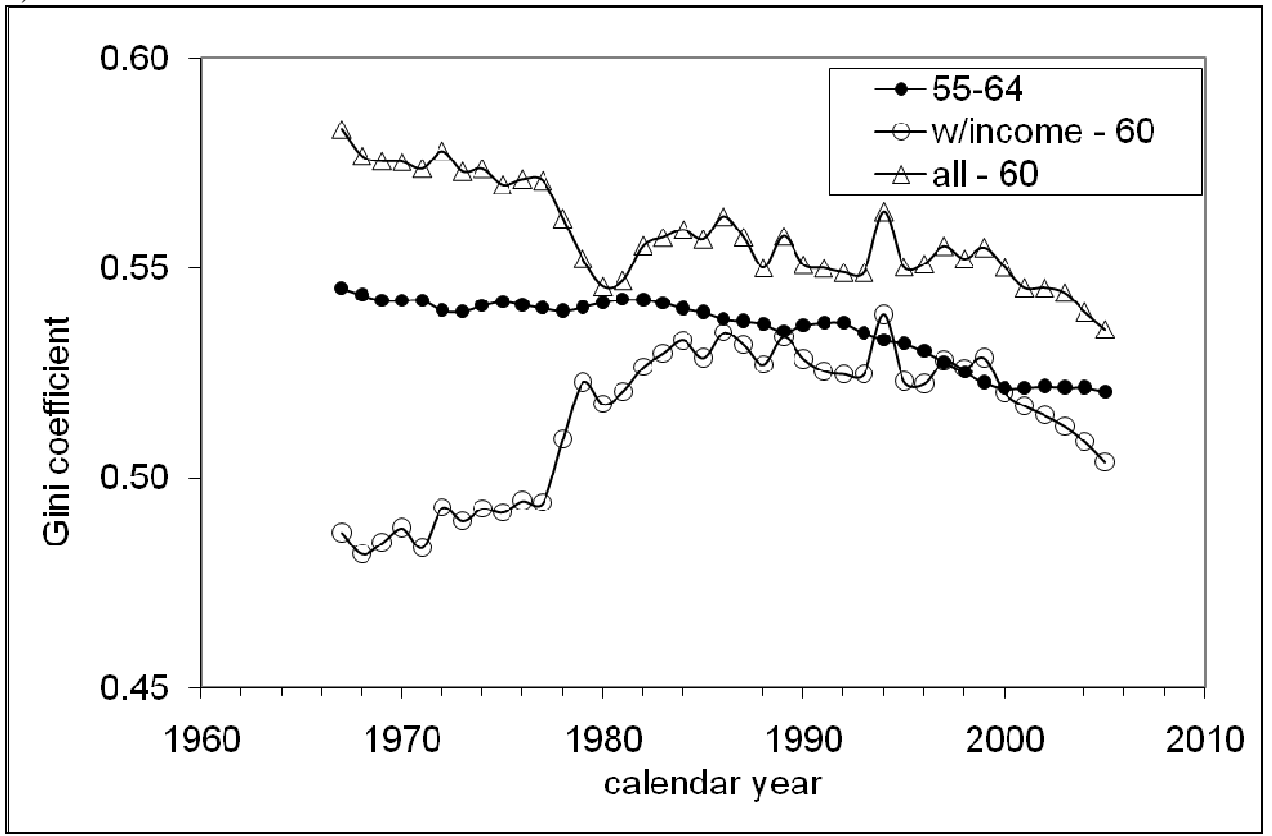

e)

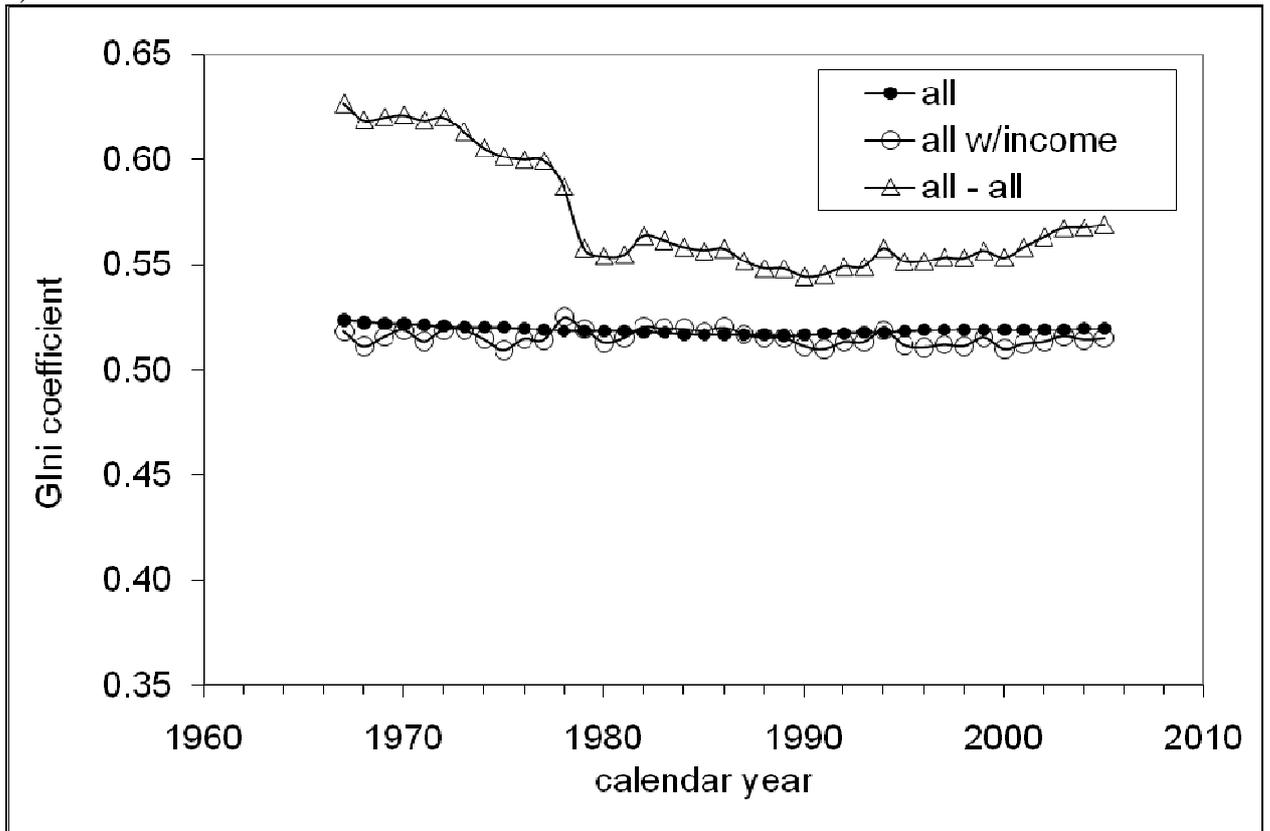

Figure 1.9.9. Comparison of predicted and empirical Gini coefficient in various age groups for the period between 1967 and 2005. In all cases $k=-1.35$.

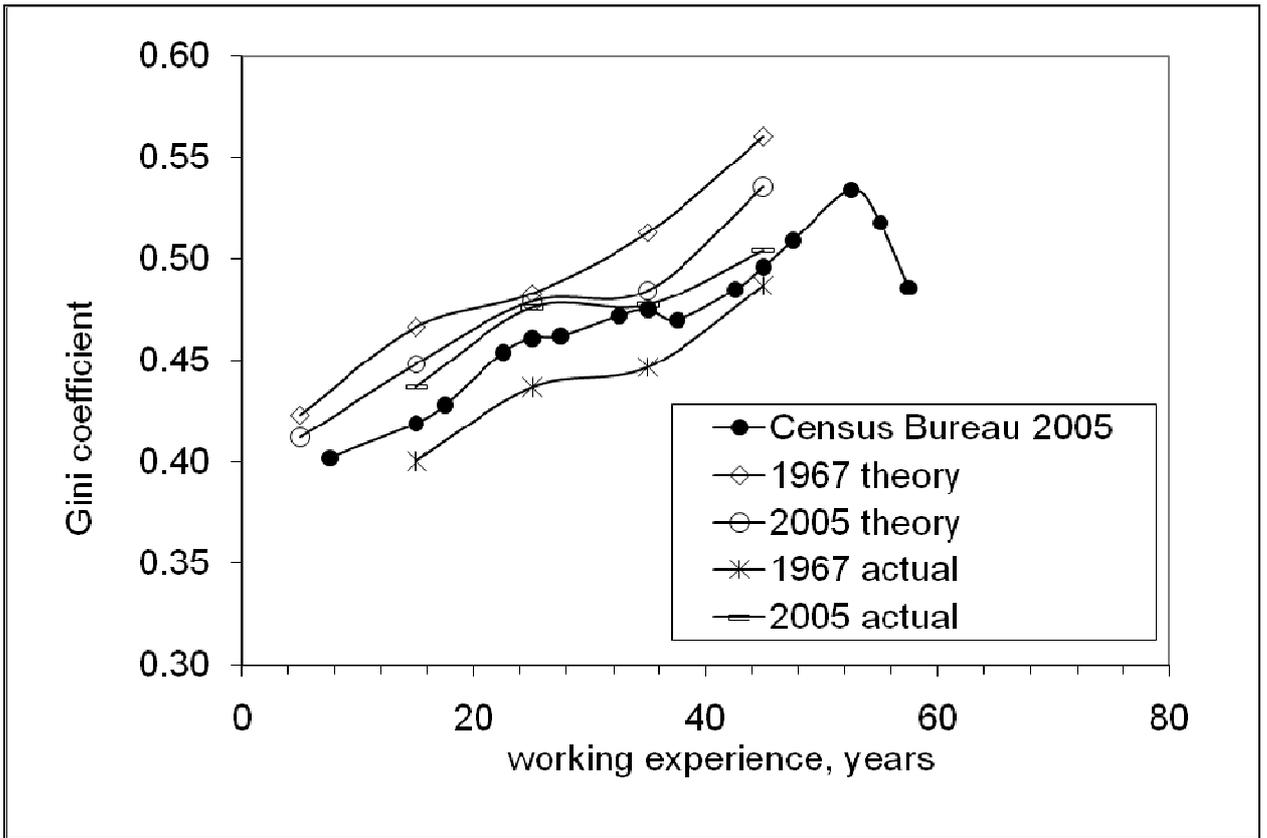

Figure 1.9.10. Comparison of Gini coefficient dependence on age, as estimated by the U.S. Census Bureau and in this study from personal income distributions in 1967 and 2005 (curves marked – actual). The Gini coefficients predicted by our model for 1967 and 2005 are also shown.

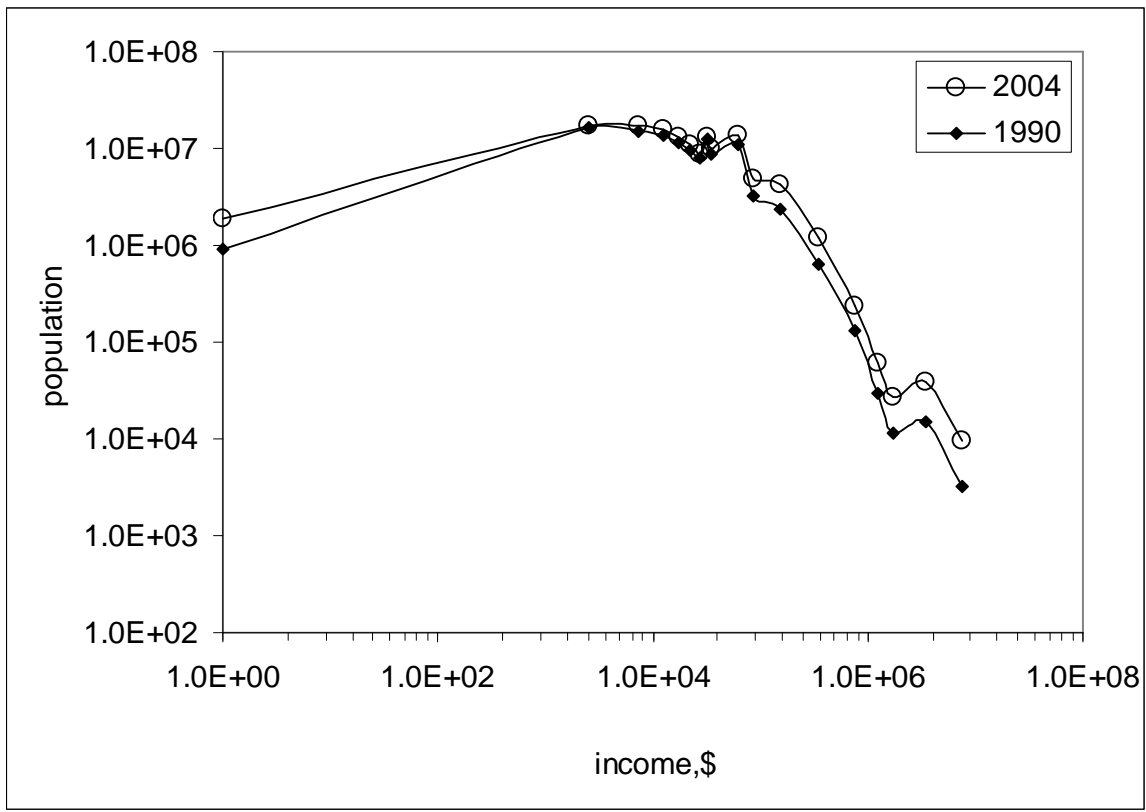

Figure 1.10.1. Comparison of the taxable income distribution reported by the IRS in 1990 and 2004. Income bins are characterized by increasing width. Enumerated populations are assigned to the centers of corresponding bins. Notice the log-log coordinates. The lowest income bin corresponds to zero and negative (loss) reported incomes, i.e. to people without positive income. The bin with incomes above $10,000,000 is not shown because of the absence of mean income estimate.

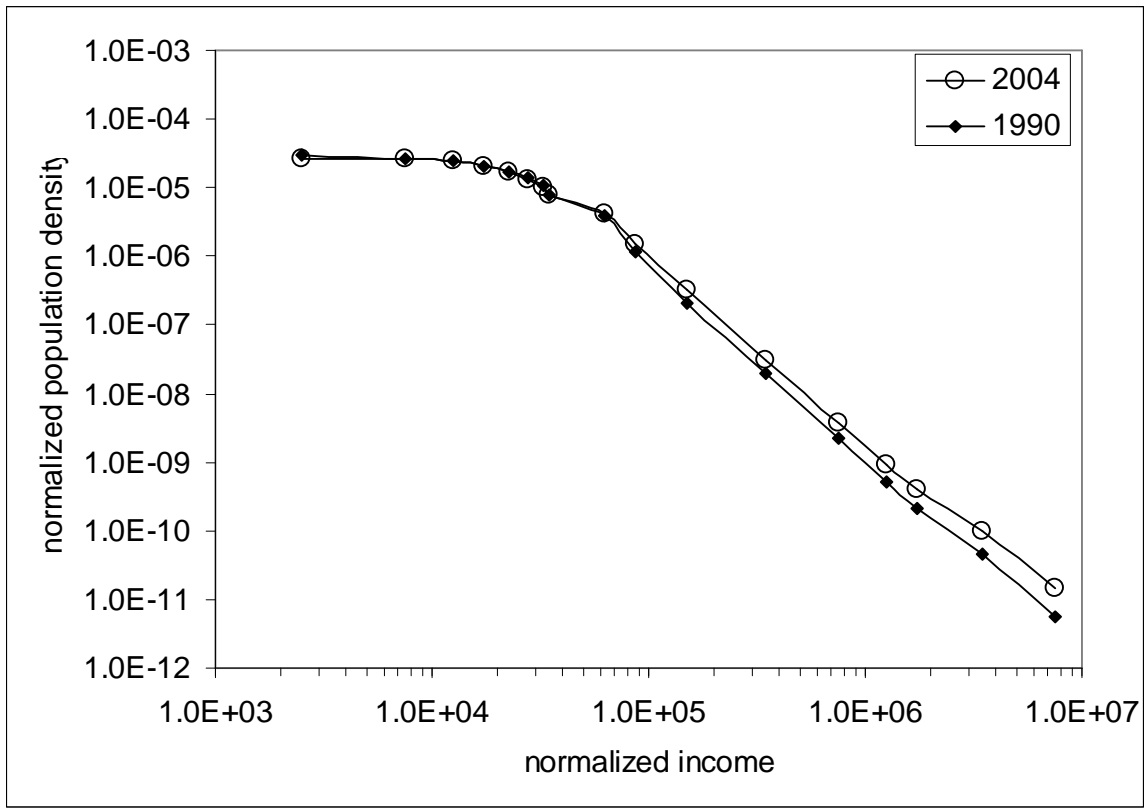

Figure 1.10.2. The readings in Figure 1.10.1 are normalized to total population with income and divided by width of corresponding income bins. Resulting population density distributions are plotted as a function of income. The first (zero width) and the last (open-ended) income bins are not presented. The curves almost coincide below $62.500 and then diverge with increasing income. Seemingly, income inequality increased as the number of people with higher incomes grew faster than that with low incomes.

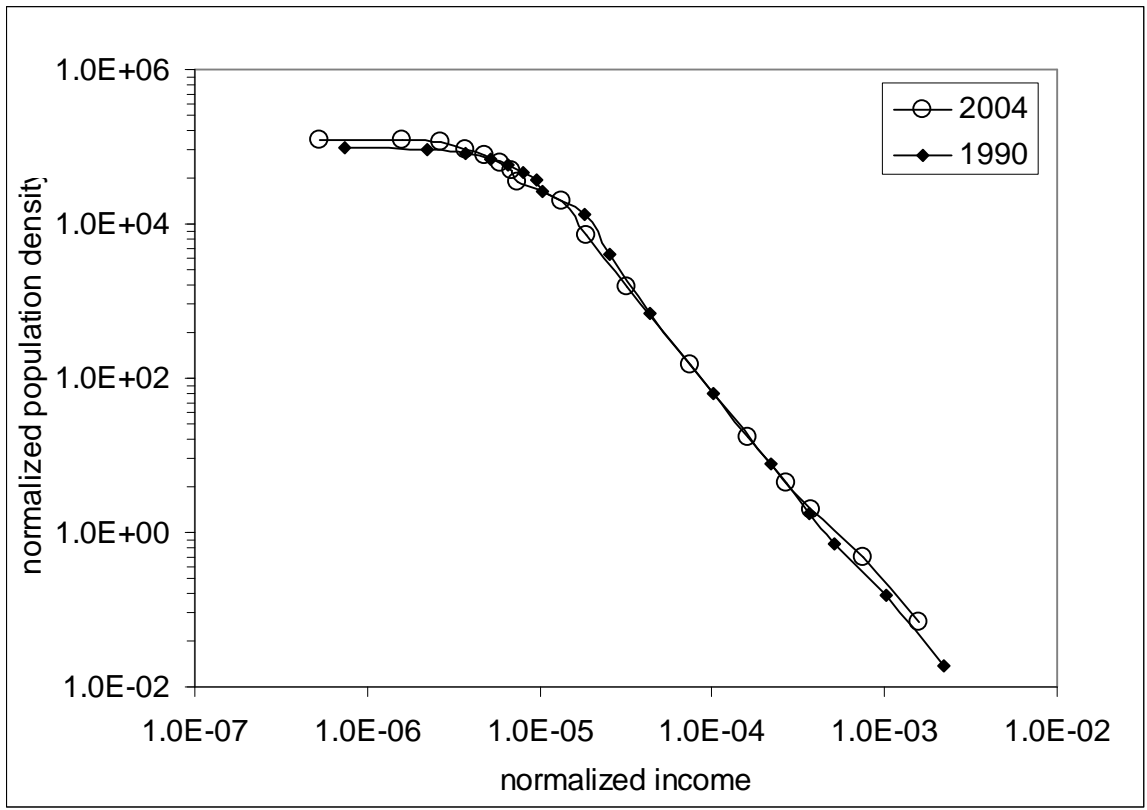

Figure 1.10.3. The curves in Figure 1.10.2 are additionally normalized to gross personal income, i.e. to $4.70E+12 in 2004 and $3.41E+12 in 1990. Relevant income scales are also normalized to these incomes and represent dimensionless portions of total income. As a result, widths of the income bins also become different since the incomes scale in 2004 and 1990 are contracted by different factors. In turn, the centers of the same original income bins in 1990 and 2004 are shifted against each other. Effectively, the curves in Figure 1.10.2 are contracted by different factors and shifted against each other.

The curves now represent density of population as a function of dimensionless income. They practically coincide at high incomes and diverge at low incomes. Therefore, density of population at higher incomes, as measured in dimensionless portions of total income, is practically the same in 1990 and 2004. In the low-income range, density of population is relatively higher is 2004.

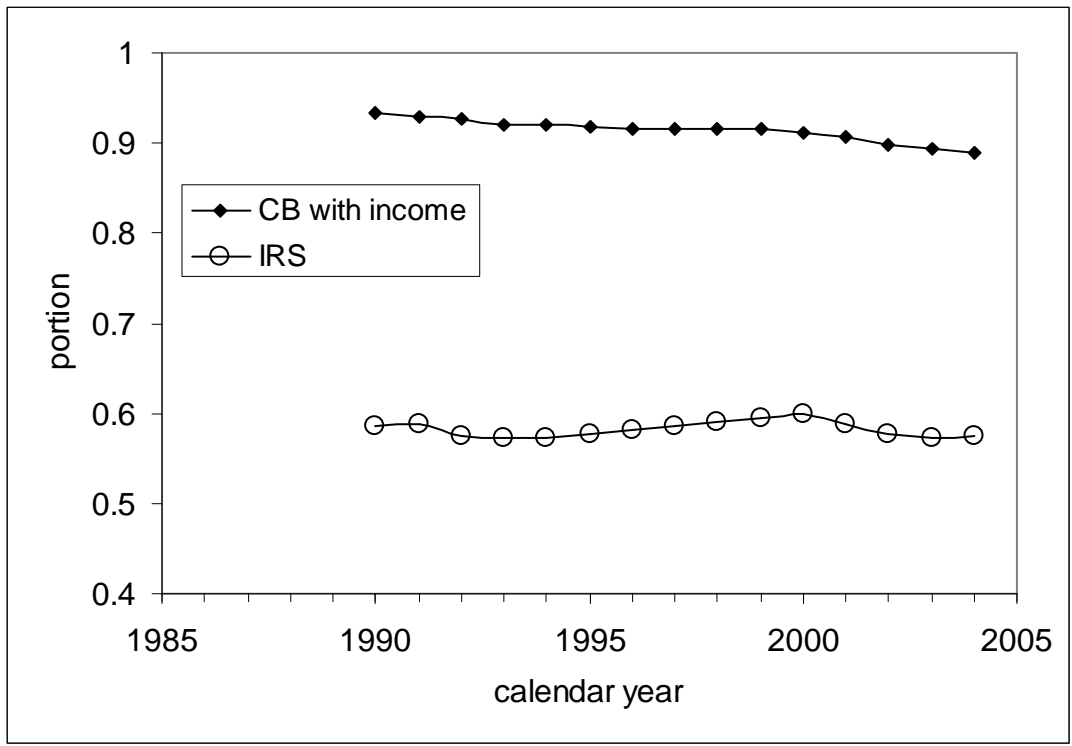

Figure 1.10.4. Portion of working age population with income, as reported by the Census Bureau and IRS. The former provides a more reliable definition of income with smaller variations over time and larger portion of working age population with income. Because of very high sensitivity of the number of low income persons to corresponding definition of income the IRS is likely not able to provide a reliable estimate. About 40 percent of working age population is beyond the IRS definition of income.

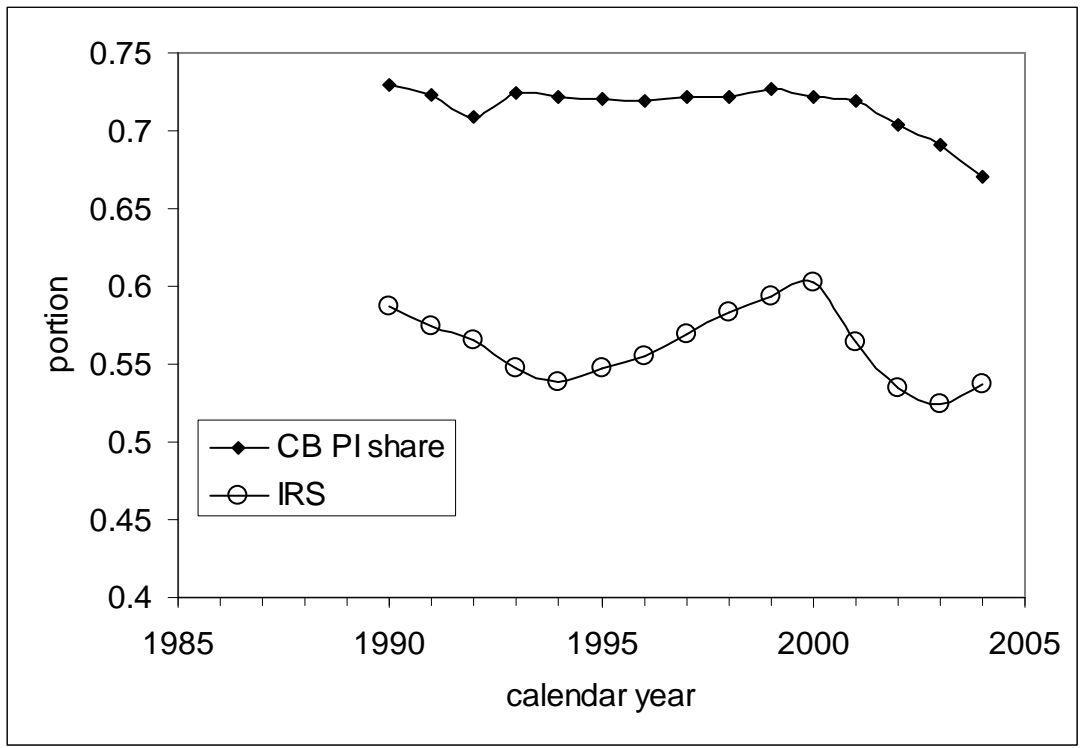

Figure 1.10.5. Gross personal income normalized to real GDP. According to the IRS, only about a half of GDP is transformed into personal income. The CB reports about 70% of real GDP as personal incomes. Volatility of the IRS estimates is very high.